%% file: thesis.tex
\documentclass[12pt,twoside]{report}
\usepackage{amsmath}
\usepackage{amssymb}
\usepackage{amsthm}
\usepackage{graphicx}
\usepackage{slashed}
\usepackage{cite}
\usepackage{fancyhdr}
\usepackage[utf8]{inputenc}
\usepackage[colorlinks=true]{hyperref}
\hypersetup{urlcolor=blue, citecolor=blue, linkcolor=red}
\let\tempmargin\oddsidemargin
\let\oddsidemargin\evensidemargin
\let\evensidemargin\tempmargin

\DeclareMathOperator{\sign}{sign}
\DeclareMathOperator{\im}{Im}
\DeclareMathOperator{\minmod}{minmod}
\DeclareMathOperator{\tr}{tr}
\DeclareMathOperator{\diag}{diag}
\title{Holography in Quark-Gluon Plasma and Neutron Stars}
\author{Govert Nijs\\\\Utrecht University\\\\Promotor: Raimond Snellings\\Copromotor: Umut G\"ursoy}
\date{Defended on July 6, 2020}
\begin{document}
\maketitle
\begin{abstract}
In this thesis, QCD is studied from three different directions, with one overarching theme: holography.
The holographic duality allows certain strongly coupled QFTs to be described in terms of much simpler classical gravity in one dimension more.
The first direction from which QCD is studied in this thesis is by examining the effects of an external magnetic field on a particular holographic model of QCD, yielding interesting qualitative insight.
The second approach examines how, in the same model, one can describe dense baryonic configurations, providing a new way to study the matter composing neutron stars.
Indeed, the equation of state produced in this way is subsequently used to compute several neutron star properties which are observable, or will be in the near future.
The last direction contains no holographic computations per se, but does incorporate several qualitative insights from holography into a new heavy ion code called \emph{Trajectum}\@.
This will in the near future be used to perform a Bayesian analysis, whereby it is hoped that these qualitative insights from holography can be tested on experimental data, to see how well the ideas coming from holography match up with experiment.
\end{abstract}
\newpage
\pagestyle{empty}
\null\newpage
\pagestyle{fancy}
\tableofcontents
\chapter*{Publications}
\addcontentsline{toc}{chapter}{Publications}
This thesis is based on the following publications:
\begin{itemize}
\item Umut G\"ursoy, Ioannis Iatrakis, Matti J\"arvinen and Govert Nijs, \\
\emph{Inverse Magnetic Catalysis from improved Holographic QCD in the Veneziano limit}, \\
\href{https://doi.org/10.1007/JHEP03(2017)053}{\emph{JHEP} \textbf{03} (2017) 053}, [\href{https://arxiv.org/abs/1611.06339}{1611.06339}].
\item Umut G\"ursoy, Matti J\"arvinen and Govert Nijs, \\
\emph{Holographic QCD in the Veneziano Limit at a Finite Magnetic Field and Chemical Potential}, \\
\href{https://doi.org/10.1103/PhysRevLett.120.242002}{\emph{Phys.~Rev.~Lett.} \textbf{120} (2018) 242002}, [\href{https://arxiv.org/abs/1707.00872}{1707.00872}].
\item Umut G\"ursoy, Matti J\"arvinen, Govert Nijs and Juan F.~Pedraza, \\
\emph{Inverse Anisotropic Catalysis in Holographic QCD}, \\
\href{https://doi.org/10.1007/JHEP04(2019)071}{\emph{JHEP} \textbf{04} (2019) 071}, [\href{https://arxiv.org/abs/1811.11724}{1811.11724}].
\item Takaaki Ishii, Matti J\"arvinen and Govert Nijs, \\
\emph{Cool baryon and quark matter in holographic QCD}, \\
\href{https://doi.org/10.1007/JHEP07(2019)003}{\emph{JHEP} \textbf{07} (2019) 003}, [\href{https://arxiv.org/abs/1903.06169}{1903.06169}].
\item Christian Ecker, Matti J\"arvinen, Govert Nijs and Wilke van der Schee, \\
\emph{Gravitational Waves from Holographic Neutron Star Mergers}, \\
\href{https://doi.org/10.1103/PhysRevD.101.103006}{\emph{Phys.~Rev.~D} \textbf{101} (2020) 103006}, [\href{https://arxiv.org/abs/1908.03213}{1908.03213}].
\end{itemize}
\chapter{Introduction}
\input{chapters/introduction}
\chapter{(Inverse) Magnetic Catalysis in Holographic QCD}\label{ch:imc}
\input{chapters/imc}
\null\newpage
\pagestyle{plain}
\null\newpage
\pagestyle{fancy}
\chapter{Holographic Baryons and Neutron Stars}\label{ch:holographicns}
\input{chapters/holographicns}
\null\newpage
\pagestyle{plain}
\null\newpage
\pagestyle{fancy}
\chapter{Simulation of Heavy Ion Collisions with \emph{Trajectum}}\label{ch:trajectum}
\input{chapters/trajectum}
\null\newpage
\pagestyle{plain}
\null\newpage
\pagestyle{fancy}
\chapter{Discussion and Outlook}
\input{chapters/discussion}
\chapter*{Acknowledgements}
\pagestyle{plain}
\addcontentsline{toc}{chapter}{Acknowledgements}
\input{chapters/acknowledgements}
\chapter*{Samenvatting}
\addcontentsline{toc}{chapter}{Samenvatting}
In dit proefschrift wordt QCD bestudeerd uit drie verschillende richtingen, met \'e\'en overkoepelend thema: holografie.
De holografische dualiteit maakt dat sommige sterk gekoppelde kwantumveldentheori\"en beschreven kunnen worden in termen van veel eenvoudigere klassieke zwaartekracht in \'e\'en dimensie extra.
De eerste richting van waaruit QCD bestudeerd wordt in dit proefschrift is door de effecten van een extern magnetisch veld op een specifiek holografisch model van QCD te bestuderen, wat interessant kwalitatief inzicht geeft.
De tweede richting bestudeert hoe, in hetzelfde model, het mogelijk is om baryonische configuraties met grote dichtheid te beschrijven, wat een nieuwe manier oplevert om de materie te bestuderen waar neutronensterren uit bestaan.
De toestandsvergelijking die op deze manier verkregen wordt wordt vervolgens inderdaad gebruikt om verscheidene eigenschappen van neutronensterren te berekenen die geobserveerd kunnen worden, of geobserveerd zullen kunnen worden in de nabije toekomst.
De laatste richting bevat op zichzelf geen holografische berekeningen, maar bevat wel verscheidene kwalitatieve inzichten vanuit holografie die worden toegepast in een nieuwe zware-ionen code genaamd \emph{Trajectum}\@.
Dit zal in de nabije toekomst gebruikt worden om een Bayesiaanse analyse te doen, waar gehoopt wordt dat deze kwalitatieve inzichten uit holografie getest kunnen worden op experimentele data, om te zien hoe goed de idee\"en uit holografie met het experiment in overeenstemming zijn.
\appendix
\null\newpage
\null\newpage
\pagestyle{fancy}
\chapter{V-QCD potentials}\label{ch:potentials}
\input{chapters/potentials}
\null\newpage
\pagestyle{plain}
\null\newpage
\pagestyle{fancy}
\addcontentsline{toc}{chapter}{Bibliography}
\bibliographystyle{JHEP}
\bibliography{refs}
\end{document}

%% file: chapters/introduction.tex
The work described in this thesis is all centered around one goal: understanding the theory of the strong interaction, QCD\@.
Looking at the Lagrangian that defines it, this theory is simple and elegant.
Yet this simple fundamental description results in a rich phenomenology, because the theory is strongly coupled in many regimes of interest.
This leads to two reasons why studying QCD is interesting.
On the one hand, the strongly coupled nature of many of the objects of study in QCD provides us with a playground in which we can learn how non-perturbative physics works.
On the other hand, many outstanding problems in QCD are the main obstacles to understanding other problems.
As an example of this, many properties of neutron stars require an equation of state (EoS) to compute, and to obtain this EoS one has to solve a QCD problem.
In a way, these two reasons for studying QCD go hand in hand.
Returning to the example of neutron stars, as our knowledge of the QCD equation of state grows, so does our knowledge of neutron stars, and on the other hand, as more measurements on neutron stars are done, we can use those measurements to learn something about QCD, and hence about strong coupling.

During my PhD, I have worked towards the goal of understanding QCD from three directions, corresponding to the remaining chapters in this thesis, excluding the conclusion.
Each of these chapters can be read mostly independently, as only minor details should be unclear from reading a chapter by itself.
Wherever this occurs I reference where the details can be looked up for the interested reader.
In the sections below I give an introduction to the concepts used throughout the remaining chapters, starting with an QCD itself, its main features and its quantities of interest.
\section{QCD}\label{sec:intro:qcd}
QCD is the non-Abelian gauge theory of $SU(3)$, which is minimally coupled to a number of quark flavors $N_f$\@.
In the standard model, there are of course 6 flavors.
However, the three heaviest flavors are too heavy to be of importance for many observables, and can be safely neglected.\footnote{Heavy quarks serve as excellent probes for energy loss in a quark-gluon plasma though, as they retain their identities on the timescales of a heavy ion collision, and hence serve as experimentally clean probes.}
In this thesis, we need a slight generalization of QCD, namely to that of a gauge group $SU(N_c)$, where now $N_c$ and $N_f$ can be freely chosen as theory parameters.
The Lagrangian for this generalized QCD is
\begin{equation}
\mathcal{L}_\text{QCD} = -\frac{1}{4}\tr\left[G_{\mu\nu}G^{\mu\nu}\right] + \sum_{i=1}^{N_f}\bar q_i\left(i\slashed{D} - m_i\right)q_i,\label{eq:intro:qcdlagrangian}
\end{equation}
with
\[
G_{\mu\nu} = \partial_\mu A_\nu - \partial_\nu A_\mu + g\left[A_\mu,A_\nu\right], \qquad i\slashed{D}q_i = \gamma^\mu\left(i\partial_\mu + gA_\mu\right)q_i,
\]
with $A_\mu \in \mathfrak{su}(N_c)$, and $m_i$ the mass of quark flavor $i$\@.

\begin{figure}[ht]
\centering
\includegraphics[width=0.7\textwidth]{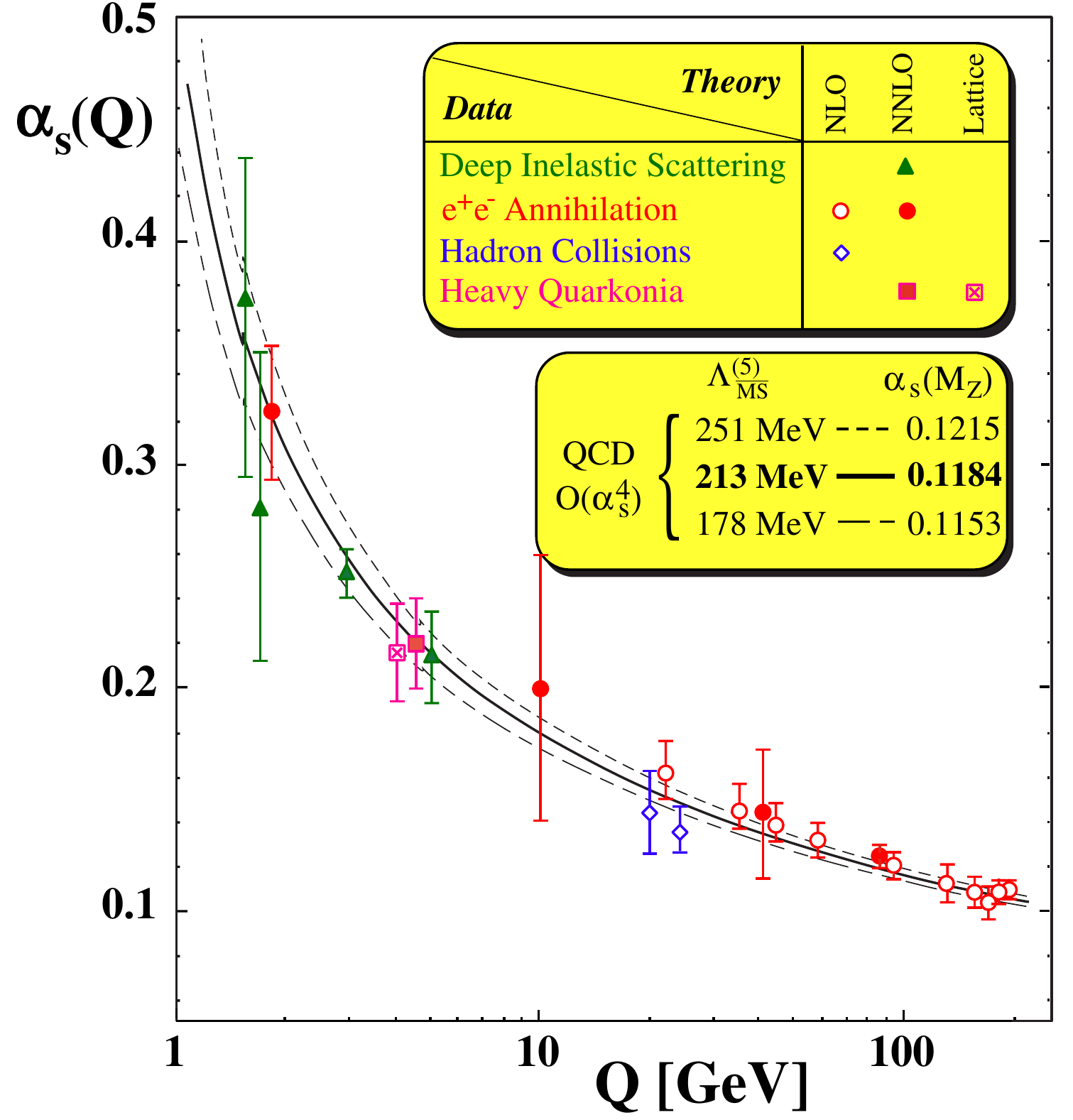}
\caption{\label{fig:intro:introlatticealpharunning}Strong coupling constant as a function of the energy scale $Q$\@. One can see that for energies smaller than about $1\,\text{GeV}$, the coupling becomes $\mathcal{O}(1)$\@. Figure taken from \cite{Bethke:2000ai}\@.}
\end{figure}

This theory has the property that the beta function of the coupling $\alpha_s = g^2/4\pi$ is negative to first order in perturbation theory provided that $N_f/N_c < 11/2$ \cite{Gross:1973id,Politzer:1973fx}\@.
A consequence of this negative beta function is that the coupling constant decreases towards higher energies, a phenomenon known as asymptotic freedom, and increases towards lower energies.
In figure \ref{fig:intro:introlatticealpharunning}, one can see that this is indeed also seen in experiments.
It can be seen that around $1\,\text{GeV}$, the coupling constant becomes $\mathcal{O}(1)$, and the theory can no longer be accurately described by perturbation theory.
This is a huge obstacle in the way of understanding QCD at low energy scales.
One method by which one can still compute certain observables in the non-perturbative regime is lattice QCD\@.
This method discretizes QCD on a Euclidean lattice, enabling the computation of non-dynamical observables by Monte Carlo integration of the euclidean path integral.
Lattice QCD is a reliable method to compute not only thermal properties of QCD, but also hadron masses, which have been favorably compared to experimental values.
It is not without its downsides though, as the Euclidean formalism makes the computation of dynamical processes extremely challenging.
Also, for similar reasons, it turns out to be rather difficult to introduce a finite baryon chemical potential, an issue known as the \emph{sign problem} \cite{Aarts:2015tyj}\@.
A detailed discussion of lattice QCD is beyond the scope of this thesis, but excellent introductions can be found in \cite{Smit:2002ug,Gattringer:2010zz}\@.

\begin{figure}[ht]
\centering
\includegraphics[width=0.49\textwidth]{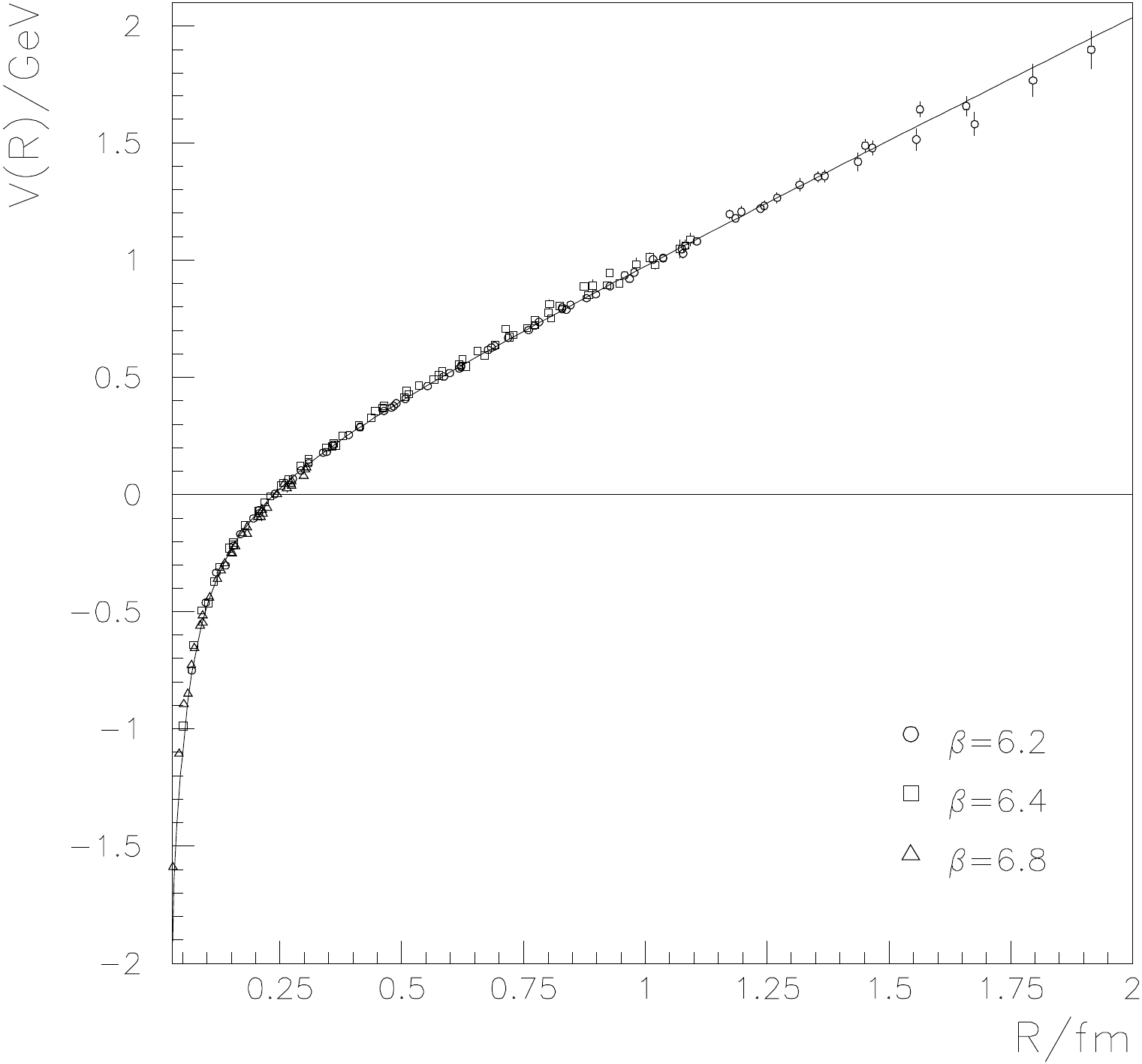}
\includegraphics[width=0.49\textwidth]{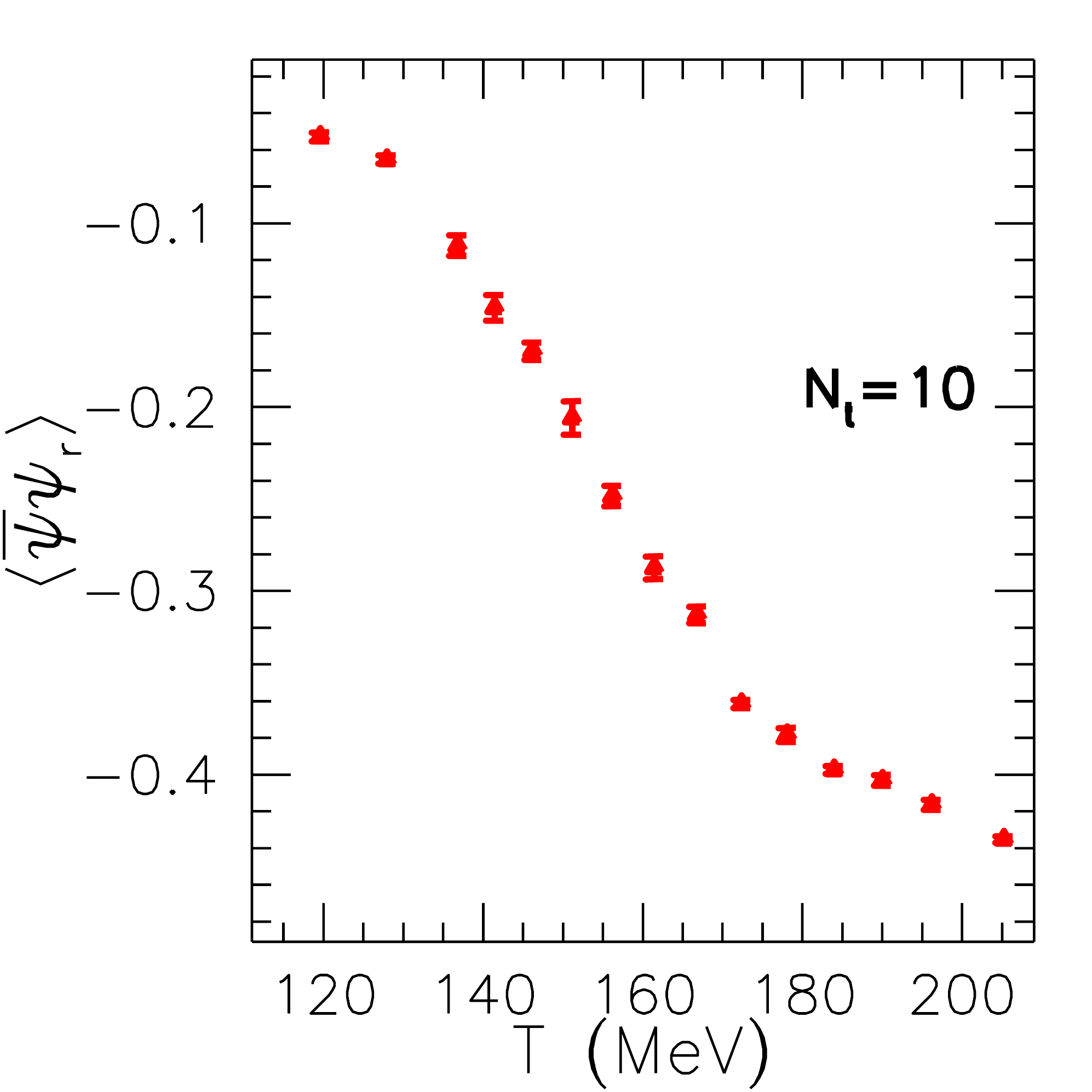}
\caption{\label{fig:intro:introlatticeqqpotential}Left: Quark-antiquark potential computed using lattice QCD in the quenched approximation. Figure taken from \cite{Bali:1993zj}\@. Right: renormalized chiral condensate $\langle\bar\psi\psi\rangle$ (labeled $\langle\bar qq\rangle$ in the main text), renormalized such that $\langle\bar\psi\psi\rangle(T=0) = 0$\@. Figure taken from \cite{Endrodi:2011gv}\@.}
\end{figure}

Two features of QCD which appear in the low energy regime are confinement and chiral symmetry breaking.
Confinement is the phenomenon that states have to be color neutral, implying that particles carrying color charge, such as quarks and gluons, can not occur as isolated particles.
One example of this which can be computed in lattice QCD is the quark-antiquark potential, which is shown in the left panel of figure \ref{fig:intro:introlatticeqqpotential} \cite{Bali:1993zj}\@.
This quark-antiquark potential quantifies the potential energy between a heavy quark paired with an antiquark of the same flavor, and can be seen to grow linearly at large separation.
Assuming that all quarks in the theory are infinitely massive, this means that one would have to spend an infinite amount of energy to separate the quark-antiquark pair.
In the case of realistic quark masses, instead this implies that once the quarks are separated far enough, the potential energy stored in the gluon field will be large enough such that a new quark-antiquark pair can be created.
The new quarks then each pair with one of the original quarks to create two color neutral mesons.

Chiral symmetry breaking refers to the approximate global chiral symmetry
\[
SU(N_f)_L \times SU(N_f)_R \times U_1(L) \times U_1(R),
\]
which acts on \eqref{eq:intro:qcdlagrangian} such that the $SU(N_f)_L \times U(1)_L$ generators act on the left-handed quark components by multiplication, while the $SU(N_f)_R \times U(1)_R$ generators act on the right-handed components.
This symmetry is only approximate in \eqref{eq:intro:qcdlagrangian}, but becomes exact in the limit where the quarks are massless.
The QCD vacuum, however, breaks this approximate symmetry further spontaneously.
The order parameter of this chiral symmetry breaking is the chiral condensate operator $\langle\bar q_iq_i\rangle$, which can be defined for each flavor $i$\@.
In the right panel of figure \ref{fig:intro:introlatticeqqpotential}, one can see the chiral condensate as a function of temperature, computed using lattice QCD \cite{Endrodi:2011gv}\@.
Note that the renormalized chiral condensate is shown, which is defined by the subtraction of a constant such that the chiral condensate at zero temperature vanishes.
One can clearly see that indeed the order parameter $\langle\bar q_iq_i\rangle$, which is small at large temperatures, grows for small temperatures.

\begin{figure}[ht]
\centering
\includegraphics[width=0.49\textwidth]{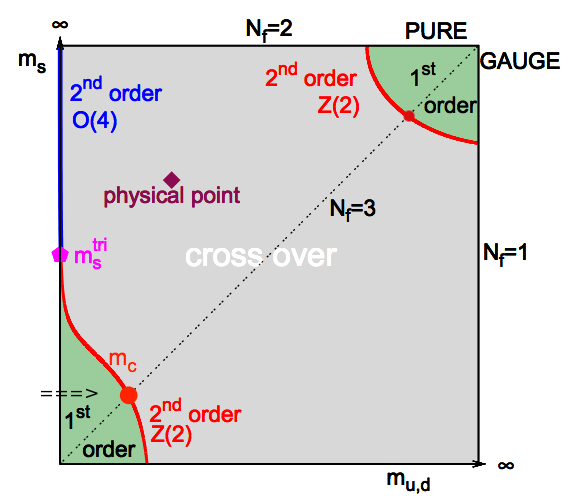}
\includegraphics[width=0.49\textwidth]{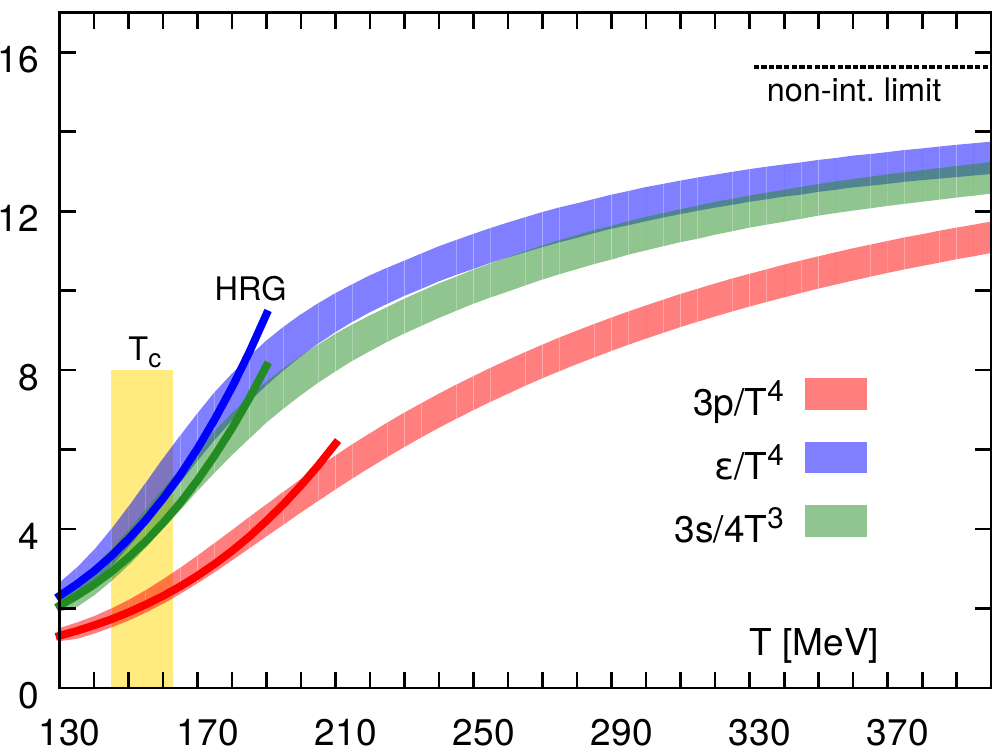}
\caption{\label{fig:intro:crossover}Left: Columbia plot, showing for which values of the up and down quark masses $m_\text{u,d}$ on the horizontal axis and the strange quark mass $m_s$ on the vertical axis, QCD has a first order phase transition or a cross-over. Physical values for the quark masses are indicated, yielding a cross-over. Figure taken from \cite{deForcrand:2017cgb}\@. Right: Equation of state at vanishing baryon chemical potential as a function of temperature, showing the pressure $p$, the energy density $e$, and the entropy $s$\@. The transition can clearly be seen to be a cross-over. Figure taken from \cite{Bazavov:2014pvz}\@.}
\end{figure}

A subsequent question one can ask is whether the transition between a chirally symmetric phase without confinement, known as the quark-gluon plasma, at high temperatures and the chirally broken confining vacuum are separated by a cross-over or a first order phase transition.
In the Columbia plot \cite{deForcrand:2017cgb}, shown in the left panel of figure \ref{fig:intro:crossover}, one can see that the answer to this question depends on the masses of the quarks.
One can see that for the physical values of the quark masses, the transition is a cross-over.
In the right panel of figure \ref{fig:intro:crossover}, one can see the equation of state as a function of temperature for vanishing baryon chemical potential \cite{Bazavov:2014pvz}\@.
Here too, it is apparent that the transition is a cross-over.

\begin{figure}[ht]
\centering
\includegraphics[width=0.8\textwidth]{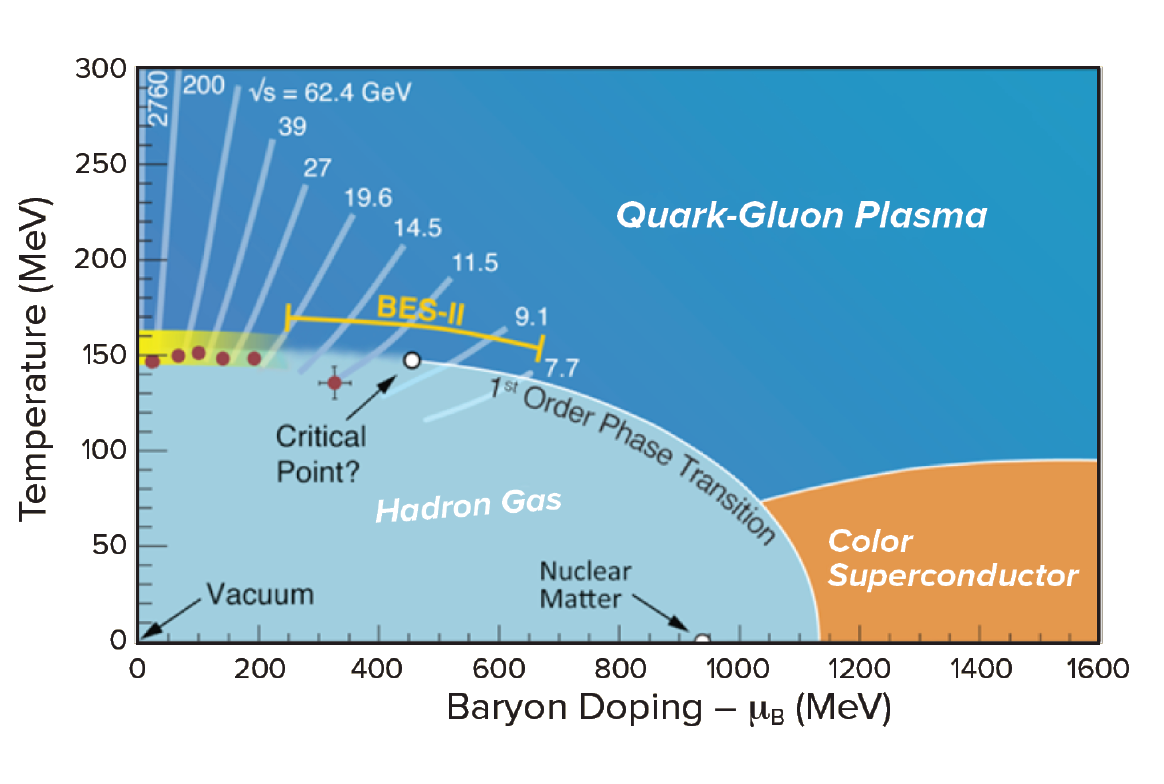}
\caption{\label{fig:intro:phasediagram}Sketch of the expected phase diagram of QCD as a function of temperature and chemical potential. The regions probed by heavy ion collisions are also shown, with the corresponding center-of-mass energies indicated. Figure taken from \cite{2015LRPNS}\@.}
\end{figure}

In the presence of a finite baryon chemical potential, the situation may be different.
In figure \ref{fig:intro:phasediagram}, a sketch is shown of what the phase diagram is expected to look like as a function of both temperature and baryon chemical potential.
Also, the regions probed by heavy ion collision experiments at RHIC and LHC are shown, with center-of-mass energies indicated.
The cross-over seen in figure \ref{fig:intro:crossover} is seen in figure \ref{fig:intro:phasediagram} along the vertical axis.
As one moves to larger chemical potentials, the cross-over is expected to turn into a first order phase transition at a critical endpoint.
The search for such an endpoint is the purpose of the Beam Energy Scan program at RHIC \cite{Luo:2015doi}\@.
If we look at the low temperature, large chemical potential region of the phase diagram, we see nuclear matter indicated.
At chemical potential values beyond this value, still at low temperatures, we enter the regime in which the matter making up neutron stars exists.
It is not known whether the densities inside a neutron star are large enough to probe a potential phase transition as indicated in figure \ref{fig:intro:phasediagram}, but at some large density it is expected that yet a new state of matter forms, known as a color superconductor \cite{Alford:2007xm}\@.
One of the main reasons why these features in the phase diagram are as of yet unknown is the aforementioned sign problem, which precludes a first principles calculation of these features.
The two experimental probes into the phase diagram, namely heavy ion collisions and neutron stars, are both described by relativistic hydrodynamics, which is what we will describe next.
\section{Relativistic hydrodynamics}
Relativistic hydrodynamics is an effective theory which describes the behavior of fluids in local thermal equilibrium.
It can be described by the conservation of conserved quantities that the underlying theory has.
This always includes the conservation of the stress-energy tensor, but can also include other conserved currents, such as baryon number density.\footnote{In principle, one can write down as many conserved currents as desired. We will restrict ourselves to just the baryon number density.}
Let us take this theory as an example.
We then have
\[
\partial_\mu T^{\mu\nu} = 0, \qquad \partial_\mu J^\mu = 0,
\]
where $J^\mu$ is the conserved current associated to baryon number density.
This set of equations cannot be solved though, which can be seen by a simple counting argument.
Indeed, we have 10 independent components in the stress-energy tensor, and 4 in the baryon current, whereas we have only 5 equations to constrain them.
Further input is therefore needed.
This comes as no surprise, as it would be rather strange if the behavior of the conserved quantities were completely determined by the conservation laws themselves, and had no dependence on the underlying microscopic theory.

The extra input used to close the system of equations is called the constitutive relations, which determine $T^{\mu\nu}$ and $J^\mu$ in terms of the temperature $T$, baryon chemical potential $\mu$ and fluid velocity $u^\mu$\@.
The constitutive relations can framed in terms of an expansion in derivatives of $T$, $\mu$ and $u^\mu$\@.
Below, we will discuss three examples of such relations, namely that of ideal hydrodynamics with a conserved baryon number density, first order Israel-Stewart theory without a conserved baryon number density, and second order hydrodynamics, also without any conserved quantities other than the stress-energy tensor.
These three cases correspond exactly to the three cases which will be used in the remainder of this thesis.
We will however restrict ourselves to just a description of these theories, giving just the basic idea of the ingredients used to derive them.
There are many different detailed derivations available in the literature, see e.g.~\cite{Romatschke:2009im,Banerjee:2012iz,Kovtun:2012rj}.
Before moving on to the examples of constitutive relations however, note that one can easily couple the equations of hydrodynamics to those of general relativity by using the hydrodynamic stress-energy tensor defined by the constitutive relations in the Einstein equations.

Let us now look at the first example of constitutive relations, namely that of ideal hydrodynamics with a conserved baryon number density.
In this case, we have
\begin{equation}
T^{\mu\nu} = e(T,\mu)u^\mu u^\nu - P(T,\mu)\Delta^{\mu\nu}, \qquad J^\mu = n(T,\mu)u^\mu,\label{eq:intro:idealhydro}
\end{equation}
where $\Delta^{\mu\nu} = g^{\mu\nu} - u^\mu u^\nu$ is a projector satisfying $u_\mu\Delta^{\mu\nu} = 0$, and $e$, $P$ and $n$ are for now arbitrary functions of the temperature and chemical potential.
Note also that the metric $g_{\mu\nu}$ follows the mostly minus convention, in accordance with most literature on hydrodynamics.
Also, since the metric can in principle be something other than Minkowski, all derivatives in this section can be assumed to be covariant unless stated otherwise.
One can check that \eqref{eq:intro:idealhydro} are the most general expressions for $T^{\mu\nu}$ and $J^\mu$ which do not involve derivatives of $T$, $\mu$ or $u^\mu$\@.
For this reason this constitutive relation is zeroth order in the derivative expansion.
Now let us examine the above expression in the fluid rest frame, in which $u^\mu = (1,0,0,0)$\@.
We then have
\[
T^{\mu\nu} = \diag(e,P,P,P), \qquad J^\mu = (n,0,0,0),
\]
which we can compare to the known result for a fluid at rest to deduce that we should interpret the arbitrary functions $e$, $P$ and $n$ as energy density, pressure and baryon number density, respectively.
Relating these three quantities through the equation of state, we can close the system of equations, rendering it solvable.
These constitutive relations are used for the neutron star merger simulations in chapter \ref{ch:holographicns}, where they are solved together with the Einstein equations.

For the remainder of this section, we will disregard $J^\mu$, and consider a theory with only a conserved stress-energy tensor.
We will also add the first order in derivative corrections.
To this end, let us first introduce the following derivatives:
\[
\nabla^\nu \equiv \Delta^{\mu\nu}\partial_\nu, \qquad D \equiv u^\mu\partial_\mu,
\]
where $\nabla^\nu$ is the gradient in the fluid rest frame, and $D$ is the time derivative in the fluid rest frame.
We now write the first equation of \eqref{eq:intro:idealhydro} as
\begin{equation}
T^{\mu\nu} = eu^\mu u^\nu - (P(e) + \Pi)\Delta^{\mu\nu} + \pi^{\mu\nu}.\label{eq:intro:viscoushydro}
\end{equation}
Here we have removed the dependence on $\mu$ as there is no more baryon number density, and added the bulk pressure $\Pi$ and the shear tensor $\pi^{\mu\nu}$, where $\pi^{\mu\nu}$ is traceless ($\pi_\mu^\mu = 0$) and orthogonal ($u_\mu\pi^{\mu\nu} = 0$)\@.
We have also rewritten $e(T)$ and $P(T)$ as a single function $P(e)$\@.
For the bulk pressure and shear tensor we have the following expression in terms of derivatives:
\begin{equation}
\Pi = -\zeta(e)\nabla\cdot u, \qquad \pi^{\mu\nu} = 2\eta(e)\sigma^{\mu\nu},\label{eq:intro:piPifirstorder}
\end{equation}
which is the most general expression at first order in derivatives which satisfies the second law of thermodynamics \cite{Kovtun:2012rj}\@.
Here $\zeta(e)$ and $\eta(e)$ are the bulk viscosity and shear viscosity, respectively.
Both $\eta(e)$ and $\zeta(e)$ are required to be positive to respect the second law of thermodynamics \cite{Kovtun:2012rj}\@.
The $\sigma^{\mu\nu}$ tensor is a symmetric tensor satisfying the same tracelessness and orthogonality conditions as $\pi^{\mu\nu}$:
\begin{equation}
\sigma^{\mu\nu} = \nabla^{\langle\mu}u^{\nu\rangle} = \frac{1}{2}\left(\nabla^\mu u^\nu + \nabla^\nu u^\mu\right) - \frac{1}{3}\Delta^{\mu\nu}\nabla\cdot u.\label{eq:intro:sigmadef}
\end{equation}
Here we define the angled brackets as symmetrizing a tensor, and at the same time removing the trace.

There is one big problem with these constitutive relations though, namely that they allow for superluminal propagation, thereby violating causality.
This can be solved in the following way, which may seem ad hoc, but can be derived in several different ways \cite{Romatschke:2009im}\@.
The solution is to replace the identifications in \eqref{eq:intro:piPifirstorder} by the following differential equations, called the Israel-Stewart equations \cite{Israel:1979wp}:
\begin{align}
D\Pi & = -\frac{1}{\tau_\Pi(e)}\left[\Pi + \zeta(e)\nabla\cdot u\right],\label{eq:intro:isbulk}\\
\Delta^\mu_\alpha\Delta^\nu_\beta D\pi^{\alpha\beta} & = -\frac{1}{\tau_\pi(e)}\left[\pi^{\mu\nu} - 2\eta(e)\sigma^{\mu\nu}\right],\label{eq:intro:isshear}
\end{align}
where the projectors in front of $D\pi^{\alpha\beta}$ ensure that the differential equation preserves tracelessness and orthogonality, and the positive functions $\tau_\pi(e)$ and $\tau_\Pi(e)$ are called the shear relaxation time and bulk relaxation time, respectively.
Summarizing, the Israel-Stewart equations give us four parameters, called transport coefficients:
\[
\eta(e), \qquad \zeta(e), \qquad \tau_\pi(e), \qquad \tau_\Pi(e),
\]
where the dependence on the energy density, or equivalently the temperature, is indicated.
These transport coefficients depend on the microscopic details of the theory, and hence encode information about the underlying theory.
Since they also enter the equations governing the hydrodynamical evolution, they also have an influence on macroscopic observables, which can in principle be measured experimentally.

Second order hydrodynamics is a generalization of the above discussion.
The stress-energy tensor is still described by \eqref{eq:intro:viscoushydro}, but the relaxation equations for the bulk pressure \eqref{eq:intro:isbulk} and shear stress \eqref{eq:intro:isshear} are expanded to the following form:
\begin{align}
D\Pi & = -\frac{1}{\tau_\Pi(e)}\left[\Pi + \zeta(e)\nabla\cdot u + \delta_{\Pi\Pi}(e)\nabla\cdot u\Pi\right.\label{eq:intro:secondorderbulk}\\
& \qquad - \left.\lambda_{\Pi\pi}(e)\pi^{\mu\nu}\sigma_{\mu\nu}\right],\nonumber\\
\Delta^\mu_\alpha\Delta^\nu_\beta D\pi^{\alpha\beta} & = -\frac{1}{\tau_\pi(e)}\left[\pi^{\mu\nu} - 2\eta(e)\sigma^{\mu\nu} + \delta_{\pi\pi}(e)\pi^{\mu\nu}\nabla\cdot u\right.\label{eq:intro:secondordershear}\\
& \qquad - \left.\phi_7(e)\pi_\alpha^{\langle\mu}\pi^{\nu\rangle\alpha} + \tau_{\pi\pi}(e)\pi_\alpha^{\langle\mu}\sigma^{\nu\rangle\alpha} - \lambda_{\pi\Pi}(e)\Pi\sigma^{\mu\nu}\right].\nonumber
\end{align}
We can see that the second order terms add the following transport coefficients:
\[
\delta_{\Pi\Pi}(e), \qquad \lambda_{\Pi\pi}(e), \qquad \delta_{\pi\pi}(e), \qquad \phi_7(e), \qquad \tau_{\pi\pi}(e), \qquad \lambda_{\pi\pi}(e).
\]
These transport coefficients can, just as the first order transport coefficients, in principle be derived from the microscopic theory, and they also can potentially be measured experimentally.
Both the first and second order constitutive relations will be used in chapter \ref{ch:trajectum}, where they will be used to describe the quark-gluon plasma stage of simulations of heavy ion collisions.
\section{Heavy Ion Collisions}
At large temperatures, QCD matter undergoes a transition to a quark-gluon plasma (QGP) phase, as can be seen in figure \ref{fig:intro:phasediagram}\@.
Such large temperatures can be achieved experimentally by depositing extreme amounts of energy inside a small volume, and letting the system equilibrate towards a thermal state.
For this system to be able to reach a near-equilibrium state though, the spatial extent of the system must be sufficient such that the energy density dissipates away faster than the system takes to equilibrate.
If this condition is met, a significant portion of the system's evolution will be described by a QGP, which can be described using hydrodynamics as discussed above.
Systems in which this is possible are heavy ion collisions.
In a heavy ion collision experiment such as those conducted at RHIC and the LHC, two atomic nuclei are accelerated in opposite directions, and collide inside a particle detector.
The resulting matter produced in the collision `hydrodynamizes' on a timescale of less than $1\,\text{fm}/c$, which is much smaller than the spatial extent of the resulting plasma if the colliding nuclei are large and collide `head-on'\@.
An interesting question is indeed how small a collision system can be for it to still form a QGP (See \cite{Nagle:2018nvi} for a recent review.)\@.
In the following paragraphs, we will describe in some detail the physical processes occuring during a heavy ion collision.
We will subsequently end this section with a discussion of experimental observables.
See also \cite{Busza:2018rrf} for a recent review of this topic.

\begin{figure}[ht]
\centering
\includegraphics[width=0.54\textwidth]{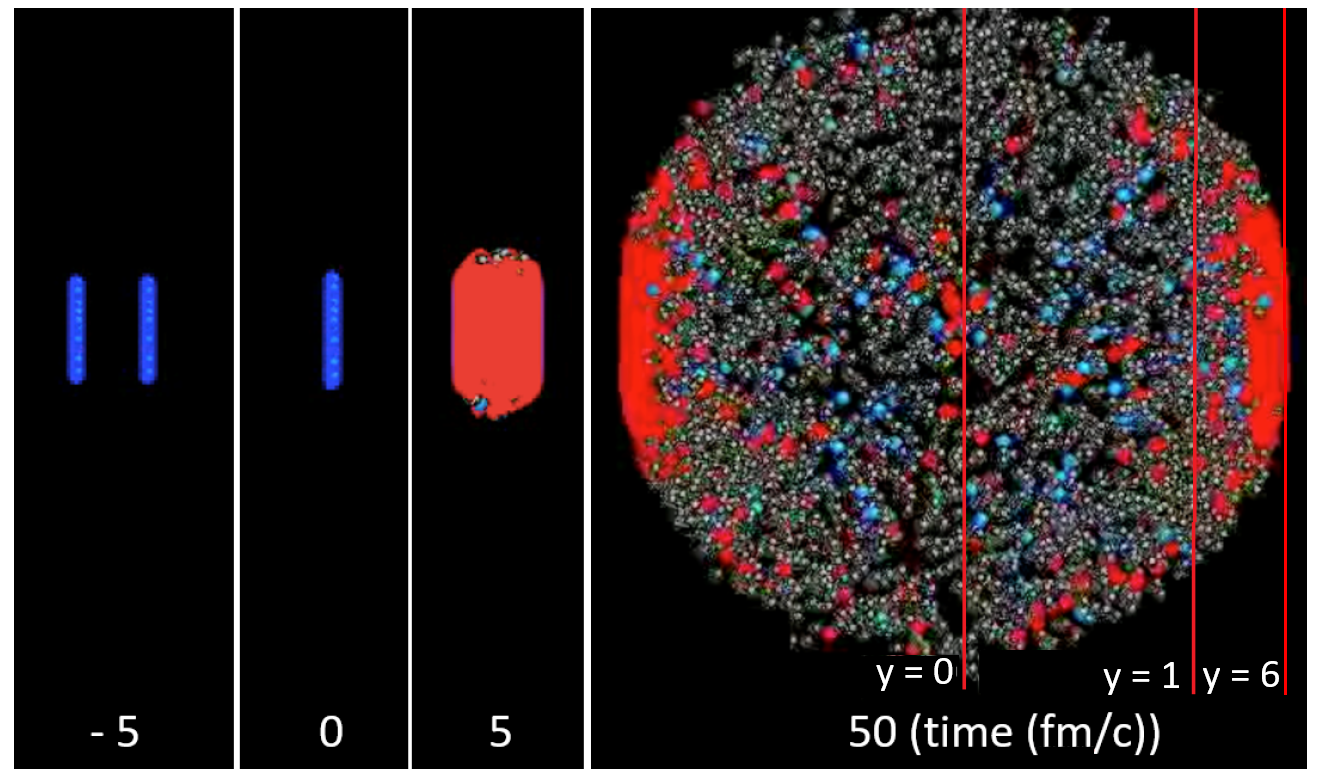}
\includegraphics[width=0.44\textwidth]{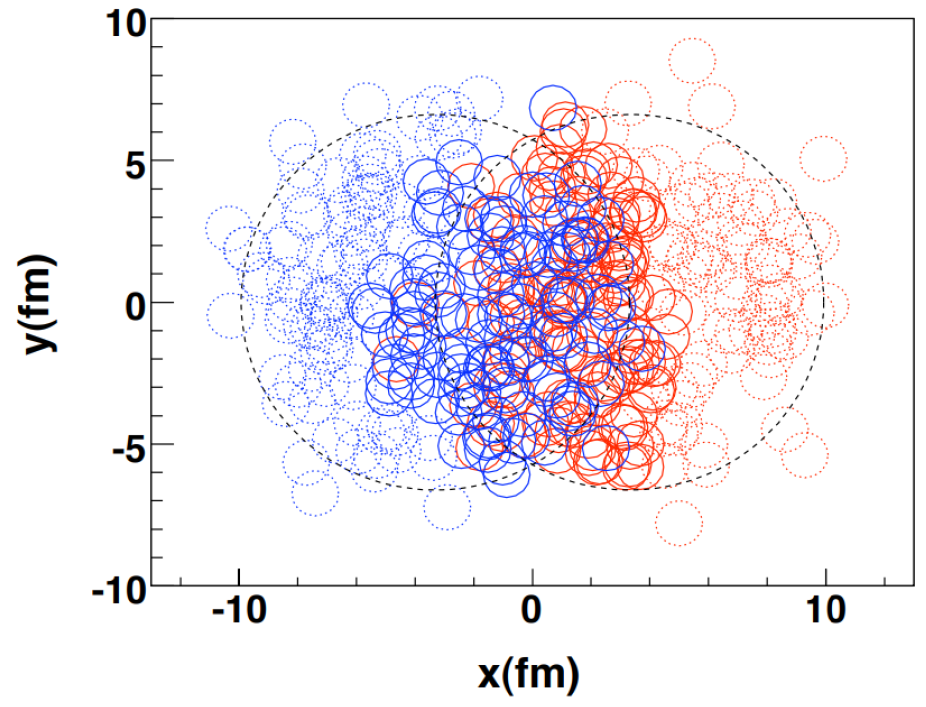}
\caption{\label{fig:intro:introheavyioncollision}Left: Snapshots from an animation of a heavy ion collision taken at $5\,\text{fm}/c$ before the collision, at the moment of collision, and at $5$ and $50\,\text{fm}/c$ after the collision, where the collision is viewed from the side, and where quark-gluon plasma is indicated in red. In the final snapshot the approximate rapidity $y$ (see \eqref{eq:intro:rapiditydef}) of particles is shown for $y = 0,1,6$\@. Figure taken from \cite{Busza:2018rrf}, which was adapted from \cite{LHCanmation}\@. Right: Location of nucleons participating in the collision, viewed in the plane transverse to the beam. Nucleons from one nucleus are shown in blue, the ones from the other are shown in red. Nucleons participating in the collision are shown as solid circles, while dotted circles indicate `spectator' nucleons. Figure taken from \cite{Alver:2008aq}\@.}
\end{figure}

For the discussion of the processes occuring during a heavy ion collision, let us focus on a specific example.
At the LHC, lead-208 nuclei are collided at center-of-mass energy per nucleon of $2.76\,\text{TeV}$ and $5.02\,\text{TeV}$\@.
Different snapshots of an animation of such a collision can be seen in the left panel of figure \ref{fig:intro:introheavyioncollision}\@.
Here the numbers in the bottom of each snapshot indicate the time in $\text{fm}/c$, with $t = 0$ being defined as the moment the collision occurs, and the collision is viewed from the side, i.e.~the beam passes through the figure from left to right.
Because the nuclei each have a very large energy, in the lab frame they appear extremely Lorentz contracted, as can be seen in the snapshot at $t = -5\,\text{fm}/c$\@.
The nuclei then pass through each other, interacting and leaving matter in their wake.
This matter is what we will be discussing the evolution of below.

Before continuing the discussion on the different stages this matter goes through, one important point is that the two nuclei need not collide head-on, as can be seen in the right panel of figure \ref{fig:intro:introheavyioncollision}\@.
In that figure, we are looking in the direction of the beam, i.e.~the two dimensions shown are transverse to the beam.
Since the nuclei are very small compared to the size of the beam, the so-called `impact parameter', or the transverse distance between the centers of the colliding nuclei, is essentially random.
As a consequence of this, heavy ion collisions as measured in an experiment are not all of the same type, as collisions with a small impact parameter (called central events) are very different from those with a large impact parameter (called peripheral or off-central events)\@.
One difference is that the number of participants in the collision correlates with the number of particles measured in the final state, which causes central events to have more particles in their final states.
Another difference is in the initial geometry.
Lead-208 is spherical on average, and therefore central events are also to a good approximation spherical.
Off-central events like the one shown in the right panel of \ref{fig:intro:introheavyioncollision} instead are quite elongated.

Let us now consider with the discussion of what happens after the collision.
As was mentioned, when the nuclei pass through each other, they leave matter in their wake.
This matter is produced, to a good approximation, in a way which is invariant under boosts in the beam direction.
At some time after the initial collision, the resulting matter can be described by hydrodynamics.
This process is called `hydrodynamization'\@.
Note that this is different from thermalization, as the matter is at this stage not yet in equilibrium, which shows itself in the fact that the matter is not homogeneous, and large gradients of the stress-energy tensor exist.
The process by which this hydrodynamization happens is poorly understood, and even how fast this happens is subject to debate.
Kinetic theory suggests that hydrodynamization occurs after roughly $1\,\text{fm}/c$ \cite{Kurkela:2015qoa}\@.
Holography, which will be discussed in section \ref{sec:intro:holography}, predicts even earlier values of perhaps $0.35\,\text{fm}/c$ \cite{vanderSchee:2013pia}\@.
Furthermore, different models for this pre-equilibrium stage describing this hydrodynamization process give qualitatively different answers for the state of the hydrodynamic fluid immediately after hydrodynamization.

\begin{figure}[ht]
\centering
\includegraphics[width=\textwidth]{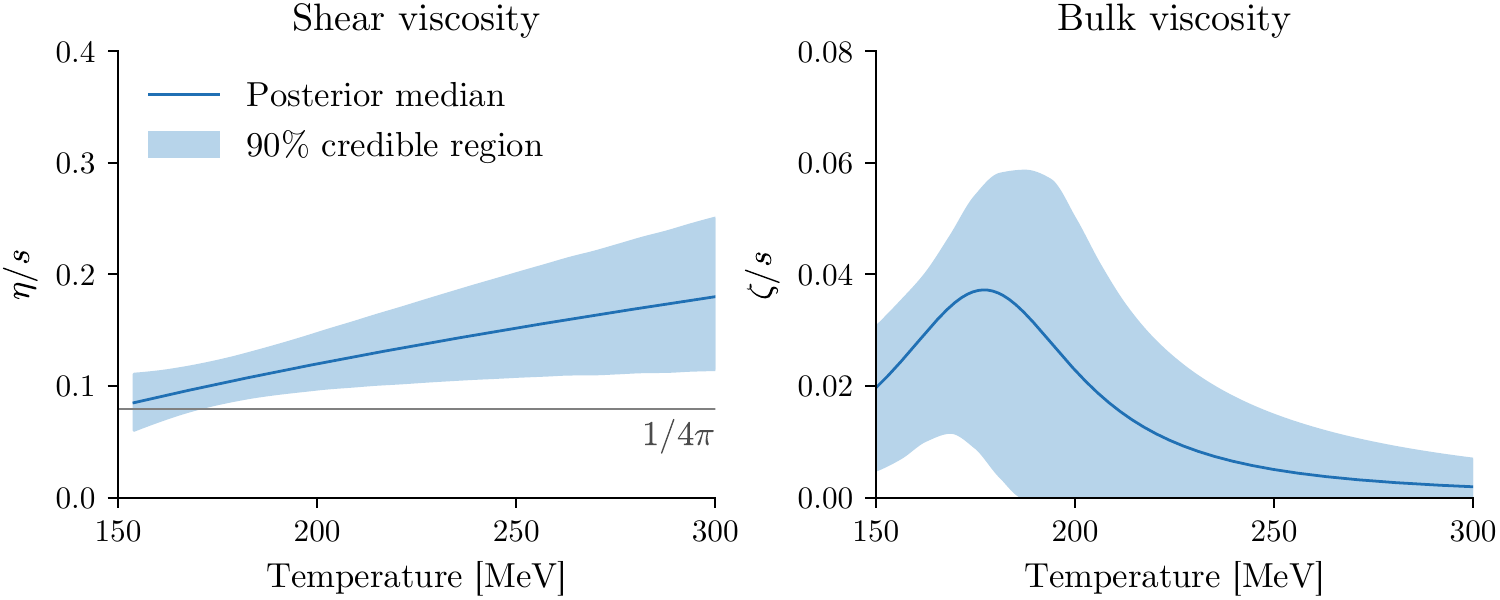}
\caption{\label{fig:intro:introviscosities}Left: Shear viscosity divided by entropy density as a function of temperature. Right: Bulk viscosity divided by entropy density as a function of temperature. In both figures the 90\% credible interval is shown. Also shown is the holographic result $\eta/s = 1/4\pi$ for the shear viscosity. Both figures taken from \cite{Bernhard:2018hnz}\@.}
\end{figure}

The next stage in the description of a heavy ion collision is the hydrodynamical evolution of the fluid created in the hydrodynamization process.
For this stage, the physical description is well understood, namely viscous hydrodynamics.
What is less well understood are the values of the various transport coefficients entering the evolution through (\ref{eq:intro:secondorderbulk}--\ref{eq:intro:secondordershear})\@.
Of the transport coefficients listed, the ones with the most pronounced effect on the experimental observables are the shear and bulk viscosities.
This makes sense, because hydrodynamics is a derivative expansion, where higher order derivatives are assumed to be less important for the evolution.
The shear and bulk viscosities are the only first order coefficients in this expansion, expressing the fact that they have the most influence on the evolution of the fluid and hence on the final experimental observables.
In figure \ref{fig:intro:introviscosities}, the results from a Bayesian analysis are shown, in which among other quantities both these viscosities were fitted to experimental data.
Of theoretical interested is that these viscosities can be obtained by means of holography, which will be discussed in section \ref{sec:intro:holography}\@.
In particular, \cite{Maldacena:1997re,Policastro:2001yc,Kovtun:2004de,CasalderreySolana:2011us} obtained a surprisingly small value for the ratio of the shear viscosity to entropy density ratio:
\[
\frac{\eta}{s} = \frac{1}{4\pi},
\]
where one should note that the assumption of infinite coupling strength is an important ingredient in the holographic computation.
As can be seen in figure \ref{fig:intro:introviscosities}, this value is compatible with the results from the Bayesian analysis for values near the QCD cross-over, where the effective coupling strength is expected to be large.
This lends credibility to the idea that holography can be used to at least give qualitative insight into QCD\@.

At some point after the collision, the fluid has cooled and diluted enough so that the interactions can no longer maintain local hydrodynamic equilibrium, and hydrodynamics no longer provides a good description of the fluid.
Theoretical models reflect this change by switching to a particle description at a certain temperature called the freeze-out temperature $T_\text{fr}$\@.
Note here that this change in description depends not so much on the time, but instead on the temperature.
This means that even though the language in this section conveys this process as occuring sequentially in time, the time at which freeze-out occurs is not the same for different regions of the plasma.

After the system has cooled enough so that it is no longer described by hydrodynamics, there are still interactions between the particles, which can be well described by solving a Boltzmann equation.
However, as the system expands further, at some point the particles become far enough separated that they no longer interact.
After this time, except for the decay of unstable particles, no further interactions occur, and the particles travel in straight trajectories until they are detected.
In fact, these final particles are all that can be measured.
None of the other processes mentioned can be directly observed, so all conclusions about the processes described above have to be inferred from the final state particles and their correlations.
As one can imagine, this is an enormous obstacle towards understanding the processes involved, because the final state typically depends on all of the physical processes involved in the collision.

Before discussing the various observables one can define in terms of the final state particles, let us mention one more physical process, which will be neglected in the rest of this thesis, but should nevertheless be mentioned.
During the initial collision, it is possible that two nucleons undergo a hard scattering, creating high-momentum particles.
These particles form jets, which subsequently propagate through the medium.
This process contains a wealth of information about the medium, but as we will not simulate jets in chapter \ref{ch:trajectum}, we will not discuss them in detail.

Let us now move on to the discussion of the observables which can be experimentally measured.
This discussion will necessarily be limited to a small subset, but this should give a good impression of the main types of observables and the type of information they carry about the physical processes mentioned above.
Here, note that this will be a general discussion, and the precise definitions of observables computed in this thesis and their comparison to experimental data will be done in chapter \ref{ch:trajectum}\@.
To start, let us note that the spatial extent of the QGP is only a couple of femtometers, which is too small to be able to measure any spatial information.
Hence all observables are defined in terms of the momenta of the final state particles, where subtle differences between observables can be made based on which particles to count, such as conditions on the momenta and particle species.

For these definitions, let us decompose the transverse momentum of each particle in the following way:
\[
p_x = p_T\cos\phi, \qquad p_y = p_T\sin\phi,
\]
where $p_T$ is the transverse momentum and $\phi$ is the azimuthal angle.
In addition, we define the rapidity $y$ and the pseudorapidity $\eta$:
\begin{equation}
y = \frac{1}{2}\log\left(\frac{E + p_z}{E - p_z}\right), \qquad \eta = \frac{1}{2}\log\left(\frac{|p| + p_z}{|p| + p_z}\right),\label{eq:intro:rapiditydef}
\end{equation}
where $E$ is the energy of the particle, and the $z$-component of the momentum points along the beam axis.
Note that in the case of massless particles, we have $y = \eta$, and also note that while computing $y$ requires knowledge of a particle's mass, $\eta$ is a pure angle.
Most observables are defined in terms of only particles satisfying certain constraints on their momenta.
The main reason for this is that experimentally, detectors are not 100\% efficient in detecting every single particle from an event, where efficiencies vary depending on particularly $p_T$ and $\eta$\@.
One could try and correct for this, but it is easier to just exclude particles from the most inefficient regions from the analysis.
Indeed, for theorists it is easy to simply apply the same cuts, and this allows for a cleaner comparison.

With the momentum decomposition in hand, let us now define centrality.
As was mentioned in the beginning of this section, the amount of overlap of the initial nuclei is very important, where central events with a small impact parameter produce many particles in a roughly spherical manner, while more peripheral events produce fewer particle in a more anisotropic way.
Unfortunately, there is no way to experimentally determine the impact parameter, and therefore the `centrality' is defined in a different way.
Since we know that central events produce more particles than peripheral ones, it makes sense to use the number of particles (most often the number of charged particles to be precise) produced by an event as a proxy for the impact parameter.
In this way, we determine for each event how many particles it produced, and sort all of them from many particles to few.
Then we define centrality by percentiles, i.e.~the event with the most particles is by definition 0\% central, while the event with the fewest is by definition 100\% central, and the other events interpolate between these extremes.

Using the above discussion, we can already define a few observables, such as the number of particles produced per unit pseudorapidity $dN/d\eta$ and the mean transverse momentum $\langle p_T\rangle$\@.
As it turns out, the number of particles produced correlates well with the entropy produced in the initial stage of the collision, because viscous corrections in the hydrodynamical evolution are too small to generate appreciable amounts of entropy, and the final state entropy is proportional to the number of particles.
The momenta of the particles produced in the final state are to some approximation those of a boosted thermal ensemble.
Because of this, the mean transverse momentum is mostly sensitive to the freeze-out temperature and the velocity of the fluid at the freeze-out surface.

It was mentioned above that the initial geometry of the plasma is generically anisotropic.
It turns out that this initial spatial anisotropy is translated by the hydrodynamic evolution into anisotropy in momentum space, specifically in the azimuthal distribution of the momenta of the final state particles.
In particular, one can perform a Fourier decomposition of the particle distribution $dN/d\phi$ in an event:
\[
\frac{dN}{d\phi} = \frac{N}{2\pi}\left(1 + \sum_{n=1}^\infty v_n\cos\left[n(\phi - \Psi_n)\right]\right),
\]
where $v_n$ are called the anisotropic flow coefficients, and $\Psi_n$ are the event plane angles.
Averaged over a large number of events, the flow coefficients show a pronounced dependence on the centrality and hence on the impact parameter.
The reason for this is that the $v_n$, but also correlations between different $\Psi_n$, inherit information about the initial geometry, and this depends strongly on the impact parameter.
This is however not the complete story.
The viscosities tend to smooth out spatial structure.
As such, large values of the viscosities tend to lower the final anisotropy present, making $v_n$ a probe of especially the shear viscosity.

These observables will be discussed in more detail in section \ref{sec:trajectum:analyze}, along with figures of these experimental results compared to theoretical predictions described in section \ref{sec:trajectum:collide}\@.
Next, we will examine a different corner of the QCD phase diagram, namely neutron stars.
\section{Neutron stars}
When a star burns through its supply of hydrogen, it reaches the end of its life.
In a sequence in which it starts burning ever heavier elements, it eventually sheds its outer layers to leave behind a compact remnant.
The nature of this remnant is determined mainly by the mass of the progenitor star.
For stars like our sun, the remnant will be a so-called white dwarf: an object with a mass in the order of magnitude of one solar mass ($1\,M_\odot$) and a radius comparable with that of the earth.
Unlike an ordinary main sequence star, a white dwarf is not held in static equilibrium by thermal pressure of gas resisting gravitational collapse.
Instead, the degeneracy pressure due to the Pauli exclusion principle of the electrons is what resists further gravitational collapse.
There is however a limit to how much mass such an object can have before electron degeneracy pressure becomes insufficient to maintain hydrostatic equilibrium.
This is called the Chandrasekhar limit, and is equal to about $1.4\,M_\odot$\@.

Indeed, if the progenitor star is too massive, the resulting compact remnant is no longer a white dwarf, but a neutron star instead.
For small pressures, neutrons are unstable, as they can undergo beta decay into a proton, an electron and an anti-electronneutrino.
At extreme densities, however, it is thermodynamically favorable for the protons and electrons inside ordinary matter to merge and form neutrons and neutrinos.
This explains the name neutron star, as a neutron star is extremely neutron-rich.
Observationally, it is known that the masses of known neutron stars are typically about $1.4\,M_\odot$, but with masses of around $2\,M_\odot$ also occuring.
The radius depends on the mass, as we see below, and recent experimental constraints by NICER put the radius of a typical $1.4\,M_\odot$ neutron star at around $13\,\text{km}$ \cite{Raaijmakers:2019qny}\@.
Finally, we note that neutron stars are cold as far as QCD is concerned.
This may seem like an odd statement given the fact that they have temperatures on the order of $100\,\text{eV}$ \cite{Vigano:2013lea}\@.\footnote{Newly formed neutron stars are much hotter, but this phase does not last very long.}
However, since neutron stars consist of densely packed neutron-rich matter, for which the relevant physics is QCD, one should compare this temperature to the energy scale of QCD, which is around $1\,\text{GeV}$\@.
In this unit, we can safely neglect the effects of temperature on the structure of the neutron star and assume $T = 0$\@.\footnote{Note though that for non-QCD processes, like the emission of thermal energy in the form of light, the temperature can most definitely not be neglected. Note also that during a binary neutron star merger event, the temperatures cannot be neglected.}
This means that, given their enormous density and cold temperature, neutron stars occupy the low temperature, large chemical potential region of the QCD phase diagram.

Let us now show that indeed the mass and radius are related.
Assuming a non-rotating neutron star, one can assume spherical symmetry.
This, in combination with the assumption of hydrostatic equilibrium ($T^{\mu\nu} = \diag(e,P,P,P)$), allows us to write down the Tolman-Oppenheimer-Volkov (TOV) equations:
\[
\frac{dP}{dr} = -\frac{(e + P)(Gm(r) + 4\pi r^3GP)}{r(r - 2Gm(r))},
\]
where $r$ is the radial coordinate, $G$ the gravitational constant and where the mass $m(r)$ enclosed within radius $r$ satisfies
\[
\frac{dm}{dr} = 4\pi r^2e.
\]
Supplying these two equations with the boundary conditions at the center of the star at $r = 0$ that $m(0) = 0$ and that the central density is some specified value $\rho^*$, the differential equations can be integrated to yield $P(r)$ and $m(r)$\@.
From the solution, we can then identify the radius $R$ of the star as the point where $P(R) = 0$, and subsequently we can also find the mass $M = m(R)$\@.\footnote{Note that there is a subtlety here, namely that in general relativity one has to define what one means by radius and mass. We define the radius $R$ to be in Schwarzschild coordinates, which implies that the area of a star of radius $R$ equals $4\pi R^2$\@. For the mass, we define that the gravitational mass, as measured by examining the Schwarzschild metric outside the star, is the same as that of a black hole of mass $M$\@.}
In this way, we obtain $R(\rho^*)$ and $M(\rho^*)$ as parametric functions of the central density $\rho^*$\@.

\begin{figure}[ht]
\centering
\includegraphics[width=0.49\textwidth]{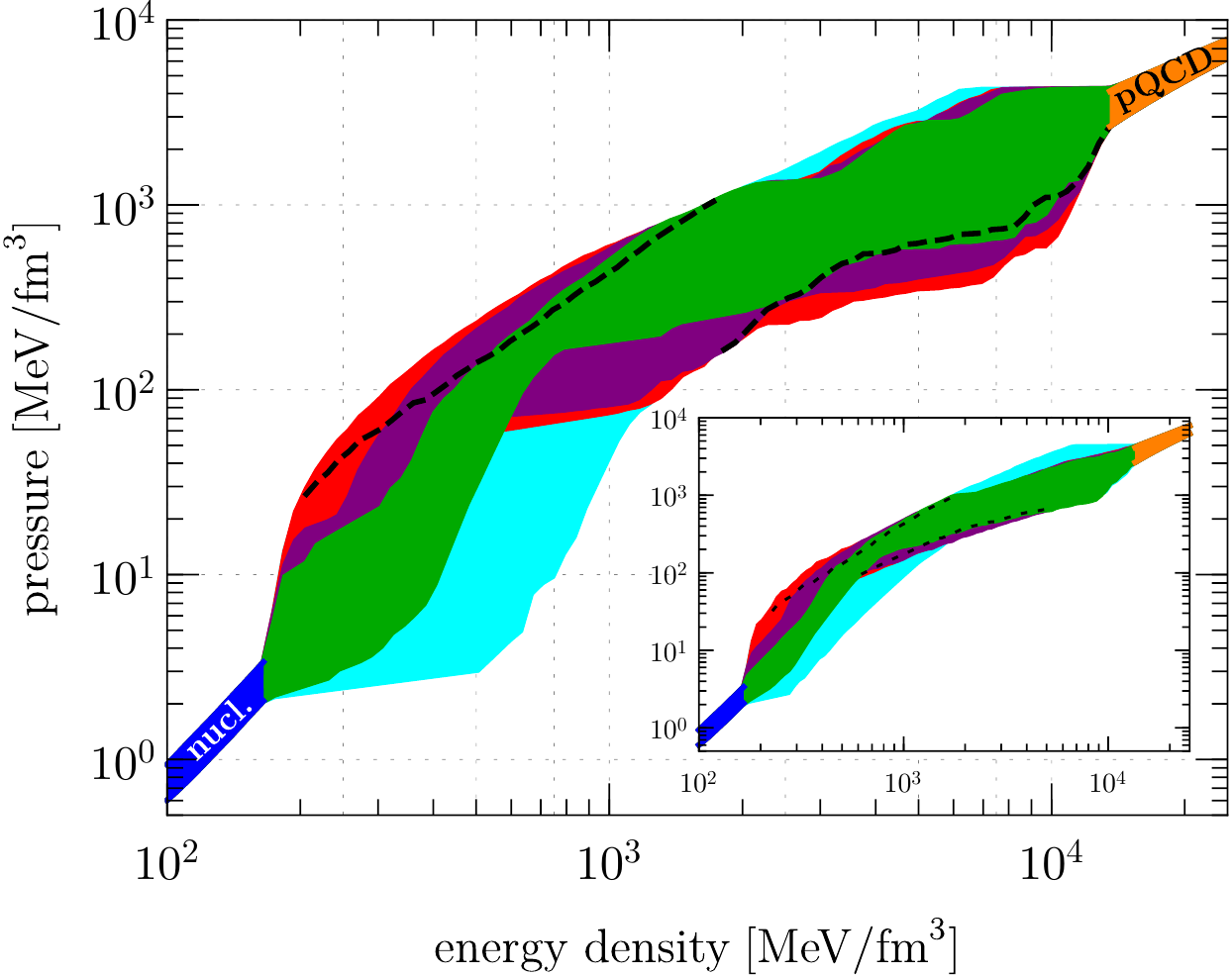}
\includegraphics[width=0.49\textwidth]{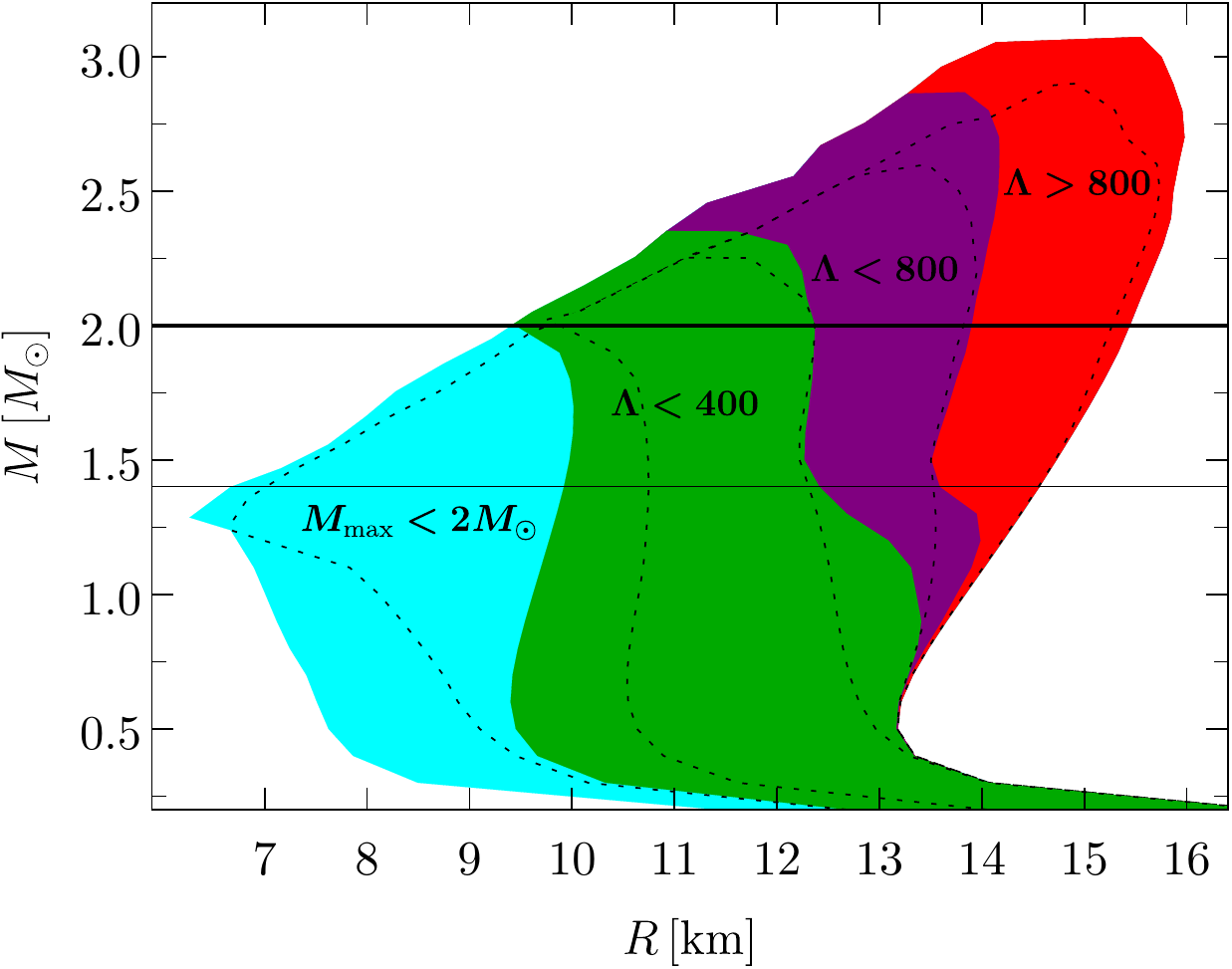}
\caption{\label{fig:intro:introneutronstareos}Left: Family of equations of state compatible with QCD constraints at both small and large densities. The green and purple bands lead to neutron stars compatible with known observational constraints. The inset shows a similar family of equations of state constructed in a slightly different setup. Right: Mass to radius relation for the same equations of state shown in the left panel. The color coding is the same. Both figures taken from \cite{Annala:2017llu}\@.}
\end{figure}

Note though that the TOV equations can only be solved given an equation of state.
In \cite{Annala:2017llu}, a large number of equations of state was generated, which are shown in the left panel of figure \ref{fig:intro:introneutronstareos}\@.
Here only equations of state were used which are causal, i.e.~the speed of sound is less than the speed of light, and which simultaneously satisfy constraints from nuclear matter models at low densities and from perturbative QCD at high densities.
In the right panel, the resulting mass to radius relations are shown for the same equations of state.
One can see several important features.
First of all, each equation of state has a maximum allowed mass for neutron stars it supports, in much the same way as we saw above for white dwarfs.
It is observationally not precisely known what this maximum mass is precisely, but it is known that a neutron star named J0348+0432 has a precisely measured mass of $2.01 \pm 0.04\,M_\odot$ \cite{Antoniadis:2013pzd}\@.\footnote{An even more massive star, J0740+6620, was detected after the publication of \cite{Annala:2017llu}, with a mass of $2.14_{-0.09}^{+0.10}\,M_\odot$ \cite{Cromartie:2019kug}\@.}
This means that all the equations of state colored blue in figure \ref{fig:intro:introneutronstareos} are excluded by this observation, as these equations of state do not support a $2\,M_\odot$ star.
Similarly, the equations of state colored red are also excluded by observational constraints, this time by the tidal deformability $\Lambda$ of the neutron stars involved in the binary neutron star merger GW170817 \cite{TheLIGOScientific:2017qsa}\@.

\begin{figure}[ht]
\centering
\includegraphics[width=\textwidth]{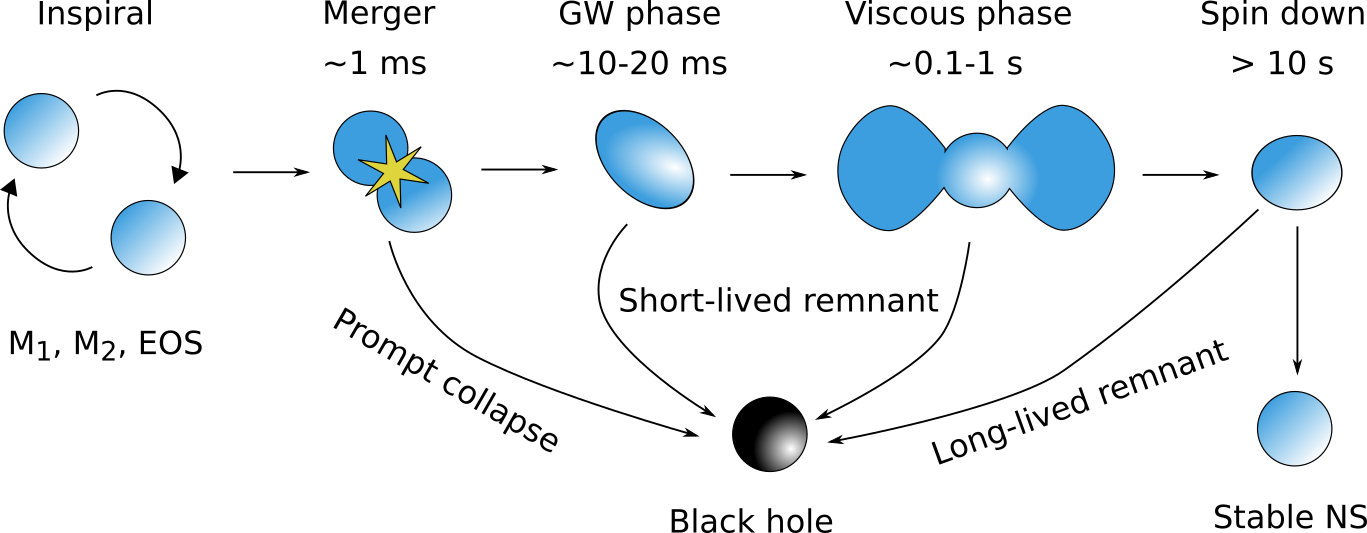}
\caption{\label{fig:intro:intromergerfates}Overview of the different stages of a binary neutron star merger. Arrows indicate the possible scenarios, where the outcome depends mostly on the masses of the progenitors and the equation of state. Figure taken from \cite{Radice:2020ddv}\@.}
\end{figure}

Let us next discuss neutron star mergers.
In 2017, the first neutron star merger, GW170817, was discovered using gravitational waves \cite{TheLIGOScientific:2017qsa}, and was accompanied by an electromagnetic counterpart \cite{Goldstein:2017mmi,GBM:2017lvd,Monitor:2017mdv}\@.
In the remainder of this section, we will discuss the various stages involved in such a merger, and how QCD enters the problem.
For more detailed reviews, see \cite{Baiotti:2016qnr,Radice:2020ddv}\@.
In figure \ref{fig:intro:intromergerfates}, a schematic overview is given of the different stages of a merger event.
The first stage of the merger is by far the longest lasting, as it covers the inspiral.
This inspiral can quite literally take billions of years, as gravitational wave emission circularizes the orbit and slowly but surely shrinks the size of the orbit.
As the orbits shrink, the orbital period does too, and the amplitude of the emitted gravitational waves increases.
Only in the last minute or so does this happen to a large enough extent such that observatories such as LIGO and VIRGO can detect the gravitational waves emanating from the source.
During the merger, the two stars exert a tidal force on one another, which slightly deforms the stars.
This produces an imprint in the gravitational wave emission, the size of which depends on the equation of state through the tidal deformability $\Lambda$ \cite{Read:2013zra}\@.

When the stars touch, two things can happen.
Either the stars are heavy enough that the densities immediately exceed what the equation of state can support, and they collapse to a black hole.
In this case, very little material will be ejected, producing only a small electromagnetic counterpart.
Also, gravitational wave emission dies down quickly, as the resulting black hole will ring down with its characteristic quasinormal mode frequencies.
The other option is that the stars merge to form a highly deformed object, which loses energy by gravitational wave emission, and ejects a substantial amount of neutron-rich matter.
This matter, no longer under enormous pressure, decays to form heavy elements, and emits electromagnetic radiation in the process.
The gravitational waves emitted during this phase are characteristic of the equation of state.

As the deformed object circularizes over a timescale of around $10\,\text{ms}$, gravitational wave emission dies down, there is again the possibility of gravitational collapse to a black hole.
If this does not happen, the merger remnant will slowly lose angular momentum due to various processes over the course of a few seconds.
The angular momentum effectively contributes partly to the pressure preventing the star from collapsing, and therefore as the star spins down, there is again the possibility of collapse to a black hole.
If the mass of the merger remnant is below the maximum allowed mass however, the remnant will be a heavier neutron star.

\begin{figure}[ht]
\centering
\includegraphics[width=\textwidth]{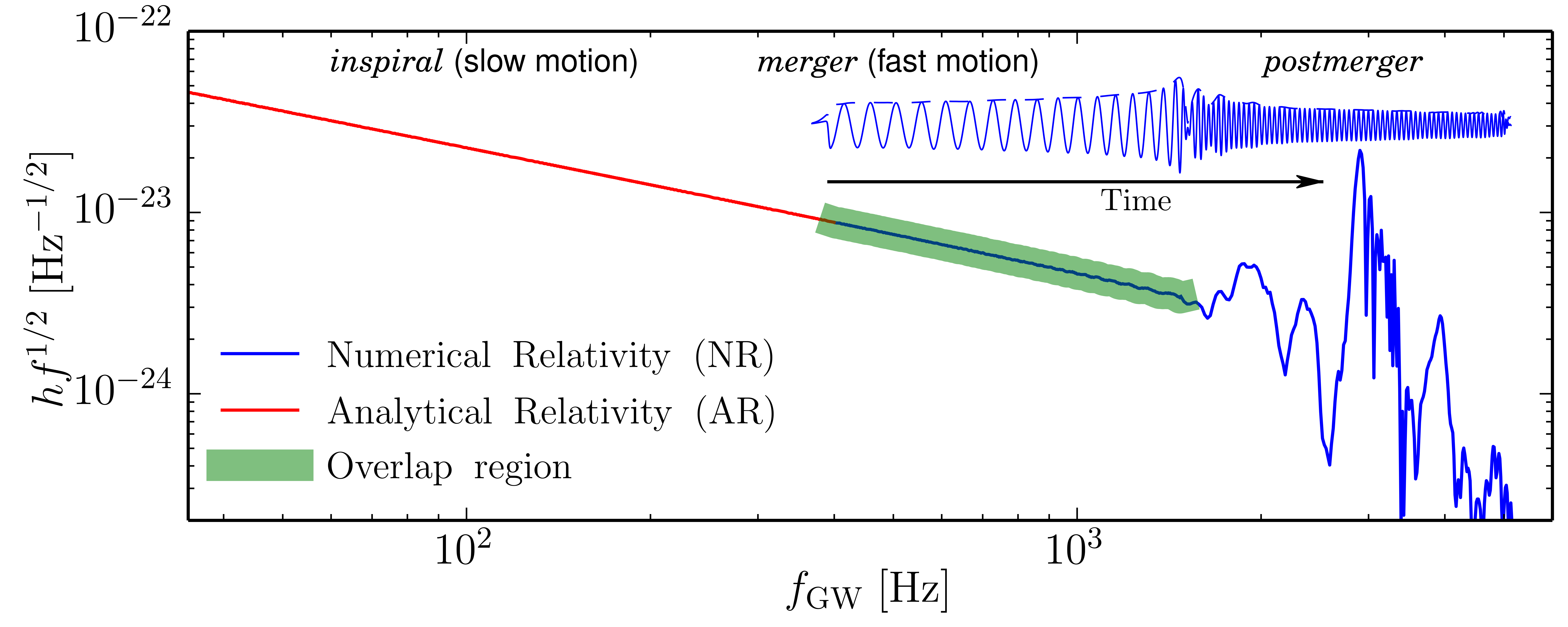}
\caption{\label{fig:intro:introgwcomputation}Amplitude of gravitational waves emitted as a function of frequency. Methods of computation are indicated, and a part of the waveform is shown as an inset. Figure taken from \cite{Radice:2020ddv}\@.}
\end{figure}

To end this section, let us discuss the methods used to theoretically compute a waveform.
Neutron star mergers are described theoretically by relativistic hydrodynamics coupled to general relativity.
In figure \ref{fig:intro:introgwcomputation}, one can see the amplitude of gravitational waves emitted as a function of frequency.
The low frequencies are mostly produced during the inspiral phase, while the peaks at high frequencies are mostly produced during the post-merger phase.
Also indicated are the methods used, namely analytical methods for most of the inspiral, and numerical relativity for the merger part, where there is an overlap interval in which both methods are applicable.
Numerical relativity is immensely computationally expensive, so it would be impractical to have to compute the inspiral using numerical relativity.
In this way, the two methods neatly complement each other.
Explaining either of these methods in detail is beyond the scope of this thesis, however good introductions can be found in \cite{Will:2014kxa,Baumgarte:2010ndz}\@.
\section{Holography}\label{sec:intro:holography}
Holography is a duality between two at first sight very different classes of theories.
To illustrate this, let us focus on the first constructed example, namely that of $\mathcal{N} = 4$ super Yang-Mills (SYM) theory living in four spacetime dimensions.
For the purpose of this section, this can be seen as a highly supersymmetric version of QCD, where we take the number of colors $N_c$ to be infinite.
In \cite{Maldacena:1997re}, a convincing argument was made that this theory is the same as type-IIB string theory living in an anti-de Sitter (AdS) background with five spacetime dimensions.
In other words, the string theory lives in one dimension more than the gauge theory that it is dual to.

Furthermore, what makes this duality particularly interesting, is that the string theory side of the duality simplifies if in addition to $N_c \rightarrow \infty$ we also take the 't Hooft coupling $\lambda \equiv g^2N_c$ to be infinite, where $g$ is the gauge theory coupling constant.
When taking this limit, known as the 't Hooft limit, two things happen:
The string coupling on the string theory side of the duality goes to zero, leaving us with a classical string theory.
Additionally, also the string length vanishes, reducing the classical string theory further to a theory of point-like particles, which in this example is classical type IIB supergravity.

In this way, we obtain a duality between on one side a strongly coupled quantum field theory, which we say lives on the boundary, and on the other side a classical gravitational theory which lives in one dimension extra, which we say lives in the bulk.
This is extremely useful, because this duality relates something difficult, namely strongly coupled QFT, to something relatively easy, namely classical general relativity.
Note that in \cite{Maldacena:1997re}, the duality was only introduced for $\mathcal{N} = 4$ SYM theory and related theories, and that there is no known general way to obtain a holographic dual for an arbitrary QFT\@.
It is expected though, both from the large $N_c$ expansion in gauge theory \cite{tHooft:1973alw} and from black hole thermodynamics \cite{Bekenstein:1973ur,tHooft:1993dmi,Susskind:1994vu}, that the class of theories with holographic duals is larger.
We will touch further upon this problem of constructing holographic duals in section \ref{sec:intro:bottomup}\@.
Before doing so however, let us discuss some more how the duality precisely works.
Indeed, for the two theories on either side of the duality to be equal, one needs a precise dictionary for how to relate quantities and problems on one side to the corresponding quantities and problems on the other side.
Such a dictionary has been developed over the years, and in the following paragraphs we will discuss a selection of this dictionary, introducing only the quantities that will be used in this thesis.
This discussion will just describe the dictionary without going into the derivations.
An excellent review which goes in more detail can be found in \cite{CasalderreySolana:2011us}\@.

Let us start the discussion of the dictionary with a few thermodynamical quantities.
The first of these is the free energy.
The starting point for this is the observation that the partition functions of both sides of the duality are equal \cite{Gubser:1998bc,Witten:1998qj,Aharony:1999ti}\@.
After performing a Wick rotation, and using the fact that the bulk theory is classical, one obtains that the free energy of the boundary theory $F$ obeys
\[
F = -TS,
\]
where $T$ is the temperature, and $S$ is the on-shell action of the bulk theory.
Note here that the on-shell bulk action is divergent towards the AdS boundary.
This problem is similar in origin to the UV divergences originating in QFTs, and the solution is similar, namely holographic renormalization \cite{deHaro:2000vlm}\@.
In holographic renormalization, one regularizes the divergence by introducing a cutoff $\epsilon$ in the bulk spacetime integral for the action.
Subsequently, one then compares the desired action to that of a reference solution, after which one can take the limit $\epsilon \rightarrow 0$ for the difference of the two regularized actions.
In this way one can compute free energies up to an overall constant, which is for most purposes enough.

Let us next discuss temperature and entropy.
In a bulk geometry with a horizon, such as one with a planar horizon called a black brane, one can obtain the temperature as the Hawking temperature of the horizon, which can be expressed in terms of the local metric at the horizon by requiring that the Wick rotated geometry has no conical singularity at the horizon \cite{Gibbons:1976ue}\@.
The entropy can also be obtained purely from horizon data, namely by use of the Bekenstein-Hawking formula \cite{Bekenstein:1973ur}:
\[
S = \frac{A}{4},
\]
where now $S$ is the entropy, and $A$ is the area of the black hole.
Note that in the case of a black brane solution, the black hole is infinite in extent, in which case it makes more sense to divide out the volume on the boundary.
Indeed, when one examines the entropy density, the result is still finite.

The next important element in the holographic dictionary that we will need is the field-operator correspondence \cite{Gubser:1998bc,Witten:1998qj}\@.
Imagine an operator $\mathcal{O}(x)$ in the boundary theory that we want to compute, and imagine introducing a source for that operator $\phi(x)$\@.
Then the field-operator correspondence tells us that in the corresponding bulk theory there is a field $\Phi(x,r)$, with $r$ the bulk coordinate, where $r = 0$ corresponds to the boundary of AdS\@.
This bulk field $\Phi$ then has the following near-boundary expansion:
\[
\Phi(x,r) = \phi(x)r^{4-\Delta} + \frac{1}{2\Delta - 4}\langle\mathcal{O}\rangle(x)r^\Delta,
\]
where $\Delta$ is the scaling dimension of the operator $\mathcal{O}$\@.
For most operators the first term will be non-normalizable, and the second will be normalizable.
One can see that in this way, one obtains a way to evaluate expectation values of operators in the boundary theory by an equivalent computation in the bulk theory, namely by extracting the subleading behavior of the corresponding bulk field.
This result also allows for the computation of Green's functions.
For example, by considering the appropriate space-time dependent metric fluctuation $\delta g^{\mu\nu}$ as the source for the stress-energy tensor $T^{\mu\nu}$, one can obtain the Green's function for the stress-energy tensor, leading to the famous result mentioned earlier, namely that the shear viscosity divided by the entropy density of a holographic fluid is equal to $1/4\pi$ \cite{Policastro:2001yc,Kovtun:2004de,CasalderreySolana:2011us}\@.

Let us next move on to two non-local operators, namely the Polyakov loop correlator and the entanglement entropy.
In QCD, the Wilson line operator
\[
\tr\mathcal{P}\exp\left[i\int_\mathcal{C}dx^\mu A_\mu(x)\right],
\]
where $\mathcal{P}$ denotes path ordering and $\mathcal{C}$ denotes a closed path, contains information on among other things confinement.
The reason for this is that if one takes $C$ to run in the time direction from $-\infty$ to $\infty$, and if one then takes two such loops at a constant distance $L$ from each other, the expectation value of this Polyakov loop correlator is equal to the potential energy stored in the gluon field separating a heavy quark-antiquark pair.
The holographic dual of this operator is the on-shell action of an open string in the bulk attached to the path $\mathcal{C}$ on the boundary \cite{Maldacena:1998im,Rey:1998ik}\@.
As the action of a string is just the area measured in the string frame metric, the complicated non-perturbative problem of evaluating the expectation value of the Polyakov loop correlator itself is therefore replaced in the holographic dual by the much easier task of finding a minimal surface.

A computationally related quantity to the Polyakov loop correlator is the entanglement entropy.
In a QFT, if we imagine dividing the spacetime into a region $A$ and its complement $A^c$, we can partition the Hilbert space as $H = H_A \otimes H_{A^c}$, and define the reduced density matrix for a pure state $\Psi \in H$ by $\rho_A = \tr_{A^c}(|\Psi\rangle\langle\Psi|)$\@.
We can subsequently define the entanglement entropy as
\[
S_A = -\tr_A(\rho_A\log\rho_A).
\]
For static spacetimes, \cite{Ryu:2006bv} proposed that the holographic dual of entanglement entropy is, similarly to the Wilson loop, a minimal surface in the bulk with its ends attached to the boundary of the region $A$\@.
Important differences with the Wilson loop are that in the case of entanglement entropy, the minimal surface is a codimension 2 surface in the bulk, whereas in the case of the Wilson loop, the minimal surface is a dimension 2 surface.
Also, for the entanglement entropy we use the Einstein frame metric, whereas for the Wilson loop, one had to use the string frame metric.
The proposition was later generalized to non-static spacetimes in \cite{Hubeny:2007xt}, and both propositions were proven in \cite{Lewkowycz:2013nqa} and \cite{Dong:2016hjy}, respectively.

Lastly, let us briefly discuss baryons, which are important if one aims for a holographic description of neutron stars, as at least up to some depth these are composed of mostly baryons.
In \cite{Witten:1998xy,Gross:1998gk}, it was shown in the $\mathcal{N} = 4$ SYM example which was also used above, that baryons in the boundary theory can be identified with D5-branes wrapping the 5 compact dimensions in the bulk, which we previously neglected.
In this way the baryon appears in the bulk as a small pointlike topological defect, i.e.~a soliton.
This analysis was later extended to other holographic models obtained from string theory, such as the Witten-Sakai-Sugimoto (WSS) model \cite{Witten:1998zw,Sakai:2004cn,Sakai:2005yt}, with similar conclusions \cite{Hong:2007kx,Hata:2007mb,Hong:2007dq,Hashimoto:2008zw,Kim:2008pw,Cherman:2009gb,Cherman:2011ve,Bolognesi:2013nja,Rozali:2013fna,Kaplunovsky:2012gb,deBoer:2012ij,Kaplunovsky:2015zsa,Preis:2016fsp,BitaghsirFadafan:2018uzs,Bergman:2007wp,Rozali:2007rx,Ghoroku:2012am,Li:2015uea,Elliot-Ripley:2016uwb}\@.
\subsection{Bottom-up holography: IHQCD and V-QCD}\label{sec:intro:bottomup}
One issue that we have so far glossed over is the fact that even though holography allows for an enormous simplification of certain computations, the theories discussed so far are not QCD\@.
For example, even though $\mathcal{N} = 4$ SYM theory is in essence `just' QCD with a large number of colors and a lot of supersymmetry, phenomenologically the two theories are quite different, most notably in the fact that $\mathcal{N} = 4$ SYM theory is conformal, whereas QCD is not.
In the construction of holographic models to describe QCD, there are two general classes of models.
On the one hand, there are the `top-down' approaches, which includes $\mathcal{N} = 4$ SYM theory, but also the previously mentioned WSS model.
In a top-down approach, one starts from a string theoretical construction, and in that way arrives at a precise holographic theory.
This has the obvious advantage that in this approach, the holographic dictionary is precisely known, and the general amount of control over the computations is larger.
The main disadvantage of such theories is that, like $\mathcal{N} = 4$ SYM, the phenomenological resemblance to QCD is not very good.

An alternative approach is the so-called `bottom-up' approach, where one tries to construct a holographic model without a derivation from string theory, where the aim is to make the model as phenomenologically accurate as possible.
Early examples of this approach are the `hard-wall' models \cite{Erlich:2005qh,DaRold:2005mxj}, which were followed by the `soft-wall' model introduced in \cite{Karch:2006pv}\@.
In this subsection, we will focus on the IHQCD model, including its extension V-QCD, as this is the model we will be using in later chapters.

In Improved Holographic QCD (IHQCD), a holographic theory is constructed for the fields dual to the $\tr F^2$ and $T^{\mu\nu}$ operators in QCD\@.
These fields are the dilaton $\Phi$ and the metric $g^{\mu\nu}$, respectively.
Note though that in this thesis we will write $\Phi = \log\lambda$\@.
The IHQCD action is the following \cite{Gursoy:2007cb,Gursoy:2007er}:
\begin{equation}
S_g = M^3N_c^2\int d^5x\sqrt{-g}\left(R - \frac{4}{3}\frac{(\partial\lambda)^2}{\lambda^2} + V_g(\lambda)\right),\label{eq:intro:Sg}
\end{equation}
with $R$ the Ricci scalar, and $V_g(\lambda)$ a potential function.
The choice of a non-trivial potential $V_g$ allows for breaking of conformality in IHQCD\@.
One can see this as follows:
The metric which solves the IHQCD action is, near the AdS boundary, of the form
\[
ds^2 = e^{2A(r)}\left(-dt^2 + dr^2 + dx_1^2 + dx_2^2 + dx_3^2\right),
\]
where the scale factor $\exp A(r) = 1/r$ can be interpreted as the renormalization scale.
On the other hand, $\lambda$ can be interpreted as the QCD coupling strength.
With the appropriate choice for the small $\lambda$ expansion of $V_g(\lambda)$, one can then make sure that $d\lambda/dA$ is equal to the QCD $\beta$-function in the UV\@.

Another demand on the potential fixes the large $\lambda$ behavior of the potential as well.
By requiring that the theory is confining and simultaneously has a linear glueball spectrum, the large $\lambda$ behavior of $V_g$ is restricted to be of the form
\[
V_g \sim \lambda^{4/3}\sqrt{\log\lambda}.
\]
The intermediate behavior of the potentials can still be freely chosen, but can in principle be fixed by computing observables also computed on the lattice.
Doing this results in the conclusion that IHQCD can match very well lattice results for pure Yang-Mills \cite{Gursoy:2008za,Gursoy:2009jd}\@.

IHQCD does not contain quarks.
For this reason, it has been extended to include $N_f$ flavor $D4$ and $\bar D4$ branes, yielding V-QCD \cite{Sen:2003tm,Bigazzi:2005md,Casero:2006pt,Casero:2007ae,Jarvinen:2011qe,Alho:2012mh}\@.
The V in the name stands for Veneziano, as we take $N_f$ to be large, with $x_f \equiv N_f/N_c$ fixed, a limit known as the Veneziano limit \cite{Veneziano:1979ec}\@.
We then obtain the following action in addition to the one for IHQCD \eqref{eq:intro:Sg}:
\begin{align}
S_\mathrm{DBI} & = -\frac{1}{2} M^3 N_c\,  {\mathbb Tr} \int d^5x\label{eq:intro:DBI}\\
& \quad \times \left(V_f(\lambda,T^\dagger T)\sqrt{-\det {\bf A}^{(L)}}+V_f(\lambda, TT^\dagger)\sqrt{-\det {\bf A}^{(R)}}\right),\nonumber
\end{align}
with
\begin{align*}
{\bf A}_{MN}^{(L)} & = g_{MN} + w(\lambda,T) F^{(L)}_{MN} + {\kappa(\lambda, T) \over 2 } \left[(D_M T)^\dagger (D_N T) + (D_N T)^\dagger (D_M T)\right],\\
{\bf A}_{MN}^{(R)} & = g_{MN} + w(\lambda,T) F^{(R)}_{MN} + {\kappa(\lambda, T) \over 2 } \left[(D_M T) (D_N T)^\dagger + (D_N T) (D_M T)^\dagger\right],
\end{align*}
and where the covariant derivative for the tachyon $T$ is given by
\[
D_M T = \partial_M T + i  T A_M^L- i A_M^R T.
\]
Here $A^L_M$ and $A^R_M$ are gauge fields corresponding to the global $U(N_f)_L \times U(N_f)_R$ flavor symmetry, and $F^{(L)}_{MN}$ and $F^{(R)}_{MN}$ are the corresponding field strength tensors.
This action contains 3 new phenomenological potentials: $V_f$, $\kappa$ and $w$, which we will discuss shortly.
While \eqref{eq:intro:DBI} is required in chapter \ref{ch:holographicns}, in chapter \ref{ch:imc} we can make the simplifying assumption that the non-Abelian parts of the gauge fields are zero, and that $T = \tau(r)\mathbb{I}_{N_f}$, which simplifies $S_\mathrm{DBI}$ to the following expression:
\begin{align}
S_f & = -x_fM^3N_c^2\int\,d^5xV_f(\lambda,\tau)\label{eq:intro:Sf}\\
& \qquad \times \sqrt{-\det\left[g_{\mu\nu} + w(\lambda)V_{\mu\nu} + \kappa(\lambda)\partial_\mu\tau\partial_\nu\tau\right]},\nonumber
\end{align}
which we will call the diagonalized action.
Here $V^{\mu\nu}$ is the Abelian component of both the left and right gauge fields, which are also assumed to be equal.

As was done for IHQCD, the potentials are chosen to satisfy phenomenological properties of QCD\@.
For the $V_f$-potential, the UV (small $\lambda$) behavior is fixed by requiring that the beta function matches that of QCD for different values of $x_f$\@.
The UV behavior of the $\kappa$-potential is determined by the RG flow of the quark mass \cite{Jarvinen:2011qe}, as well as the behavior at large quark mass \cite{Jarvinen:2015ofa}\@.
In the IR, the potentials are constrained to reproduce phenomenologically reasonable features in the phase diagram, as well as the properties of meson spectra \cite{Arean:2012mq,Arean:2013tja,Alho:2012mh,Alho:2013hsa,Iatrakis:2010zf,Iatrakis:2010jb,Alho:2015zua}\@.
In \cite{Jokela:2018ers}, the potentials were fitted to lattice data, resulting in a holographic model of QCD which matches with known phenomenological constraints as much as possible.

%% file: chapters/imc.tex
Magnetic fields play an important role in two widely studied QCD systems, namely heavy ion collisions, and neutron stars.
In peripheral heavy ion collisions, the spectator nucleons, which are charged and are moving close to the speed of light, induce a magnetic field of $10^{15}\,T$ which, in the appropriate units of the pion mass squared, gives around $eB/m_\pi^2 \approx 10$ \cite{Skokov:2009qp,Tuchin:2010vs,Voronyuk:2011jd,Deng:2012pc,Tuchin:2013ie,McLerran:2013hla,Gursoy:2014aka}\@.
In the context of neutron stars, magnetars exhibit magnetic fields of potentially $10^{11}\,T$ \cite{Duncan:1992hi}, which in units of the pion mass squared is about $eB/m_\pi^2 \approx 10^{-3}$\@.
Given that one of the salient features of QCD at vanishing magnetic field is its phase structure, it makes sense to study the effect of the magnetic field on the phase structure.
In particular, one can study how the phase transition temperatures, as well as the associated order parameters, change as one applies a magnetic field.
It was precisely in this context that a surprising effect was discovered.

As was discussed in section \ref{sec:intro:qcd}, at low temperatures QCD spontaneously breaks chiral symmetry.
At low temperatures, one expects that the order parameter of this chiral symmetry breaking, the chiral condensate, increases as one applies a magnetic field \cite{Miransky:2015ava,Gusynin:1994re,Gusynin:1994xp,Gusynin:1994va}\@.
This phenomenon is called `magnetic catalysis', and the reason for this is that Landau quantization leads to an effective reduction from $3+1$ to $1+1$ dimensions.
In lower dimensions, the gauge theory IR dynamics are stronger, leading to a strengthening of the chiral condensate.

\begin{figure}[ht]
\centering
\includegraphics[width=0.49\textwidth]{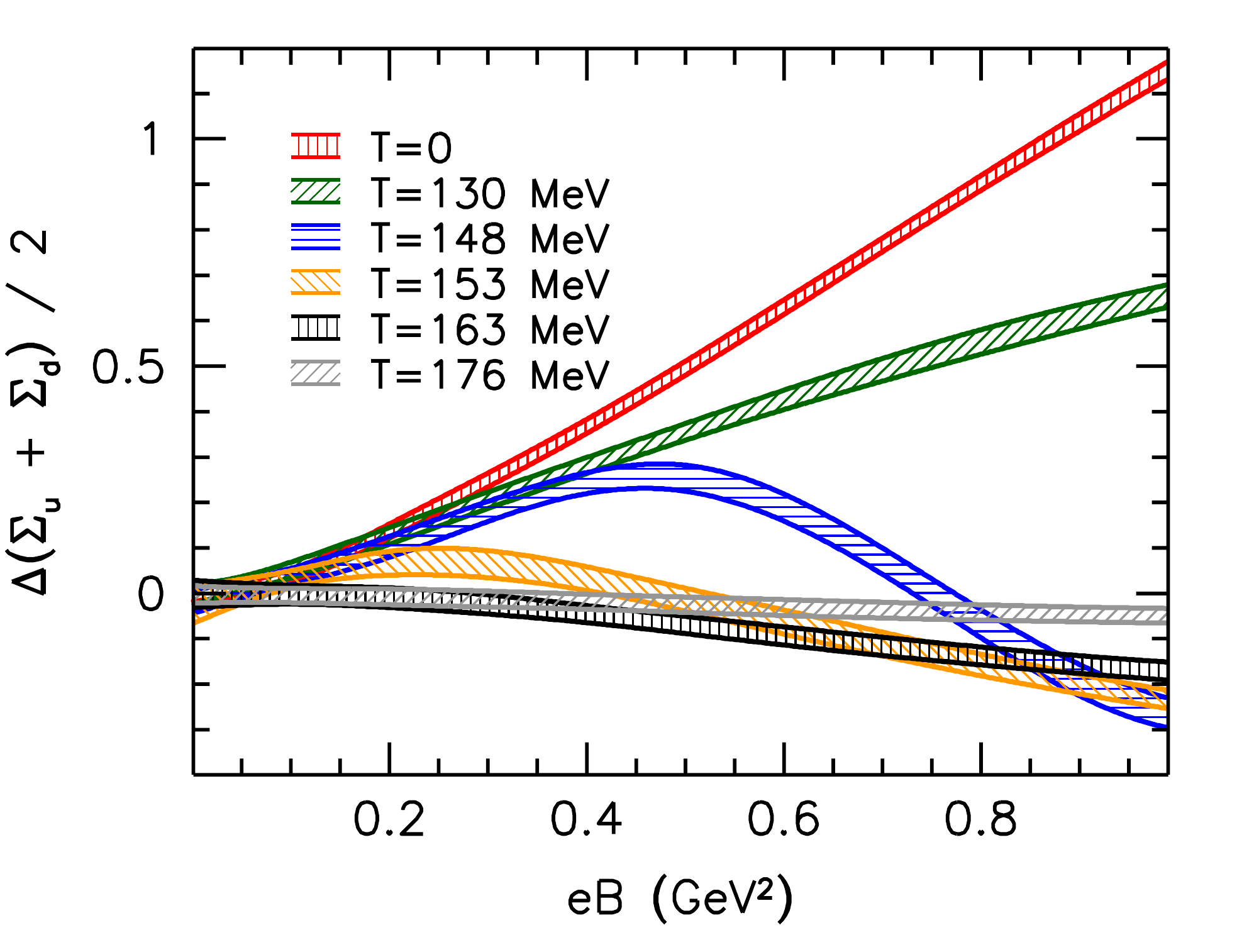}
\includegraphics[width=0.49\textwidth]{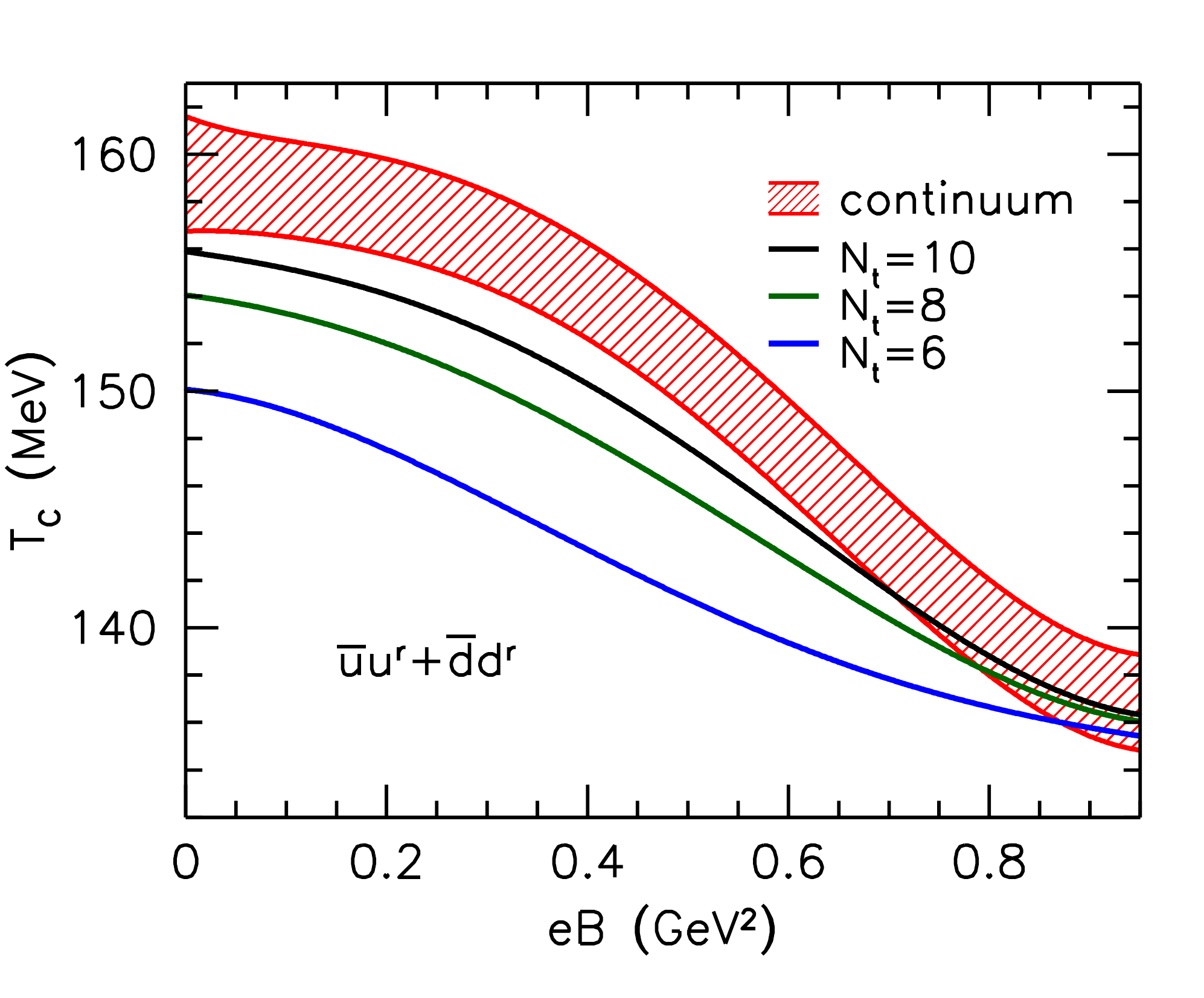}
\caption{\label{fig:imc:introimc}Left: Chiral condensate as a function of the magnetic field $B$, using the normalization as in (\ref{eq:imc:sigma}--\ref{eq:imc:deltasigma})\@. Figure taken from \cite{Bali:2012zg}\@. Right: Chiral cross-over temperature as a function of the magnetic field $B$\@. Figure taken from \cite{Bali:2011qj}\@.}
\end{figure}

However, when lattice studies were done, surprisingly the opposite effect was seen in around the crossover temperature, and this effect was named `inverse magnetic catalysis' (IMC) \cite{Bali:2011qj,Bali:2011uf,Bali:2012zg,DElia:2012ems}\@.
In figure \ref{fig:imc:introimc}, two such lattice results are shown.
In the left panel, one can see the chiral condensate as a function of $B$ for fixed temperature.
One can see that for small temperatures, one sees magnetic catalysis, whereas for larger temperatures the condensate instead decreases with $B$\@.
In the right panel, a related quantity is shown, namely the cross-over temperature as a function of $B$\@.
One can see that the cross-over temperature decreases with $B$, signalling inverse magnetic catalysis.
Given that IMC seems to require strong coupling to exhibit itself, it is natural to study this effect in holography, where we can hope to obtain a qualitative understanding of the mechanism leading to IMC\@.

In this chapter, which is based on \cite{Gursoy:2016ofp,Gursoy:2017wzz,Gursoy:2018ydr} as well as upcoming work with the authors of \cite{Gursoy:2018ydr}, we will address these questions.
While the questions asked in each of these papers are different, the model and the methods used are quite similar.
So similar in fact, that it is possible to write down a `master' model, which contains each of the models used in the papers that this chapter is based on by taking appropriate limits.
In section \ref{sec:imc:model}, we will discuss this master model, as well as how to obtain useful information from it.
This should allow for a streamlined treatment of computations that would otherwise have to return in slightly different setups in sections \ref{sec:imc:b} through \ref{sec:imc:ba}\@.
The strategy for solving the model parallels \cite{Alho:2013hsa}, where of course the discussion is slightly modified because our master model is more general than the one considered there.
Also, in a few places, it was necessary to make non-trivial adjustments to the analysis.
Wherever this occurs this will be clearly stated.
\section{Analysis of V-QCD model in the presence of a magnetic field and anisotropy}\label{sec:imc:model}
In this section, we will go through the computations necessary to obtain the relevant observables for sections \ref{sec:imc:b} through \ref{sec:imc:ba}\@.
This section is written with the aim of providing the reader with a practical guide as to how to perform these computations.
As such, it necessarily contains a lot of technical details, which are required for the computations.
The rest of this chapter has been written to only use the results from this section, and not the computations themselves, so it should be possible to follow the rest of this chapter without having read this section.
\subsection{Extending V-QCD to incorporate a magnetic field and anisotropy}
To study magnetic fields in V-QCD, we consider the diagonalized action \ref{eq:intro:Sf}\@.
A magnetic field in the $x_3$-direction can then be introduced by changing the ansatz for the Abelian gauge field to \cite{Gursoy:2016ofp,Gursoy:2017wzz}
\begin{equation}
A = \Phi(r)\,dt + Bx_1\,dx_2,\label{eq:imc:gaugefieldansatz}
\end{equation}
where we recall that $\Phi$ was dual to the baryon chemical potential.
This ansatz makes one important assumption, namely that all quark flavors have are identical, and in particular, that they have the same electric charge.
In nature, this is of course not the case, but this assumption allows us to consider only the Abelian part of the DBI action, greatly simplifying the analysis.

In addition to a magnetic field, we will also be adding an axion field $\chi$ to the action \cite{Giataganas:2017koz,Gursoy:2018ydr}, where we will use the following ansatz:
\begin{equation}
\chi = a_\perp x_2 + a_\parallel x_3.\label{eq:imc:axionansatz}
\end{equation}
In this way, we can introduce an anisotropy to the system in a way that is different from a magnetic field.
With this ansatz, the axion is dual to a space-dependent theta term.
As the axion will only appear in the action through a derivative, the presence of the axion does not break translation symmetry.
Instead, it breaks rotational symmetry.
To determine the remaining symmetry left over from rotations, we can distinguish two cases: one in which the axion is parallel to the magnetic field ($a_\parallel \neq 0$, $a_\perp = 0$) or where there is no magnetic field, and all other configurations.
In the first case, the remaining symmetry is given by axial symmetry around the $x_3$ axis, whereas in the latter case the rotational symmetry is completely broken.
As will become clear below, it turns out that to be able to use a diagonal ansatz for the metric, we have to choose either $a_\perp = 0$ or $a_\parallel = 0$ if a non-zero magnetic field is present.
Also, for notational convenience, whenever no magnetic field is present $a_\perp$ and $a_\parallel$ will both be denoted $a$, since in this case the orientation of the axion is irrelevant.

With the addition of the magnetic field and the axion, the V-QCD action becomes
\[
S = S_g + S_f,
\]
with
\begin{align}
S_g & = M^3N_c^2\int\,d^5x\sqrt{-g}\left(R - \frac{4}{3}\frac{(\partial\lambda)^2}{\lambda^2} + V_g(\lambda) - \frac{1}{2}Z(\lambda)(\partial\chi)^2\right),\label{eq:imc:Sg}\\
S_f & = -x_fM^3N_c^2\int\,d^5xV_f(\lambda,\tau)\label{eq:imc:Sf}\\
& \qquad \times \sqrt{-\det\left[g_{\mu\nu} + w(\lambda)V_{\mu\nu} + \kappa(\lambda)\partial_\mu\tau\partial_\nu\tau\right]},\nonumber
\end{align}
where $V_{\mu\nu}$ is the electromagnetic field strength tensor for the gauge field given by \eqref{eq:imc:gaugefieldansatz}, and the potentials $V_g$, $V_f$, $\kappa$, $w$ and the axion potential $Z$ are given in appendix \ref{sec:potentials:imc}\@.
We will keep using these potentials throughout this chapter.
In the rest of this section, it will be explained how this model can be solved to obtain black hole solutions which are dual to a QGP-like phase.
It is sufficient to focus on solutions containing a black hole, as horizonless solutions must always be obtained from a black hole solution where a limit is taken that lets the horizon shrink to zero size.
This requirement ensures that the IR singularity contained in such a horizonless geometry is of the `good' type, as discussed in more detail in \cite{Gubser:2000nd}.
\subsection{Equations of motion and boundary conditions}\label{sec:imc:eomandbc}
To obtain the equations of motion, we first choose the following ansatz for the metric:\footnote{Note that we use the Minkowski signature here. In principle one has to perform a Wick rotation for the thermodynamical observables, but since we only consider time-independent solutions to the equations of motion this is trivial.}
\begin{equation}
ds^2 = e^{2A(r)}\left(\frac{dr^2}{f(r)} - f(r)\,dt^2 + dx_1^2 + e^{2U(r)}dx_2^2 + e^{2W(r)}\,dx_3^2\right),\label{eq:imc:metricansatz}
\end{equation}
which contains the anisotropy factors $U$ and $W$\@.
These are necessary in order for the Einstein equations to be consistent.\footnote{Note that if $a_\perp = 0$ we have $U = 0$ as well, and if $a_\parallel = B = 0$ we have $W = 0$\@.}
Note that this is also the reason that if a magnetic field is present, either $a_\perp$ or $a_\parallel$ must vanish, because otherwise one of the Einstein equations cannot be satisfied.
This problem could be remedied by choosing instead a more general metric ansatz.
However this would greatly complicate the analysis, while the additional physical insight from allowing for a general angle between $a$ and $B$ would probably be limited.
Note further that the metric contains a blackening factor $f$\@.
This allows for the existence of a black hole horizon at the location where $f = 0$\@.

Before stating the Einstein equations and the equations of motion for the dilaton, tachyon and $\Phi$ field, note that in principle $B$ and $a$ also have equations of motion, so we are not completely free to choose any ansatz for them that we want.
It is important therefore that we check that our ans\"atze \eqref{eq:imc:gaugefieldansatz}, \eqref{eq:imc:axionansatz} are consistent with these equations of motion, and it turns out that this is indeed the case.
Using the metric ansatz \eqref{eq:imc:metricansatz} we can write the Einstein equations as follows:
\begin{align}
0 & = \ddot A + \dot A\left(3\dot A + \dot U + \dot W + \frac{\dot f}{f}\right) - \frac{e^{2A}V_g(\lambda)}{3f}\label{eq:imc:AEoM}\\
& \quad + \frac{e^{2A}x_fV_f(\lambda,\tau)}{6QGf\sqrt{1 + K}}\left((1 + K)Q^2 + G^2((2 + K)Q^2 - 1)\right),\nonumber
\end{align}
\begin{align}
0 & = \ddot U + \dot U\left(3\dot A + \dot U + \dot W + \frac{\dot f}{f}\right) + \frac{e^{2A}Y_2Z(\lambda)}{2f},\nonumber\\
0 & = \ddot W + \dot W\left(3\dot A + \dot U + \dot W + \frac{\dot f}{f}\right)\nonumber\\
& \quad - \frac{e^{2A}Gx_fV_f(\lambda,\tau)\left(Q^2 - 1\right)}{2\sqrt{1 + K}Qf} + \frac{e^{2A}Y_3Z(\lambda)}{2f},\nonumber\\
0 & = \ddot f + (3\dot A + \dot U + \dot W)\dot f + \frac{x_fV_f(\lambda,\tau)e^{2A}G}{Q\sqrt{1 + K}}\left[1 - (1 + K)Q^2\right],\nonumber\\
0 & = \frac{2}{3}\frac{\dot\lambda^2}{\lambda^2} - 6\dot A^2 - 3\dot A\left(\dot U + \dot W\right) - \dot U\dot W - \frac{\dot f}{2f}\left(3\dot A + \dot U + \dot W\right)\label{eq:imc:constraint}\\
& \quad + \frac{e^{2A}V_g(\lambda)}{2f} - \frac{e^{2A}YZ(\lambda)}{4f} - \frac{e^{2A}x_fV_f(\lambda,\tau)Q\sqrt{1 + K}}{2fG},\nonumber
\end{align}
where we use a dot for derivatives with respect to $r$, a convention we will keep throughout this chapter.
We also define
\[
Q = \sqrt{1 + w^2(\lambda)B^2e^{-4A - 2U}}, \qquad G = \sqrt{1 + e^{-2A}f\kappa(\lambda)\dot\tau^2},
\]
\[
K = \frac{\hat n^2}{e^{6A + 2U + 2W}Q^2x_f^2V_f^2(\lambda,\tau)w^2(\lambda)},
\]
\[
Y_2 = a_\perp^2e^{-2A - 2U}, \qquad Y_3 = a_\parallel^2e^{-2A - 2W}, \qquad Y = Y_2 + Y_3,
\]
where $\hat n$ is an integration constant that arises from integrating the $\Phi$ equation of motion.
Using the above definitions, the $\Phi$ equation of motion can now be written as what is essentially a simple integral.
\[
\dot\Phi = -\frac{e^{2A}G\sqrt{K}}{w(\lambda)\sqrt{1+K}}.
\]
Even though $\Phi$ can be integrated out in favor of the integration constant $\hat n$, we still need to integrate its equation of motion, because as we will see below the value of the chemical potential equals the difference of $\Phi$ evaluated at the boundary and at the horizon, necessitating that we evaluate this integral.
Lastly, to complete the system, we also have equations of motion for $\lambda$ and $\tau$:
\begin{align*}
0 & = \ddot\lambda - \frac{\dot\lambda^2}{\lambda} + \dot\lambda\left(3\dot A + \dot U + \dot W + \frac{\dot f}{f}\right) + \frac{3e^{2A}\lambda^2}{8f}\left(\partial_\lambda V_g(\lambda) - \frac{Y}{2}\partial_\lambda Z(\lambda)\right) \\
& \quad + \frac{3e^{2A}x_fV_f(\lambda,\tau)\lambda^2}{8\sqrt{1 + K}f}\left(-GQ\partial_\lambda\log V_f(\lambda,\tau)\right.\\
& \qquad + \frac{Q(1 - G^2)(1 + K)}{2G}\partial_\lambda\log\kappa(\lambda) \\
& \qquad + \left.\frac{G(1 - Q^2 + KQ^2)}{Q}\partial_\lambda\log w(\lambda)\right),
\end{align*}
\begin{align}
0 & = (1 + K)\ddot\tau - \frac{e^{2A}G^2}{f\kappa(\lambda)}\partial_\tau\log V_f(\lambda,\tau)\label{eq:imc:taueomr}\\
& \quad + G^2\dot\tau\left[\left(1 + \frac{(G^2 - 1)(1 + K)}{G^2} + \frac{2}{Q^2}\right)\dot A\right.\nonumber\\
& \qquad + \frac{\dot U}{Q^2} + \dot W + \frac{(1 + G^2)(1 + K)}{2G^2}\frac{\dot f}{f}\nonumber\\
& \qquad + \dot\lambda\left(\partial_\lambda\log V_f(\lambda,\tau) + \frac{(G^2 + 1)(1 + K)}{2G^2}\partial_\lambda\log\kappa(\lambda)\right.\nonumber\\
& \qquad \quad + \left.\left.\left(\frac{Q^2 - 1}{Q^2} - K\right)\partial_\lambda\log w(\lambda)\right)\right].\nonumber
\end{align}
Observe that we have 8 equations of motion for 7 degrees of freedom, so in principle this system could be overconstrained.
One can check however that \eqref{eq:imc:constraint} is a constraint, by taking the derivative of the right hand side, and using the other equations of motion to show that the derivative of \eqref{eq:imc:constraint} is automatically zero.
This implies that if \eqref{eq:imc:constraint} is satisfied for one particular $r$, it is satisfied for any $r$\@.
Therefore this equation will be trivially solved, provided that we choose the proper boundary conditions.

Before stating the boundary conditions, it is useful to write the equations of motion in another form.
The reason why this is useful is that near the boundary, which in $r$-coondinates is located at $r = 0$, $A(r)$ grows like $\log r$\@.
Numerically this poses a problem, as this behavior makes it difficult to satisfy the boundary conditions at the AdS boundary to a good accuracy, and as we shall see below, the observables that we are interested in require the boundary conditions to be precisely met.
The solution for this is to use $A$ as the independent variable instead of $r$\@.
We can do this as long as $A$ is a monotonic function of $r$ throughout the bulk.
Interestingly, this is not always the case.
For this reason, it is prudent to use $r$ as the independent variable near the horizon, and to do a coordinate transformation at some point in the bulk in order to use $A$ as the independent variable near the boundary.

Since $A$ satisfies a second order differential equation in $r$-coordinates, to perform the transformation to $A$-coordinates one has to introduce $q \equiv e^Adr/dA$\@.
Using this definition, one obtains for the Einstein equations:
\begin{align*}
0 & = 4 - \frac{q'}{q} + U' + W' + \frac{f'}{f} - \frac{q^2V_g(\lambda)}{3f} \\
& \quad + \frac{q^2x_fV_f(\lambda,\tau)}{6QGf\sqrt{1 + K}}\left((1 + K)Q^2 + G^2\left((2 + K)Q^2 - 1\right)\right), \\
0 & = U'' + U'\left(4 - \frac{q'}{q} + U' + W' + \frac{f'}{f}\right) + \frac{q^2Y_2Z(\lambda)}{2f}, \\
0 & = W'' + W'\left(4 - \frac{q'}{q} + U' + W' + \frac{f'}{f}\right) \\
& \quad - \frac{q^2Gx_fV_f(\lambda,\tau)\left(Q^2 - 1\right)}{2\sqrt{1 + K}Qf} + \frac{q^2Y_3Z(\lambda)}{2f}, \\
0 & = f'' + \left(4 - \frac{q'}{q} + U' + W'\right)f' + \frac{q^2x_fV_f(\lambda,\tau)G}{Q\sqrt{1 + K}}\left[1 - (1 + K)Q^2\right], \\
0 & = \frac{2}{3}\frac{\lambda'^2}{\lambda^2} - 6 - 3\left(U' + W'\right) - U'W' - \frac{f'}{2f}\left(3 + U' + W'\right) \\
& \quad + \frac{q^2V_g(\lambda)}{2f} - \frac{q^2YZ(\lambda)}{4f} - \frac{q^2x_fV_f(\lambda,\tau)Q\sqrt{1 + K}}{2fG},
\end{align*}
where $G$ is now given by
\[
G = \sqrt{1 + \frac{f\kappa(\lambda)\tau'^2}{q^2}},
\]
and where we introduce the notational convention, which will be used for the rest of this chapter, that a prime denotes a derivative with respect to $A$\@.
For the remaining equations of motion, one obtains:
\[
\Phi' = -\frac{e^AqG\sqrt{K}}{w(\lambda)\sqrt{1 + K}},
\]
\begin{align*}
0 & = \lambda'' - \frac{\lambda'^2}{\lambda} + \lambda'\left(4 - \frac{q'}{q} + U' + W' + \frac{f'}{f}\right) + \frac{3q^2\lambda^2}{8f}\left(\partial_\lambda V_g(\lambda) - \frac{Y}{2}\partial_\lambda Z(\lambda)\right) \\
& \quad + \frac{3q^2x_fV_f(\lambda,\tau)\lambda^2}{8\sqrt{1 + K}f}\left(-GQ\partial_\lambda\log V_f(\lambda,\tau) + \frac{Q(1 - G^2)(1 + K)}{2G}\partial_\lambda\log\kappa(\lambda)\right. \\
& \qquad + \left.\frac{G(1 - Q^2 + KQ^2)}{Q}\partial_\lambda\log w(\lambda)\right),
\end{align*}
\begin{align}
0 & = (1 + K)\tau'' - \frac{q^2G^2}{f\kappa(\lambda)}\partial_\tau\log V_f(\lambda,\tau)\label{eq:imc:taueomA}\\
& \quad + G^2\tau'\left[2 + K - \frac{1 + K}{G^2}\frac{q'}{q} + \frac{2}{Q^2} + \frac{U'}{Q^2} + W' + \frac{(G^2 + 1)(1 + K)}{2G^2}\frac{f'}{f}\right.\nonumber\\
& \qquad + \lambda'\left(\partial_\lambda\log V_f(\lambda,\tau) + \frac{(G^2 + 1)(1 + K)}{2G^2}\partial_\lambda\log\kappa(\lambda)\right.\nonumber\\
& \qquad \quad + \left.\left.\left(\frac{Q^2 - 1}{Q^2} - K\right)\partial_\lambda\log w(\lambda)\right)\right].\nonumber
\end{align}

In addition to satisfying the equations of motion, solutions must also obey boundary conditions.
We will first discuss the boundary conditions that need to be imposed at the horizon, before moving on the the boundary conditions at the boundary.
The first of these boundary conditions is of course that $f_h = 0$, where a subscript $h$ will from now on always denote a quantity evaluated at the horizon.
This first boundary condition simply follows from the definition of a black hole horizon, as this makes an observer stationary at the horizon move on a lightlike trajectory.
We will also assume $A_h = W_h = U_h = 0$ and $\dot f = 1$\@.\footnote{Here, we swapped the usual definition of $r$ by a minus sign so that $r_b > r_h$\@. This means that a few signs are different from other texts, but this is more convenient when computing the solutions.}
While it may seem strange to just assume this, it turns out one can do this without loss of generality.
The reason for this is that the solutions are invariant under symmetries which can be used to rescale the solutions to satisfy the boundary conditions at the boundary.
For this reason it is irrelevant which choice we make for these assumptions, as any different choice will later be absorbed by these symmetries.
This will be discussed in more detail in the next subsection.
The last remaining boundary conditions are consequences of the horizon being just a coordinate singularity.
This can be imposed by requiring that all variables are smooth at the horizon.
Taking equation \eqref{eq:imc:AEoM} as an example, this means that we must require in particular that $\ddot A$ is finite.
Given that $f_h = 0$ by definition, to make sure that $\ddot A$ is finite we must make sure that all terms inversely proportional to $f$ cancel.
This leads to the following condition:
\[
\dot A_h = \frac{V_g(\lambda_h)}{3} - \frac{x_fV_f(\lambda_h,\tau_h)\left((3 + 2K_h)Q_h^2 - 1\right)}{6Q_h\sqrt{1 + K_h}}.
\]
Similar arguments for the other equations of motion yield
\begin{align}
\dot U_h & = -\frac{a_\perp^2Z(\lambda_h)}{2},\nonumber\\
\dot W_h & = \frac{x_fV_f(\lambda_h,\tau_h)\left(Q_h^2 - 1\right)}{2Q_h\sqrt{1 + K_h}} - \frac{a_\parallel^2Z(\lambda_h)}{2},\nonumber\\
\dot\lambda_h & = \frac{3\lambda_h^2}{8}\left(-\partial_\lambda V_g(\lambda_h) + \frac{Y_h\partial_\lambda Z(\lambda_h)}{2}\right.\nonumber\\
& \quad + \frac{x_fV_f(\lambda_h,\tau_h)}{\sqrt{1 + K_h}}\left(Q_h\partial_\lambda\log V_f(\lambda_h,\tau_h)\right.\nonumber\\
& \qquad - \left.\left.\frac{1 - Q_h^2 + K_hQ_h^2}{Q_h}\partial_\lambda\log w(\lambda_h)\right)\right),\nonumber\\
\dot\tau_h & = \frac{\partial_\tau\log V_f(\lambda_h,\tau_h)}{\kappa(\lambda_h)(1 + K_h)}.\label{eq:imc:tauhbc}
\end{align}
Note here that even though \eqref{eq:imc:constraint} does not contribute a non-trivial equation of motion in the bulk, it \emph{is} important to take it into account in the boundary conditions, so in particular the constraint on $\dot\lambda_h$ comes from \eqref{eq:imc:constraint}\@.
The last boundary condition at the horizon that is required, is that $\Phi_h = 0$\@.
This is needed because $\Phi_h$ is a component of the gauge field, and in the Euclidean geometry the gauge field would not be continuous unless $\Phi_h = 0$ \cite{DeWolfe:2010he}\@.
Applying these boundary conditions one is left with the freedom to choose $\lambda_h$ and $\tau_h$\@.
These, together with $B$, $\hat n$, and either $a_\perp$ or $a_\parallel$, determine the entire parameter space of allowed solutions.

For the boundary conditions near the boundary, one needs to consider the asymptotic behavior of the variables near the boundary.
This asymptotic behavior essentially boils down to that to leading order the geometry is AdS, and it turns out that one can analytically expand around this ansatz for small $r$\@.
For the $f$, $U$ and $W$, the result of this procedure is that these variables approach constant values.
Subleading corrections come in at $\mathcal{O}(r^4)$, and for all works described in this thesis approximating them as constants is good enough.
In order to make sure that on the boundary the $t$, $x_i$ coordinates agree with the familiar coordinates for Minkowski space, we require that
\[
f_b = 1, \qquad U_b = 0, \qquad W_b = 0,
\]
where a subscript $b$ will from now on always denote a quantity evaluated at the boundary.

The near-boundary expansion for $A$ and $\lambda$ are a bit more complicated, they are given by \cite{Alho:2012mh}:
\begin{align*}
A(r) & = -\log\frac{r}{\mathcal{L}_\text{UV}} + \frac{4}{9\log(r\lambda)}\\
& \quad + \frac{\frac{1}{162}\left[95 - \frac{64V_2}{V_1^2}\right] + \frac{1}{81}\log\left[-\log(r\Lambda)\right]\left[-23 + \frac{64V_2}{V_1^2}\right]}{\log(r\Lambda)^2} + \mathcal{O}\left(\frac{1}{\log(r\Lambda)^3}\right),
\end{align*}
\begin{equation}
V_1\lambda(r) = -\frac{8}{9\log(r\Lambda)} + \frac{\log\left[-\log(r\Lambda)\right]\left[\frac{46}{81} - \frac{128V_2}{81V_1^2}\right]}{\log(r\Lambda)^2} + \mathcal{O}\left(\frac{1}{\log(r\Lambda)^3}\right),\label{eq:imc:lambdaofr}
\end{equation}
where $\mathcal{L}_\text{UV}$, $V_1$ and $V_2$ are determined by the potentials as discussed in appendix \ref{ch:potentials}, and where $\Lambda$ is an overall energy scale.
It turns out that all quantities one might want to compute scale with $\Lambda$ to some power, so in practice we just put $\Lambda = 1$ and if desired we can rescale later.
For the next subsections, it turns out that it is convenient to combine both of these equations, to write $A$ as a function of $\lambda$\@.
Doing this, one obtains \cite{Alho:2013hsa}:
\begin{equation}
A(\lambda) = \log\left(\mathcal{L}_\text{UV}\Lambda\right) + \frac{1}{b_0\lambda} + \frac{b_1}{b_0^2}\log\left[b_0\lambda(A)\right] + \mathcal{O}(\lambda),\label{eq:imc:Aoflambda}
\end{equation}
where $b_0$ and $b_1$ are given by the potentials and can be found in appendix \ref{ch:potentials}.
The last near-boundary expansion that we will need is that of the tachyon \cite{Alho:2012mh}:
\begin{align}
\frac{1}{\mathcal{L}_\text{UV}}\tau(r) & = m_qr\left[-\log(r\Lambda)\right]^{-\gamma_0/b_0}\left[1 + \mathcal{O}\left(\frac{1}{\log(r\Lambda)}\right)\right]\nonumber\\
& \quad + \langle\bar qq\rangle r^3\left[-\log(r\Lambda)\right]^{\gamma_0/b_0}\left[1 + \mathcal{O}\left(\frac{1}{\log(r\Lambda)}\right)\right],\label{eq:imc:tachyonboundary}
\end{align}
where $m_q$ is the quark mass and $\langle\bar qq\rangle$ is the chiral condensate.
$\gamma_0/b_0$ is given by the potentials, and can be found in appendix \ref{ch:potentials}\@.
In the following we will consider massless quarks, so we will be imposing $m_q = 0$ as the UV boundary condition for the tachyon.
How this can be achieved will be detailed in section \ref{sec:imc:constraints}.

With this, we now have a complete list of all the equations of motion, as well as the boundary conditions that we will need.
In the next section, we will describe a set of symmetries of these equations of motion that are necessary to compute solutions.
After that all the ingredients are set to describe the algorithm for obtaining the geometries, and finally the observables that one is after.
\subsection{Symmetries of the equations of motion}\label{sec:imc:symmetries}
Numerically, one of the simplest and best known methods to solve ODEs such as the ones describing this model is to initialize a solver for a specific value of the independent variable (in this case $r$ or $A$), and then integrate the equation from that value to the entire domain one is interested in.
However, we have boundary conditions on both the horizon and the boundary, and initializing the system of equations at one of the two locations by no means guarantees that the boundary conditions at the other location will also be satisfied.
While there are methods available for solving such problems, it turns out we can use symmetry properties of the equations of motion to mostly overcome this issue.
This will allow us to initialize the system of equations at the horizon, integrate towards the boundary, and rescale the solution such that the boundary conditions at the boundary are also met.

The symmetries of the equations of motion are essentially the diffeomorphism invariance that is left over after choosing the metric ansatz \eqref{eq:imc:metricansatz}\@.
It can easily be verified that the following five transformations leave the equations of motion invariant:
\begin{itemize}
\item Shift of $r$:
\[
r \mapsto r + \delta_r.
\]
\item Shift of $A$:
\[
A \mapsto A + \delta_A, \qquad r \mapsto re^{-\delta_A}, \qquad \hat n \mapsto \hat ne^{3\delta_A}, \qquad \Phi \mapsto \Phi e^{\delta_A},
\]
\[
B \mapsto Be^{2\delta_A}, \qquad a_\perp \mapsto a_\perp e^{\delta_A}, \qquad a_\parallel \mapsto a_\parallel e^{\delta_A}.
\]
\item Shift of $U$:
\[
U \mapsto U + \delta_U, \qquad B \mapsto Be^{\delta_U}, \qquad \hat n \mapsto \hat ne^{\delta_U}, \qquad a_\perp \mapsto a_\perp e^{\delta_U}.
\]
\item Shift of $W$:
\[
W \mapsto W + \delta_W, \qquad \hat n \mapsto \hat ne^{\delta_W}, \qquad a_\parallel \mapsto a_\parallel e^{\delta_W}.
\]
\item Scaling of $f$:
\[
f \mapsto \frac{f}{\delta_f^2}, \qquad q \mapsto \frac{q}{\delta_f}, \qquad r \mapsto \frac{r}{\delta_f}, \qquad \Phi \mapsto \frac{\Phi}{\delta_f}.
\]
\end{itemize}
Together, $\delta_r$, $\delta_A$, $\delta_U$, $\delta_W$ and $\delta_f$ will denoted `symmetry parameters' for the remainder of this chapter.

Note first that these symmetries indeed justify our assumptions that $A_h = W_h = U_h = 0$ and $\dot f = 1$, as these choices can just be absorbed into the various deltas defined above.
Next, observe that after generating a solution that satisfies the horizon boundary conditions, one can choose $\delta_f$, $\delta_U$, $\delta_W$ and $\delta_r$ to ensure that $f_b = 1$, $U_b = 0$, $W_b = 0$, and that $r_b = 0$\@.
In equation \eqref{eq:imc:Aoflambda}, only the left hand side transforms under these transformations, and it is easy to see that by using the appropriate $\delta_A$, one can make sure that \eqref{eq:imc:Aoflambda} is satisfied.
Summarizing, we can satisfy all boundary conditions except that of the tachyon, namely that the quark mass vanishes.
This issue will be addressed in section \ref{sec:imc:constraints}\@.
Lastly, note that there is a price to pay for using these symmetries.
Quantities like $B$, $a$ and $\hat n$ enter in the transformations.
This implies that while we are guaranteed to get a solution that satisfies the correct boundary conditions on both sides, we have no direct control over for instance the value of the magnetic field we want to compute the solution of.
This need not necessarily be a bad thing though, because we are guaranteed that together with $\lambda_h$, all possible values of $B$, $\hat n$ and $a$ before any rescaling happens span the space of all possible solutions.
If it is feasible to produce solutions which explore this entire parameter space, then we will be guaranteed to find all possible solutions for all possible rescaled values of $B$, $\hat n$ and $a$ as well.
In some of the setups that will be described below this is indeed the case, but in other cases it is necessary to fine tune the unrescaled values to produce the desired rescaled values.
The procedure for doing so will also be described in section \ref{sec:imc:constraints}.
\subsection{Computing observables}\label{sec:imc:observables}
Now that the equations of motion, their symmetries, and the boundary conditions are established, we can describe how to obtain the solutions, and how to extract useful information from these solutions.
As it turns out, many quantities of interest can be expressed purely in terms of quantities defined at the horizon, and the symmetry parameters described in the previous section.
The reason for this is that these quantities do not require the solving of any additional differential equations or integrals on top of the ones already mentioned in section \ref{sec:imc:eomandbc}\@.\footnote{Note that for example the magnetization does require solving an integral, but this only has to be done once, so it is done at the same time as solving the equations of motion for the metric, dilaton and tachyon.}
The quantities for which this is the case will from now on be called `background' observables, and their computation will be discussed first.
There are also observables which \emph{do} require the solving of additional equations of motion.
In principle, one could follow the same computation scheme as for the background observables, but as this would require solving the same background equations of motion multiple times with the same boundary conditions, it is more efficient to solve the background once, and then solve the additional equations afterward.
The computations of these non-background observables will be discussed towards the end of this subsection.
\subsubsection{Background observables}
In this subsection, we will focus on the background observables, where for convenience we will introduce the notation that quantities with a tilde denote quantities before the symmetries are used to impose the boundary conditions at the AdS boundary, and quantities without a tilde do correspond to the quantities with the proper boundary conditions imposed.
Before moving on to the computation of these obervables, observe that the property of only needing symmetries and horizon data is extremely useful.
Because the required symmetry transformations only depend on the non-rescaled solution that one has computed at the boundary, there is actually no need to retain the information about the bulk geometry at all, and therefore it can be immediately discarded.
This does away with the need for reading/writing a lot of data to memory, making the computation faster.
Of course, if one is interested in one of the quantities that do not have this property, this shortcut cannot be taken, but since the quark-antiquark potential and the entanglement entropy will only be computed for zero temperature solutions, it turns out that for a majority of the computations in this chapter, the shortcut is possible.

From the above discussion, the strategy for computing observables becomes clear.
Choose $\lambda_h$, $\tau_h$, $\tilde n$, $\tilde B$ and $\tilde a$, and then initialize according to the horizon boundary conditions described above.
Subsequently integrate the equations of motion, switching from $r$- to $A$-coordinates once close enough to the boundary, and then integrate further up to some large enough value of $A$\@.
Then extract the symmetry parameters $\delta_A$, $\delta_U$, $\delta_W$ and $\delta_f$, and use them to evaluate the desired observables.\footnote{It turns out that $\delta_r$ is not needed for any of the observables that will be listed below.}
The next few paragraphs will describe how to compute the following observables in this way:
\begin{itemize}
\item Temperature $T$,
\item Entropy density $s$,
\item Baryon chemical potential $\mu$,
\item Magnetic field $B$,
\item Anisotropies $a_\parallel$ and $a_\perp$,
\item Baryon number density $n$,
\item Magnetization $M$,
\item `Anisotropization' $M_a$\@. This is the analog of magnetization for the anisotropy;
\item Quark mass $m_q$,
\item Chiral condensate $\langle\bar qq\rangle$.
\end{itemize}
The first 5 of these are straightforward, whereas the latter 5 require a bit more work.
We will now go through each of these observables in order, after which there will be a discussion on how to accurately determine the required symmetry parameters.

The temperature is given by the Hawking temperature associated to the horizon.
Using the metric ansatz \ref{eq:imc:metricansatz}, this can be expressed as
\[
\frac{T}{\Lambda} = \frac{\dot f_h}{4\pi\Lambda} = \frac{e^{\delta_A}}{4\pi\delta_f\Lambda}\frac{\mathrm{d}\tilde f(\tilde r)}{\mathrm{d}\tilde r} = \frac{e^{\delta_A}}{4\pi\delta_f\Lambda},
\]
where the $\Lambda$-dependence enters because of $\delta_A$, as we will see below the discussion of the observables that we can only extract $\delta_A - \log\Lambda$ from the model.
Moving on to the entropy density, note that the entropy is given by the area of the black brane.
Once again using the metric ansatz \ref{eq:imc:metricansatz} and dividing out the overall volume factor, one obtains the entropy density
\begin{align}
\frac{s}{\Lambda^3} & = \frac{\exp(3A_h + U_h + W_h)}{\Lambda^3} = \frac{\exp(3\tilde A_h + 3\delta_A + \tilde U_h + \delta_U + \tilde W_h + \delta_W)}{\Lambda^3},\nonumber\\
& = \frac{\exp(3\delta_A + \delta_U + \delta_W)}{\Lambda^3},\label{eq:imc:entropydensity}
\end{align}
where one may note that this result is different from literature by a factor 4\@.
This amounts to choosing $M^3N_c^2 = 1/4\pi$ in the action (\ref{eq:imc:Sg}, \ref{eq:imc:Sf}), and is done for notational convenience.
If desired, desired values of $M^3N_c^2$ can be reinstated by multiplying $s$, $n$, $M$ and $M_a$ by appropriate factors.
The chemical potential is given by the value of $\Phi$ at the boundary.
As $\Phi$ can be shifted by a constant due to gauge symmetry, naively one would say that this is ill-defined.
However, as discussed before in section \ref{sec:imc:eomandbc}, $\Phi_h = 0$ to ensure continuity of the gauge field at the horizon.
Therefore we obtain:
\[
\frac{\mu}{\Lambda} = \frac{\Phi_b}{\Lambda} = \frac{e^{\delta_A}\tilde\Phi_b}{\delta_f\Lambda}.
\]
The magnetic field and the anisotropy can be obtained as follows:
\[
\frac{B}{\Lambda^2} = \tilde B\frac{e^{2\delta_A + \delta_U}}{\Lambda^2}, \qquad \frac{a_\perp}{\Lambda} = \tilde a_\perp\frac{e^{\delta_A + \delta_U}}{\Lambda}, \qquad \frac{a_\parallel}{\Lambda} = \tilde a_\parallel\frac{e^{\delta_A + \delta_W}}{\Lambda}.
\]

Next, we move on to compute the number density, magnetization and anisotropization.
These quantities have in common that they are computed by taking derivatives of the on-shell action.
Recall that
\begin{equation}
S = -\frac{\Omega}{T},\label{eq:imc:Omegadef}
\end{equation}
where $\Omega$ is the grand potential.
By definition, the number density is given by
\begin{equation}
nV = -\left.\frac{d\Omega}{d\mu}\right|_T,\label{eq:imc:ndef}
\end{equation}
which means that we can express $n$ in terms of the action.
To make this more concrete, consider the variation of the action with respect to $\Phi$:
\begin{equation}
\delta S = \frac{V}{T}\int_{r_b}^{r_h}dr\frac{\delta\mathcal{L}}{\delta\dot\Phi}\delta\dot\Phi = \frac{V}{T}\int_{r_b}^{r_h}dr\frac{d}{dr}\left(\frac{\delta\mathcal{L}}{\delta\dot\Phi}\delta\Phi\right),\label{eq:imc:varactionPhi}
\end{equation}
where the last equality holds because the action is on-shell, and where the $V/T$ factor comes from integration over $t$ and $x_i$\@.\footnote{Note that the action only depends on $\dot\Phi$\@. This is ultimately the reason why we could immediately integrate the equation of motion for $\Phi$.}
As it turns out, the integration constant that we defined as $\hat n$ obeys
\[
\hat n = -\frac{1}{4\pi}\frac{\delta\mathcal{L}}{\delta\dot\Phi},
\]
allowing us to write
\[
\delta S = \frac{\hat nV\delta\Phi_b}{4\pi T},
\]
where we keep $\delta\Phi_h = 0$, and recognize $\delta\Phi_b$ as being an infinitesimal change in baryon chemical potential.
Combining the last equation with \eqref{eq:imc:Omegadef} and \eqref{eq:imc:ndef}, one obtains that $n = \hat n/4\pi$\@.
Applying the appropriate rescalings, one then obtains
\[
n = \frac{\hat n}{4\pi} = \frac{\tilde n\exp(3\delta_A + \delta_U + \delta_W)}{4\pi} = \frac{\tilde ns}{4\pi},
\]
where in the last step we used \eqref{eq:imc:entropydensity}\@.

For the magnetization, defined by $MV = -d\Omega/dB$, one can perform a similar computation to obtain the following integral:
\begin{align*}
\frac{M}{\Lambda^2} & = \frac{B}{4\pi\Lambda^2}\int_{r_b}^{r_h}dr\frac{e^{A - U + W}x_fV_f(\lambda,\tau)w(\lambda)^2G}{Q\sqrt{K + 1}}, \\
& = \frac{e^{2\delta_A + \delta_W}}{4\pi\delta_f\Lambda^2}\tilde B\int_{\tilde r_h}^{\tilde r_b}d\tilde r\frac{e^{\tilde A - \tilde U + \tilde W}x_fV_f(\lambda,\tau)w(\lambda)^2G}{Q\sqrt{K + 1}}.
\end{align*}
In this case there is no way to analytically evaluate the integral, so it has to be integrated along with the other equations of motion.
Note however that $M$ factorizes into a factor containing all the rescalings, and an integral that depends only on unrescaled variables.
This allows us to still perform the integral first, without the need to retain any intermediate quantities.
One further point of interest is that $M$ is a divergent quantity, as the integrand diverges near the boundary.
This means that in practice we can only compute differences in $M$ between solutions which have the same $B$, by cutting off the integral at some non-zero value of $A$, and comparing the differences.\footnote{Note here that it is important to cut off the integral at some prescribed value of $A$, not $\tilde A$\@.}
Also, defining the anisotropization $M_aV = -d\Omega/da_\parallel$, we can obtain in a similar fashion that
\begin{align*}
\frac{M_{a_\perp}}{\Lambda^3} & = \frac{a_\perp}{4\pi\Lambda^3}\int_{r_b}^{r_h}dr\,e^{3A - U + W}Z(\lambda) = \frac{e^{3\delta_A + \delta_W}a_\perp}{4\pi\delta_f\Lambda^3}\int_{\tilde r_b}^{\tilde r_h}d\tilde r\,e^{3\tilde A - \tilde U + \tilde W}Z(\lambda),\\
\frac{M_{a_\parallel}}{\Lambda^3} & = \frac{a_\parallel}{4\pi\Lambda^3}\int_{r_b}^{r_h}dr\,e^{3A - W}Z(\lambda) = \frac{e^{3\delta_A}a_\parallel}{4\pi\delta_f\Lambda^3}\int_{\tilde r_b}^{\tilde r_h}d\tilde r\,e^{3\tilde A - \tilde W}Z(\lambda),
\end{align*}
where we note that $U = 0$ in the second case, whereas in the first case both a non-trivial $U$ and $W$ can occur if both $a_\perp$ and $B$ are non-vanishing.

The last quantities we would like to extract are the quark mass and the chiral condensate.
Both of these can be extracted from \eqref{eq:imc:tachyonboundary}\@.
Given that the term containing $m_q$ becomes dominant near the boundary, it can be extracted by simply ignoring the term containing $\langle\bar qq\rangle$, and rearranging:
\begin{equation}
\frac{m_q}{\Lambda} = \lim_{A \rightarrow \infty}\tau\mathcal{L}_\text{UV}^{-2}e^{\tilde A + \delta_A}\left(\tilde A + \delta_A - \log(\mathcal{L}_\text{UV}\Lambda)\right)^{\gamma_0/b_0}.\label{eq:imc:mqextract}
\end{equation}
The procedure for extracting the chiral condensate is slightly more involved.
The problem is of course that the $\langle\bar qq\rangle$ term gets smaller relative to the $m_q$ term as one gets closer to the boundary.
In \cite{Gursoy:2016ofp}, we developed a relatively simple way to solve this problem.
The trick is to divide \eqref{eq:imc:tachyonboundary} by $r[-\log(r\Lambda)]^{-\gamma_0/b_0}$, so that the $m_q$-term is to leading order constant.
Subsequently taking a derivative on both sides and rearranging, one obtains the following expression:
\begin{align}
\frac{\langle\bar qq\rangle}{\Lambda^3} & = \lim_{A\rightarrow\infty}\frac{\exp(2A)\left(A - \log(\mathcal{L}_\text{UV}\Lambda)\right)^{-\gamma_0/b_0}}{2\mathcal{L}_\text{UV}^3\left(\frac{\gamma_0}{b_0} - A + \log(\mathcal{L}_\text{UV}\Lambda)\right)}\label{eq:imc:qqbarextract}\\
& \quad \times \left[\exp(A)\left(\frac{\gamma_0}{b_0} + A - \log(\mathcal{L}_\text{UV}\Lambda)\right)\tau + \mathcal{L}_\text{UV}(-A + \log(\mathcal{L}_\text{UV}\Lambda))\tau'\right].\nonumber
\end{align}
This works well enough to extract the chiral condensate reliably, but one has to be careful not to do the extraction \emph{too} close to the boundary.
This is because \eqref{eq:imc:qqbarextract} essentially computes the difference of two numbers, where that difference tends to zero as one approaches the boundary.
For this reason, one eventually runs into accuracy issues.

To conclude this subsection, we will discuss how to obtain the symmetry parameters required for the computation of the observables discussed above.
As an example, consider $\delta_f$\@.
The blackening factor $f$ does not change under any of the symmetry transformations listed in section \ref{sec:imc:symmetries} except for the one parameterized by $\delta_f$\@.
This implies that after having computed an unrescaled solution by shooting from the horizon, one can extract $\delta_f$ solely by using information from the blackening factor.
Since $f$ approaches a constant value with subleading corrections only entering at $\mathcal{O}(r^4)$ as one approaches the boundary, one can simply divide the unrescaled $\tilde f$ and rescale such that it becomes zero:
\[
\delta_f = \frac{1}{\sqrt{\tilde f}}.
\]
As the anisotropy factors $U$ and $W$ approach constants in the same way $f$ does, $\delta_U$ and $\delta_W$ can be extracted in an analogous way by demanding that $U_b$ and $W_b$ both vanish to obtain
\[
\delta_U = \tilde U, \qquad \delta_W = \tilde W.
\]

The remaining symmetry parameter that is important for computing observables, $\delta_A$, is slightly more complicated to obtain.
The key is to use \ref{eq:imc:Aoflambda}, which, by substituting $A = \tilde A + \delta_A$, can be written in the following form:
\begin{equation}
\delta_A - \log\Lambda = -\tilde A(\lambda) + \log\mathcal{L}_\text{UV} + \frac{1}{b_0\lambda} + \frac{b_1}{b_0^2}\log\left[b_0\lambda(A)\right] + \mathcal{O}(\lambda).\label{eq:imc:deltaAextraction}
\end{equation}
Using this equation, we can find $\delta_A - \log\Lambda$ as long as we evaluate the right hand side close enough to the boundary.
Note that this is the reason why in the observables computed above every power of $\exp\delta_A$ came with a factor of $1/\Lambda$\@.
We can now either choose to compute observables in units of $\Lambda$, which is something we will mostly do in this chapter, or we can try to use a physically reasonable value for $\Lambda$, which is something we will do in chapter \ref{ch:holographicns}\@.

We will conclude this section with an improvement to the computation of $\delta_A$ as described above, developed in \cite{Gursoy:2016ofp}.
Note that in \eqref{eq:imc:deltaAextraction} the subleading corrections go like $\lambda$\@.
Also recall that $\lambda$ goes to zero on the boundary by equation \eqref{eq:imc:lambdaofr}\@.
Taking these two facts together, one can see that if we linearly extrapolate $\delta_A$ as computed by \eqref{eq:imc:deltaAextraction} to $\lambda = 0$, then the subleading linear correction should cancel.
We tested that this works in practice, against the code used in \cite{Alho:2013hsa}\@.
We compared specifically that with the extrapolation, where one extrapolates linearly using $A = 18$ and $A = 20$ to compute two values of $\delta_A$, one gets a $0.7\%$ deviation from a solution computed up to $A = 400$ without extrapolation.
Furthermore, the deviation is an overall deviation which is always the same regardless of input parameters like $\lambda_h$\@.
This means that such a deviation is acceptable, as it effectively amounts to a slight redefinition of $\Lambda$\@.
The linear extrapolation procedure presents an immediate advantage in computation time because it is computationally a lot cheaper.
However, as it turns out, for \eqref{eq:imc:qqbarextract}, the accuracy issues one always eventually runs into as one gets closer to the boundary imply that for this observable one \emph{has} to do the extraction at a value of $A$ for which subleading corrections are sizable.
For this reason, the extrapolation procedure is also done for \eqref{eq:imc:mqextract} and \eqref{eq:imc:qqbarextract}, as both of these have subleading behavior which is linear in $\lambda$\@.
\subsubsection{Non-background observables}
We will now continue with what we defined to be `non-background' observables.
These observables will all be computed on top of a background metric, dilaton field and tachyon, where we can assume that the backgrounds have been properly rescaled using the symmetries, so that all the boundary conditions are satisfied.
We will discuss the following observables:
\begin{itemize}
\item Helicity 2 glueballs,
\item Quark-antiquark potential,
\item Entanglement entropy.
\end{itemize}

Let us start by discussing the helicity 2 glueballs.
To compute the spectral density associated to these glueballs, one needs to examine the behavior of the $\delta g_{12}$ metric perturbation:
\[
\delta g_{12} = \delta g_{21} = e^{2 A(r) }e^{iq_\mu x^\mu}h(r),
\]
where by an appropriate coordinate transformation we may assume $q^\mu = (\omega,0,0,q)$, and where the sum in the plane wave term goes over the time and space indices.
The linearized Einstein's equations then become
\[
\ddot h(r) +3\dot A(r)\dot h(r) + \dot W(r)\dot h(r) - q^2 e^{-2 W(r)} h(r) +\omega ^2 h(r) = 0.
\]
These equations can also be put in the Schr\"odinger form, by defining $h(r) = e^{-3A(r)/2 - W(r)/2}\psi(r)$:
\[
-\psi''(r) + V_s(r) \psi(r) = (\omega ^2- q^2 e^{-2 W(r)}  )\psi(r),
\]
\begin{equation}
V_S(r) = \frac{1}{2} \left(3 A''(r)+W''(r)\right)+\frac{1}{4}\left(3 A'(r)+W'(r)\right)^2.\label{eq:imc:schrodinger}
\end{equation}
Here we can easily determine whether the spectrum is discrete by looking at the asymptotics of $V_s$\@.
In the UV, $V_S$ diverges as $A \sim -\log r$, and if in the IR $V_S$ diverges as well, the spectrum will be discrete, otherwise it will be continuous.

To extract the spectral density, we recall that it can be obtained from the correlator of the energy-momentum tensort $T_{12}$, which we can obtain from $h(r)$ by the following near-boundary expansion:
\[
h(r) = 1 + \mathcal{O}(r^2) + G(\omega,q) r^4\left[1+\mathcal{O}\left(\frac{1}{\log r}\right)\right].
\]
Here $h$ has to satisfy infalling boundary conditions in the IR $\psi(r) \propto e^{i\omega r}$\@.
We then want to extract the imaginary part of $G(\omega,q)$, which is equal to the spectral density.
In practice, numerically, it is easier to define $h(r) = e^{k(r)}$, and solve the corresponding equations of motion for $k$\@.
This simplifies the IR boundary conditions to $k(r) = i\omega r$, and more importantly it removes the oscillatory behavior that $h$ will usually have.
Furthermore, the $\mathcal{O}(r^2)$ terms in the near-boundary expansion drop out as they are real, and allow one to write:
\[
\im\left[G(\omega,q)\right] = \frac{\im(k')}{4r^3},
\]
where we note that this expression has the added advantage of being insensitive to $h$ having the correct normalization to 1 at the boundary.
Note lastly that because the equations of motion for $h$ are linear, we can easily demand that the constant term in the near-boundary expansion equals 1\@.

As mentioned in the introduction, the quark-antiquark potential indicates whether it is possible to pull a quark-antiquark pair apart.
In holography, this quantity is computed by evaluating the on-shell Nambu-Goto action of a static string hanging in the bulk from two points on the boundary \cite{Bak:2007fk}\@.
In principle, this is a divergent quantity.
However, the difference of such a solution with a solution of two strings extending infinitely deep into the IR with non-holographic coordinates fixed is finite.
By varying the distance between the endpoints, one can then examine the free energy, and hence the potential, associated to each quark-antiquark distance.
Throughout this discussion, it is important that we use the string frame metric for the computation, as this is what the string `feels'\@.
This amounts ro replacing the $A$ metric factor by $A_S = A + \frac{2}{3}\log\lambda$\@.

As it turns out, it is not necessary to solve equations of motion for the Nambu-Goto action, as it has been worked out that the on-shell action satisfies \cite{Gursoy:2007er,Kinar:1998vq}\@:
\begin{align*}
\frac{V_\parallel(r_F)}{T_f} & = e^{2A_S(r_F)+W(r_F)}L(r_F)\\
& \quad + 2\int_0^{r_F}\frac{dr}{e^{W(r)}}\sqrt{e^{4A_S(r)+2W(r)} - e^{4A_S(r_F)+2W(r_F)}} - 2 \int_0^{\infty}dr\, e^{2 A_S(r)},\\
L(r_F) & = 2\int_0^{r_F}\frac{dr}{e^{W(r)}}\frac{1}{\sqrt{e^{4A_S(r)+2W(r)-4A_S(r_F)-2W(r_F)} - 1}},
\end{align*}
where $T_f$ is the string tension, and $r_F$ is the turning point of the string in the bulk.
This allows us to compute $V_\parallel(L)$ as a parametric function by choosing a value for $r_F$ and integrating.
Also note that while the integrals above are given for $V_\parallel$, one can obtain the integral for $V_\perp$ by setting $W = 0$\@.

The last observable we will be examining is the entanglement entropy of the following two regions:
\begin{itemize}
\item A region, $A$, defined by $0 < x_3 < L$, where we note that $x_3$ is in the parallel to both $a_\parallel$ and $B$\@.
We denote the entanglement entropy of this region by $S_{E,\parallel}$\@.
\item A region, $B$, defined by $0 < x_1 < L$, where we note that $x_1$ is perpendicular to any source of anisotropy, be it $a$ or $B$\@.
We denote the entanglement entropy of this region by $S_{E,\perp}$\@.
\end{itemize}
The entanglement entropy can be computed in holography by finding a minimal surface with its endpoints fixed to the boundary of region $A$ or $B$, respectively.
Such a surface is called a Ryu-Takayanagi (RT) surface \cite{Ryu:2006bv}\@.
Given the symmetries of both regions considered, this results in a similar computation to that for the quark-antiquark potential, with the difference that for the entanglement entropy we need to minimize the surface in the Einstein frame instead of the string frame.
Using the same techniques as for the quark-antiquark potential, one can find both $S_{E,\parallel}$ and $L$ as functions of the turning point in the bulk $r_F$\@:\footnote{This regularization defines the entanglement entropy of the entire boundary as 0.}
\begin{align*}
\frac{S_{E,\parallel}(r_F)}{4\pi A_VM^3N_c^2} & = e^{3A(r_F)+W(r_F)}L(r_F)\\
& \quad + 2\int_0^{r_F}\frac{dr}{e^{W(r)}}\sqrt{e^{6A(r)+2W(r)} - e^{6A(r_F)+2W(r_F)}}\\
& \quad - 2\int_0^{\infty}dr \,e^{3A(r)}, \\
L(r_F) & = 2\int_0^{r_F}\frac{dr}{e^{W(r)}}\frac{1}{\sqrt{e^{6A(r)+2W(r)-6A(r_F)-2W(r_F)} - 1}},
\end{align*}
with $A_V$ an infinite factor arising becaus the region is spatially infinite in two dimension.
Analogously, one obtains for $S_{E,\perp}$ and $L$:
\begin{align*}
\frac{S_{E,\perp}(r_F)}{4\pi A_VM^3N_c^2} & = e^{3A(r_F)+W(r_F)}L(r_F)\\
& \quad + 2\int_0^{r_F}dr\,\sqrt{e^{6A(r)+2W(r)} - e^{6A(r_F)+2W(r_F)}}\\
& \quad - 2\int_0^{\infty}dr \,e^{3A(r)+W(r)}, \\
L(r_F) & = 2\int_0^{r_F}dr\,\frac{1}{\sqrt{e^{6A(r)+2W(r)-6A(r_F)-2W(r_F)} - 1}}.
\end{align*}
\subsection{Satisfying constraints}\label{sec:imc:constraints}
With the above discussion, it is now possible to obtain a solution and extract observables given the inputs $\lambda_h$, $\tau_h$, $\tilde B$, $\tilde n$ and either $\tilde a_\perp$ or $\tilde a_\parallel$\@, which together parameterize the available space of solutions.
We will now move up a level of abstraction, treating the entire discussion above as a function which takes in these inputs and outputs the observables listed in section \ref{sec:imc:observables}\@.
Here we must also emphasize that for some of these inputs this function will fail to return an answer, because it is not guaranteed that every set of inputs correspond to a valid solution.
All of these inputs are given at the horizon, and there is no guarantee that the geometry near the horizon smoothly connects to a near-boundary geometry.
For example, for some of these solutions $A$ might grow to negative infinity instead of positive infinity, or some of the other variables might hit poles at some point in the bulk.

One issue that arises now is that while we have an efficient way of evaluating this function, we would rather have control over some of the outputs.
In particular, even one of the boundary conditions, namely $m_q = 0$, is not satisfied automatically.
Also, it is sometimes needed to for example look at solutions at fixed $B$ or $a$, which requires to fine-tune the inputs to reproduce the desired outputs.
In this subsection, we will explain the solutions to both of these problems, starting with fixing $m_q = 0$\@.

Requiring that $m_q = 0$ can be done in two ways, namely setting $\tau_h = 0$, and finding a non-zero $\tau_h$ for which $m_q = 0$\@.
While this seems like a trivial distinction, making this distinction immediately leads to finding out whether the chiral condensate is zero or not, which is one of the order parameters we're interested in.
In the case where the chiral condensate is non-zero, it can immediately be concluded that such a solution spontaneously breaks chiral symmetry since $m_q = 0$, which is one of the main features of V-QCD\@.
First, note that setting $\tau_h = 0$ immediately implies that $\dot\tau_h = 0$ by \eqref{eq:imc:tauhbc}, which by the equation of motion for $\tau$, \eqref{eq:imc:taueomA}, implies that $\tau = 0$ everywhere.
Then from the near-boundary expansion \eqref{eq:imc:tachyonboundary}, it immediately follows that $m_q$ and $\langle\bar qq\rangle$ both vanish.

Finding a non-zero $\tau_h$ for which $m_q = 0$, keeping the other inputs fixed, is a more complicated procedure.
The algorithm for doing so must determine where a solution with $m_q = 0$ exists at all.
If it does, it must choose the largest such value of $\tau_h$, and then accurately determine the outputs for that value of $\tau_h$\@.
The reason for choosing the largest such value is that one needs to find, given $T$, $\mu$, $B$ and $a$, the value with the smallest grand potential.
It turns out that the value for $\tau_h$ which corresponds to this most stable solution is the one with the largest $\tau_h$ \cite{Jarvinen:2011qe,Alho:2012mh}\@.
The procedure to find this value is done using the steps described below, where any failure in any of the steps will lead to the conclusion that no solution with $m_q = 0$ exists for a non-zero $\tau_h$\@.\footnote{Note that this algorithm, and also the algorithm for constraining other output values like $B$ and $a$ to be described later, are obtained empirically. There are cases in which it fails to find legitimate solutions or when the solution it finds does not have the largest $\tau_h$ possible, but occurances of this are rare, and when they do occur the result stands out as being wrong. Therefore the algorithm can safely be used.}
This algorithm is mostly the same as in \cite{Alho:2013hsa}, except for the last step, which is modified in a non-trivial way.
Also, several minor things like the amount of iterations before an algorithm terminates have been changed.
These changes turn out to have little effect on the accuracy of the algorithm, but have a large effect on execution time, as every time we generate a bulk geometry and extract the observables takes a few milliseconds.
For this reason iterating over this process is very computationally expensive, and lowering the amount of iterations saves a lot of computation time.
\begin{enumerate}
\item The first step involves finding any $\tau_h$ for which a bulk solution exists, without any requirements on $m_q$\@.
This is done by starting with $\tau_h = 1$, and doubling $\tau_h$ until a solution is found, which we call $\tau_{h,\text{exist}}$\@.
If no solution is found after 30 iterations, and the procedure terminates.
\item We next need to find a solution $\tau_{h,\text{min}}$ for which $m_q < 0$\@.
Having found at least one solution $\tau_{h,\text{exist}}$, we check the sign of $m_q(\tau_{h,\text{exist}})$\@.
If it is negative, this step is complete.
If it is positive, we bisect the interval $(0, \tau_{h,\text{exist}})$\@.
If $m_q < 0$ for this solution, we found $\tau_{h,\text{min}}$ and continue to the next step.
If $m_q > 0$, we continue searching in the interval $(0, \tau_{h,\text{exist}}/2)$\@.
If for this value of $\tau_h$ there exists no solution, we continue searching in the interval $(\tau_{h,\text{exist}}/2, \tau_{h,\text{exist}})$\@.
We keep bisecting in this way until we find $\tau_{h,\text{min}}$, or, if we haven't found $\tau_{h,\text{min}}$ after 30 iterations, we terminate the procedure, and conclude that no solution with $m_q = 0$ exists.
\item After finding a value of $\tau_h$ for which $m_q < 0$, we need to find a value $\tau_{h,\text{max}}$ for which $m_q > 0$\@.
Doing so guarantees, by continuity, that we will at least find \emph{one} solution with $m_q = 0$, so after this step, the algorithm can no longer terminate with a failure to find a solution.
If $m_q(\tau_{h,\text{exist}}) > 0$, this step is trivially completed, and otherwise we keep doubling $\tau_{h,\text{exist}}$ until a solution with positive $m_q$ is found.
As in the previous steps, if after 30 iterations no acceptable solution is found, the algorithm terminates with a failure.
\item The next step is meant to find values of $\tau_h$ bracketing the largest value of $\tau_h$ for which $m_q = 0$\@.
This means that we have to search for zeroes between the zero that we already found and some large value of $\tau_h$\@.
It is known that for large values of $\tau_h$ the quark mass increases asymptotically \cite{Alho:2012mh}, so what we do is to increase $\tau_h$ stepwise until we see this asymptotic behavior.
These steps need to be small enough so as to not miss zeroes, but must also not be too small, as this would slow down the algorithm considerably.
A good compromise is to take the current highest $\tau_h$ value, named $\tau_{h,\text{last}}$, and compute $m_q$ at $\tau_{h,\text{last}}$, $\tau_{h,\text{last}} + \Delta\tau_h$ and $\tau_{h,\text{last}} + 2\Delta\tau_h$, for an appropriate $\Delta\tau_h$\@.
If $m_q$ changes sign between any of these three values, we found a zero, and the $\tau_h$-value larger than where the zero is located becomes the new $\tau_{h,\text{last}}$\@.
If $m_q$ does not change sign, we take $\tau_{h,\text{last}} + 2\Delta\tau_h$ to become the new $\tau_{h,\text{last}}$\@.

To completely describe this step, we need two more things, namely what to choose for $\Delta\tau_h$, and when to terminate the search.
At the start of this step, $\Delta\tau_h$ is initialized as
\[
\Delta\tau_h = \tau_{h,\text{min}}\left[\left(\frac{\tau_{h,\text{max}}}{\tau_{h,\text{min}}}\right)^{10^{-4}} - 1\right].
\]
Also, $\Delta\tau_h$ is updated every time we find a new $\tau_{h,\text{last}}$\@.
If we found a new $\tau_{h,\text{last}}$ because we found a new zero, then we set $\Delta\tau_h$ to the difference between the last found zero and $\tau_{h,\text{min}}$, where we divide by 5 so as not to miss any zeroes.
If we found a new $\tau_{h,\text{last}}$ without a new zero, i.e.~because $m_q(\tau_{h,\text{last}} + n\Delta\tau_h)$ has the same sign for $n = 0, 1, 2$, then we construct a parabola through these three points, and look at its zeroes.
If it has zeroes, $\Delta\tau_h$ is set to be the difference between the two zeroes, if there are no zeroes $\Delta\tau_h$ is doubled from its last value.

Lastly, the search is terminated if the following three requirements are met at the same time:
\begin{itemize}
\item $m_q$ has not decreased in the last iteration.
\item $m_q > 0$.
\item $m_q$ is more than 100 times larger than the largest $m_q$ found in all previous iterations.
\end{itemize}
These three conditions correspond to the ones described in \cite{Alho:2013hsa}\@.
Sometimes, however, this never terminates, so for these cases I have added the condition that the search is also terminated if $\tau_h > 10^5$, which is large enough that the search can be safely terminated.
\item Now that we have found values of $\tau_h$ which bracket the largest zero of $m_q$, it is possible to iteratively find the zero.
For this, \cite{Alho:2013hsa} uses Brent's method, which combines the best properties of Newton's method, which doesn't always converge, but converges fast if it does, and the bisection method, which is guaranteed to find a zero, but does so more slowly.
In \cite{Gursoy:2017wzz}, we improved this algorithm slightly.
Instead of using Brent's method to reach some small enough value of $m_q$, we use Brent's method to reach a much larger interval.
We then compute at 10 intermediate values within this interval, and perform a least square fit to a linear function, of which we can then find the zero analytically.
This enables us to not only obtain the desired zero of $m_q$, but also obtain an error estimate for quantities like $T$ and $s$\@.
We shall see in the rest of this section that this is crucial to obtain quantities like the speed of sound numerically.
\end{enumerate}

With the above discussion, one can obtain a solution satisfying the boundary condition $m_q = 0$, either by setting $\tau_h = 0$ to find a chirally symmetric solution, or by following the algorithm described above to obtain a chirally broken solution.
To conclude this section, let us go yet one level of abstraction higher, and view the result of the discussion above as two functions (one for chirally symmetric and another for chirally broken solutions) which take in $\lambda_h$, $\tilde n$, $\tilde B$ and either $a_\perp$ or $a_\parallel$, and output the desired observables for a solution which satisfies $m_q = 0$\@.
The discussion below will explain how we can iteratively call these functions to fix $B$ or $a$ to some specified value by fine-tuning $\tilde B$ and $\tilde a$, respectively.
As the algorithm works the same for both chirally symmetric and chirally broken solutions, I will not distinguish between the two.

The algorithm is similar in setup, albeit simpler, than the one described above.
One starts by finding two values of the input parameters which bound the desired solution.
This is simpler than the procedure for finding $\tau_h$ such that $m_q = 0$, as generically $B(\tilde B)$ and $a(\tilde a)$ are single-valued functions.
For $B$, this is done by starting from $\tilde B = 0$, then $\tilde B = 1$, and then doubling until two values bracketing the desired solution are found.
For $a$, one starts from $\tilde a = 0$, and increments $\tilde a$ by 0.1 until either one finds values bracketing the solution.
For both these cases, a $\tilde B$ or $\tilde a$ for which no solution exists will be interpreted as potentially bracketing the true solution if its `neighbor' in the algorithm does exist.
Empirically, it turns out that this scheme usually is able to find values bracketing the true solution.

Once two values which potentially bracket the true solution are found, one once again uses Brent's method to shrink the size of these brackets.\footnote{Note that due to missing values being interpreted as potentially bracketing the true solution, there may be multiple such brackets per requested solution.}
As soon as Brent's method has shrunk the brackets to a small enough size, we once again perform a fit over intermediate values in this interval, and this is where the value of being able to compute an error bar on observables in determining $\tau_h$ shows its value.
As it turns out, the accuracy of observables varies by as much as an order of magnitude for different values of $\tau_h$ and $\tilde B$ or $\tilde a$\@.
This also means that the accuracy of observables obtained from the fit to $m_q = 0$ can vary by an order of magnitude.
Now, having error bars, one can perform a weighted fit, in which the `bad' points will receive a lower weight, enabling us to obtain a more accurate result.
This turns out to be crucial to determine quantities like the speed of sound, which ultimately depend on quantities like $dT/d\lambda_h$\@.
For such an observable to not be completely dominated by numerical noise, one needs to be able to determine $T$ and other such observables to as high an accuracy as possible.
\subsection{Obtaining a phase diagram}
The above discussion enables us to construct both chirally symmetric solutions and chirally broken solutions, both satisfying all the required boundary conditions.
We are also able to constrain either $B$ or $a$ to have a specified value, and we can extract all the required observables from the solutions.
Only one thing remains now in order for us to determine what the observables are as functions of $T$, $\mu$, $B$ and $a$, which together parameterize the phase diagram of the theory.
In principle, this is an easy task; one just has to evaluate the grand potential of every solution, and if there are multiple solutions with the same $T$, $\mu$, $B$ and $a$, choose the one with the smallest grand potential.

In practice, however, this is more complicated.
The grand potential diverges like $A^4$ near the boundary.
In principle one could try to perform holographic renormalization like one does for the magnetic field, but this turns out to be very hard to do numerically.
What works better numerically is to use the first law of thermodynamics
\[
d\Omega/V = -s\,dT - n\,d\mu - M\,dB - M_a\,da,
\]
and integrate along a family of solutions where one is able to continuously vary the input parameters.
Of course, this brings with it an integration constant, which we choose such that the thermal gas geometry has grand potential equal to zero.
This thermal gas geometry is obtained by taking $\lambda_h \rightarrow \infty$, which makes the horizon shrink to zero size.\footnote{This limit of taking $\lambda_h \rightarrow \infty$ is well-defined because for any $\mu$ this corresponds to the same solution. For different $B$ and $a$, the solution is not the same, but in those cases the holographic renormalization of $M$ and $M_a$, respectively, are defined with respect to these different horizonless solutions, which hence defines the grand potentials of these solutions to be the same.}
This corresponds to a horizonless geometry with a `good singularity', as discussed in \cite{Gubser:2000nd}\@.

By integrating the first law in this way, one can obtain the grand potential for every solution.
Note that this is a non-trivial statement, as it requires the whole parameter space to be connected, and in particular it requires that the chirally broken solutions have a limit in which they approach a corresponding chirally symmetric solution.
Both of these statements turn out to be correct \cite{Alho:2012mh}\@.

After one has computed the grand potential throughout the parameter space in this way, one knows which geometry is the dominant one for every $T$, $\mu$, $B$ and $a$\@.
To then label the resulting phase diagram appropriately, in this chapter we will use the following order parameters:
\begin{itemize}
\item Chiral condensate $\langle\bar qq\rangle$: this distinguishes chirally symmetric from chirally broken solutions.
\item Confinement: this is defined in this chapter as the quark-antiquark potential having a linear branch.
Usually this is equivalent with the geometry being horizonless.
In section \ref{sec:imc:a}, we will find that in the presence of anisotropy this is no longer always true.
This has some interesting consequences, which will be explored there.
\end{itemize}

This concludes the introduction into the computations needed throughout the rest of this chapter.
In the next sections, we will explore various limits of the general model described so far, starting with one where we have only a magnetic field, and hence set $\mu = 0$ and $a = 0$\@.
\section{Inverse magnetic catalysis due to a magnetic field}\label{sec:imc:b}
In the beginning of this chapter, the puzzle of inverse magnetic catalysis (IMC) was introduced.
An interesting possible explanation was put forward in \cite{Bruckmann:2013oba,Bruckmann:2013ufa}, namely that there are two competing effects at work, called `valence' and `sea' quark contributions.
The valence quark contribution corresponds to the $\bar qq$ operators appearing in the path integral.
This effect is found to increase the condensate as $B$ increases.
The sea quark contribution, instead, arises from the quark determinant.
This contribution tends to decrease the condensate as $B$ increases.
The appearance of magnetic catalysis or its opposite are then explained as follows: in regions of the phase diagram where valence quark effects are dominant, one finds magnetic catalysis, whereas in regions where the sea quark effects dominate, one finds inverse magnetic catalysis.
Note that the sea quark contribution is a backreaction effect.
To see that this is true, consider the quenched approximation, or the approximation of infinitely massive quarks.
In this approximation, the quark determinant becomes a constant, and hence the sea quark contribution vanishes.
In other words, when one takes the quarks as non-dynamical probes, there is no sea quark contribution.

In this section, which is based on \cite{Gursoy:2016ofp}, we study this problem in the holographic model introduced in section \ref{sec:imc:model}\@.
The aim of this study is twofold: to reproduce IMC in holography, and to investigate whether in holography one can similarly isolate two competing effects in analogy to the findings in \cite{Bruckmann:2013oba,Bruckmann:2013ufa}\@.
In earlier studies, holographic gauge theories in the presence of a magnetic field have been investigated either in the absence of flavors, or with $N_f \ll N_c$ \cite{Zayakin:2008cy,Filev:2011mt,Erdmenger:2011bw,Preis:2012fh,Mamo:2015dea,Dudal:2015wfn,Rougemont:2015oea,Evans:2016jzo}, or with smeared backreacted flavor branes in the Veneziano limit, which leads to a different flavor global symmetry group \cite{Jokela:2013qya}\@.
In our model, we have the correct flavor symmetry group, as well as a fully backreacted flavor sector in the Veneziano limit.
The model can be obtained from the `master' model introduced in section \ref{sec:imc:model} by setting $a_\perp = a_\parallel = \mu = 0$\@.
The baryon chemical potential can be set to zero by choosing $\tilde n = 0$\@.
\subsection{Varying the $w$ potential}
An important ingredient to reproduce IMC is the choice of potentials.
As briefly mentioned in section \ref{sec:imc:model}, the potentials used in this chapter can be found in appendix \ref{sec:potentials:imc}\@.
These potentials are to a large extent the same as the ones used in \cite{Alho:2013hsa}, with one important difference.
The coupling of the magnetic field to the flavor sector depends on the $w$-potential.
In \cite{Alho:2013hsa}, the choice is made that $w(\lambda) = \kappa(\lambda)$\@.
This is a natural choice, as it is expected that $\kappa$ and $w$ have similar asymptotics in both the UV and the IR\@.
Indeed, this assumption is consistent with the flavor vector current two point function and the asymptotics of the meson spectra \cite{Iatrakis:2010jb,Arean:2013tja}\@.
Also, QGP conductivity and its diffusion constant \cite{Iatrakis:2014txa}, as well as thermal photon emission during the QGP phase of heavy ion collisions \cite{Iatrakis:2016ugz}, are well described by assuming this assumption.
A simple modification to $w(\lambda) = \kappa(\lambda)$ is to insert a multiplicative constant $b$ such that $w(\lambda) = b\kappa(\lambda)$\@.
Looking at the appearance of $w$ in the action \eqref{eq:imc:Sf} however, such a modification would just lead to a trivial redefinition of $B$\@.
Another simple modification which keeps the asymptotics of $\kappa$ and $w$ the same is the one we will use, namely
\[
w(\lambda) = \kappa(c\lambda),
\]
with $c$ a constant.

\begin{figure}[ht]
\centering
\includegraphics[width=0.49\textwidth]{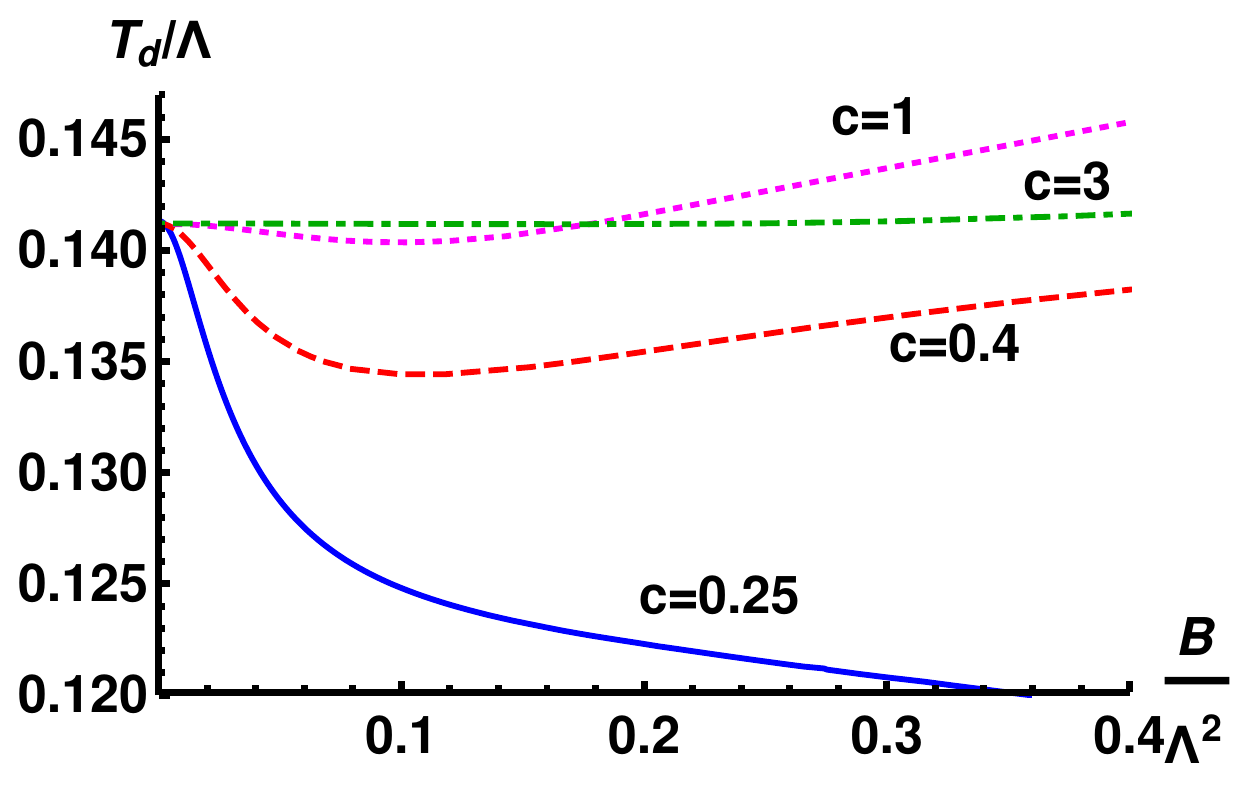}
\includegraphics[width=0.49\textwidth]{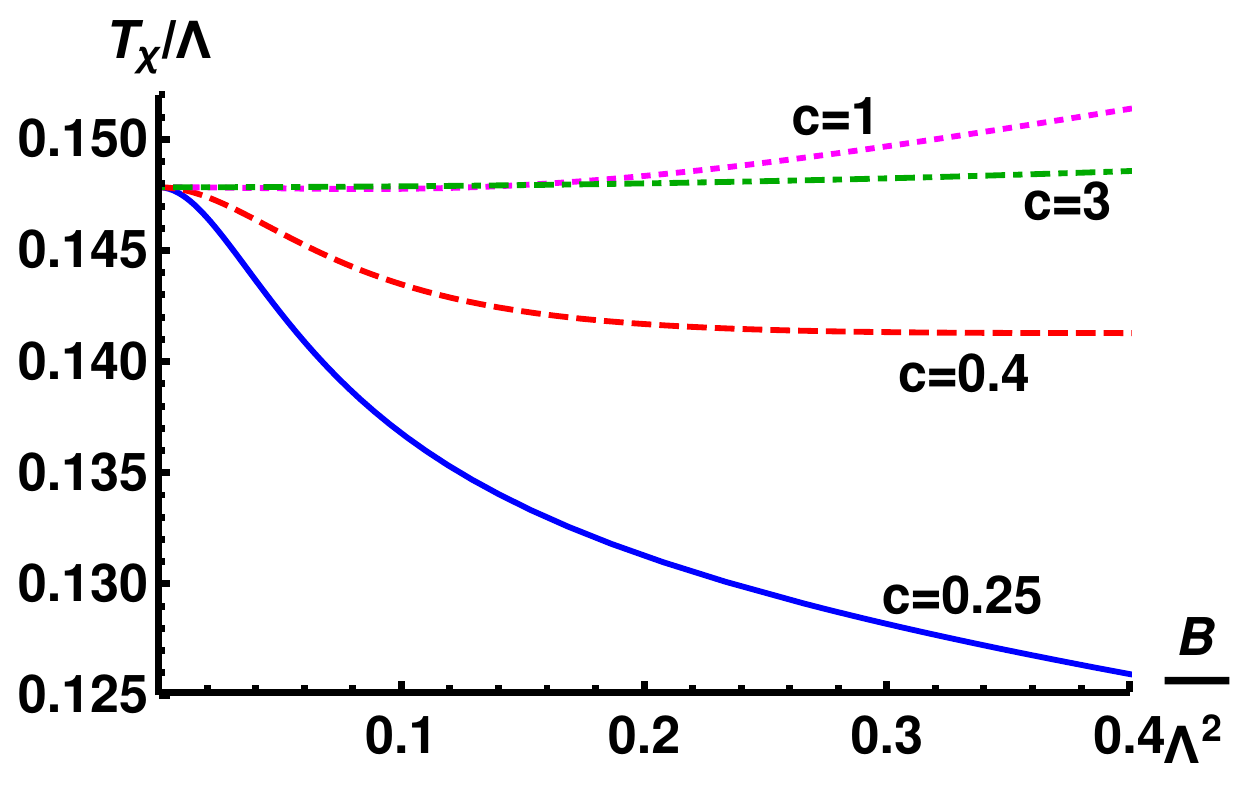}
\caption{\label{fig:imc:imc1TdvsB}Deconfinement transition temperature $T_d$ (left) and the chiral transition temperature (right) as a function of the magnetic field $B$ for different choices of $c$, with $x_f$ set to 1\@.}
\end{figure}

A first interesting question is then how various observables depend on $c$\@.
In figure \ref{fig:imc:imc1TdvsB}, the magnetic field dependence of both the deconfinement transition temperature and the chiral transition temperature is shown for different choices of $c$\@.
Here we have kept the number of flavors constant by setting $x_f \equiv N_f/N_c = 1$\@.
An interesting observation here is that both phase transition temperatures decrease with $B$ for smaller values of $c$, whereas for $c \simeq 1$ the transition temperatures generally grow with $B$\@.
Given that the chiral transition is second order in this model, this means that for smaller values of $c$ and temperatures slightly below the chiral transition, the chiral condensate must continuously decrease to zero, signaling inverse magnetic catalysis.\footnote{We assume massless quarks, which implies that the condensate must vanish at the chiral transition.}
An interesting sidenote is that larger values of $w$, which correspond to smaller values of $c$ in our model, also match better the electric conductivity of the QGP in the absence of a magnetic field \cite{Iatrakis:2016ugz}\@.

\begin{figure}[ht]
\centering
\includegraphics[width=0.9\textwidth]{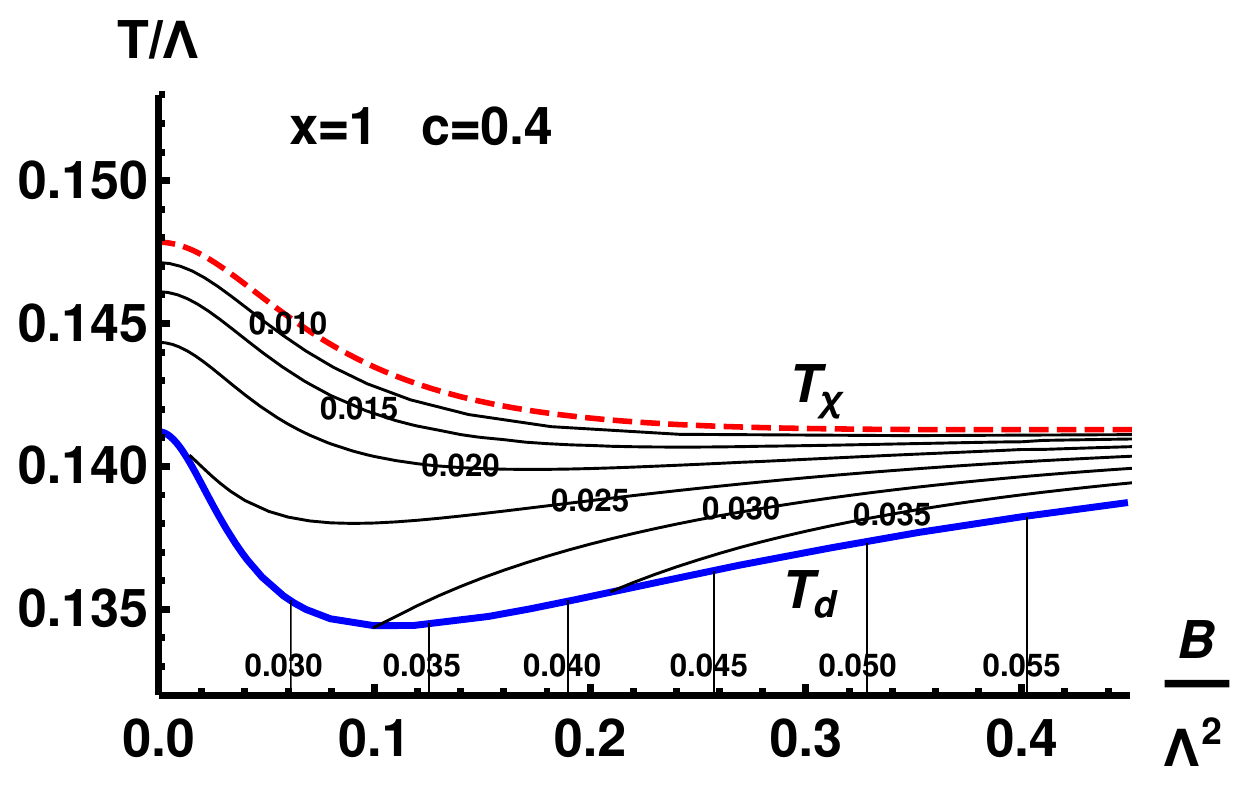}
\caption{\label{fig:imc:imc1TcandcondensatevsBx1c04}Chiral condensate as a function of $T$ and $B$, shown as isolines of $\langle\bar qq\rangle/\Lambda^3$, with $x_f = 1$ and $c = 0.4$ fixed. Shown in red and blue are the chiral transition and the deconfinement transition, respectively.}
\end{figure}

\begin{figure}[ht]
\centering
\includegraphics[width=0.49\textwidth]{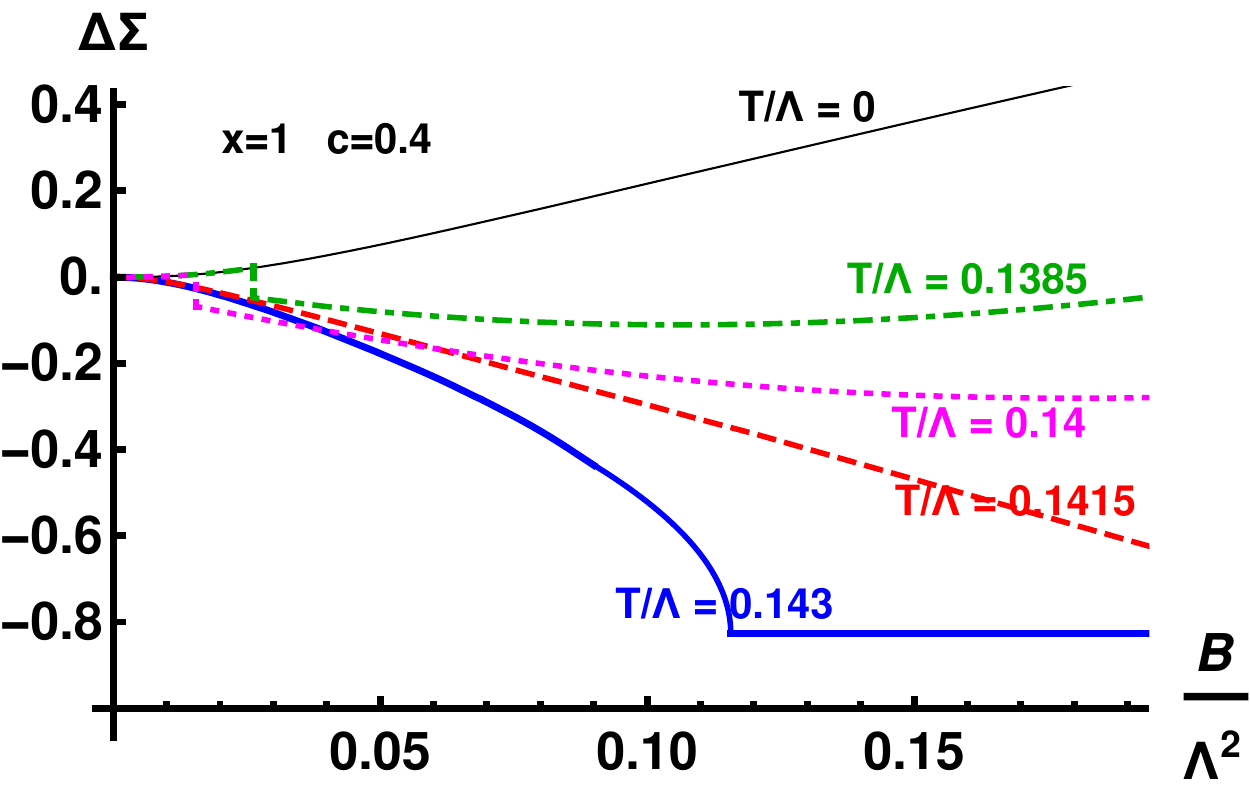}
\includegraphics[width=0.49\textwidth]{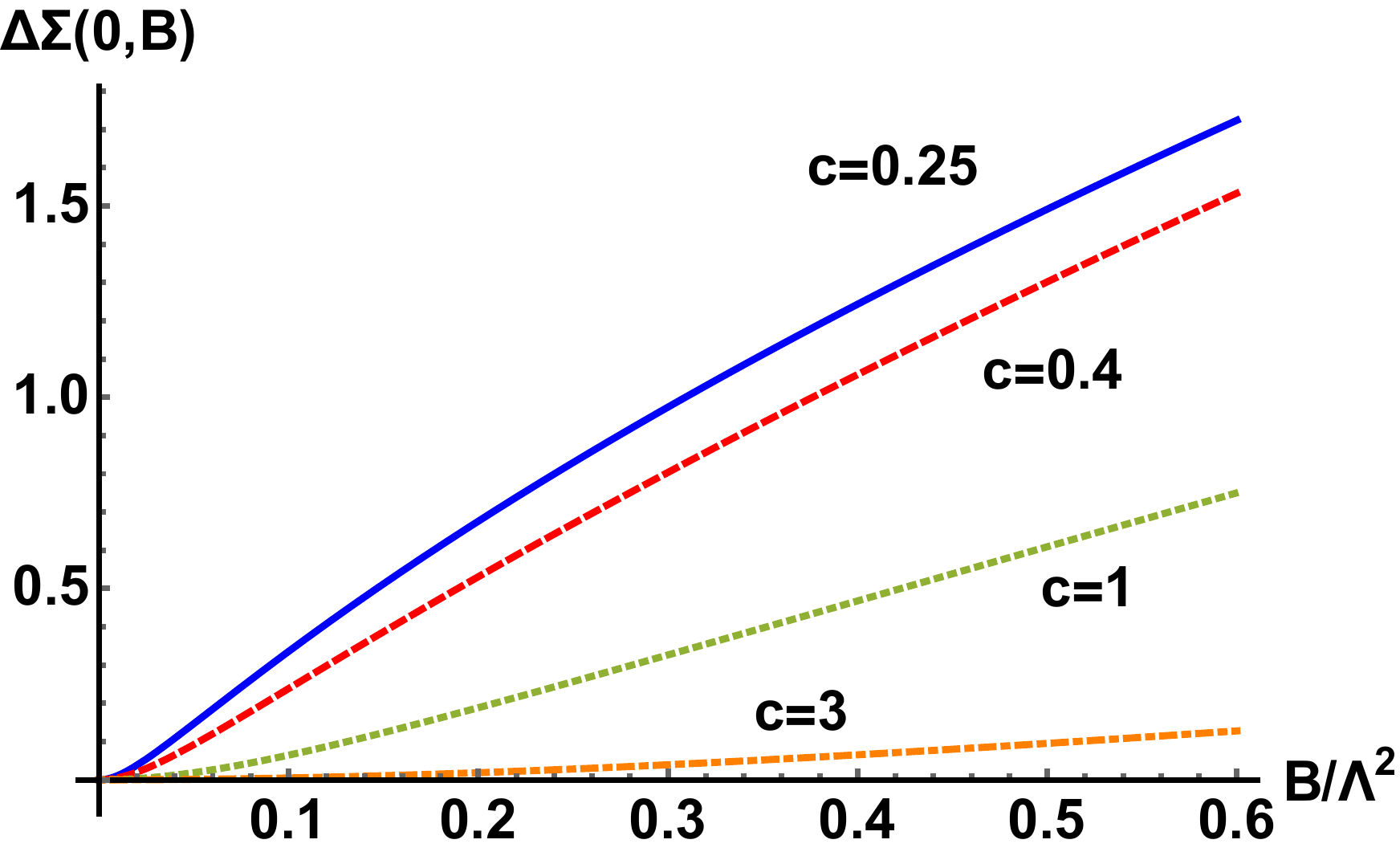}
\caption{\label{fig:imc:imc1qqconfx1}Left: $\Delta\Sigma$ as defined in \eqref{eq:imc:deltasigma} as a function of $B$ for constant $T$, $x_f = 1$ and $c = 0.4$\@. Right: $\Delta\Sigma$ for different $c$ for $T = 0$ and $x_f = 1$.}
\end{figure}

The appearance of inverse magnetic catalysis can be seen more explicitly in figure \ref{fig:imc:imc1TcandcondensatevsBx1c04}\@.
Here $\langle\bar qq\rangle/\Lambda^3$ is shown as a function of $T$ and $B$ for $x_f = 1$ and $c = 0.4$\@.\footnote{Note that $x_f$ is labeled $x$ in the figure.}
Between the two phase transitions for small enough values of the magnetic field, it can indeed be seen that the chiral condensate decreases with increasing values of the magnetic field.
In the low temperature thermal gas phase however, the condensate increases with the magnetic field.
These two observations show good agreement with the discussion at the beginning of this chapter.
In the left panel of figure \ref{fig:imc:imc1qqconfx1}, the same chiral condensate is shown in a different way, namely by computing the renormalization group invariant
\begin{equation}
\Sigma(T,B) = \frac{\langle\bar qq\rangle(T,B)}{\langle\bar qq\rangle(0,0)},\label{eq:imc:sigma}
\end{equation}
and subsequently looking at the difference
\begin{equation}
\Delta\Sigma(T,B) = \Sigma(T,B) - \Sigma(T,0).\label{eq:imc:deltasigma}
\end{equation}
By looking specifically at this quantity, it is possible to compare to lattice results, and indeed we find qualitative agreement \cite{Bali:2012zg}\@.
At small temperatures, the condensate increases monotonically as a function of $B$, while at temperatures around the phase transitions the condensate first increases, then jumps, and then decreases.
For even larger temperatures, the condensate decreases monotonically.
This is in qualitative agreement with \cite{Bali:2012zg}, with the notable exception that real QCD has no first order deconfinement transition, whereas the holographic model does.\footnote{This artefact is due to the fact that in the holographic model we take the large $N_c$ limit, which leads to a first order phase transition.}
In the right panel of figure \ref{fig:imc:imc1qqconfx1}, we show $\Delta\Sigma$ at vanishing temperature for different choices of $c$\@.
It can be seen that the condensate behaves like
\[
\Delta\Sigma(0,B) = D_{\bar qq}(c)B^2 + \mathcal{O}(B^3),
\]
where $D_{\bar qq}$ depends on $c$\@.
It is clear that $D_{\bar qq}$ decreases as $c$ increases.

\begin{figure}[ht]
\centering
\includegraphics[width=0.9\textwidth]{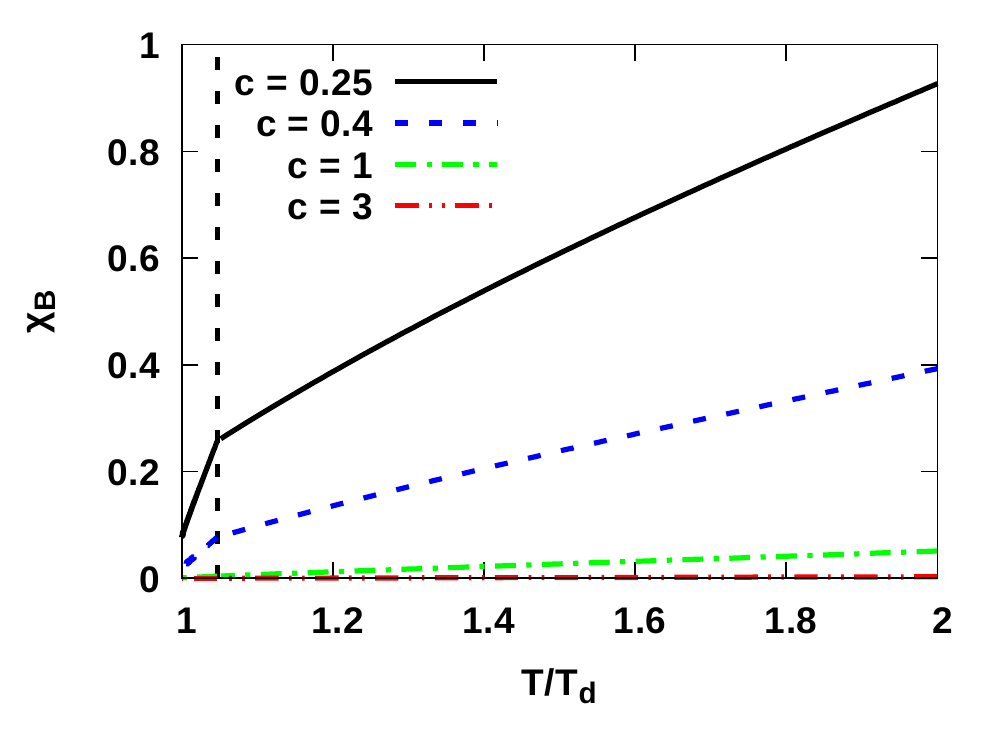}
\caption{\label{fig:imc:imc1susceptibilitiesx1}Magnetic susceptibility as a function of $T$ for $x_f = 1$ and different values of $c$\@. The chiral transition is shown by the vertical dashed line.}
\end{figure}

One final quantity we examine is the magnetic susceptibility at $B = 0$ as a function of $T$ for different values of $c$, which is shown in figure \ref{fig:imc:imc1susceptibilitiesx1}\@.
This is a useful quantity, as it can be used to determine the behavior of the deconfinement transition around $B = 0$\@.
This can be seen as follows:
Using the first law of thermodynamics, and using that the thermal gas has $s = 0$ and $M_B = 0$, one can derive that
\[
\frac{dT_d}{dB} = -\frac{M_B(B)}{s(B)},
\]
where $M_B$ and $s$ are now taken to be observables in the deconfined phase.
Together with the observation that
\[
M_B(B) = \chi_BB + \mathcal{O}(B^3), \qquad s(B) = s(0) + \mathcal{O}(B^2),
\]
this leads to
\begin{equation}
\left.\frac{d^2T_d}{dB^2}\right|_{B=0} = -\frac{\chi_B}{s(0)}.\label{eq:imc:dTddB}
\end{equation}
Knowing that the $w$-potential does not influence the entropy density for vanishing $B$, we can determine the dependence on $c$ of the behavior of the phase transition as a function of $B$ purely from the dependence of the magnetic susceptibility on $c$\@.
In this way, we can conclude that the deconfinement transition will decrease more sharply with $B$ for small values of $c$\@.
\subsection{Varying the number of flavors}
\begin{figure}[ht]
\centering
\includegraphics[width=0.9\textwidth]{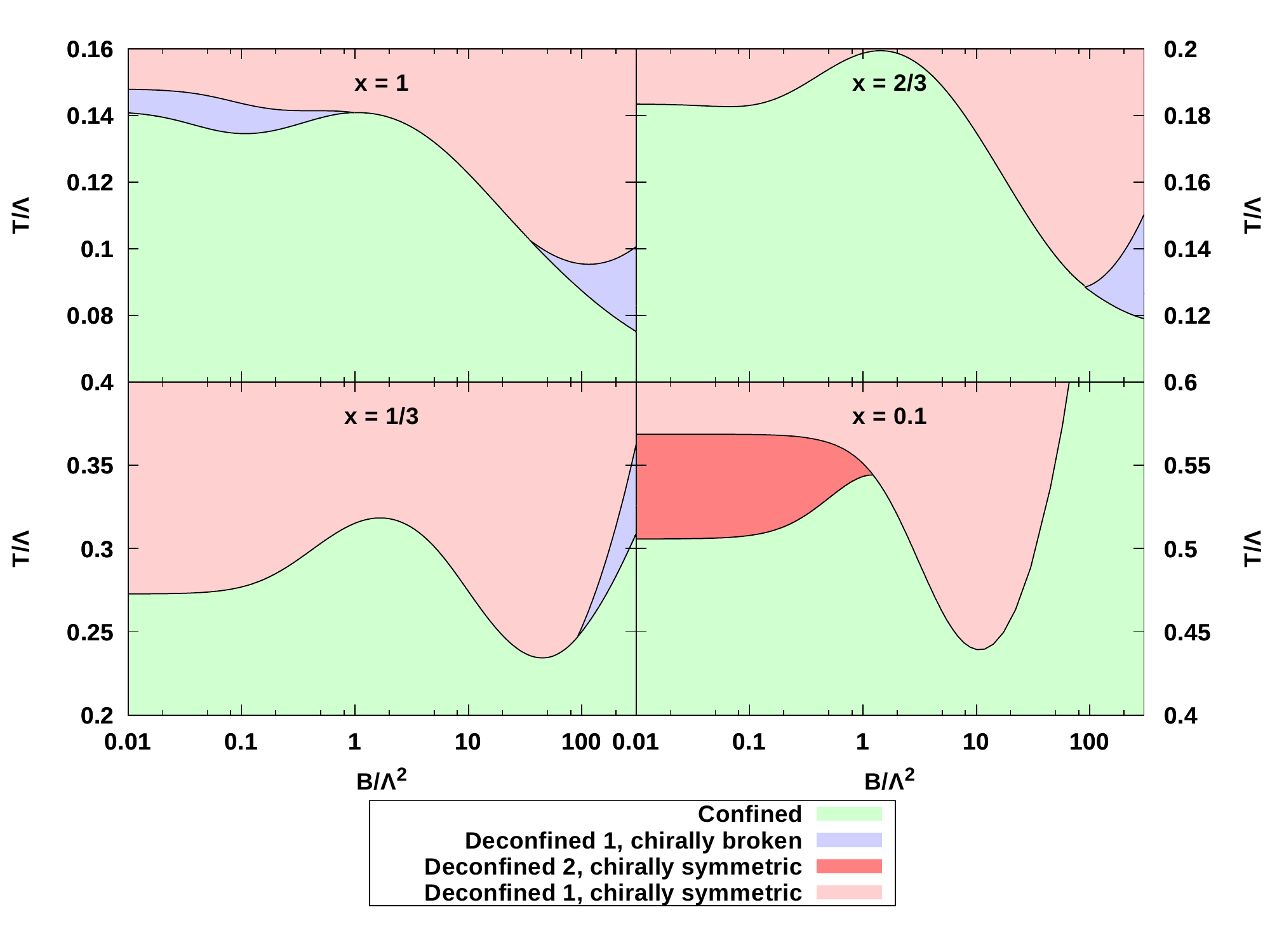}
\caption{\label{fig:imc:imc1phases}Phase diagrams for $c = 0.4$ and different values of $x_f$, which is labeled $x$ in the figure.}
\end{figure}

Now that we have established that it is possible to obtain IMC in V-QCD by setting $c < 1$, we can further investigate the mechanism behind its appearance.
To do this, we vary the number of flavors by tuning $x_f \equiv N_f/N_c$\@.
This has the effect of varying the amount of backreaction that the quark sector has on the gluon sector.
Indeed, examining \eqref{eq:imc:Sf}, one can see that $x_f$ multiplies the entire quark sector of the action, so that tuning $x_f \ll 1$ turns the quark sector into a probe, whereas $x_f \sim 1$ allows for significant backreaction.
One of the first things one can investigate is how the phase structure changes for different values of $x_f$\@.
In figure \ref{fig:imc:imc1phases}, this phase structure is shown for $c = 0.4$, where we also go to larger values of $B$ as compared to figure \ref{fig:imc:imc1TcandcondensatevsBx1c04}\@.
The phase diagram in figure \ref{fig:imc:imc1phases} contains the following four phases:
\begin{itemize}
\item A quark-gluon plasma phase at large temperatures.
This phase exhibits chiral symmetry and is not confining.
\item An intermediate temperature phase, exhibiting chiral symmetry breaking, but, like the QGP-like phase, not confining.
In real QCD, where the phase transition is a crossover, both chiral symmetry breaking and confinement set in in a smooth way.
In our holographic model, since the chiral transition and the deconfinement transition are second\footnote{Note that the chiral transition is first order whenever it concides with the deconfinement transition. It is only second order when it is a distinct phase transition.} and first order, respectively, these transitions can occur at distinct temperatures depending on the choice of potentials.
\item A low temperature thermal gas phase, exhibiting confinement and chiral symmetry breaking.
\item At $x_f = 0.1$, another deconfined chirally symmetric phase exists at intermediate temperatures, separated from the one mentioned above by a first order phase transition.
This is the same phase that was previously found at $B = 0$ in \cite{Alho:2012mh}\@.
\end{itemize}
In addition to the phase structure, it can be seen that around $B/\Lambda^2 \sim 1$, for all choices of $x_f$ we see inverse magnetic catalysis, chararcterized by a decrease in the chiral transition temperature.
Also for all considered values of $x_f$, we see that for still larger values of $B$, the chiral transition temperature increases, signaling magnetic catalysis.
For small values of $B/\Lambda^2$, the behavior of the chiral transition depends on $x_f$, where we see that for $x_f = 1$ there is inverse magnetic catalysis, whereas for $x_f = 2/3$ there is only slight inverse magnetic catalysis, and for smaller values of $x_f$ the dip in transition temperature decreases to a point that it is no longer detectable.\footnote{The dip can never really go away, as $\chi/s > 0$ for all deconfined solutions, which by \eqref{eq:imc:dTddB} implies $d^2T_d/dB^2 < 0$\@. Since for $x \lesssim 2/3$ the chiral transition coincides with the deconfinement transition, this implies that there sould always be slight inverse magnetic catalysis.}
Another interesting feature is the reappearance of the chirally broken deconfined plasma phase at large $B/\Lambda^2$, which seems to appear for all considered $x_f$, but for $x_f = 0.1$ this could not be shown conclusively due to numerical inaccuracies.

\begin{figure}[ht]
\centering
\includegraphics[width=0.9\textwidth]{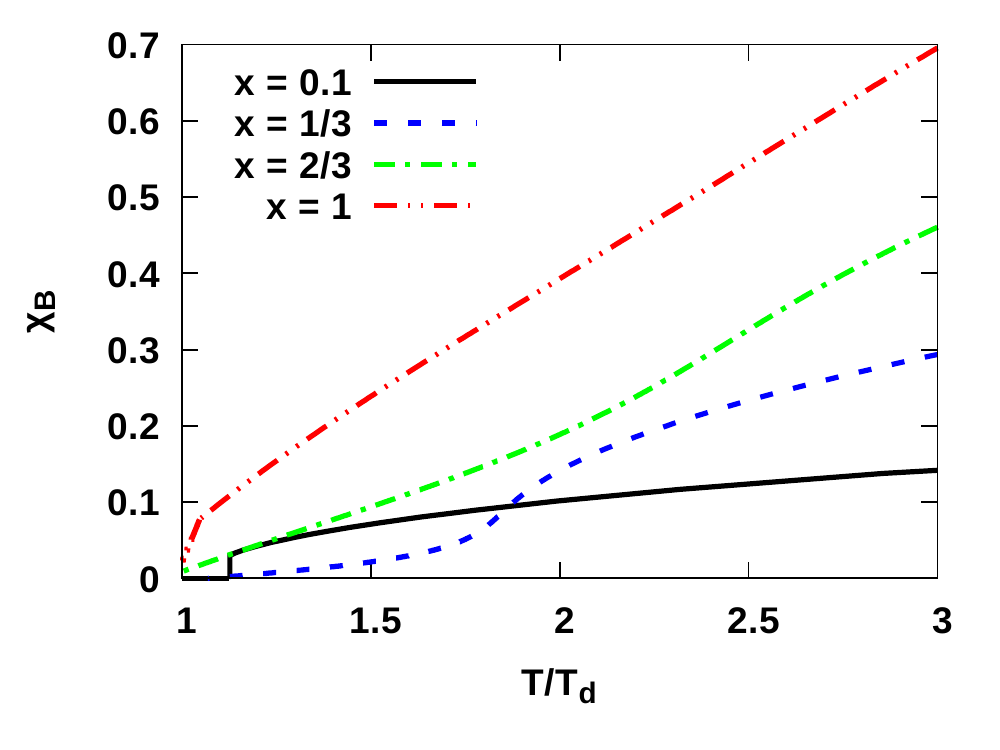}
\caption{\label{fig:imc:imc1susceptibilitiesc0-4}Magnetic susceptibility as a function of temperature for $c = 0.4$ and for different values of $x_f$, which is labeled as $x$ in the figure.}
\end{figure}

In figure \ref{fig:imc:imc1susceptibilitiesc0-4}, the magnetic susceptibility for $c = 0.4$ for the same values of $x_f$ considered in figure \ref{fig:imc:imc1phases} is shown.
Because in this case the entropy density at the deconfinement transition is not the same for all $x_f$, as opposed to the situation for figure \ref{fig:imc:imc1susceptibilitiesx1}, we can not conclude anything about the behavior of the deconfinement transition from the magnetic susceptibility.
Nevertheless, interesting features can be discerned.
Firstly, the susceptibility becomes larger as $x_f$ is increased.
This reflects the observation that the coupling between the magnetic field and the gluon sector becomes stronger as $x_f$ is increased.
Also, as one decreases $x_f$, an inflection point can be seen to appear, which eventually forms a first order phase transition somewhere between $x_f = 0.1$ and $x_f = 1/3$\@.

\begin{figure}[ht]
\centering
\includegraphics[width=0.49\textwidth]{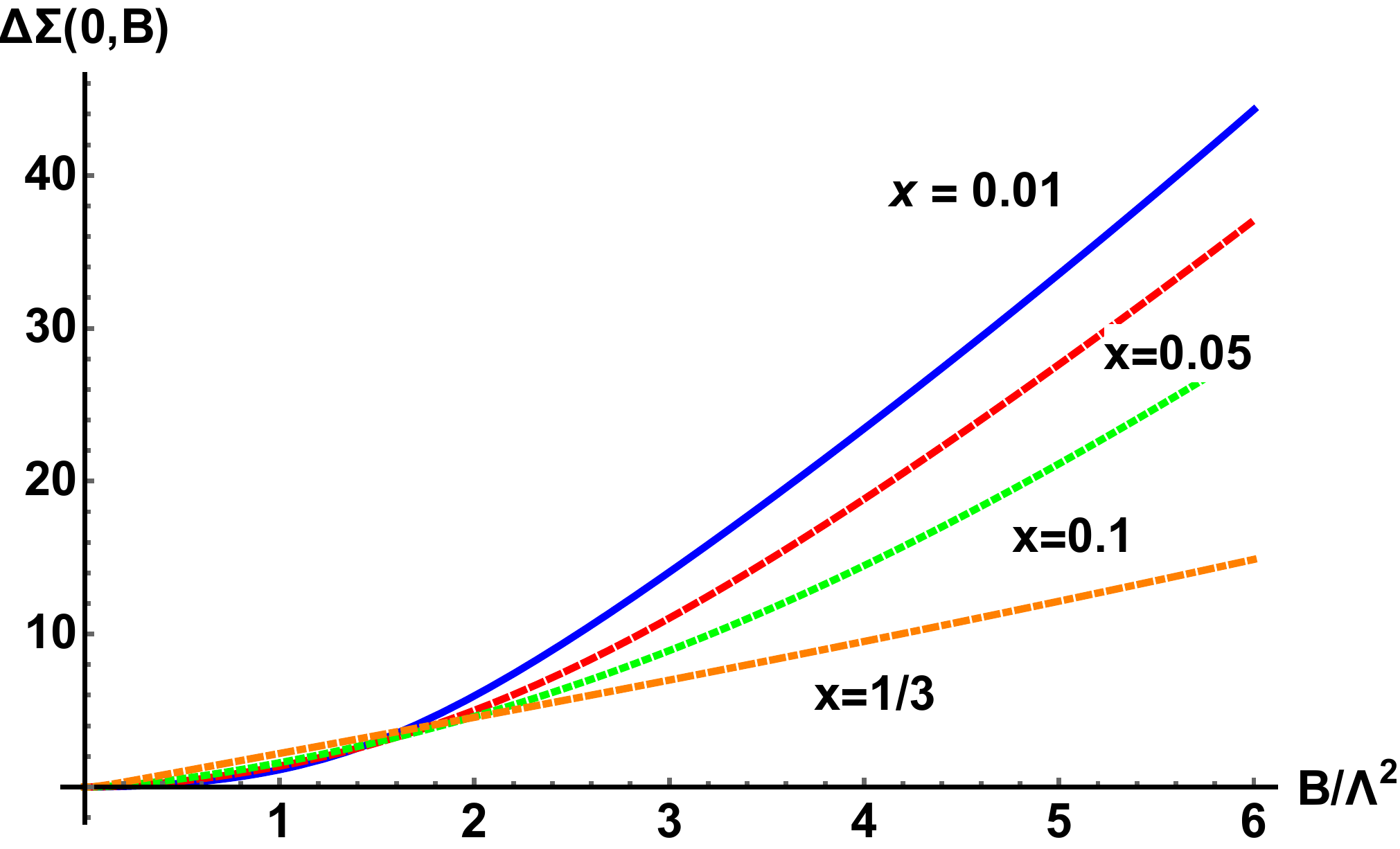}
\includegraphics[width=0.49\textwidth]{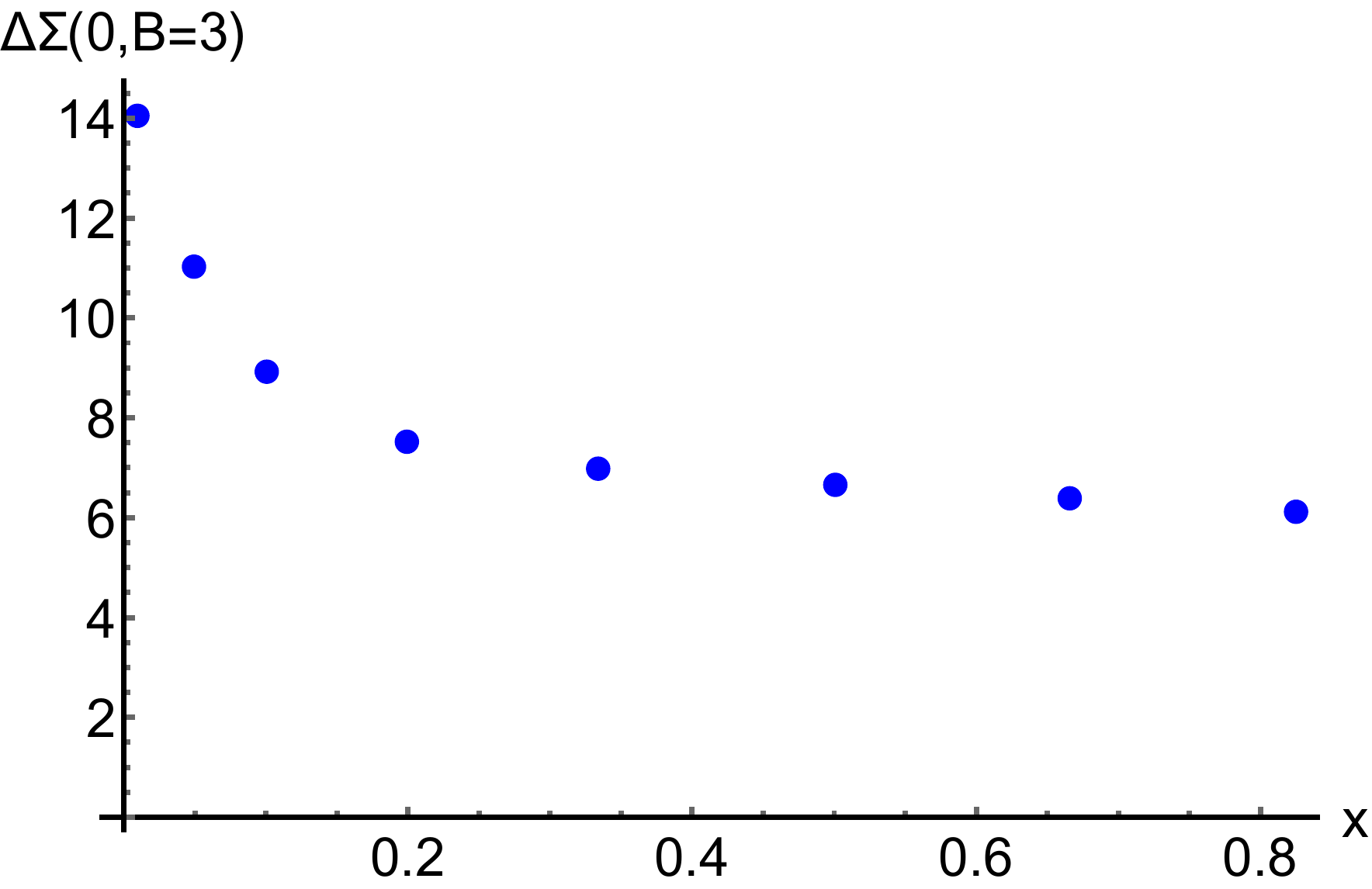}
\caption{\label{fig:imc:imc1qqConfinedx}Left: $\Delta\Sigma$ as defined in \eqref{eq:imc:deltasigma} as a function of $B$ at zero temperature for $c = 0.4$ and different $x_f$\@. Right: $\Delta\Sigma$ for $T = 0$, $B/\Lambda^2 = 3$, $c = 0.4$ and different $x_f$\@. Note that in both panels $x_f$ is denoted as $x$\@.}
\end{figure}

The last observable we will discuss is the chiral condensate.
It turns out that by examining the $x_f$-dependence of the chiral condensate, we can isolate two competing effects in analogy to \cite{Bruckmann:2013oba,Bruckmann:2013ufa}\@.
The `valence' quark effect will be identified with the direct coupling of the magnetic field to the tachyon field, which is dual to the chiral condensate.
The `sea' quark effect, in contrast, is identified with the indirect coupling of the magnetic field to the tachyon field through the metric.
To investigate what effect these two contributions have on the chiral condensate, we need to find a way to change their relative contributions.
Looking at the tachyon equation of motion \eqref{eq:imc:taueomr}, one can see that the magnetic field only enters through $Q$, which, for large $B$, behaves as $Q \sim e^{-2A}w(\lambda)B$\@.
One can then check that the explicit dependence of \eqref{eq:imc:taueomr} on $B$ vanishes for large $B$\@.
In other words, the `valence' quark effect becomes constant for large $B$\@.
On the other hand, we can tune the magnitude of the `sea' quark effect by changing the amount of backreaction of the flavor sector onto the gluon sector.
This can easily be achieved by changing $x_f$\@.
With this in mind, in the left panel of figure \ref{fig:imc:imc1qqConfinedx}, we examine $\Delta\Sigma$ as defined in \eqref{eq:imc:deltasigma} as a function of $B$, for zero temperature, $c = 0.4$ and different values of $x_f$\@.
Interestingly, for large $B$, we can see that even though the condensate is larger than for $B = 0$, the difference $\Delta\Sigma$ is smaller for larger $x_f$ than for smaller $x_f$\@.
This can be seen more clearly in the right panel of figure \ref{fig:imc:imc1qqConfinedx}, where $\Delta\Sigma$ is shown for different $x_f$, $T = 0$, $B = 3$, with $c = 0.4$\@.
Since at large $B$ the `valence' quark contribution is constant, we find that indeed the `sea' quark effect, which we control through $x_f$, tends to lower the condensate.
This corroborates the findings of \cite{Bruckmann:2013oba,Bruckmann:2013ufa}\@.

To conclude this section, we find that we can indeed reproduce inverse magnetic catalysis in a holographic model, provided that we take backreaction of the flavor sector onto the gluon sector into account.
Furthermore, we find, similarly to the lattice QCD studies \cite{Bruckmann:2013oba,Bruckmann:2013ufa}, that there are two competing effects that influence the chiral condensate.
Of these, the backreaction effect associated with `sea' quarks, tends to decrease the condensate, whereas the direct `valence' quark, increases it.
In the next section, we will discuss what happens to IMC in the presence of a finite baryon chemical potential, which, due to the sign problem, is a hard problem to address with lattice QCD\@.
\section{Inverse magnetic catalysis in the presence of a nonzero chemical potential}\label{sec:imc:mub}
In this section, which is based on \cite{Gursoy:2017wzz}, we extend the analysis of the previous section to also include a finite baryon chemical potential.
There are several reasons why such an extension is interesting.
Of these, the most important one is that this region of the phase diagram is explored by the two systems mentioned at the beginning of this chapter.
Indeed, low-energy heavy ion collisions such as those performed at RHIC occur in a region of both non-negligible baryon chemical potential \cite{Inghirami:2019muf} and magnetic field \cite{Gursoy:2014aka}, whereas neutron stars occupy the low temperature, high baryon density region of the phase diagram, and in particular magnetars have substantial magnetic fields.\footnote{Even for magnetars the magnetic field is still small in comparison to the relevant energy scales of QCD, but nevertheless observables can perhaps be defined which are sensitive to such effects.}
Additionally, it is of theoretical interest what effect the chemical potential can have on inverse magnetic catalysis \cite{Andersen:2014xxa}\@.
The latter is a question that cannot be addressed using lattice QCD, due to the sign problem \cite{Aarts:2015tyj}\@.
Because of this, attempts to explore this question have been made both with effective models \cite{Miransky:2015ava} and using holography \cite{Maldacena:1997re,Ballon-Bayona:2017dvv,Jarvinen:2011qe,Drwenski:2015sha,Jokela:2013qya,Dudal:2015wfn,Mamo:2015dea,Evans:2016jzo}\@.

Both the results of the previous section, as well as lattice studies \cite{Bruckmann:2013oba,Bruckmann:2013ufa}, show that the effects of backreaction are important to capture the physics behind inverse magnetic catalysis.
Therefore, we choose to study this problem in the same setup as in the previous section, with the addition of the baryon chemical potential.
In terms of the `master' model introduced in section \ref{sec:imc:model}, this amounts to setting $a_\perp = a_\parallel = 0$\@.
As in the rest of this chapter, we will use the potentials from appendix \ref{sec:potentials:imc}\@.
Throughout this section, we will additionally fix $c = 0.4$, and we keep the number of flavors fixed to $x_f \equiv N_f/N_c = 1$\@.
\subsection{Phase diagram and thermodynamics}
\begin{figure}[ht]
\centering
\includegraphics[width=0.9\textwidth]{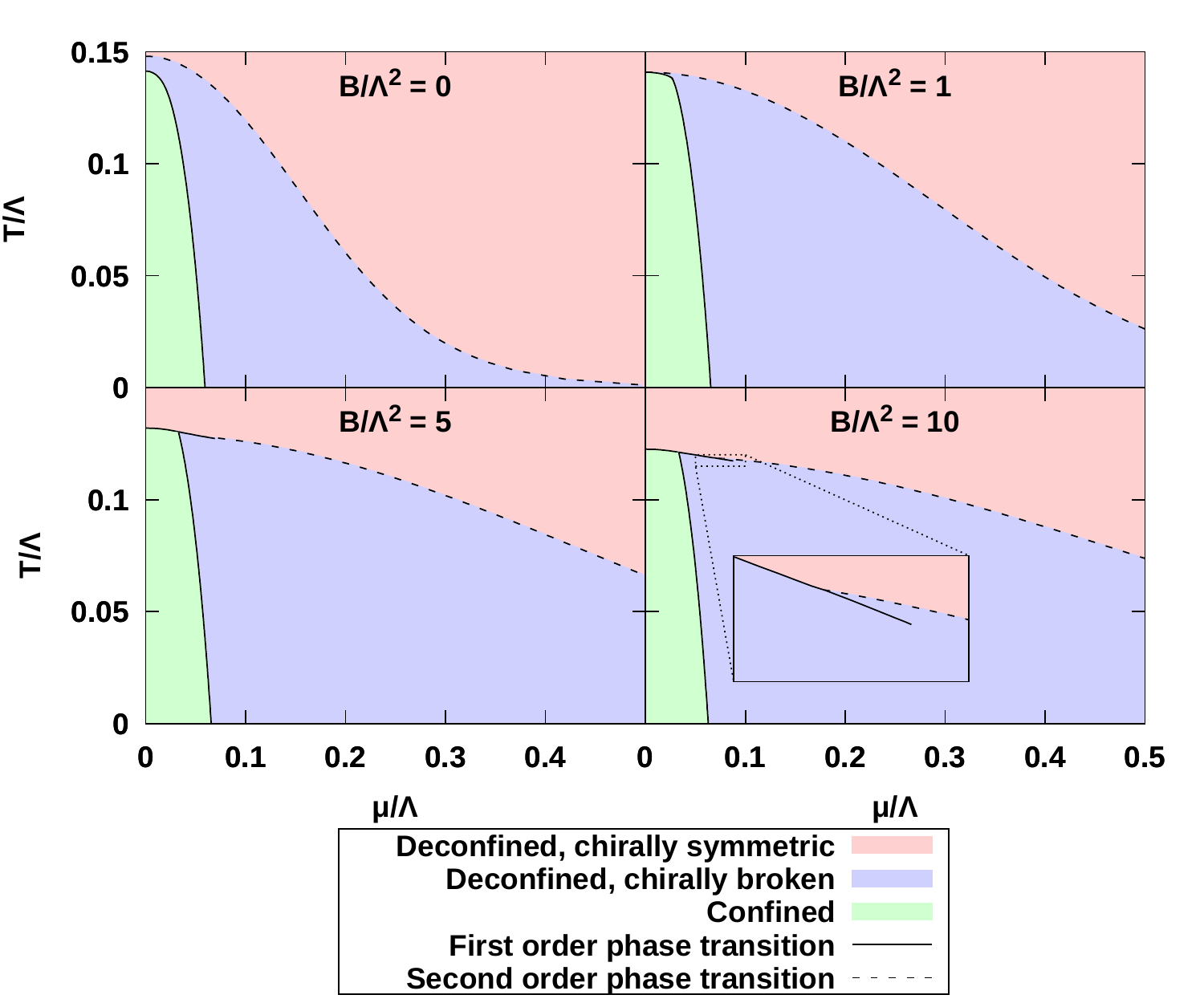}
\caption{\label{fig:imc:imc2phasediagram}Phase diagram of V-QCD as a function of $T$ and $\mu$, shown for different choices of $B$, $c = 0.4$ and $x_f = 1$\@.}
\end{figure}

As before, we start the discussion of the results by examining the phase diagram, which is shown in figure \ref{fig:imc:imc2phasediagram}\@.
The phases featured in the phase diagram are the same as in section \ref{sec:imc:b}, with the exception that the extra chirally symmetric phase that appears for $x_f = 0.1$ is not present, as we fix $x_f = 1$ in this section.
The phase structure shows an interesting dependence on the magnetic field.
Firstly, the deconfined, chirally symmetric phase can be seen to move down for small chemical potential values, whereas it increases for larger $\mu$\@.
The behavior of the chiral transition will be examined in more detail in section \ref{sec:imc:imcatmu}\@.
In contrast to the chiral transition, the deconfinement transition does not exhibit a large dependence on the magnetic field, with one notable exception.
From section \ref{sec:imc:b}, we already know that in the absence of chemical potential around $B/\Lambda^2 \approx 1$, the two phase transitions join.
As the deconfined, chirally symmetric phase continues to exist for larger values of $B$, there is a triple point.
Interestingly, as this triple point forms around $B/\Lambda^2 \approx 1$, the second order transition between both deconfined phases turns first order.
After this, as can be seen in the inset in figure \ref{fig:imc:imc2phasediagram}, another triple point forms, after which the first order transition becomes a transition between two deconfined, chirally broken phases, before ending in a critical point.

\begin{figure}[ht]
\centering
\includegraphics[width=0.8\textwidth]{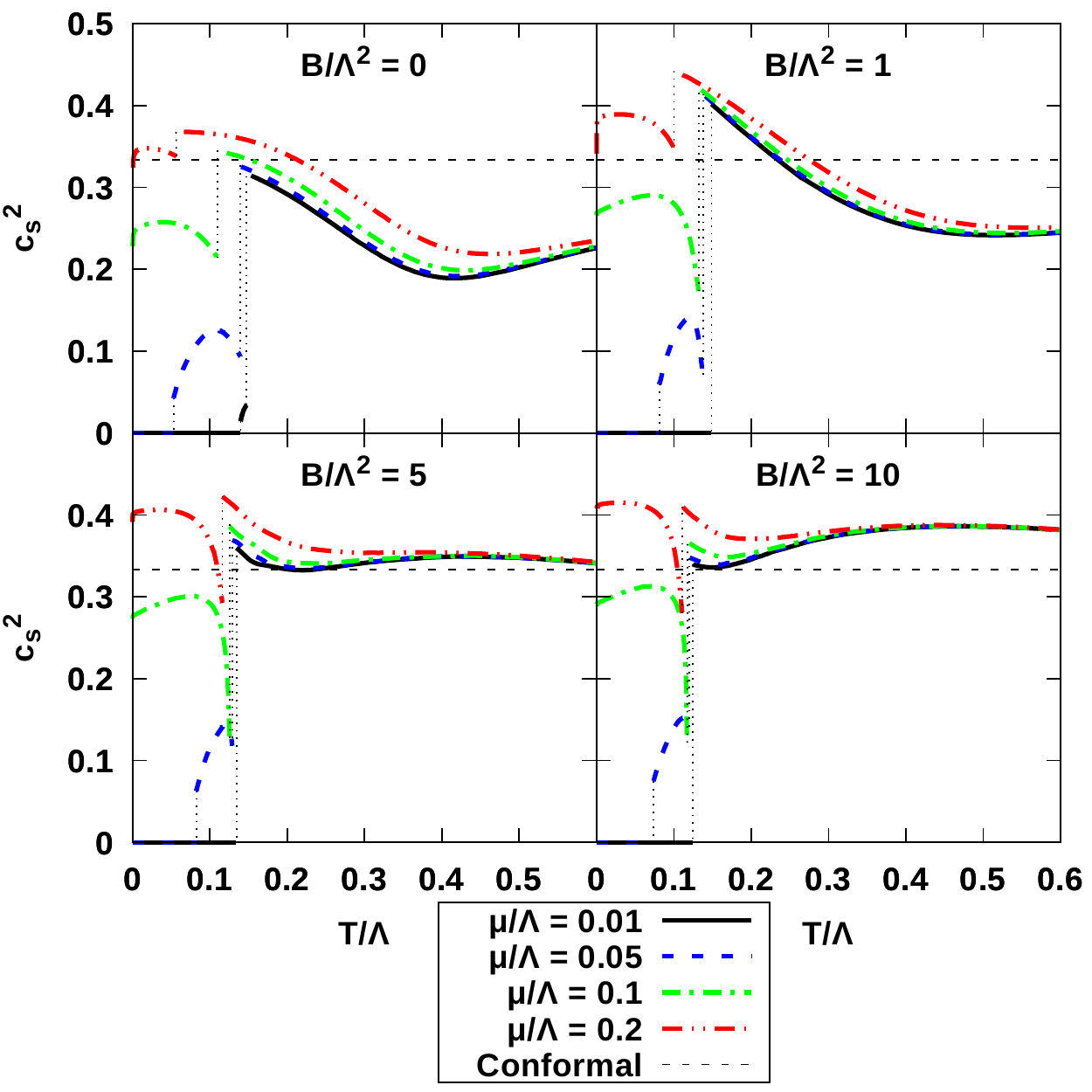}
\caption{\label{fig:imc:imc2soundspeed}The speed of sound squared $c_s^2$ as a function of temperature for different values of chemical potential and magnetic field.}
\end{figure}

An interesting observable which encodes much of the thermodynamical information is the speed of sound.
In figure \ref{fig:imc:imc2soundspeed}, we show the speed of sound in the direction of the magnetic field.\footnote{Due to the anisotropy induced by the magnetic field the speed of sound can take a different value perpendicular to $B$\@.}
The speed of sound $c_s$ is given by
\[
c_s^2 = \left.\frac{s\,dT + n\,d\mu}{T\,ds + \mu\,dn + B\,dM}\right|_{n/s,B}.
\]
Note that since $c_s$ depends on derivatives of thermodynamic variables, it is crucial to obtain accurate estimates for these quantities, as any numerical noise will cause the derivative to be unusable.
For this reason, the fitting procedure described at the end of section \ref{sec:imc:constraints} is essential to obtain the speed of sound, although even with the improved accuracy obtained from this method the speed of sound still has some numerical noise, which was removed artificially from the curves in figure \ref{fig:imc:imc2soundspeed}\@.
Rather surprisingly, the speed of sound squared is, for certain values of the chemical potential and magnetic field, larger than the conformal value of $1/3$, which is surprising given \cite{Hohler:2009tv,Cherman:2009tw}, though it does not contradict their findings.
Finally, note that at large $T$, outside the range of figure \ref{fig:imc:imc2soundspeed}, the speed of sound approaches the conformal value from below, consistent with \cite{Hohler:2009tv,Cherman:2009tw}\@.
\subsection{Chiral condensate and inverse magnetic catalysis}\label{sec:imc:imcatmu}
\begin{figure}[ht]
\centering
\includegraphics[width=0.49\textwidth]{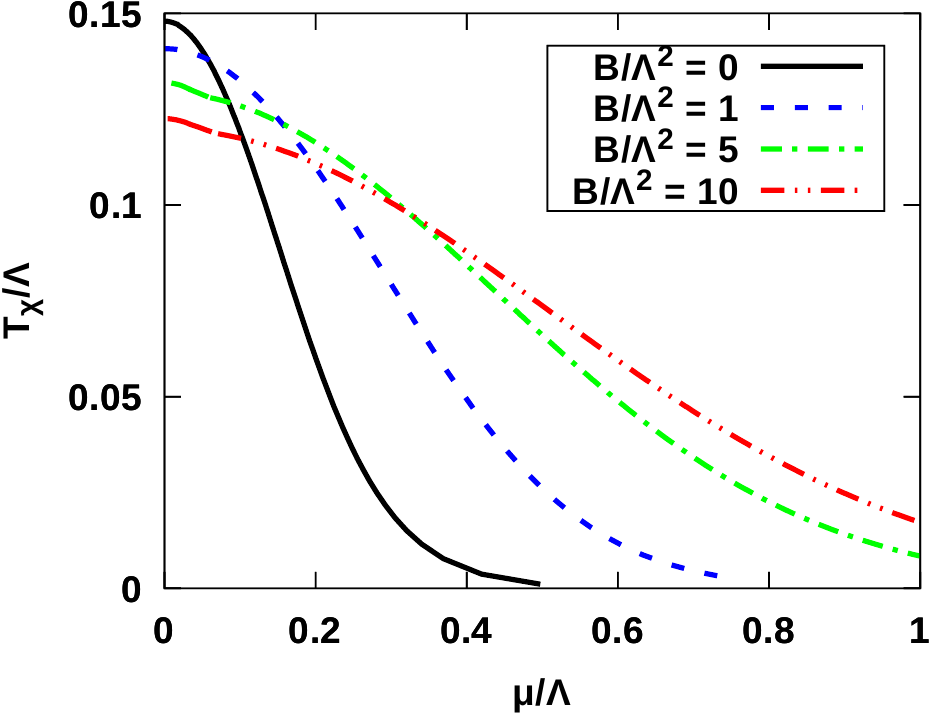}
\includegraphics[width=0.49\textwidth]{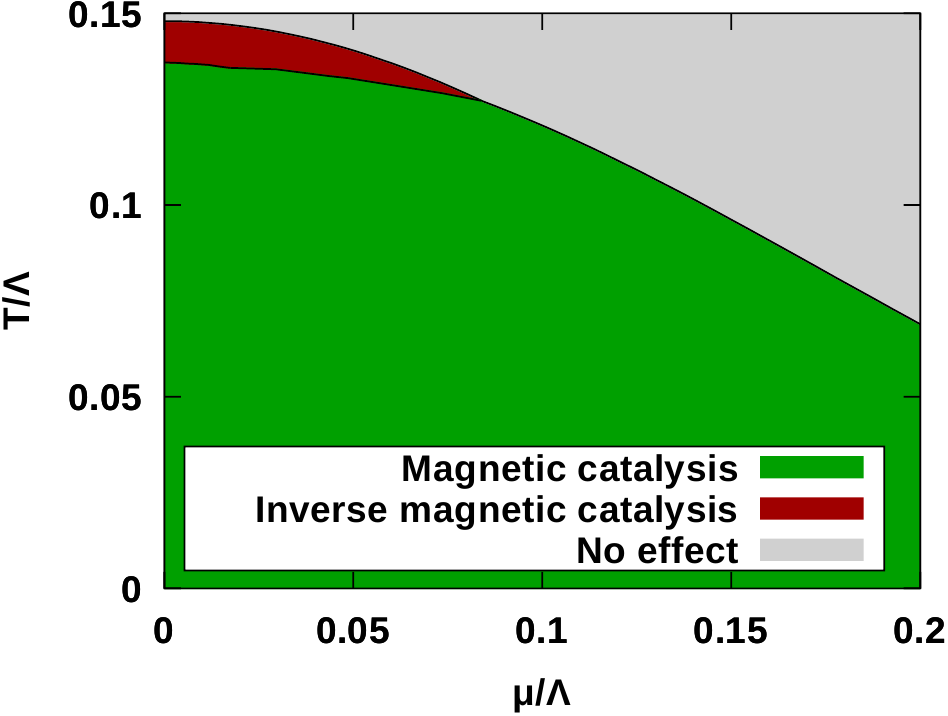}
\caption{\label{fig:imc:imc2chiraltransition}Left: The chiral transition temperature $T_\chi$ as a function of chemical potential for different values of the magnetic field. Right: Region in the $T$--$\mu$ plane around $B = 0$ where (inverse) magnetic catalysis occurs, where we define by looking at the sign of $\langle\bar qq\rangle_{B/\Lambda^2=0.1} - \langle\bar qq\rangle_{B/\Lambda^2=0}$\@.}
\end{figure}

As the chiral transition is second order in a large part of the phase diagram, we can use the behavior of the chiral transition temperature as a function of chemical potential and magnetic field to study inverse magnetic catalysis.
In the left panel of figure \ref{fig:imc:imc2chiraltransition}, the chiral transition temperature is shown as a function of $\mu$ for different values of $B$\@.
For small values of the chemical potential, the chiral transition temperature decreases as a function of $B$ for the values of $B$ considered in the left panel of figure \ref{fig:imc:imc2chiraltransition}\@.
This signals inverse magnetic catalysis, and is in line with expectation from section \ref{sec:imc:b}\@.
Interestingly, for larger values of the chemical potential, the exact opposite behavior is seen, signaling magnetic catalysis instead.

Looking explicitly at the change in the condensate around $B = 0$, we can examine in more detail where we have (inverse) magnetic catalysis.
We do this by looking at the sign of
\[
\langle\bar qq\rangle_{B/\Lambda^2=0.1} - \langle\bar qq\rangle_{B/\Lambda^2=0},
\]
where a positive sign signals magnetic catalysis, a negative sign signals inverse magnetic catalysis, and zero signals no change.\footnote{The latter mostly happens in the chirally symmetric phase, where the constant vanishes independently of $B$\@.}
Using this quantity, we obtain the right panel of figure \ref{fig:imc:imc2chiraltransition}\@.
We can clearly see now that for temperatures well below the phase transitions, magnetic catalysis occurs, and that for a small region below the chiral transition inverse magnetic catalysis occurs, but only for small enough chemical potential.

\begin{figure}[ht]
\centering
\includegraphics[width=0.8\textwidth]{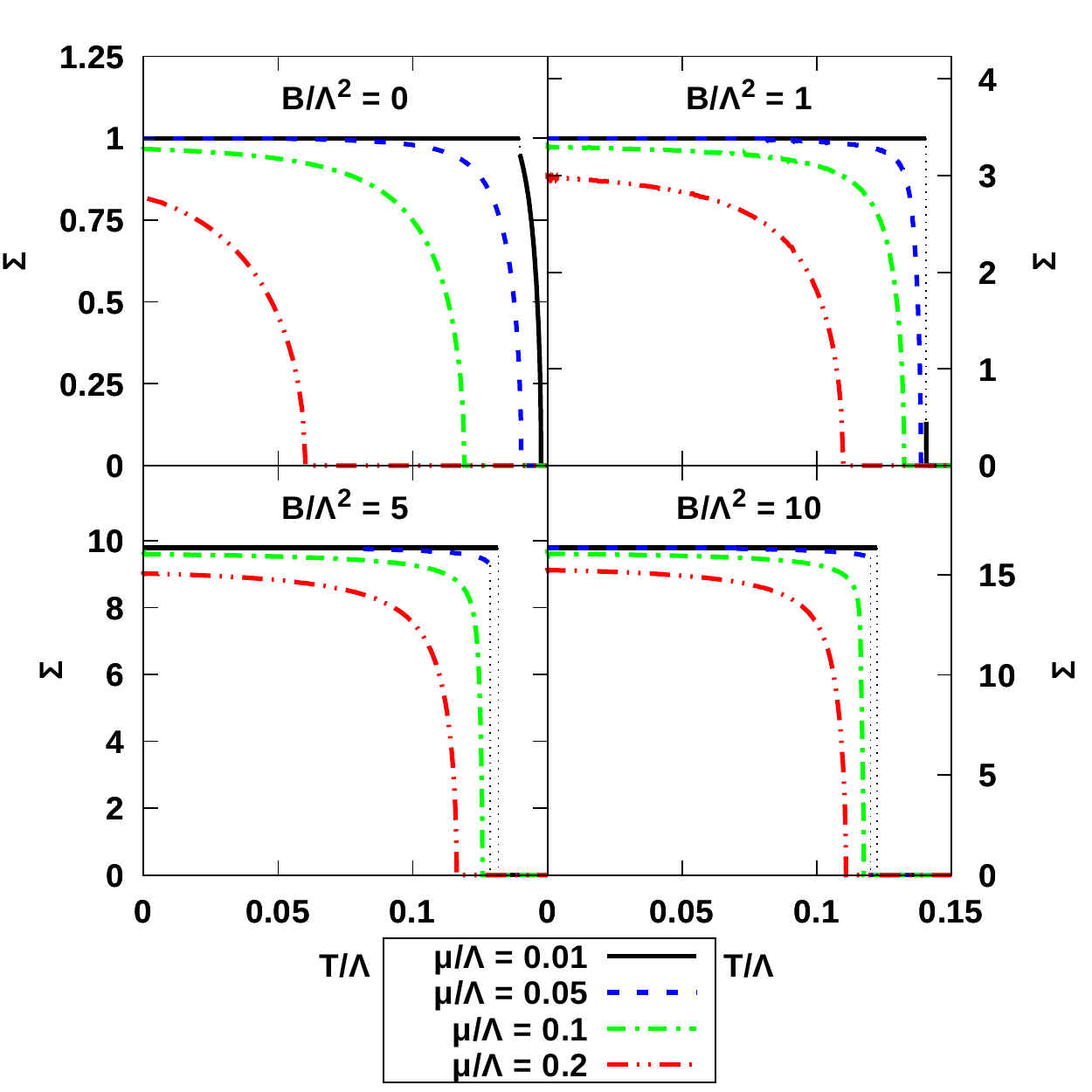}
\caption{\label{fig:imc:imc2chiralcondensate}Chiral condensate, normalized as in \eqref{eq:imc:sigma}, as a function of temperature for different values of chemical potential and magnetic field.}
\end{figure}

For larger values of $B$, this behavior continues.
This can be seen both from the left panel of figure \ref{fig:imc:imc2chiraltransition} and from figure \ref{fig:imc:imc2chiralcondensate}, where $\Sigma$ is shown, as defined in \eqref{eq:imc:sigma}\@.
One can see that the chiral condensate always decreases with temperature, and for most values of $T$ and $\mu$, the chiral condensate increases with $B$\@.
However, for certain values of $T$, $\mu$ and $B$, the condensate decreases.
One can check that indeed that for small $B$ this is consistent with the right panel of figure \ref{fig:imc:imc2chiraltransition}, but also for larger values $B$, one can find examples of inverse magnetic catalysis.
Indeed, one can see that for example for $T\Lambda = 0.12$, $\mu/\Lambda = 0.05$, the chiral condensate has a finite value for $B/\Lambda^2 = 5$, whereas the condensate vanishes for $B/\Lambda^2 = 10$\@.

\begin{figure}[ht]
\centering
\includegraphics[width=0.8\textwidth]{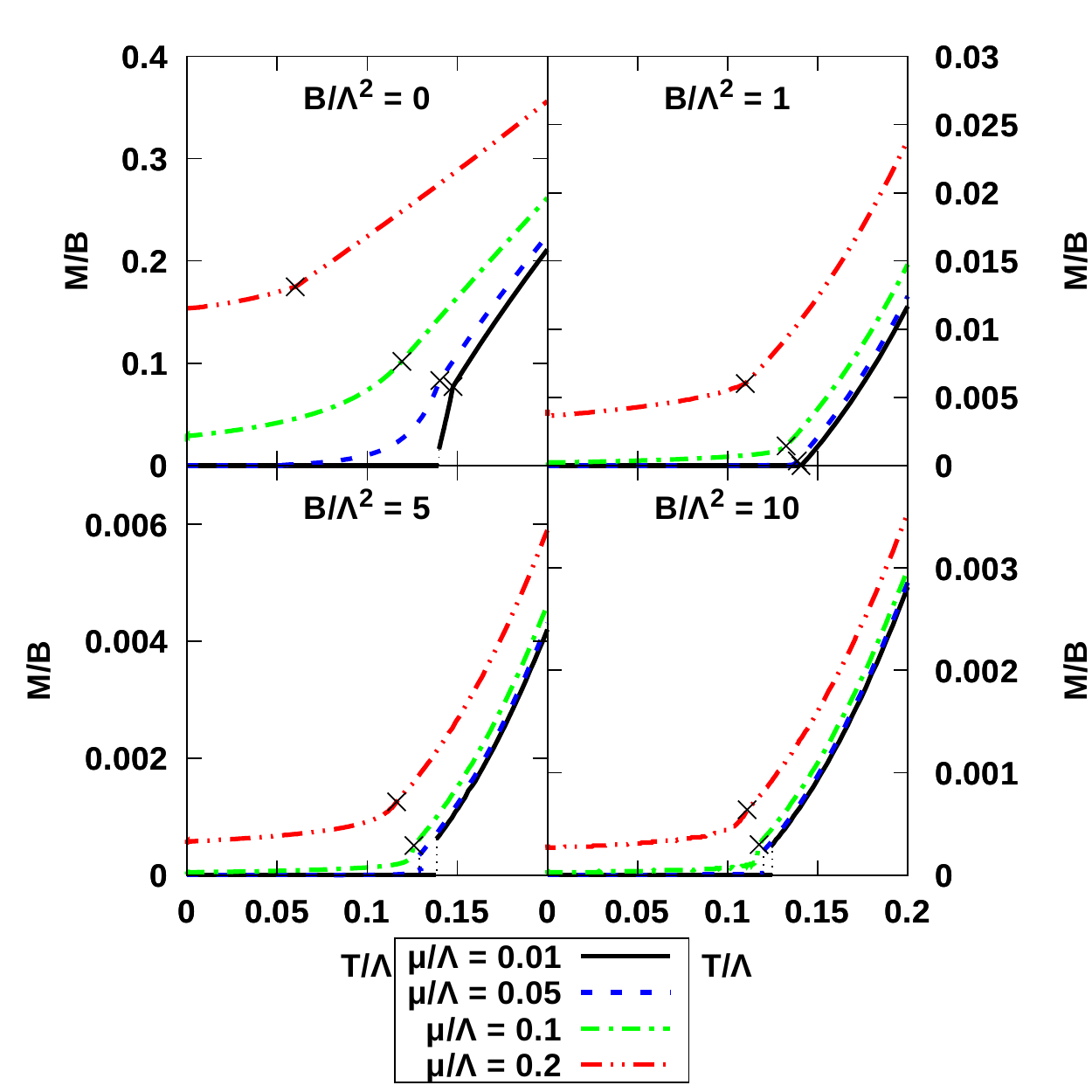}
\caption{\label{fig:imc:imc2magnetization}Magnetization divided by the magnetic field as a function of $T$, $\mu$ and $B$\@. The crosses denote the locations of the chiral transition.}
\end{figure}

In figure \ref{fig:imc:imc2magnetization}, we show the magnetization divided by the magnetic field as a function of temperature for different values of the chemical potential and magnetic field.
In section \ref{sec:imc:b}, we saw that for the first order deconfinement transition, one can predict the slope of the transition temperature $dT_d/dB$ from the sign of the jump of the magnetization across the transition.
This was studied in more detail in \cite{Ballon-Bayona:2017dvv}\@.
It turns out one can extend the argument to second order transitions as well.
In this case one looks at the difference in entropy densities $\Delta s$, which must be zero along the phase transition line.
Then by using a Maxwell relation and using that $\partial s/\partial T$ is always larger for the higher temperature phase, one arrives at the following formula,
\[
\sign\left(\frac{dT_\chi}{dB}\right) = \sign\left(\frac{dM(T_\chi + \epsilon)}{dT} - \frac{dM(T_\chi - \epsilon)}{dT}\right),
\]
where $\epsilon$ approaches zero from above.
In other words, from the direction of the kinks of $M(T)$, we can infer whether the chiral transition moves up or down.
Indeed, this formula correctly predicts the behavior of the chiral transition temperature, even though many of the kinks in figure \ref{fig:imc:imc2magnetization} are too small to be visible.

In conclusion, we find that in our holographic model, at finite chemical potential inverse magnetic catalysis persists up to some value of the chemical potential, after which magnetic catalysis sets in.
Also, the region of the phase diagram covered by deconfined, chirally broken matter, grows in size at finite $B$\@.
In the next section, we will take a different look at inverse magnetic catalysis.
Instead of applying an external magnetic field, we turn on a different source of anisotropy, and we examine how different this is from applying a magnetic field.
\section{Inverse anisotropic catalysis due to an anisotropy}\label{sec:imc:a}
When we apply an external magnetic field to a quark-gluon plasma, this breaks rotational invariance of the system, as the magnetic field is a vector.
It is then an interesting question to what extent phenomena associated to magnetic fields, like (inverse) magnetic catalysis, are in fact more general phenomena.
In other words, are these phenomena caused by anisotropy being introduced into the system, or does that anisotropy need to specifically be a magnetic field in order for these phenomena to appear?
In this section, which is based on \cite{Gursoy:2018ydr}, we will introduce anisotropy by means of the anisotropy parameter $a$ introduced in section \ref{sec:imc:model}, which is dual to a space-dependent $\theta$-term.
Note that this way of introducing anisotropy is very different from a magnetic field, as $a$ couples to the gluon sector directly, while the magnetic field couples indirectly through the quark sector.

In terms of the `master' model from section \ref{sec:imc:model}, we set the magnetic field $B$ to zero, while also setting $\tilde n = 0$ to ensure a vanishing chemical potential.
In this section, we will not distinguish between $a_\parallel$ and $a_\perp$\@.
This is possible because in this section we set the magnetic field to zero, which means that there is no other source of anisotropy, and we can choose the $x_3$-axis to align with $a$\@.
The choice of potentials is the same as in previous sections, namely those in appendix \ref{sec:potentials:imc}\@.
Most of these potentials have been used in previous sections, and their asymptotics have been motivated in section \ref{sec:intro:bottomup}\@.
This is not true for the $Z$ potential though.
We choose it to be of the same form as what was used in \cite{Gursoy:2012bt,Drwenski:2015sha}, where we have set the constant term \cite{Gursoy:2007cb} to 1, as an overall factor can always be absorbed into $a$\@.
In the IR, we must have $\lambda \sim \lambda^4$ to ensure a linear glueball spectrum for the $0^{-+}$ glueballs \cite{Gursoy:2012bt,Gursoy:2007er,Kiritsis:2009hu,Gursoy:2010fj}, and the choice
\[
Z(\lambda) = 1 + \frac{\lambda^4}{\alpha}
\]
is the simplest choice that satisfies these constraints.
We subsequently choose $\alpha = 10$ \cite{Arean:2016hcs}\@.

Before discussing physical observables, will next discuss the IR asymptotics of the geometry, as the geometry in the presence of $a$ is drastically different from the geometry at vanishing $a$\@.
Next, we will discuss the thermodynamics, and we will conclude by examining various observables, many of which turn out to have interesting properties as a consequence of the different geometry at non-zero $a$\@.
\subsection{IR behavior}\label{sec:imc:IRbehavior}
To examine the IR behavior of the geometry, we choose $f = 1$ so that there will not be a horizon capping the geometry off.
We also set the tachyon $\tau$ to zero, as the tachyon decouples from the other variables in the IR \cite{Jarvinen:2011qe,Arean:2013tja}\@.
Before continuing, it is also convenient to define $\lambda = e^\phi$, $\frac{dr}{dA} e^A = q = -e^p$, and $\widetilde W = W + A$\@.
In terms of these quantities, the equations of motion become
\begin{align*}
8\phi'^2 & = e^{2p} \left(3 a^2 e^{-2 \widetilde W} Z(\phi )-6 V_g(\phi )\right)+36 \left(\widetilde W'+1\right), \\
p' & = \frac{1}{6} \left(-2 e^{2 p} V_g(\phi )+6 \left(\widetilde W'-1\right)+24\right),\\
\widetilde W'' & = -\frac{1}{6} e^{2 p} \left(3a^2e^{-2 \widetilde W} Z(\phi )+2  V_g(\phi ) \left(\widetilde W'-1\right)\right).
\end{align*}
It is not difficult to see that these equations admit fixed point solutions provided that
\begin{equation}
e^{2p_*} V_g(\lambda_*)=9, \qquad 3 a^2 Z(\lambda_*)=2 V_g(\lambda_*),\label{eq:imc:fixedpoint}
\end{equation}
with $\widetilde W$ a constant which we can set to zero, as it can be absorbed into $a$\@.
First note that this solution describes an $\text{AdS}_4 \times \mathbb{R}$ solution, as $e^A \sim 1/r$ and $e^{A+W}$ is constant.
Also, this solution turns out to be unstable.
This can be seen from the second order dilaton EoM:
\begin{align*}
12\phi''\phi' & = \frac{9}{4} e^{2 p} \phi' \left[a^2 e^{-2 \widetilde W} Z'(\phi )-2 V_g'(\phi )\right]\\
& \quad + e^{2 p} V_g(\phi ) \left[e^{2 p}\left(-\frac{3}{2}a^2e^{-2 \widetilde W} Z(\phi )+  V_g(\phi )\right)\right.\\
& \qquad + \left.\left(2 e^{2 p} V_g(\phi ) -18\right)-18\widetilde W'\right].
\end{align*}
The term in the second square brackets vanishes at the fixed point, but the term in the first square brackets does not.
Because of this a small $\phi'$ will cause $\phi''$ to be non-zero as well, which implies an instability.
The stability can be restored by requiring
\[
\frac{d}{d\phi}\log Z(\phi_* ) = 3 \frac{d}{d\phi}\log V_g(\phi_* )
\]
in addition to \eqref{eq:imc:fixedpoint}, which can only be satisfied for specific $a^*$\@.
It turns out that this fixed point is realized by a chirally symmetric vacuum that exists for $x_f = 1/3$, as we will see later.
Note that in this case one has to replace $V_g$ in the analysis by $V_\text{eff} = V_g - x_fV_{f0}$, where $V_{f0}(\lambda) = V_f(\lambda,\tau = 0)$ \cite{Jarvinen:2011qe}\@.

To find the IR behavior satisfied by the chirally broken vacuum, we need to generalize the fixed point to a `slow roll' solution around the fixed point.
In other words, we assume that deviations from the fixed point solution are small and linear in $A$\@.
To do this, we set all second derivatives to zero, which, from the argument above, requires that $Z(\phi) \propto V_g(\phi)^3$, which, for our choice potentials, holds to good enough precision.
Under these assumptions, one can obtain that
\begin{align}
\phi' & = \frac{6 \left(\frac{d}{d\phi}\log Z(\phi )-3 \frac{d}{d\phi}\log V_g(\phi )\right)}{\mathcal{D}},\label{eq:imc:asympt1}\\
\widetilde W' & = \frac{3 \left(\frac{d}{d\phi}\log V_g(\phi )-\frac{d}{d\phi}\log Z(\phi )\right) \left(3 \frac{d}{d\phi}\log V_g(\phi )-\frac{d}{d\phi}\log Z(\phi )\right)}{\mathcal{D}},\label{eq:imc:asympt2}\\
p' & = \frac{3 \frac{d}{d\phi}\log V_g(\phi ) \left(3 \frac{d}{d\phi}\log V_g(\phi )-\frac{d}{d\phi}\log Z(\phi )\right)}{\mathcal{D}},\label{eq:imc:asympt3}\\
e^{2p} & = \frac{18  \left(8-3 \frac{d}{d\phi}\log V_g(\phi ) \frac{d}{d\phi}\log Z(\phi )+2 \left(\frac{d}{d\phi}\log Z(\phi )\right)^2\right)}{ V_g(\phi)\mathcal{D}},\label{eq:imc:asympt4}\\
a^2e^{-2\widetilde W} & = \frac{2 V_g(\phi) \left(16+9 \frac{d}{d\phi}\log V_g(\phi ) \frac{d}{d\phi}\log Z(\phi )-9 \left(\frac{d}{d\phi}\log V_g(\phi )\right)^2\right)}{3 Z(\phi) \mathcal{D}},\label{eq:imc:asympt5}
\end{align}
where
\[
\mathcal{D} = 16 -3 \frac{d}{d\phi}\log V_g(\phi ) \frac{d}{d\phi}\log Z(\phi ) +3 \left(\frac{d}{d\phi}\log Z(\phi ) \right)^2.
\]
Furthermore, by substituting the asymptotic behavior of the potentials $V_g \,\propto\, V_\mathrm{IR} e^{4\phi/3}\sqrt{\phi}$, $Z \,\propto\, Z_\mathrm{IR} e^{4\phi}$, we can obtain analytically that
\begin{equation}
e^A \sim \frac{1}{r}e^{-\sqrt{(\log r)/6}-(\log\log r)/8}, \qquad e^{W+A} = e^{\widetilde W} \sim e^{\sqrt{(2\log r)/3}-(\log\log r)/8},\label{eq:imc:flowasympt1}
\end{equation}
\begin{equation}
\phi \sim \sqrt{(3\log r)/8}.\label{eq:imc:flowasympt2}
\end{equation}
This has been investigavted for more generic potentials in \cite{AriasTamargo:2018pcx}\@.
Indeed, the above asymptotics describe an approximately $\text{AdS}_4 \times \mathbb{R}$ metric.

\begin{figure}[ht]
\centering
\includegraphics[width=0.49\textwidth]{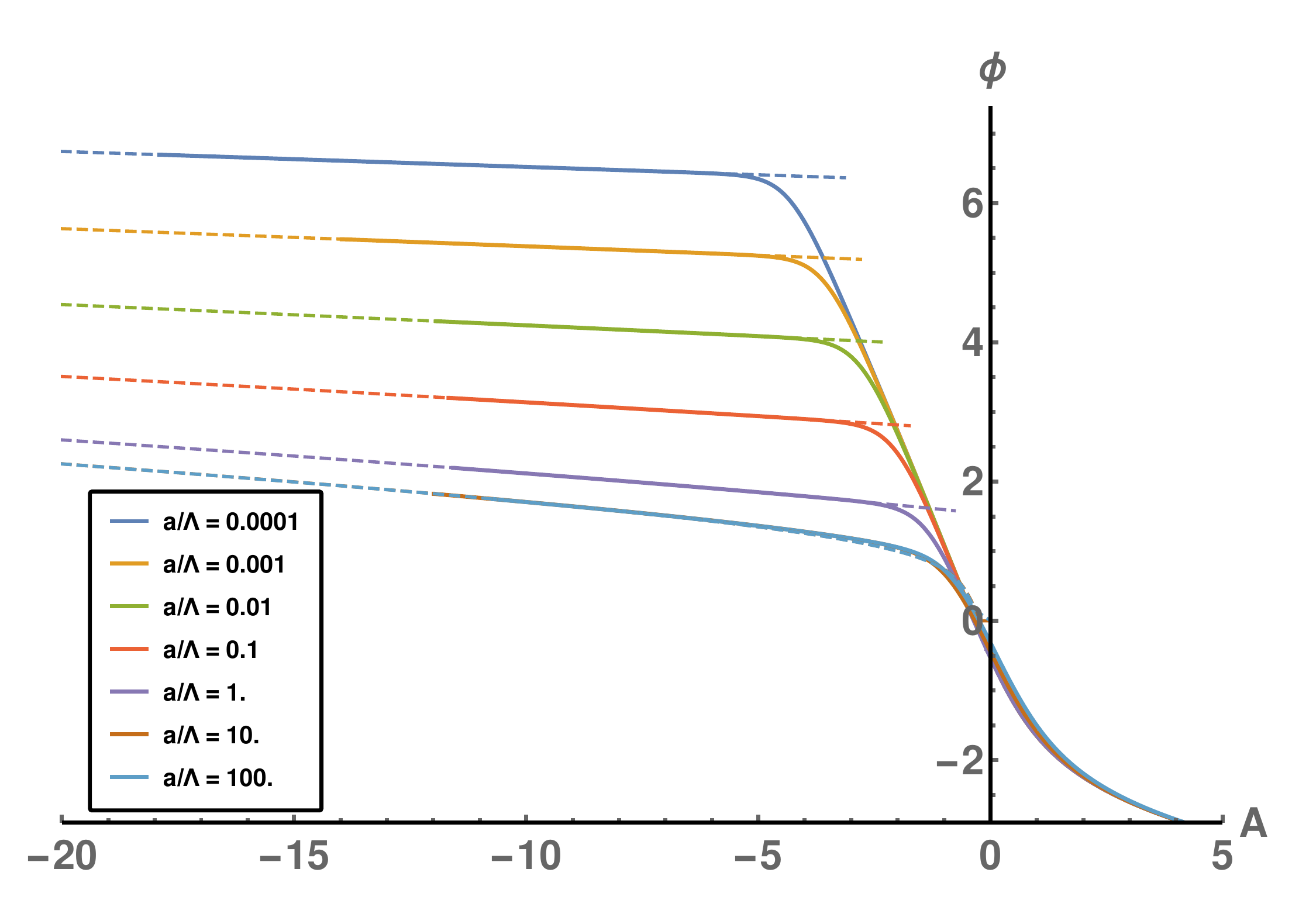}
\includegraphics[width=0.49\textwidth]{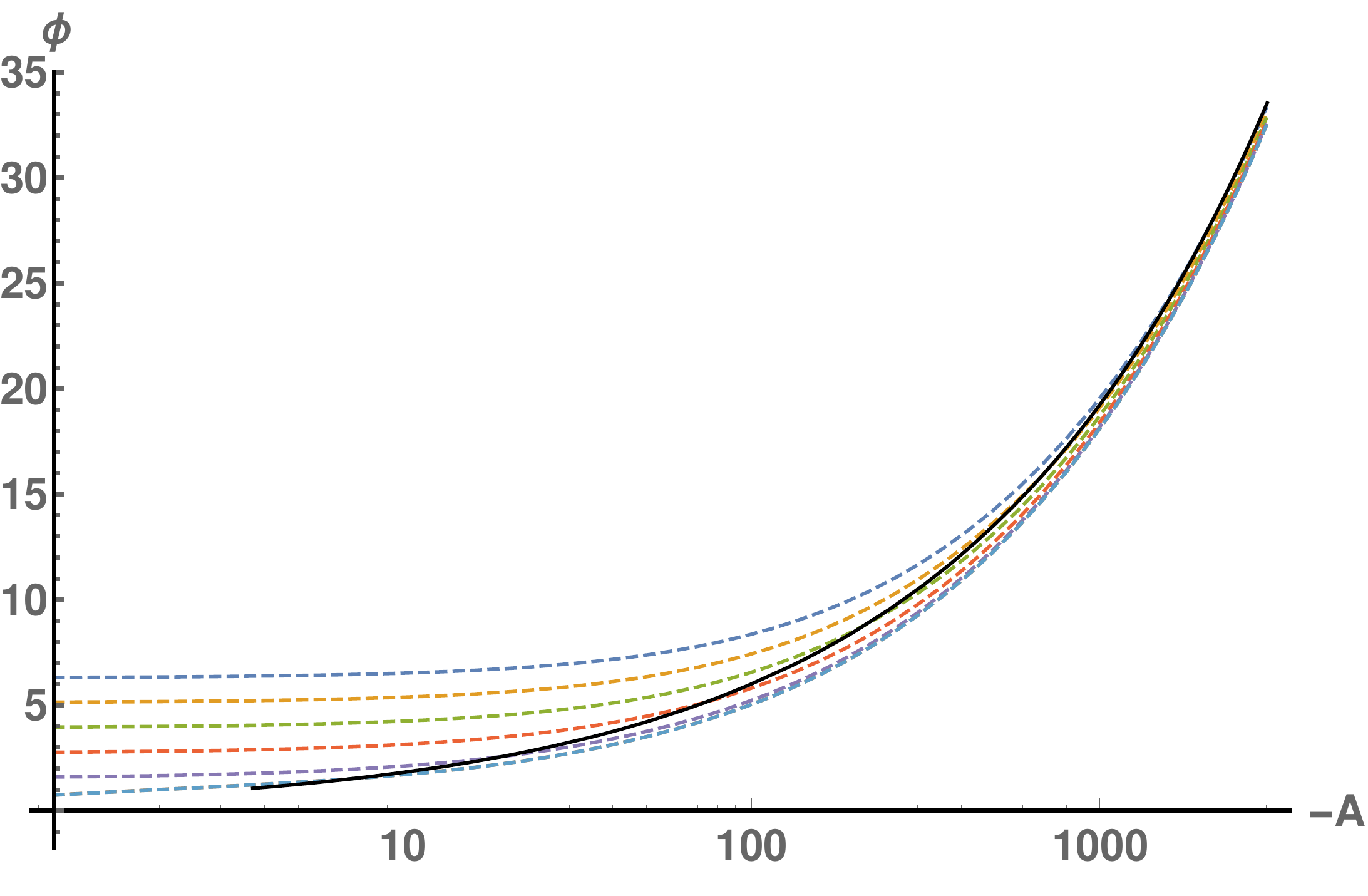}
\caption{\label{fig:imc:imc3RG_flow_x0}Left: RG flow of the coupling $\phi$ in the IR as a function of $A$ for different values of $a$, and $x_f = 0$\@. The solid curves correspond to the exact equations of motion from section \ref{sec:imc:eomandbc}, whereas the dashed curves were obtained by numerically solving (\ref{eq:imc:asympt1}--\ref{eq:imc:asympt5})\@. Right: RG flow of the coupling $\phi$ in the deep IR as a function of $A$ for different values of $a$, and $x_f = 0$, where we compare the numerical solution obtained from (\ref{eq:imc:asympt1}--\ref{eq:imc:asympt5}) to the asymptotic result in (\ref{eq:imc:flowasympt1}--\ref{eq:imc:flowasympt2})\@.}
\end{figure}

To check that these asymptotics indeed describe the IR behavior displayed by the model, we compare the solutions of (\ref{eq:imc:asympt1}--\ref{eq:imc:asympt5}) to the full numerical solution of the equations of motion given in section \ref{sec:imc:eomandbc}\@.
In the left panel of figure \ref{fig:imc:imc3RG_flow_x0}, this comparison is shown for different values of $a$, and for $x_f = 0$\@.
One can see that far enough into the IR, the solution of (\ref{eq:imc:asympt1}--\ref{eq:imc:asympt5}) agrees well with the exact numerical solution of the equations of motion given in section \ref{sec:imc:eomandbc}\@.
In the right panel of figure \ref{fig:imc:imc3RG_flow_x0}, the solution of (\ref{eq:imc:asympt1}--\ref{eq:imc:asympt5}) is compared to the analytical asymptotics obtained in (\ref{eq:imc:flowasympt1}--\ref{eq:imc:flowasympt2})\@.
These can be seen to eventually agree, but only for extremely large values of $-A$\@.
\subsection{Thermodynamics}
\begin{figure}[ht]
\centering
\includegraphics[width=0.9\textwidth]{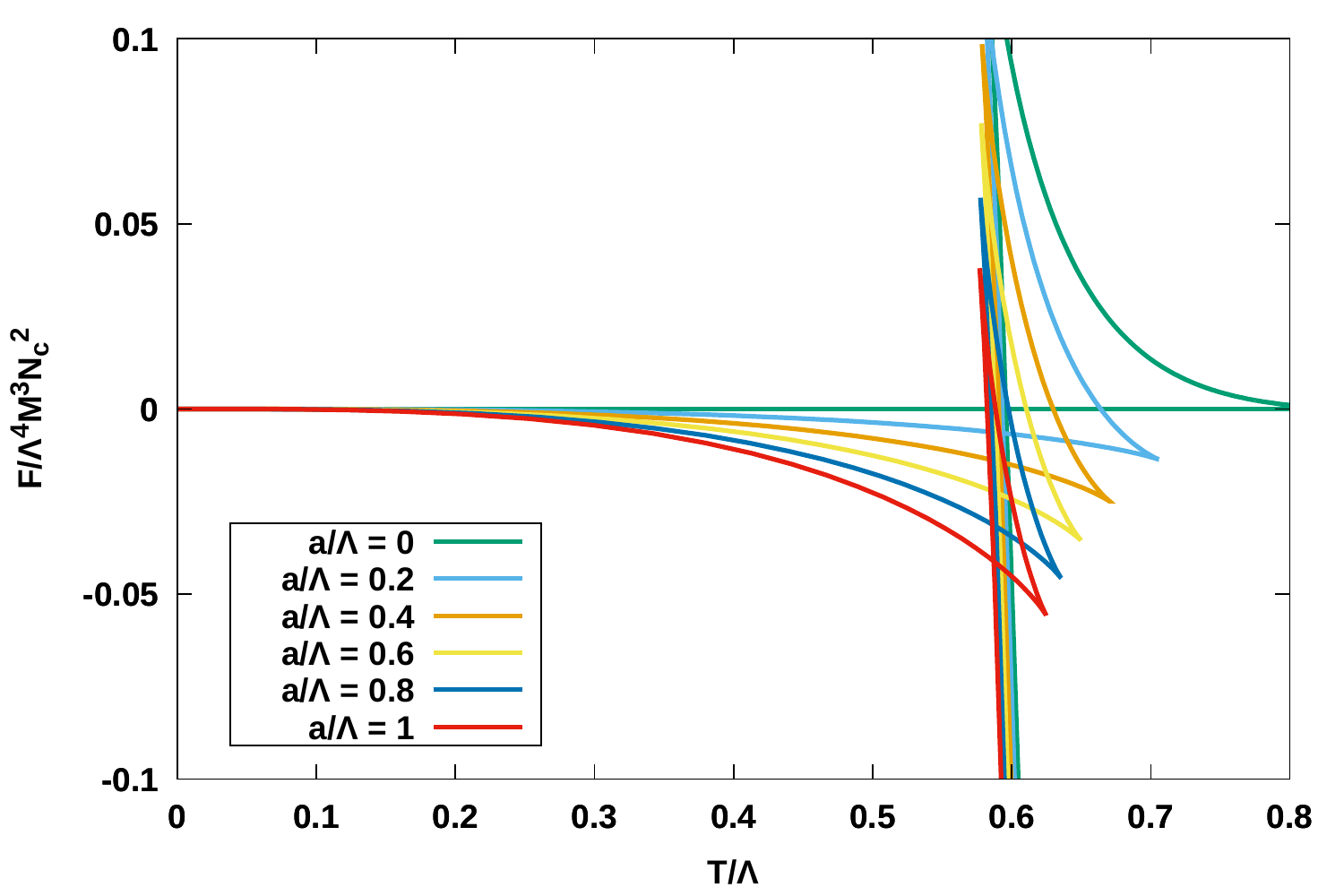}
\caption{\label{fig:imc:imc3freeenergy}Free energy $F$ as a function of $T$ for $x_f = 0$ and different values of $a$\@.}
\end{figure}

We will continue the discussion by examining various thermodynamical properties of the system at different values of $a$\@.
As was mentioned before, some of these properties are substantially altered by the modifications to the IR geometry discussed in section \ref{sec:imc:IRbehavior}\@.
In figure \ref{fig:imc:imc3freeenergy}, we show the free energy as a function of temperature for various values of $a$, and for $x_f = 0$\@.
As different values of $x_f$ show the same qualitative behavior, we will show only the result for $x_f = 0$\@.
One can see that generally there are three branches of black hole solutions: one small black hole branch (labeled $I$ in figure \ref{fig:imc:imc3freeenergy}), one big black hole branch (labeled $II$ in figure \ref{fig:imc:imc3freeenergy}), and an unstable branch (unlabeled).
If the anisotropy parameter $a$ vanishes, the result qualitatively agrees with earlier results \cite{Gursoy:2008za,Gursoy:2009jd}\@.\footnote{The result is not exactly the same because the potentials are different.}
In this case, the small black hole branch in fact has infinitesimal area, and corresponds to the horizonless thermal gas solution.
As one turns on a non-trivial $a$, however, the behavior is qualitatively different, as a black hole solution now always dominates in region $I$\@.
This is due to the $\text{AdS}_4 \times \mathbb{R}$ geometry in the IR\@.

\begin{figure}[ht]
\centering
\includegraphics[width=0.85\textwidth]{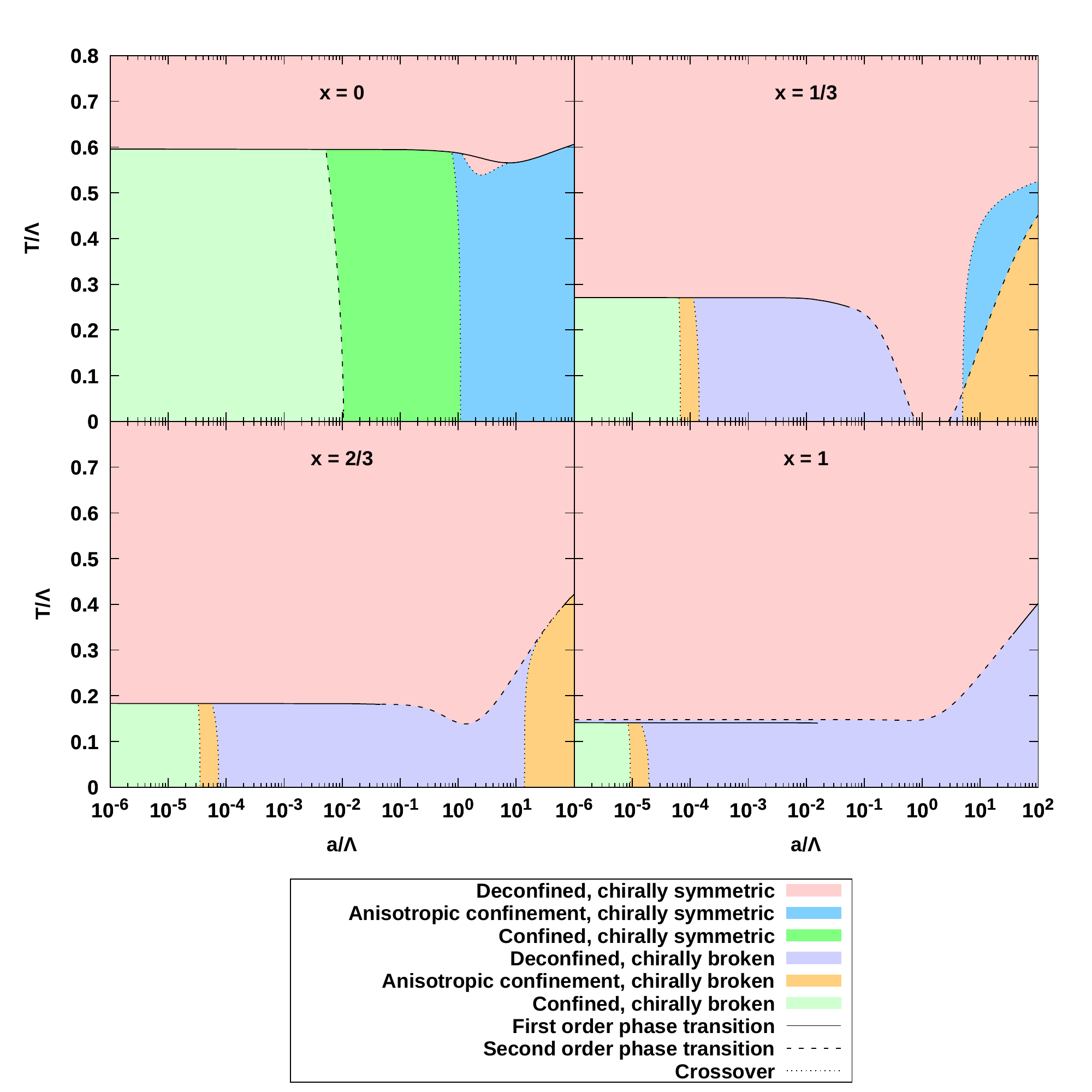}
\caption{\label{fig:imc:imc3phases}Phase diagram as a function of $T$ and $a$ for several different $x_f$\@. Note that $x_f$ is labeled $x$ in the figure.}
\end{figure}

In figure \ref{fig:imc:imc3phases}, we show the phase diagram as a function of temperature and anisotropy parameter for different values of $x_f$\@.
In this figure, as in the rest of this section, we define a phase to be confining if the quark-antiquark potential has a linear branch.
As the system is anisotropic, this need not happen in every direction.
This leads to a phase with anisotropic confinement, which is confining in the direction parallel to the magnetic field, and deconfined perpendicular to it.
Also note that with this definition of confinement, the transition between a confining geometry and a deconfined geometry is a crossover.
We will discuss the confinement properties of the system in more detail in section \ref{sec:imc:imc3observables}, but it can already be seen using this definition that for $x_f \gtrsim 1/3$, the geometries become deconfined, i.e.~the quark-antiquark potential has no linear branch.

Another interesting observation is that the first order phase transitions which exist at $a = 0$ disappear at sufficiently large $a$ for $x_f \gtrsim 1/3$\@.
In the cases of $x_f = 1/3$ and $x_f = 2/3$, the first order transition turns into a second order one, while for $x_f = 1$ the first order transition ends in a critical point.
For large values of $a$, in the case of $x_f = 1$, a first order transition reappears.
The numerics were not stable enough to determine whether this also happens for the other values of $x_f$ considered.
Also, note that for $x_f = 1/3$, the chiral transition decreases to zero temperature around $a/\Lambda \sim 1$, and at some larger $a$ the chiral transition reappears.
In between these two values, the vacuum is the chirally symmetric vacuum mentioned above.

The last feature that can be seen in figure \ref{fig:imc:imc3phases} is the behavior of the chiral transition.
Comparing to figure \ref{fig:imc:imc1phases}, we can see that in both cases the chiral transition temperature first decreases as a function of $a$ and $B$, respectively.
After this decrease, an increase follows in both cases.
This tells us that a magnetic field and an anisotropy have similar effects on the resulting chiral transition temperature.
We will examine the chiral condensate explicitly in section \ref{sec:imc:imc3observables}\@.

\begin{figure}[ht]
\centering
\includegraphics[width=0.9\textwidth]{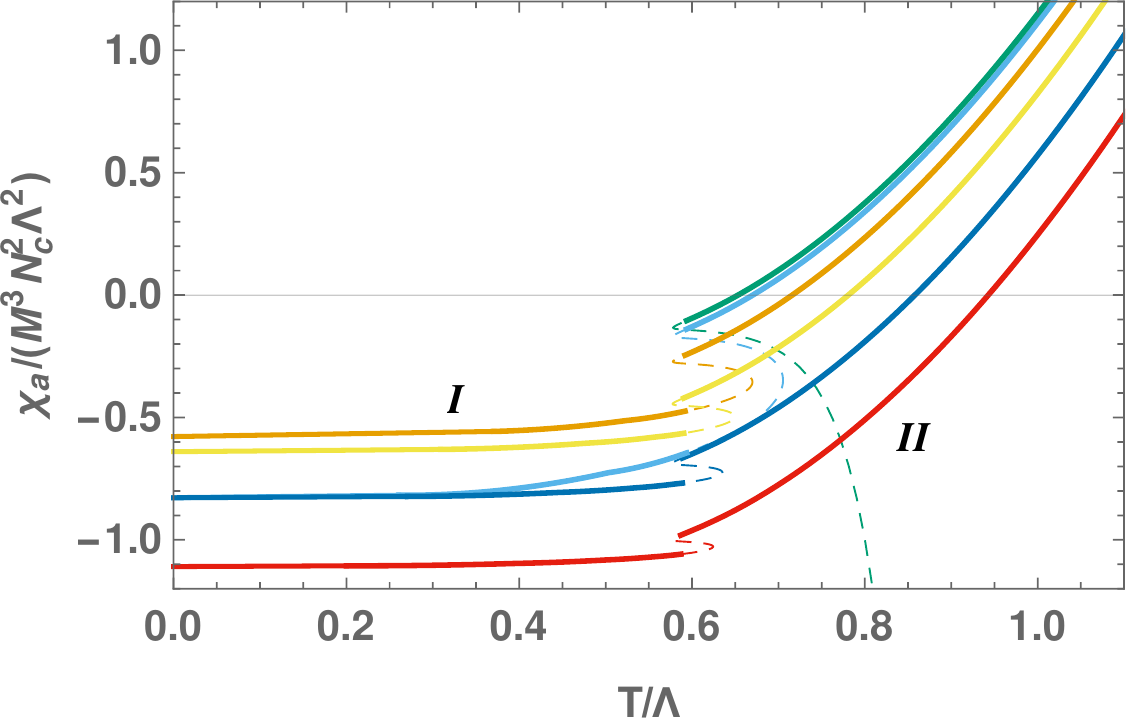}
\caption{\label{fig:imc:imc3chi_a_vs_T}Anisotropic susceptibility $\chi_a$ as a function of $T$ for $x_f = 0$ and different values of $a$\@. The color coding is the same as in figure \ref{fig:imc:imc3freeenergy}\@.}
\end{figure}

Lastly, in figure \ref{fig:imc:imc3chi_a_vs_T} we show the `anisotropic susceptibility' $\chi_a$, which we define by
\[
\chi_a = \frac{M_a}{a},
\]
with $M_a$ as defined in section \ref{sec:imc:observables}\@.
This definition agrees with the standard definition of a susceptibility in the limit where $a \rightarrow 0$\@.
In the same way as what we did in the previous two sections, we can derive for the first order transition that
\[
\frac{dT_c}{da} = -\frac{\Delta M_a}{\Delta s}.
\]
Noting that the higher temperature phase at the phase transition always has the largest entropy, we can read off from the jump of $\chi_a$ whether $T_c$ is increasing or decreasing with $a$\@.
Indeed, this agrees between figures \ref{fig:imc:imc3phases} and \ref{fig:imc:imc3chi_a_vs_T}\@.
\subsection{Observables}\label{sec:imc:imc3observables}
\begin{figure}[ht]
\centering
\includegraphics[width=0.85\textwidth]{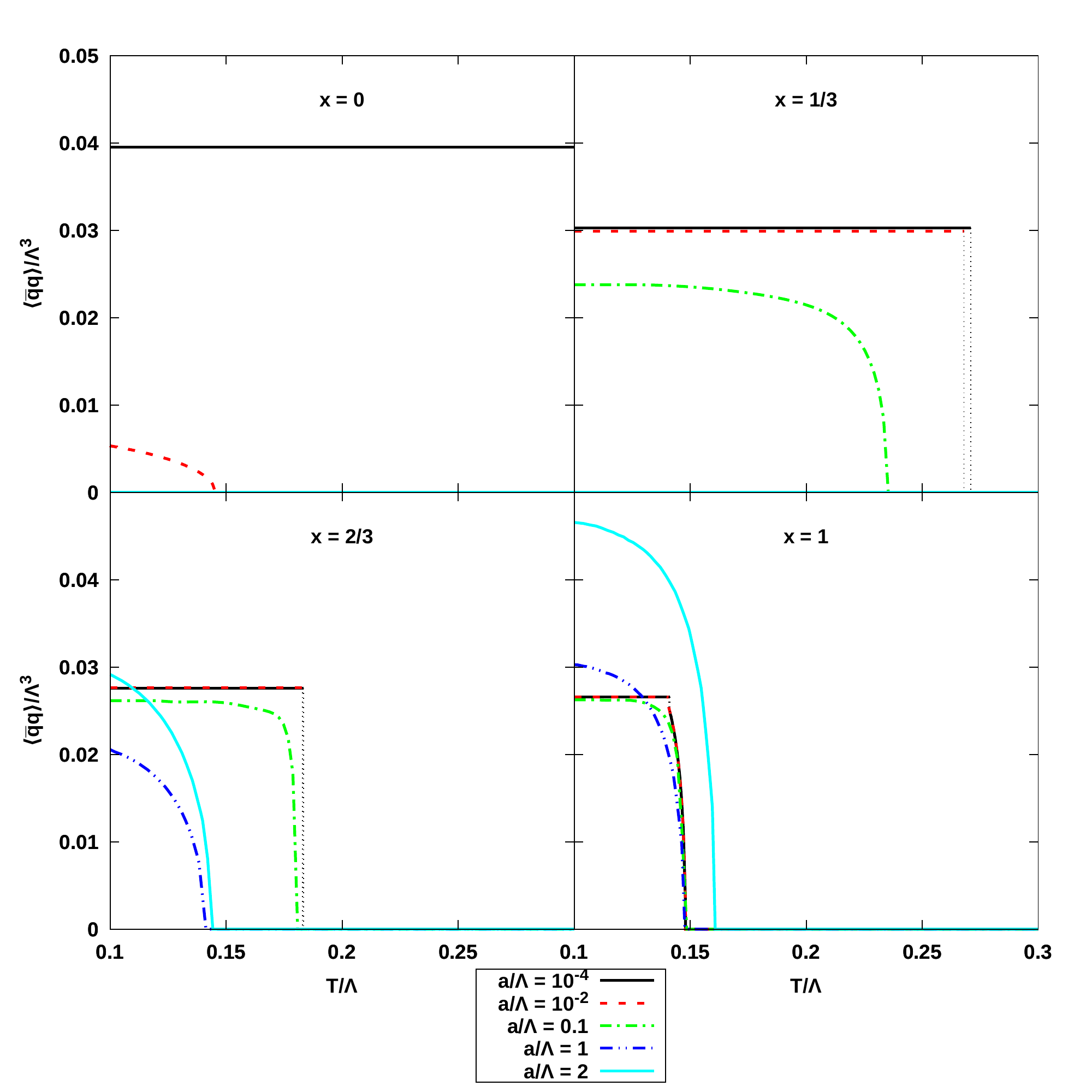}
\caption{\label{fig:imc:imc3qqbar}Chiral condensate $\langle\bar qq\rangle$ as a function of temperature for different values of $a$ and $x_f$\@. Note that $x_f$ is labeled $x$ in the figure.}
\end{figure}

\begin{figure}[ht]
\centering
\includegraphics[width=0.9\textwidth]{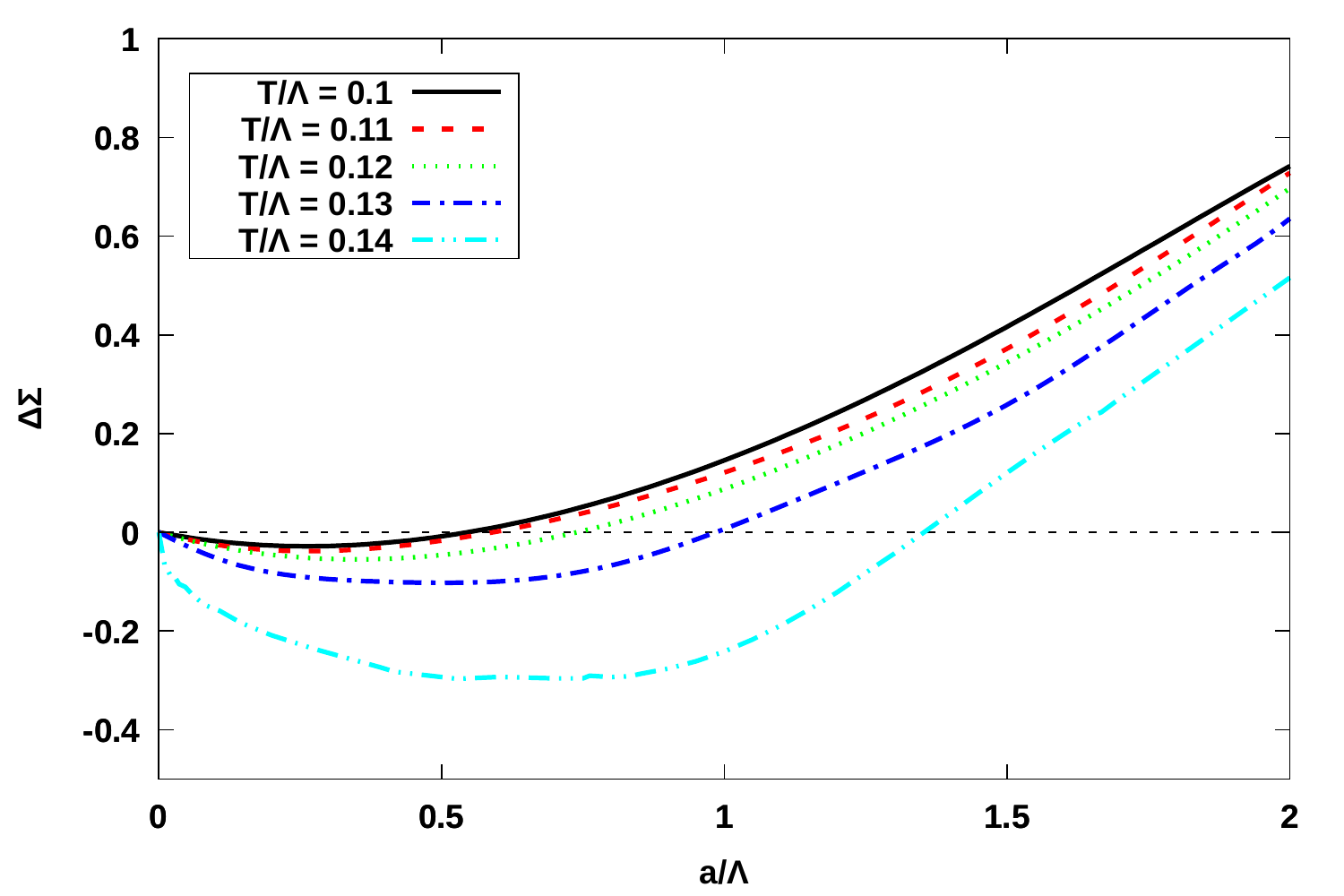}
\caption{\label{fig:imc:imc3DeltaSigma}$\Delta\Sigma$ as defined in \eqref{eq:imc:deltasigmaanis}, as a function of $a$ for various fixed $T$, with $x_f = 1$\@.}
\end{figure}

Next, we examine in some more detail the chiral condensate, after which we will look at some more consequences of the IR geometry.
The chiral condensate is shown in figure \ref{fig:imc:imc3qqbar}\@.
Just as in the previous sections, the chiral condensate decreases with temperature.
Also, at $x_f = 0$ the condensate always decreases with $a$, while for larger $x_f$, the condensate first decreases as a function of $a$, and then increases again.
The latter behavior can be studied in more detail by defining
\begin{equation}
\Sigma(T,a) = \frac{\langle\bar qq\rangle(T,a)}{\langle\bar qq\rangle(0,0)}, \qquad \Delta\Sigma(T,a) = \Sigma(T,a) - \Sigma(T,0),\label{eq:imc:deltasigmaanis}
\end{equation}
in analogy with (\ref{eq:imc:sigma}--\ref{eq:imc:deltasigma})\@.
This quantity is shown as a function of $a$ for various fixed temperatures in figure \ref{fig:imc:imc3DeltaSigma}, where we keep $x_f = 1$\@.
For all values shown, one can see first a decrease in $\Delta\Sigma$, which is then followed by an increase.
In particular, the decrease in $\Delta\Sigma$ is greatest for temperatures just below the chiral transition.
This gives evidence that indeed an anisotropy as introduced in this section leads to similar physical consequences for the chiral condensate, leading to the claim that indeed the cause of inverse magnetic catalysis may be the presence of anisotropy induced by the magnetic field.

\begin{figure}[ht]
\centering
\includegraphics[width=0.49\textwidth]{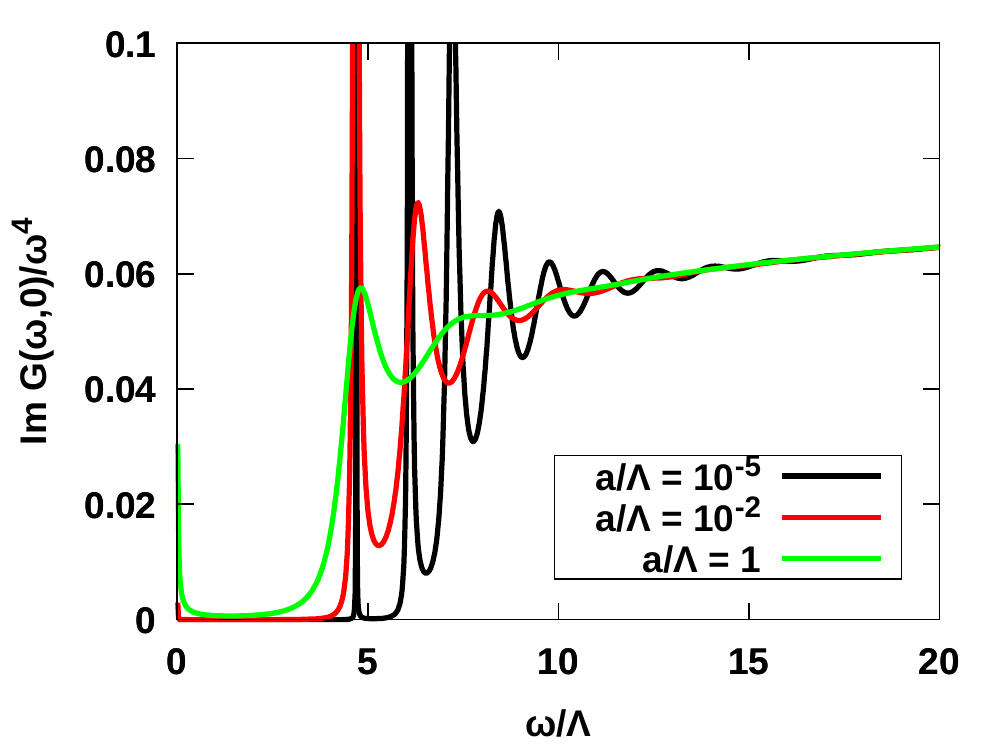}
\includegraphics[width=0.49\textwidth]{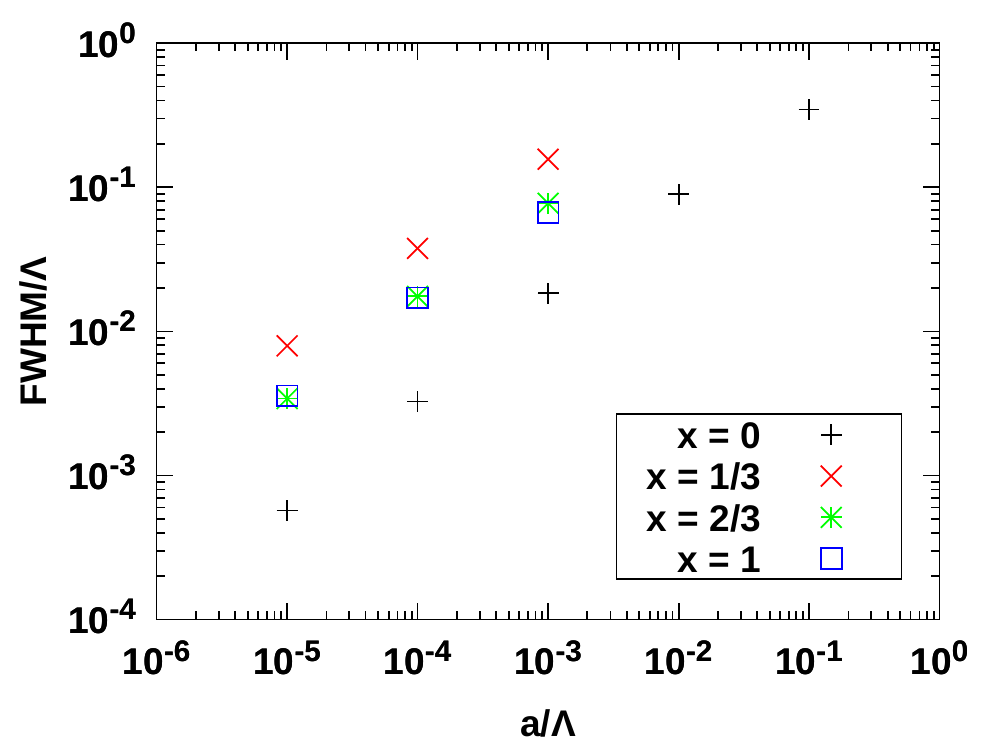}
\caption{\label{fig:imc:glueballspectrum}Left: spectral density for helicity 2 glueballs at $x_f = 0$ for different $a$\@. Right: widths of the lowest helicity two glueball states as a function of $a$ for various different $x_f$, which is labeled $x$ in the figure. The width is defined as the full width at half maximum.}
\end{figure}

Next, we will investigate the spectrum of helicity 2 glueballs.
As was mentioned in section \ref{sec:imc:observables}, we can infer from the Schr\"odinger potential \eqref{eq:imc:schrodinger} whether the spectrum is discrete or continuous.
For the geometry without the anisotropy, the Schr\"odinger potential diverges towards the IR, creating a discrete spectrum.
In the presence of $a$ though, we have $V_S(r) \sim 2/r^2$ due to the $\text{AdS}_4 \times \mathbb{R}$ geometry, leading to a continuous spectrum.
Such a dissociation of mesons due to anistropy has been studied earlier in holographic models in \cite{Chernicoff:2012bu,Avila:2016mno}\@.
In the left panel of figure \ref{fig:imc:glueballspectrum}, the resulting spectral density is shown for $x_f = 0$, normalized by $\omega^4$\@.
It can be seen that for small $a$, the glueballs have small but finite widths, whereas they would be described by delta functions at $a = 0$\@.
As one increases $a$, the widths grow larger, and the peaks melt.
Still, the mass gap which is exact at $a = 0$ is still approximately present even at $a = 1$\@.
Note that the peak appearing around $\omega = 0$ arises because due to the $\text{AdS}_4 \times \mathbb{R}$ geometry changing the power law behavior around the origin to $G \propto \omega^3$\@.
In the right panel of figure \ref{fig:imc:glueballspectrum}, the widths of the first glueball peak are shown for different values of $a$ and $x_f$\@.
The widths can be seen to be getting wider with a power-law dependence on $a$\@.

Since the helicity 2 glueballs are not stable for $a \neq 0$, it is interesting to examine another quantity related to the stability of mesons, namely the quark-antiquark potential $V$\@.
This observable quantifies the potential energy of a quark-antiquark pair at some distance, where we assume the quarks to be infinitely heavy.
If the quark-antiquark potential grows infinitely with distance, it is impossible to pull the quarks apart, and in this case they are confined into mesons.\footnote{In reality, a new quark-antiquark pair will nucleate between the original quarks, and one ends up with two mesons. However, if we assume that no light enough quarks exist for this to happen, it is impossible to pull the heavy quarks apart.}
However, if the potential is bounded from above, at some distance the force between the quarks will vanish, and in this case it is possible to pull the quarks apart.

\begin{figure}[ht]
\centering
\includegraphics[width=0.9\textwidth]{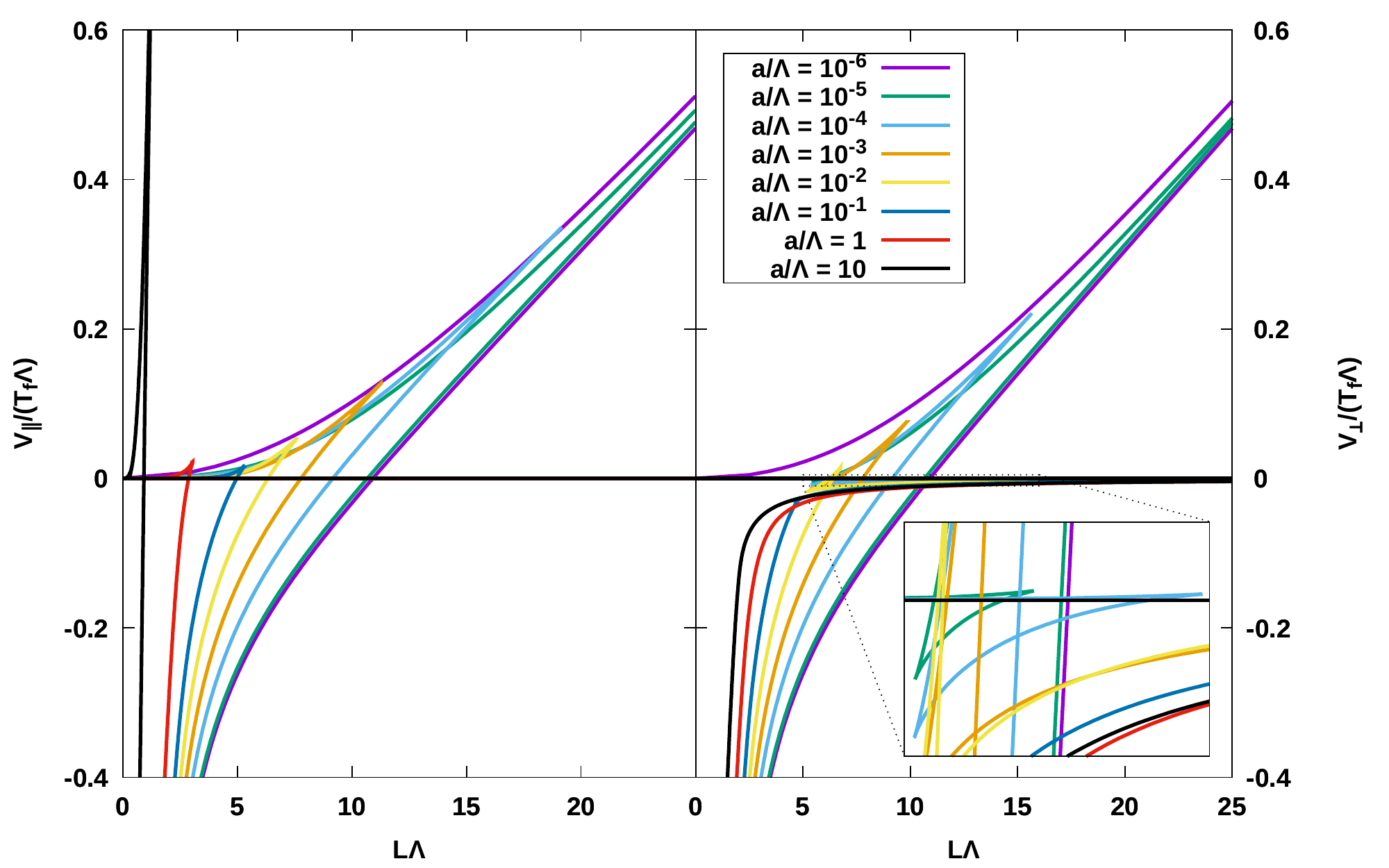}
\caption{\label{fig:imc:imc3qqpotentials}Quark-antiquark potential parallel and perpendicular to the anisotropy, denoted $V_\parallel$ and $V_\perp$, respectively, for different $a$ and $x_f = 1$\@. Note that also unstable branches are shown. Note that in the parallel case, all shown solutions have a branch equal to zero.}
\end{figure}

In \cite{Gursoy:2007er,Kinar:1998vq}, a simple criterion was found to describe whether there is a linearly growing branch of the quark-antiquark potential which, in figure \ref{fig:imc:imc3phases}, is used as the criterion to label confinement.
In our case, this implies that if $A_S \equiv A + \frac{2}{3}\log\lambda$ has a local minimum, then the quark-antiquark in the direction perpendicular to the anistropy $V_\perp$ will have a linear branch.
Similarly, if $A_S + W/2$ has a local minimum, then the quark-antiquark potential in the direction of the anisotropy $V_\parallel$ will have a linear branch.
In figure \ref{fig:imc:imc3qqpotentials}, we show the entire quark-antiquark potential for different values of $a$ and with $x_f = 1$\@.
One can see that in both the parallel and perpendicular cases, the solutions have multiple branches, with in particular the perpendicular case showing multiple swallowtail structures.
Of these branches, we are supposed to pick the smallest one, which is stable.
A consequence of this is that even if $A_S$ resp.~$A_S + W/2$ have local minima, the linear branch which is then guaranteed to exist may not be stable.
As with the previous quantities we examined, the reason behind this is the $\text{AdS}_4 \times \mathbb{R}$ geometry, which causes the $A_S$ minimum to be only a local one.
However, even if the linear branch is not stable, for small enough $a$ there is still a large potential barrier for the string to decay to the stable configuration.
This may indicate that the decay is not very fast, in line with the observations from the glueball spectrum that the glueballs are stable but long-lived.

\begin{figure}[ht]
\centering
\includegraphics[width=0.9\textwidth]{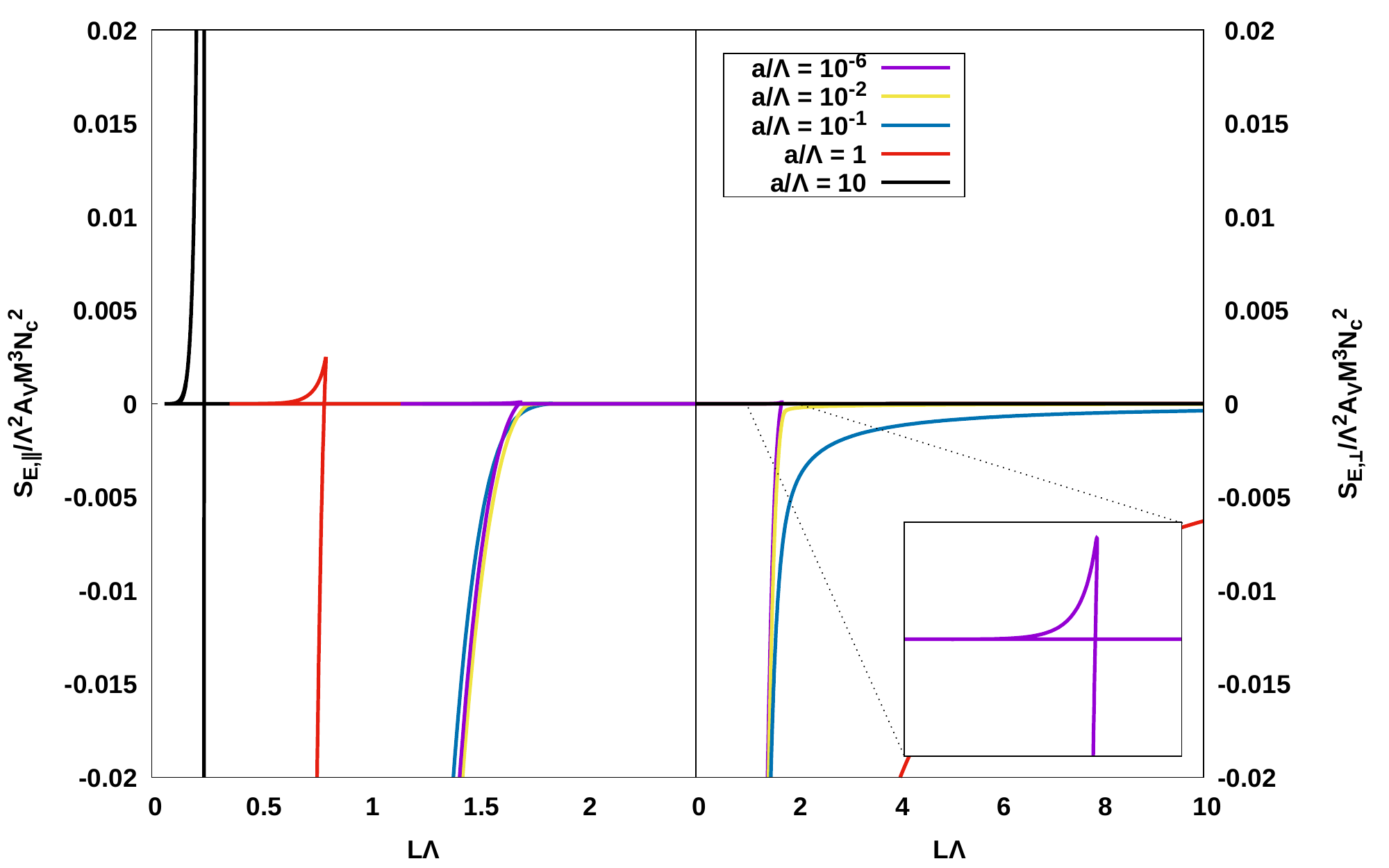}
\caption{\label{fig:imc:imc3entanglemententropy}Entanglement entropy $S_{E,\parallel}$ and $S_{E,\perp}$ for regions $A$ and $B$ defined in section \ref{sec:imc:observables}, for different values of $a$ and with $x_f = 1$\@.}
\end{figure}

The last observable we will examine is the entanglement entropy of regions $A$ and $B$ as defined in section \ref{sec:imc:observables}\@.
The result is shown in figure \ref{fig:imc:imc3entanglemententropy}\@.
First, note that there are multiple branches, the lowest of which is the stable one.
Up to about $a/\Lambda \sim 0.1$, the results are similar to isotropic results \cite{Klebanov:2007ws,Dudal:2018ztm}\@.
For large values of $L$, the result depends strongly on the orientation of the entangling region even when $a$ is small.
Specifically, for $S_{E,\perp}$, the curve of the connected surfaces always goes to $(L = 0, S_{E,\perp} = 0$, where the branch of disconnected solutions with $S_{E,\perp} = 0$ connects.
However, for $S_{E,\parallel}$ the point $(L = 0, S_{E,\parallel} = 0)$ is never reached.
Instead, the swallowtail structure present without anisotropy quickly gets smaller, and then vanishes.

As $a/\Lambda \gtrsim 1$, the result becomes very different for the different entangling regions.
$S_{E,\parallel}$ crosses zero for a smaller value of $L$, whereas $S_{E,\perp}$ moves in the opposite direction.
Such behavior has also been observed for a magnetic field \cite{Dudal:2016joz}\@.
Overall, the dependence of the entanglement entropy on the anisotropy is less pronounced than that of the quark-antiquark potential.
A possible explanation for this is that the characteristic scale of the entanglement entropy is $L\Lambda \sim 1$, causing the Ryu-Takayanagi surface to remain relatively close to the boundary, whereas the characteristic scale of the quark-antiquark potential is roughly $L\Lambda \sim 10$\@.
As the modification to the geometry due to the anisotropy is most pronounced in the IR, this implies that the quark-antiquark is more drastically modified even for small $a$, while the entanglement entropy needs rather large values of $a$ for a modification to become pronounced.
\section{Interplay between magnetic field and an\-isotropy}\label{sec:imc:ba}
In the final section of this chapter, we will investigate the interplay of both sources of anisotropy introduced before.
From the discussion in section \ref{sec:imc:model}, this gives us two options, namely to have $a$ parallel to $B$, or to have it perpendicular to $B$\@.
Recall that we were unable to put the two anisotropies at an arbitrary angle, but this is likely not an issue, as one expects to be able to already infer a lot of information from these two cases.

In terms of the `master' model introduced in section \ref{sec:imc:model}, we set $\tilde n = 0$ to ensure $\mu = 0$, and we allow $\tilde B$, $\tilde a_\parallel$ and $\tilde a_\perp$ to be non-zero.
However, $\tilde a_\parallel$ and $\tilde a_\perp$ can not both be non-zero at the same time.
We also set $x_f = 1$ throughout this section.
The potentials will be kept the same as in the previous sections, i.e.~the ones from appendix \ref{sec:potentials:imc}\@.
We do re-examine the $c$-parameter from the $w$-potential again though.
In the previous sections, we took $c = 0.4$ as this matches lattice results for the chiral transition reasonably well, while keeping $c = \mathcal{O}(1)$\@.
Even lower values match the chiral transition results better, but this may result in too large of a departure from $c = \mathcal{O}(1)$\@.

\begin{figure}[ht]
\centering
\includegraphics[width=0.49\textwidth]{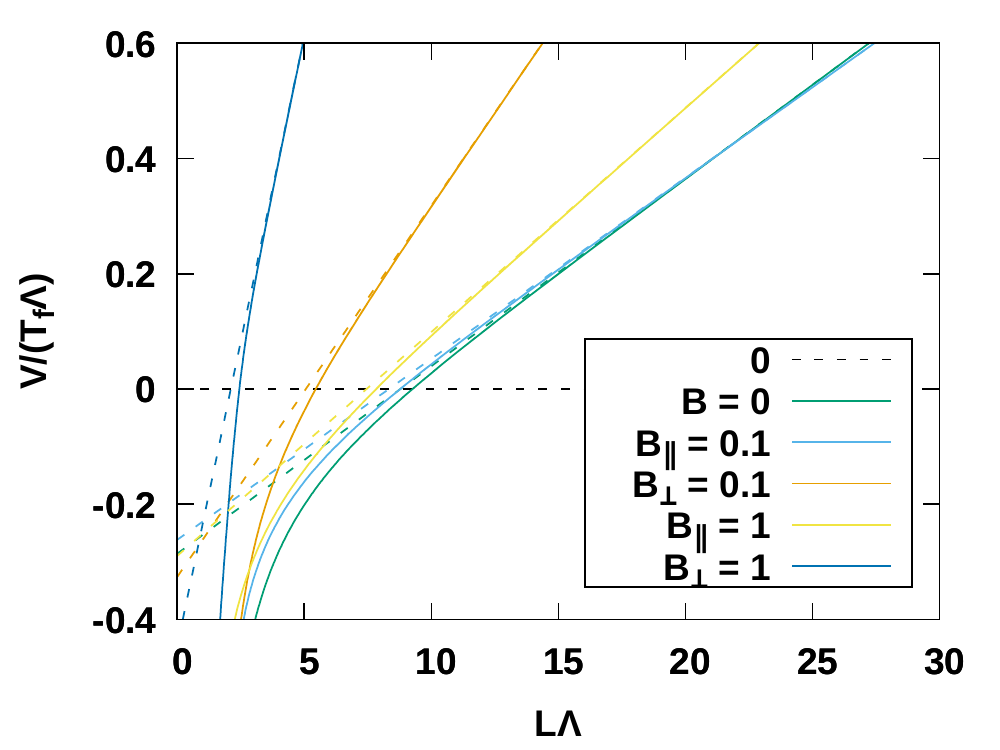}
\includegraphics[width=0.49\textwidth]{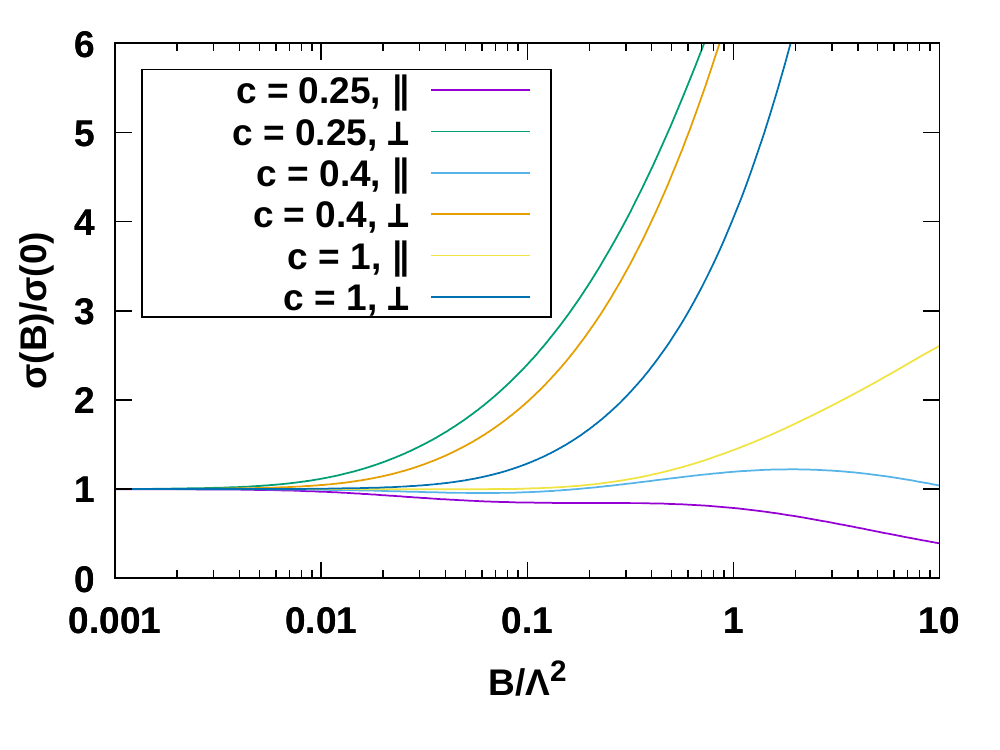}
\caption{\label{fig:imc:imc4sigmavsB}Left: Quark-antiquark potential at zero temperature for various values of $B$, with $a = 0$, $c = 0.4$ and $x_f = 1$, where the $\parallel$ and $\perp$ signs signify whether the quarks are separated parallel to the magnetic field or perpendicular to it, respectively. The dashed lines indicate the asymptotic linear behavior of the various curves at large separation. Right: String tension between the quark-antiquark pair at asymptotic distances, as a function of $B$ for different $c$, with $T = 0$, $a = 0$ and $x_f = 1$, in both the parallel and perpendicular cases.}
\end{figure}

To determine whether lower values of $c$ match other lattice results better as well, let us first examine the quark-antiquark potential at zero temperature for $x_f = 1$, $a = 0$ and $c = 0.4$, which is shown in the left panel of figure \ref{fig:imc:imc4sigmavsB}\@.
Firstly note that since we put $a = 0$, the linear branch exists for any $B$, and is always the stable branch.
One can see that in this case, at $B = 0.1$, the quark-antiquark potential gets steeper perpendicular to the magnetic field, while parallel to it the potential gets slightly shallower.
For larger values of $B$ the quark-antiquark potential gets steeper in both directions.

As the quark-antiquark potential has a linear branch in all the cases considered here, one can also compute the slope of the asymptotic linear behavior.
This slope is called the `string tension', and is shown in the right panel of figure \ref{fig:imc:imc4sigmavsB}, again for $T = 0$, $a = 0$ and $x_f = 1$, but now as a function of $B$ and for different values of $c$\@.
One can see that as $c$ decreases, the perpendicular string tension increases faster as a function of $B$, whereas the parallel string tension increases more slowly, and even decreases as a function of $B$ for a part of the range of magnetic field shown for $c = 0.4$, and for all of the range shown for $c = 0.25$\@.
In \cite{Bonati:2014ksa}, it was found on the lattice that the perpendicular string tension increases as a function of $B$, while the parallel string tension decreases.
This is most in line with the result for the holographic model for $c = 0.25$, and therefore in the rest of the section, we will show results for both $c = 0.25$ and $c = 0.4$\@.

\begin{figure}[ht]
\centering
\includegraphics[width=0.49\textwidth]{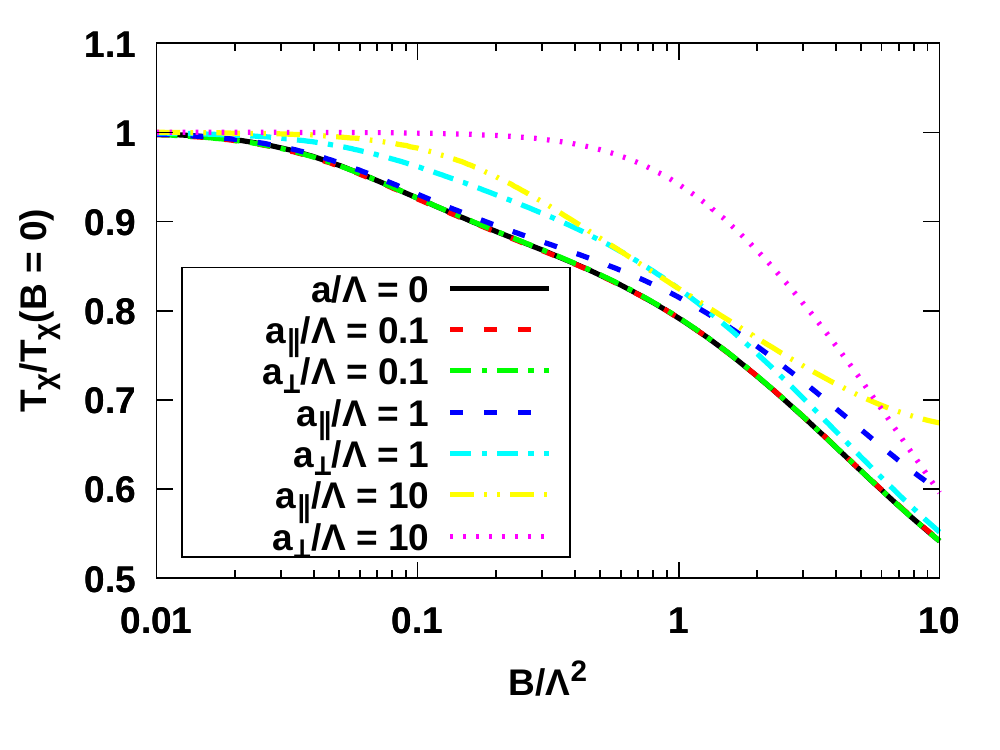}
\includegraphics[width=0.49\textwidth]{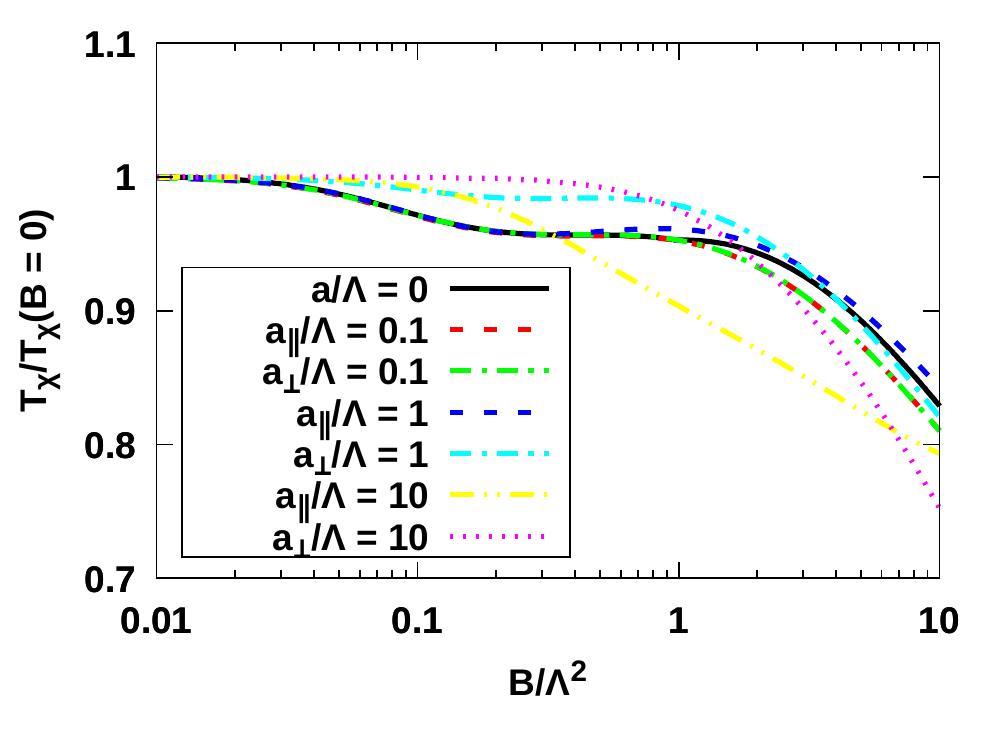}
\caption{\label{fig:imc:imc4chiraltransitions}Chiral transition temperature as a function of $B$ for different $a_\perp$, $a_\parallel$, with $x_f = 1$ and $c = 0.25$ (left) and $c = 0.4$ (right)\@.}
\end{figure}

Next, in figure \ref{fig:imc:imc4chiraltransitions}, we will examine the behavior of the chiral transition temperature as a function of $B$ for different values of $a$, where we show the result for both $a_\parallel$ and $a_\perp$\@.
At the values of $a$ shown in figure \ref{fig:imc:imc4chiraltransitions}, the chiral transition is the only feature in the phase diagram.
Therefore we do not show a separate phase diagram.
We can see several qualitative features from the chiral transitions.
Firstly, we observe that around $a \sim B$, there is equality between $T(a_\perp)$ and $T(a_\parallel)$\@.
Also, $a$ seems to effectively decrease the value of $B$ for $a > B$, which seems to be most pronounced for $a_\parallel$\@.

In conclusion, in this chapter we constructed a bottom-up holographic theory based on V-QCD which includes two different sources of anisotropy, namely the magnetic field $B$ and the axion $a$\@.
We were able to show that in this theory at zero chemical potential and zero $a$, it is possible to obtain inverse magnetic catalysis.
Furthermore, here are two contributions to the chiral condensate which have competing effects, in support of the analysis done on the lattice in \cite{Bruckmann:2013oba,Bruckmann:2013ufa}, namely that inverse magnetic catalysis is caused by `sea' quark effects, which in our model correspond to the backreaction of the gluon sector onto the flavor sector.
Next, we generalized the discussion to finite chemical potential, thereby for the first time giving insight into the fate of inverse magnetic catalysis at finite $\mu$\@.
Also, we found evidence to support the claim first made in \cite{Giataganas:2017koz} that IMC is a more general effect caused by the anisotropy induced by the magnetic field.
Lastly, we investigated the interplay between both sources of anisotropy, yielding interesting dependence on the angle between both sources.

%% file: chapters/holographicns.tex
In the study of neutron stars, some of the most basic questions are the following:
\begin{itemize}
\item Given the mass, what is the radius of the star?
\item What is the maximum mass?
\item How large is the tidal deformability?
\item What is the post-merger gravitational wave spectrum?
\item Do neutron stars contain deconfined quark matter cores?
\end{itemize}
As was discussed in the introduction, these quantities depend on two things.
The first is general relativity, which we understand well, and of which we believe it holds in the regime of neutron stars to a high degree of confidence.
The second is the neutron star equation of state, something we cannot compute from first principles in a controlled way.
This provides us with both a challenge and an opportunity.
On the one hand we are obviously challenged to find what the true EoS is.
On the other hand, neutron stars provide us with a novel laboratory to learn something about QCD\@.
As more data will be collected in the future, some candidate equations of state will be ruled out, while others will remain compatible with observations.
Subsequently one can start looking into the assumptions that went into these equations of state, to hopefully learn something about which of these assumptions are viable, and which are not.
For this to work however, it is important that a wide range of such models become available.

In this chapter, we will construct such an equation of state from holography, which hence takes non-perturbative effects into account.
In \cite{Annala:2017tqz,Jokela:2018ers,Chesler:2019osn}, holography has been used to model the deconfined phase which one expects at extreme densities, while these models have continued to rely on more traditional approaches for the nuclear matter phase, which is expected to make up the bulk of the neutron star.
In this chapter, we in addition also model most of the nuclear matter phase using holography.
The holographic model we will use for this purpose is, as in the previous chapter, V-QCD\@.
For the choice of potentials, however, we make a different choice from the previous chapter.
In \cite{Jokela:2018ers}, observables at vanishing baryon chemical potential were compared to lattice data, and the potentials were tuned to reach a good agreement.
Recall also from the introduction that various qualitative features of QCD are matched by a judicious choice of asymptotic behavior of the potentials, such as that chosen in V-QCD\@.
This means that by choosing a set of potentials from \cite{Jokela:2018ers}, we can do the computations in this chapter with a holographic model of QCD which matches real QCD to the largest extent possible at the time of this writing.

In particular, we will use potentials 7a from \cite{Jokela:2018ers}, which for convenience can be found in appendix \ref{sec:potentials:baryons}\@.
In upcoming work which is not part of this thesis, the analysis from this chapter will be repeated for several other possible choices to investigate to what extent our prediction from this chapter is a generic prediction from V-QCD as opposed to one for this specific set of potentials.

By the procedure outlined in this chapter it turns out to indeed be possible to construct an equation of state, which, in some sense, one can then see as a sort of `extrapolation' of lattice data to the neutron star regime.
Moreover, it turns out that the EoS one obtains in this way is compatible with all presently known constraints, and is hence a viable candidate.
One thing that should also be emphasised though, is that constructing this EoS can not yet be done from the V-QCD action plus the chosen potentials without making approximations.
Wherever these approximations occur they will be pointed out, but the task of improving the approximations further will be left for future work.

In the first section of this chapter, which is based on \cite{Ishii:2019gta}, we will obtain the holographic equation of state directly from the V-QCD action.
This EoS will, for reasons we will see below, be assumed to approximate the true EoS best in the regime of high density, and is expected to not work very well in the low density regime.
In the second section, based on \cite{Ecker:2019xrw}, we will introduce a matching procedure between the EoS from holography, and a nuclear matter equation of state coming from an effective Skyrme nucleon-nucleon interaction which is expected to work better in the low density regime \cite{Haensel:1993zw,Douchin:2001sv}\@.
We will then conclude this chapter by exploring the observational consequences of the resulting hybrid equation of state.
For completeness, note that section 3 of \cite{Ishii:2019gta} will not be covered, as it does not lead to reasonable neutron star physics.
\section{Baryons in V-QCD}\label{sec:holographicns:baryons}
As was discussed in the introduction, baryons arise in holography as solitonic objects living in the bulk.
In V-QCD, the fields responsible for this are the gauge fields $(A_L^\mu)^{ij}$ and $(A_R^\mu)^{ij}$, where the $i$ and $j$ are flavor indices.
These fields are dual to the globally conserved currents
\[
\bar q^i\gamma^\mu(1 \pm \gamma_5)q^j/2.
\]
In the previous chapter, the trace part of $A_L + A_R$ has already been used as the source for the conserved total baryon number, i.e.~the baryon chemical potential.
In that analysis however, the other components of $A_{L/R}$ were set to zero.
In this section, we do need to take those components into account.

Before moving on to a brief summary of the steps required for performing this analysis, let us first fix a few conventions that will be used throughout this section.
This is to keep notational consistency with \cite{Ishii:2019gta}\@.
Firstly, we will denote spacetime indices with capital latin characters, and we keep the following convention for the field strength tensors:
\[
F_{(L/R)} = dA_{L/R} - iA_{L/R} \wedge A_{L/R}\,.
\]
Also, note that when refering to an expression which applies to both left and right handed gauge fields, we will use $L/R$ to denote this.
Furthermore, since the $L$ and $R$ labels are indeed just labels, we will not be consistent in putting them up or down.
Instead they will be put in whichever place is visually convenient for a given formula.
Lastly note that since we use capital latin characters for spacetime indices, the labels for left and right handed gauge fields will be denoted with brackets as $(L)$ and $(R)$ whenever there is a possibility for confusion.

In the following subsections, we will first expand the non-Abelian part of the DBI action to first order in the gauge fields, after which we will examine the Chern-Simons terms.
After that, we will introduce a homogeneous approximation, which reduces the degrees of freedom to a single field $h$\@.
This field is discontinuous at some point in the bulk, where the discontinuity can be related to the baryon number density.
To treat the discontinuity properly, one it is therefore hard to perform the computation in the grand canonical ensemble where we derive the action.
For this reason we proceed to take a Legendre transform, and we finally derive an action involving $h$ which, when integrated on-shell, gives one the free energy density (in the canonical ensemble)\@.
We will briefly discuss how this is done numerically, and finally the result is Legendre transformed back into the grand canonical ensemble, hence giving us the phase diagram and the equation of state.
The last subsection will then be devoted to a discussion of these results.
\subsection{Expansion of the DBI action}
The starting point for this subsection is the full flavored DBI action \eqref{eq:intro:DBI} introduced in section \ref{sec:intro:bottomup}, which we repeat here for convenience:
\begin{align}
S_\mathrm{DBI} & = -\frac{1}{2} M^3 N_c\,  {\mathbb Tr} \int d^5x\,\label{eq:holo:DBI}\\
& \quad \times \left(V_f(\lambda,T^\dagger T)\sqrt{-\det {\bf A}^{(L)}}+V_f(\lambda, TT^\dagger)\sqrt{-\det {\bf A}^{(R)}}\right),\nonumber
\end{align}
where we have
\begin{align}
{\bf A}_{MN}^{(L)} & = g_{MN} + w(\lambda,T) F^{(L)}_{MN} + {\kappa(\lambda, T) \over 2 } \left[(D_M T)^\dagger (D_N T) + (D_N T)^\dagger (D_M T)\right],\nonumber\\
{\bf A}_{MN}^{(R)} & = g_{MN} + w(\lambda,T) F^{(R)}_{MN} + {\kappa(\lambda, T) \over 2 } \left[(D_M T) (D_N T)^\dagger + (D_N T) (D_M T)^\dagger\right],\label{eq:holo:senaction}
\end{align}
and the covariant derivative for $T$ is given by
\[
D_M T = \partial_M T + i  T A_M^L- i A_M^R T.
\]
Note that the non-Abelian part of this action is in principle ambiguous because the order of the gauge fields is not specified, and its full form is not completely known.
The first few terms as a series in $F$ are known precisely though \cite{Refolli:2001df,Koerber:2001uu,Grasso:2002wb,Keurentjes:2004tu}, and hence we will here also use a series expansion in $F$\@.
In fact, we will consider only the first non-Abelian correction on top of an Abelian background.
Using this restriction, the ambiguities are absent, and the trace in \ref{eq:holo:DBI} does not require a special prescription, and can be taken to be an ordinary trace.
Obviously, using an expansion in the non-Abelian field strengths assumes implicitly that these fields are small.
This is one of the points where the present analysis could potentially be improved in the future, as it would be interesting to take the next order in the expansion into account to see by how much the result changes.
An additional simplifying assumption we make is to take the tachyon to be flavor independent
\[
T = \tau(r)\mathbb{I}_{N_f},
\]
just as what was done in the diagonalized DBI action used in chapter \ref{ch:imc}\@.

The first step towards expanding in the non-Abelian part of the gauge field is to split up $A_{L/R}$ into an Abelian part, which contains the $\Phi$ field familiar from chapter \ref{ch:imc}, and a non-Abelian part.
To do this, we employ a slight abuse of notation and substitute
\begin{equation}
A_{L/R} \mapsto \mathbb{I}_{N_f}\Phi\,dt + A_{L/R},\label{eq:holo:Asubst}
\end{equation}
where we treat $A_{L/R}$ as small perturbations, while $\Phi$ is kept to all orders.
This split is artificial, however, and to make the following well-defined we have to impose the following condition on the non-Abelian part of the gauge field:
\begin{equation}
\int d^4x\, \mathbb{T}r\left(F^{(L)}_{rt} + F^{(R)}_{rt}\right) = 0.\label{eq:holo:consistencycondition}
\end{equation}
After performing the substitution \eqref{eq:holo:Asubst} into \eqref{eq:holo:senaction} and expanding for small $A_{L/R}$, we obtain
\begin{align}
{\bf A}_{MN}^{(L)} & = g_{MN} + \kappa(\lambda)\delta_M^r\delta_N^r(\tau')^2+ w(\lambda)(\delta_M^r\delta_N^t-\delta_M^t\delta_N^r)\Phi'+   w(\lambda) F^{(L)}_{MN}&\nonumber\\
& \quad + \frac{\kappa(\lambda)\tau^2}{2}\left(A_M A_N+A_NA_M\right),
\end{align}
where we define $A = A_L - A_R$\@.\footnote{In this chapter, the prime always denotes a derivative with respect to $r$.}
One can obtain a similar expression for ${\bf A}_{MN}^{(R)}$\@.
One next has to substitute this last expression into \eqref{eq:holo:DBI} to obtain an action quadratic in the non-Abelian gauge fields.
To do this, it is convenient to define the `effective metric':
\[
\tilde g_{MN} \equiv g_{MN} + \kappa(\lambda)\delta_M^r\delta_N^r(\tau')^2+ w(\lambda)(\delta_M^r\delta_N^t-\delta_M^t\delta_N^r)\Phi',
\]
which leads to the following two identities:
\begin{align*}
\left( \tilde g^{-1}\right)^{MP} {\bf A}_{PN}^{(L)} & = \delta^M_N + w(\lambda) \left( \tilde g^{-1}\right)^{MP}F^{(L)}_{PN}\\
& \quad + \frac{\kappa(\lambda)\tau^2}{2}\left( \tilde g^{-1}\right)^{MP}\left(A_P A_N+A_NA_P\right), \\
\left( \tilde g^{-1}\right)^{MN}F^{(L)}_{NM} & = 2 \Xi^{-1} e^{-4A} w(\lambda)\Phi'F^{(L)}_{rt},
\end{align*}
where
\[
\Xi = \frac{\det \tilde g}{\det g} =1 + e^{-2A}f\kappa(\lambda)(\tau')^2-e^{-4A}w(\lambda)^2(\Phi')^2.
\]
This subsequently leads to the following expression, where we keep all terms quadratic in $A_{L/R}$ and $F_{L/R}$:
\begin{align*}
\sqrt{-\det {\bf A}^{(L)}} \simeq \sqrt{-\det\tilde g}\Bigg[1 & + \Xi^{-1} e^{-4A} w(\lambda)^2\Phi'F^{(L)}_{rt}\nonumber\\
& + \frac{\kappa(\lambda)\tau^2}{2}\left( \tilde g^{-1}\right)_s^{MN}A_MA_N &\nonumber\\
&-\frac{w(\lambda)^2}{4}\left( \tilde g^{-1}\right)_s^{MN}F^{(L)}_{NP}\left( \tilde g^{-1}\right)_s^{PQ}F^{(L)}_{QM}\Bigg],
\end{align*}
where we define the diagonal part of the effective metric as
\[
(\tilde g^{-1})_s = e^{-2A}\,\mathrm{diag}\left(-f^{-1}\Xi^{-1}(1+e^{-2A}f\kappa(\lambda)(\tau')^2),1,1,1,f \Xi^{-1}\right),
\]
with the indices ordered as $(t,x_1,x_2,x_3,r)$\@.

Putting this all together, including the analogous term for ${\bf A}^{(R)}$, one obtains
\begin{align}
S_\mathrm{DBI}^{(0)} & = - M^3 N_c N_f  \int d^5x\, V_f(\lambda)\sqrt{-\det g}\nonumber\\
& \qquad \times \sqrt{1+ e^{-2A}f \kappa(\lambda)(\tau')^2-e^{-4A}w(\lambda)^2(\Phi')^2},\label{eq:holo:SDBI0}\\
S_\mathrm{DBI}^{(1)} & = - M^3 N_c  \int d^5x\, V_f(\lambda)\sqrt{-\det g}\sqrt{\Xi}\Bigg[\frac{\kappa(\lambda)\tau^2}{2}\left( \tilde g^{-1}\right)_s^{MN}\,\mathbb{T}r A_MA_N&\nonumber\\
& \quad -\frac{w(\lambda)^2}{8}\left( \tilde g^{-1}\right)_s^{MN}\left( \tilde g^{-1}\right)_s^{PQ}\,\mathbb{T}r\left(F^{(L)}_{NP}F^{(L)}_{QM}+F^{(R)}_{NP}F^{(R)}_{QM}\right) \Bigg],\label{eq:holo:SDBI1}
\end{align}
where the terms involving $F_{rt}^{(L/R)}$ cancel due to \eqref{eq:holo:consistencycondition}, and where we split up the zeroth and first order in the expansion.
With the last expression we have found the first part of the action for the non-Abelian gauge fields.
In the next subsection, we will discuss the terms coming from the Chern-Simons section of V-QCD\@.
\subsection{Obtaining baryon number from the Chern-Simons action}
Before stating the Chern-Simons (CS) action, let us first discuss why this part of the V-QCD action is important for describing baryons.
In the approximations that follow, we will not attempt to backreact the baryons onto the background geometry, which is taken to be a thermal gas.
In particular, this means that $\Phi'$ will be taken to be zero throughout the bulk in the computation of the background geometry.
This would mean that the gauge fields don't respond to a change in chemical potential, which is something they should do since they carry the correct charge.
The Chern-Simons action provide such a coupling, and indeed coupling $\Phi$ to $A_{L/R}$ is entirely appropriate, as we will see that the baryon number one can derive from the Chern-Simons action is a total derivative, and hence counts solitons, which is exactly how baryons arise.

We will now state the Chern-Simons action itself, which depends on a CP-odd potential $V_a(\lambda,\tau)$ \cite{Arean:2016hcs}\@.
This dependence is highly non-trivial though, as the explicit form of the Chern-Simons action changes with different choices of $V_a$\@.
The reason for this is that as one changes $V_a$ in a non-trivial way, gauge invariance and other essential properties are usually spoiled.
In this chapter, we choose a string motivated ansatz, namely
\begin{equation}
V_a(\lambda,\tau) = e^{-b\tau^2},\label{eq:holo:bdefinition}
\end{equation}
where the constant $b$ is introduced with the motivation that if one is to have regular IR solutions at a finite QCD $\theta$-angle, the contributions from the DBI action need to dominate over those coming from the CS action, which is achieved provided that $b > 1$ \cite{Arean:2016hcs}\@.
We will now set $b = 1$ for notational simplicity.
Later we will restore $b$ by rescaling $\tau$\@.\footnote{Note that this rescaling happens only in the CS action.}
The CS action is now given by \cite{Casero:2007ae}:
\[
S_\mathrm{CS} = \frac{iN_c}{4\pi^2}\int\Omega_5,
\]
where
\begin{align*}
\Omega_5&=\frac{1}{6}\mathbb{T}r \,e^{-\tau^2}\!\left\{ -iA_L
\wedge F^{(L)}\wedge F^{(L)} +\frac{1}{2}A_L\wedge A_L \wedge
A_L \wedge F^{(L)}\right.\\
&+\frac{i}{10} A_L\wedge A_L \wedge A_L\wedge A_L\wedge A_L
+iA_R\wedge F^{(R)}\wedge F^{(R)}\\
&-\frac{1}{2}A_R\wedge A_R
\wedge A_R \wedge F^{(R)} -\frac{i}{10} A_R\wedge A_R
\wedge A_R\wedge A_R\wedge A_R\\
&+\tau^2\Big[ iA_L\wedge F^{(R)}\wedge F^{(R)}-iA_R\wedge
F^{(L)}\wedge F^{(L)}\\
&\quad +\frac{i}{2}(A_L\!-\!A_R)\wedge
(F^{(L)}\wedge F^{(R)}+F^{(R)}\wedge F^{(L)})\\
&\quad+\frac{1}{2}A_L\wedge A_L \wedge A_L \wedge F^{(L)}-
\frac{1}{2}A_R\wedge A_R \wedge A_R \wedge F^{(R)}\\
&\quad+\frac{i}{10} A_L\wedge A_L \wedge A_L\wedge A_L\wedge A_L-\frac{i}{10} A_R\wedge A_R \wedge A_R\wedge A_R
\wedge A_R\Big]\\
&+i\tau^3\,d\tau \wedge\Big[ (A_L\wedge A_R-A_R\wedge A_L)
\wedge (F^{(L)} +F^{(R)} )\\
&\quad+i A_L \wedge A_L \wedge A_L\wedge A_R-\frac{i}{2}A_L\wedge A_R\wedge A_L\wedge A_R\\
&\quad+i A_L\wedge A_R\wedge A_R \wedge A_R \Big]\\
&\left.+\frac{i}{20}\tau^4 (A_L-A_R)\wedge (A_L-A_R)
\wedge (A_L-A_R)\wedge (A_L-A_R)\wedge (A_L-A_R)\right\}
\end{align*}

Now, to extract the coupling between $\Phi$ and the non-Abelian terms in $\Omega_5$, we make the same substitution \eqref{eq:holo:Asubst} as in the previous subsection, and collect all terms involving $\Phi$\@.
To do this, it turns out to be useful to first modify $\Omega_5$ by a total derivative, since $\Omega_5$ is only defined up to total derivatives.
The modification made is the following:
\begin{align}
 12\widetilde \Omega_5 & = 12\Omega_5 + i\, \mathbb{T}r\,d\big[\, e^{-\tau ^2} \Phi dt \wedge (4 A_L\wedge F^{(L)}+i A_L\wedge A_L\wedge A_L-4 A_R\wedge F^{(R)}&\nonumber\\
 &\ -i A_R\wedge A_R\wedge A_R)\big] +  i \,\mathbb{T}r\, d\big[\, e^{-\tau ^2} \tau ^2 \Phi dt\wedge (-2 A_L\wedge F^{(L)}-6 A_L\wedge F^{(R)}&\nonumber\\
 &\ +i A_L\wedge A_L\wedge A_L+6 A_R\wedge F^{(L)} +2 A_R\wedge F^{(R)}-i A_R\wedge A_R\wedge A_R)\big].
\end{align}
Now one can check that
\[
\widetilde \Omega_5 = \Omega_5\big|_{\Phi=0} + \frac{1}{6}\Phi\,dt \wedge H_4^{(\Phi)},
\]
with\footnote{Note that one has to use that $\tau$ and $\Phi$ both only depend on the bulk coordinate, and hence $d\tau\wedge d\Phi = 0$.}
\begin{align*}
e^{\tau ^2}H_4^{(\Phi)} &=\mathbb{T}r\,\big[ -3 i F^{(L)}\wedge F^{(L)}+3 i F^{(R)}\wedge F^{(R)}&\nonumber\\
& \quad + 6 i \tau  d\tau \wedge (A_L-A_R)\wedge (F^{(L)}+F^{(R)})&\nonumber\\
& \quad + 3 \tau ^2 (A_L-A_R)\wedge (A_L-A_R)\wedge (F^{(L)}-F^{(R)}) &\nonumber\\
& \quad + \tau ^3 d\tau \wedge (-4 i A_L\wedge F^{(R)}+4 i A_R\wedge F^{(L)}+2 A_R\wedge A_L\wedge A_L&\nonumber\\
& \qquad -2 A_R\wedge A_R\wedge A_L-2 A_L\wedge A_L\wedge A_L+2 A_R\wedge A_R\wedge A_R)\big].
\end{align*}
Using a variation of the action with respect to $\Phi$ just as in section \ref{sec:imc:observables}, one obtains the number density
\[
\varrho = -\left.\frac{\delta S_\mathrm{V-QCD}}{\delta \Phi'}\right|_\mathrm{bdry} = \int dr\, \frac{\delta S_\mathrm{V-QCD}}{\delta \Phi},
\]
where we have used the $\Phi$ equation of motion in the last equality.
The baryon number is therefore now given by:
\begin{equation}
N_c N_b = \int dr d^3x\,\frac{\delta S_\mathrm{CS}}{\delta \Phi} = \frac{iN_c}{24\pi^2} \int \ H_4^{(\Phi)},\label{eq:holo:baryonnumber}
\end{equation}
where $N_b$ is the total baryon number.
To conclude this discussion of the CS action, note that indeed, as was mentioned at the start of this subsection, that $H_4^{(\Phi)}$ is indeed exact:
\begin{align}
H_4^{(\Phi)} & =\mathbb{T}r\, d\Big[e^{-\tau ^2} \big(-3 i A_L\wedge F^{(L)}+3 i A_R\wedge F^{(R)}&\nonumber\\
& \quad + A_L\wedge A_L\wedge A_L-A_R\wedge A_R\wedge A_R &\nonumber\\
& \quad +\tau ^2 (A_L-A_R)\wedge (A_L-A_R)\wedge (A_L-A_R)&\nonumber\\
& \quad + 3 i \tau  d\tau\wedge (A_L\wedge A_R-A_R\wedge A_L)\nonumber\\
& \quad -2 i \tau ^3 d\tau\wedge (A_L\wedge A_R-A_R\wedge A_L)\big)\Big].\label{eq:holo:H4}
\end{align}

In the next subsection, we will take the next step towards including a holographic treatment of baryons in V-QCD, namely the introduction of a homogeneous approximation.
This will reduce the degrees of freedom to one single field $h$\@.
\subsection{The homogeneous approximation}
In principle, we could now attempt to construct a soliton dual to a baryon in V-QCD by solving the action defined by \eqref{eq:holo:SDBI0} and \eqref{eq:holo:SDBI1}, while using \eqref{eq:holo:baryonnumber} as a boundary condition to impose a fixed baryon number.
One could subsequently put many such solitons together to achieve a finite baryon number density.
While this has been successfully done using approximate methods in for example the Witten-Sakai-Sugimoto (WSS) model \cite{Witten:1998zw,Sakai:2004cn,Sakai:2005yt}, this is very challenging.
The reason for this is that the required solutions have non-trivial spatial profiles, and hence to obtain them one needs to solve PDEs, and not only that, the requirement of imposing a fixed number of baryons turns the problem into a particularly hard one.

To make progress, we once again look to the WSS model for inspiration, as approximations have been developed there in a controlled setup \cite{Hong:2007kx,Hata:2007mb,Hong:2007dq,Hashimoto:2008zw,Kim:2008pw,Cherman:2009gb,Cherman:2011ve,Bolognesi:2013nja,Rozali:2013fna,Kaplunovsky:2012gb,deBoer:2012ij,Kaplunovsky:2015zsa,Preis:2016fsp,BitaghsirFadafan:2018uzs,Bergman:2007wp,Rozali:2007rx,Ghoroku:2012am,Li:2015uea,Elliot-Ripley:2016uwb}\@.
The particular approach we will follow is a homogeneous approach \cite{Rozali:2007rx,Li:2015uea}, and as the name suggests, this reduces the problem to a spatially homogeneous one, which simplifies the problem to the solving of ODEs, for which we can use standard techniques.
This is not the only possible way forward though.
A similar approach to the one we will follow was developed in the WSS model in \cite{Elliot-Ripley:2016uwb}, and this approach can also be adapted for V-QCD using similar techniques to the ones that will be described below, but doing so is beyond the scope of this thesis.

The idea behind our homogeneous approach is to assume a high density of baryons.
These will energetically prefer to be located at some coordinate distance $r_c$ in the bulk.\footnote{Note that this is non-trivial. It could well happen that there exists no such $r_c$, and in that case the baryonic phase is unstable. As we shall see, in our setup the baryons \emph{are} stable, and only in the region in the phase diagram where one would expect this to happen.}
At $r = r_c$, there will therefore be a very non-trivial and highly inhomogeneous configuration of gauge fields.
Far enough away from this region, however, one can approximate the gauge fields as being homogeneous, just as when describing an electrically charged plate one does not have to worry that the plate is made up out of atoms if one is solely interested in the electric field far enough away.

The assumption we will now make is to divide the bulk into three regions: two regions away from $r_c$ where the gauge field is homogeneous, and a region around $r_c$ which we ignore, and which we take to be infinitesimally thin.
We replace instead this region around $r_c$ by a discontinuity in $h$ at $r_c$\@.
If one once again looks at the electric plate analogy, this means that if asked what for example the total energy stored in the electric field is, we choose to only include the field outside the plate, which we assume to be homogeneous, and we ignore the fields inside the plate, which we cannot easily describe.
Of course this approach is not perfect, but it is expected that one does obtain a qualitatively correct description.
In the same way, we will likely not obtain the correct on-shell action in this manner, but we can still hope that the dependence of the on-shell action on the baryon chemical potential is captured.
Of course we have no way of checking whether this actually happens without doing the full calculation, but \emph{if} this happens we can already learn valuable information about the equation of state this way.
In particular, while the equation of state would only be reliable up to an overall factor, the speed of sound would be unaffected by such a deviation.
Indeed, in section \ref{sec:holo:hybrideos}, we will make precisely this assumption, and take the speed of sound as input from the holographic model.

Concretely, the discussion from the previous chapter means that we set $N_f = 2$ and hence take
\begin{equation}
A_L^i = -A_R^i = h(r) \sigma^i,\label{eq:holo:homogeneousansatz}
\end{equation}
with $\sigma^i$ the Pauli matrices.
Here it is important to note that \eqref{eq:holo:homogeneousansatz} respects chiral symmetry and parity \cite{Pomarol:2007kr,Pomarol:2008aa}\@.
We take all integrals to mean
\[
\int_0^\infty dr \mapsto \left(\int_0^{r_c^-}+\int_{r_c^+}^\infty\right)dr \equiv \lim_{\epsilon \to 0 +}\left(\int_0^{r_c-\epsilon}+\int_{r_c+\epsilon}^\infty\right)dr.
\]
Specifically, this means that there are no delta function contributions at $r_c$, which would otherwise come from various derivatives appearing in equations of motion.

Now that the prescription is known, we can start substituting \eqref{eq:holo:homogeneousansatz} into \eqref{eq:holo:SDBI1} and \eqref{eq:holo:H4}\@.
Doing this yields
\begin{align}
S_\mathrm{DBI}^{(1)} = -12 M^3 N_c \int d^5x\,V_f(\lambda) e^{5A} \sqrt{\Xi}\bigg[&\kappa(\lambda)\tau^2e^{-2A}h^2+w(\lambda)^2e^{-4A}h^4&\nonumber\\
& +\frac{1}{4} w(\lambda)^2e^{-4A} f \Xi^{-1} \left(h'\right)^2\bigg],\label{eq:holo:homogeneousDBI1}
\end{align}
and
\begin{equation}
\int dt\wedge H_4^{(\Phi)} = 48i \int d^5 x\, \frac{d}{dr}\left[ e^{- b\, \tau(r)^2}h(r)^3(1-2b\, \tau(r)^2)\right],\label{eq:holo:homogeneousH4}
\end{equation}
respectively.
Before collecting these two results into the complete action for $h$, note that, were it not for the discontinuity at $r_c$, \eqref{eq:holo:homogeneousH4} would always evaluate to zero due to the asymptotic behavior of the tachyon, which diverges towards the IR, and $h$, which should vanish on the boundary because we do not want to introduce a source for the baryon charge.
This is not surprising given that $H_4^{(\Phi)}$ is exact, but it does show that the baryon number in the solutions is indeed sourced at the discontinuity, as is appropriate in this approximation.

We conclude this subsection by substituting \eqref{eq:holo:homogeneousH4} into \eqref{eq:holo:baryonnumber}, and then combining this result with \eqref{eq:holo:SDBI0} and \eqref{eq:holo:homogeneousDBI1}\@.
This results in the action for the homogeneous gauge field $h$:
\begin{align}
S_h &= S_\mathrm{DBI}^{(0)}+S_\mathrm{DBI}^{(1)} + S_\mathrm{CS} = - 2 M^3 N_c \int d^5x\,V_f(\lambda) e^{5A} \sqrt{\Xi}\nonumber\\
& \qquad \times \bigg[1+6\kappa(\lambda)\tau^2e^{-2A}h^2 + 6w(\lambda)^2e^{-4A}h^4+\frac{3}{2} w(\lambda)^2e^{-4A} f \Xi^{-1} \left(h'\right)^2\bigg]\nonumber\\
& \quad - \frac{2N_c}{\pi^2} \int d^5x\, \Phi  \frac{d}{dr}\left[ e^{-b\,\tau^2}h^3(1-2b\,\tau^2)\right].\label{eq:holo:homogeneoustotalaction}
\end{align}
This puts the action for the non-Abelian gauge fields in a relatively simple form, containing just a single field $h$\@.
In the next subsection, we will continue the analysis by examining the discontinuity, as we will need to decide by how much $h$ should be discontinuous.
\subsection{The Legendre transformed action}
In this subsection, we will derive a condition that needs to be satisfied at the discontinuity.
In fact, this condition will completely fix the available freedom at the discontinuity, except for its location, which we'll get back to in the next subsection.
We will also see that this condition relates the jump in $h$ at the discontinuity to the baryon number density.
This makes it hard to perform the calculation in the grand canonical ensemble, as we would need to extract the baryon number density from the solution, and simultaneously also impose it as a condition at the discontinuity.
The solution to this will be to perform a Legendre transform, after which we eliminate $\Phi$ in favor of the baryon number density.
This gives us a different action, namely one in the canonical ensemble, which we will then solve in the next subsection.

To start, we derive the baryon number density from \eqref{eq:holo:homogeneoustotalaction} in the same way as before:
\begin{align}
\rho & = -\frac{\delta S_h}{\delta \Phi'} = -\frac{V_\rho w(\lambda)^2e^{-4A}\Phi'}{\sqrt{\Xi}}\label{eq:holo:rhoofPhiprime}\\
& \qquad \times \bigg[1+6\kappa(\lambda)\tau^2e^{-2A}h^2+6w(\lambda)^2e^{-4A}h^4-\frac{3}{2} w(\lambda)^2e^{-4A} f \Xi^{-1} \left(h'\right)^2\!\bigg],\nonumber
\end{align}
where we recall that $\Xi$ depends on $\Phi'$\@.
Here we also introduced the abbreviation
\begin{equation}
V_\rho = 2 M^3 N_c V_f(\lambda) e^{5A}.\label{eq:holographicns:Vrho}
\end{equation}
From the $\Phi$ equation of motion, we can obtain the following:
\[
\rho' = - \frac{d}{dr}\frac{\delta S_h}{\delta \Phi'} = -\frac{\delta S_h}{\delta \Phi}= \frac{2N_c}{\pi^2}  \frac{d}{dr}\left[ e^{-b\,\tau^2}h^3(1-2b\,\tau^2)\right].
\]
Since we can ignore any delta function contributions at $r_c$, $\rho$ is continuous, and we can then derive
\begin{equation}
\rho = \left\{\begin{array}{lr}
               \varrho  + \frac{2N_c}{\pi^2} e^{-b\,\tau^2}h^3(1-2b\,\tau^2) \, , \qquad&(r<r_c) \\
                \frac{2N_c}{\pi^2} e^{-b\,\tau^2}h^3(1-2b\,\tau^2) \, , \qquad&(r>r_c)
              \end{array}\right.\label{eq:holo:rhoofh}
\end{equation}
where we can immediately read off $\varrho$ as the baryon number density on the boundary, which is the physical baryon number density.
Since $\rho$ is continuous, this implies that $h$ is discontinuous, and from the above equation one can derive that
\begin{equation}
\varrho = \frac{2N_c}{\pi^2} e^{-b\,\tau(r_c)^2}(1-2b\,\tau(r_c)^2)\ \mathrm{Disc}\, h^3(r_c),\label{eq:holo:varrhoofdisc}
\end{equation}
where we define the discontinuity as
\[
\mathrm{Disc}\, g(r) \equiv \lim_{\epsilon \to 0+}\left(g(r+\epsilon)-g(r-\epsilon)\right).
\]

Next, we perform the Legendre transformation to the action $\widetilde S_h$, whose on-shell value will correspond to the free energy, as follows:
\[
\widetilde S_h = S_h - \int d^4x \Phi(0) \rho(0) = S_h + \int d^5x\,  \frac{d}{dr}\left[\Phi\rho\right],
\]
where we notice that
\[
S_\mathrm{CS} = \int d^5x\, \Phi \rho'.
\]
Putting these two equations together, one obtains
\begin{equation}
\widetilde S_h = S_\mathrm{DBI} + \int d^5x\, \Phi'  \rho.\label{eq:holo:legendretransformation}
\end{equation}
To arrive at the final Legendre transformed action, there is one last non-trivial step.
We need to eliminate $\Phi'$ from the action.
The way to do this is to solve \eqref{eq:holo:rhoofPhiprime} for $\Phi'$\@.
This cannot be done analytically, but one can obtain $\Phi'$ as a series expansion, where higher order terms correspond to higher powers of $F_{L/R}$, which can by our assumptions be ignored.
Performing this calculation leads to
\begin{align}
\Phi' & = -\frac{G \rho}{V_\rho w^2 e^{-4A}\sqrt{1+\rho^2\left(V_\rho w e^{-2A}\right)^{-2}}}\label{eq:holo:Phiprimesubst}\\
& \quad \times \left[1-\frac{6\kappa\tau^2e^{-2A}h^2+6w^2e^{-4A}h^4}{1+\rho^2\left(V_\rho w e^{-2A}\right)^{-2}}+\frac{3}{2}\frac{w^2e^{-4A}f (h')^2}{G^2}\right],\nonumber
\end{align}
with
\[
G = \sqrt{1+f\kappa e^{-2A}\left(\tau'\right)^2}
\]
as in chapter \ref{ch:imc}\@.
Finally, we can substitute \eqref{eq:holo:Phiprimesubst} into \eqref{eq:holo:legendretransformation} to obtain the Legendre transformed action:
\begin{align}
\widetilde S_h & = -\int d^5x\,V_\rho G \sqrt{1+\frac{\rho^2}{\left(V_\rho w e^{-2A}\right)^2}}\label{eq:holo:legendretransformedaction}\\
& \quad \times \left[1+\frac{6w^2e^{-4A}h^4+6\kappa\tau^2e^{-2A}h^2}{1+\rho^2\left(V_\rho w e^{-2A}\right)^{-2}}+\frac{3}{2}\frac{w^2e^{-4A}f (h')^2}{G^2}\right].\nonumber
\end{align}
We have now arrived at the action that we will solve numerically.
In the next section, some important points regarding how this should be done will be given.
\subsection{Numerical method}
With \eqref{eq:holo:legendretransformedaction}, we now have an action for $h$, which, by substituting \eqref{eq:holo:rhoofh}, we can solve.
In this subsection, we will discuss step by step how this is done, and how one can finally obtain the grand potential density of the baryonic phase as a function of the baryon chemical potential.
Using this, one can construct the phase diagram and compute the equation of state.

We start the process by computing a background metric and profiles for $\lambda(r)$, $\tau(r)$ and $\Phi(r)$ in the same way as done in chapter \ref{ch:imc}\@.
For this background we will use the thermal gas background, as this is expected to be the only type of solution that will be dominant in the phase diagram.\footnote{Note that this means that $\Phi = 0$.}
Recall that we will not backreact the non-Abelian gauge fields onto these background fields, which means that we will use the same pre-computed background for all the solutions for $h$ that we will generate in the rest of this subsection.

The general idea for computing the solutions for $h$ is that we want to minimize the baryon action \eqref{eq:holo:legendretransformedaction} while keeping $\varrho$ fixed in \eqref{eq:holo:rhoofh}\@.
In the continuous sections of the bulk, this means that we simply have to satisfy the equations of motion, but this also means that we have to minimize other free parameters other than $\varrho$\@.
In particular, $h$ has boundary conditions at the boundary and in the deep IR\@.
Another remaining free parameter appears to be the location of the discontinuity.
The location, however, is completely fixed by the two boundary conditions for $h$ and \eqref{eq:holo:varrhoofdisc}\@.

At the boundary, the equations of motion imply that
\[
h(r) = C_1 + C_2r^2 + \mathcal{O}(r^3).
\]
However a non-vanishing $C_1$ would imply a source for the non-Abelian gauge field, and hence we set $C_1 = 0$\@.
This leaves us with a boundary condition at the boundary parameterized by $C_2$\@.
In the deep IR, it turns out that
\[
h \sim h_0\exp\left[-C_\tau\tau(r)^2/r^2\right],
\]
where $C_\tau$ is a constant that can be expressed in terms of the potentials.
This in principle gives us a second boundary condition $h_0$\@.
However, such solutions will be left out of the remainder of the discussion, as we have numerically checked that the value of $h_0$ which minimizes the action is consistent with zero.
Hence, for simplicity, we will just set $h(r) = 0$ for $r > r_c$\@.
If desired, the discussion below can be easily extended to reinstate $h_0$\@.
With $h_0$ set to zero, \eqref{eq:holo:varrhoofdisc} reduces to
\begin{equation}
\varrho = -\frac{2N_c}{\pi^2} e^{-b\,\tau(r_c)^2}(1-2b\,\tau(r_c)^2)h^3(r_c).\label{eq:holo:rccheck}
\end{equation}
This means that in order to find $r_c$, we just monitor \eqref{eq:holo:rccheck} as we solve for $h$ by shooting from the boundary, and when we detect a sign change, we iterate to find the $r_c$ where \eqref{eq:holo:rccheck} is satisfied.

Summarizing the previous discussion, we now need to shoot from the boundary, solving the equations of motion for $h$ until we find $r_c$ by means of \eqref{eq:holo:rccheck}\@.
All of this we have to do for different $C_2$, where in the end we must minimize the on-shell action with respect to $C_2$\@.
For this, there is one hurdle remaining, namely that the action diverges towards the boundary, implying that we need holographic renormalization to evaluate it.
There is a natural choice to use for the subtraction, and that is to subtract the on-shell action for $h = 0$\@.
This corresponds to the on-shell action of a solution without baryons, namely the thermal gas.
In this way, by subtracting the action with $h$ set to zero from the on-shell action with non-zero $h$, we find exactly the difference in free energy density between the baryonic solutions and the thermal gas solution.\footnote{Note that this subtraction, as it does not depend on $h$, does not alter any of the above discussion.}

Using all of the steps above, one is able to compute the free energy $F(\varrho,C_2)$ for any desired $\varrho$ and $C_2$\@.
In the final step towards minimizing the free energy, one performs this computation for a wide range of values for $C_2$ given a fixed $\varrho$\@.
The minimal value found in this way then yields the true free energy $F(\varrho)$, where we note that this minimal value is a stable minimum.
Finally, performing a Legendre transform on this last result enables one to obtain the grand potential, or equivalently, the pressure $p(\mu)$, as desired.
\subsection{Results}
We are now ready to discuss the results.
For this, we will fill in the value of $M$ corresponding to potentials 7a from \cite{Jokela:2018ers}, and we will also set $N_c = 3$\@.
Note though that our value for $N_f$ is not entirely consistent between the backgrounds, which are used also for the deconfined phase, and the baryonic solutions.
For the baryonic solutions we use $N_f = 2$, while the backgrounds use $N_f = 3$, which is the most appropriate given that the potentials were fitted to lattice data.
This is another source of uncertainty, that ultimately ends up modifying the overall normalization of the pressure.
Lastly, note that for all the results in this section, we use $b = 10$ for the parameter introduced in \eqref{eq:holo:bdefinition}\@.

\begin{figure}[ht]
\centering
\includegraphics[width=0.9\textwidth]{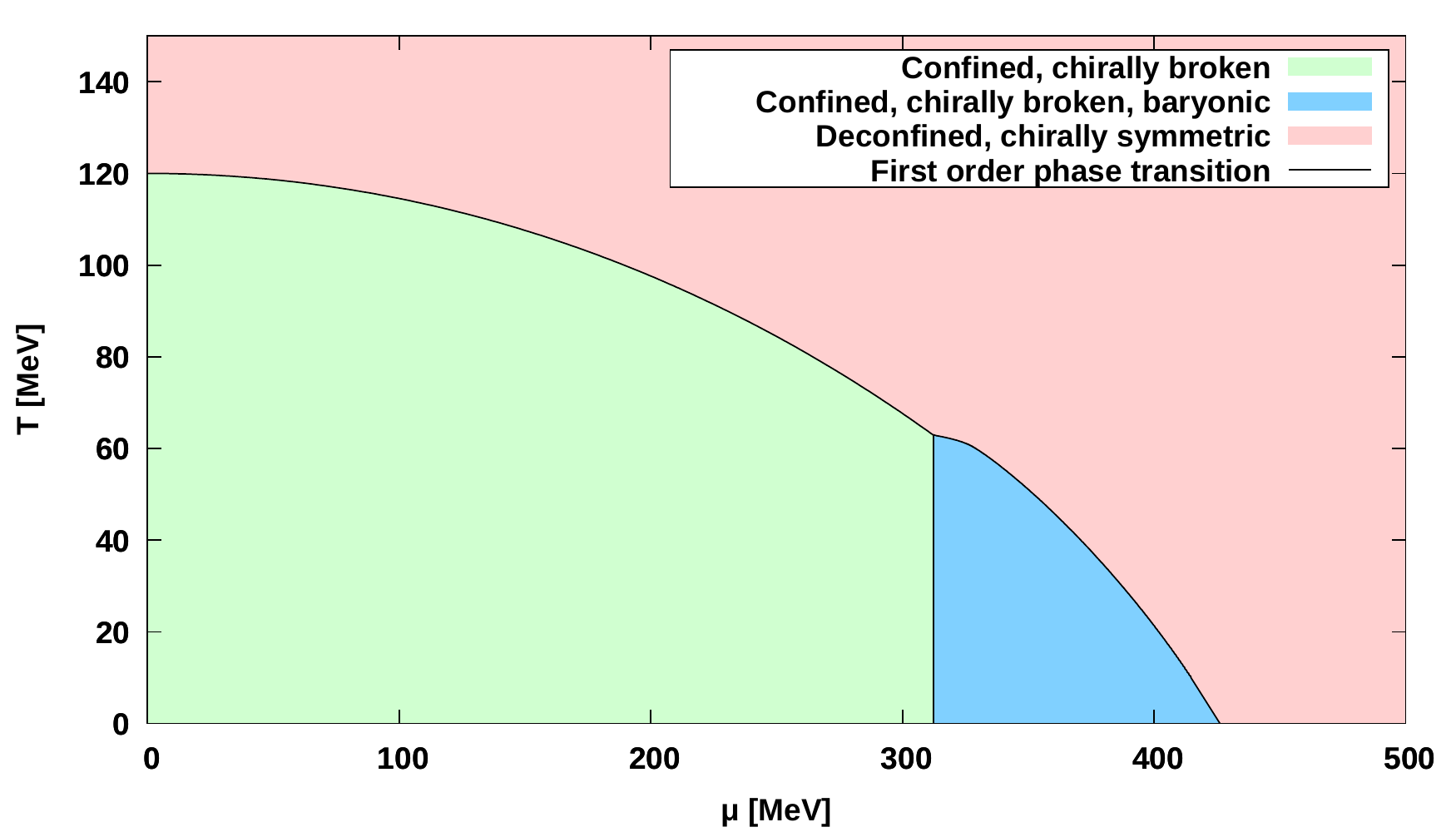}
\caption{\label{fig:holo:holo1smearedphases}Phase diagram of V-QCD including the baryonic phase, with the free parameter $b = 10$\@.}
\end{figure}

As a first step in examining the consequences of including baryons into V-QCD, let us examine the resulting phase diagram, which is shown in figure \ref{fig:holo:holo1smearedphases}\@.
Note that only the baryonic phase in this phase diagram was computed using the methods described in this chapter.
The other phases were computed using the methods described in chapter \ref{ch:imc}\@.
The phase diagram itself contains three phases:
\begin{itemize}
\item A thermal gas phase at small temperatures and chemical potentials.
This phase features confinement and chiral symmetry breaking, similar to the vacuum of real QCD\@.
This phase does not feature any temperature dependence though, as this is suppressed by powers of $N_c$ \cite{Alho:2015zua}, and hence doesn't appear in the large $N_c$ limit.
\item A baryonic phase at small temperatures and intermediate chemical potentials.
This phase shares the features of the thermal gas phase that were mentioned above, and in addition also has condensed baryons.
\item A deconfined, chirally symmetric phase at large temperatures and/or large chemical potentials.
This phase is akin to the quark-gluon plasma phase found in real QCD\@.
Due to the construction of the potentials, this phase also has an equation of state at vanishing chemical potential which is fitted to that of real QCD\@.
\end{itemize}
The resulting appears to be qualitatively reasonable within the limitations of the approach, such as the large $N_c$ limit.
The baryonic phase appears in the right place in the phase diagram, and importantly, the baryonic phase eventually yields to a deconfined phase at some large chemical potential.
This is in contrast to a similar approach in the WSS model \cite{Li:2015uea}\@.

\begin{figure}[ht]
\centering
\includegraphics[width=0.9\textwidth]{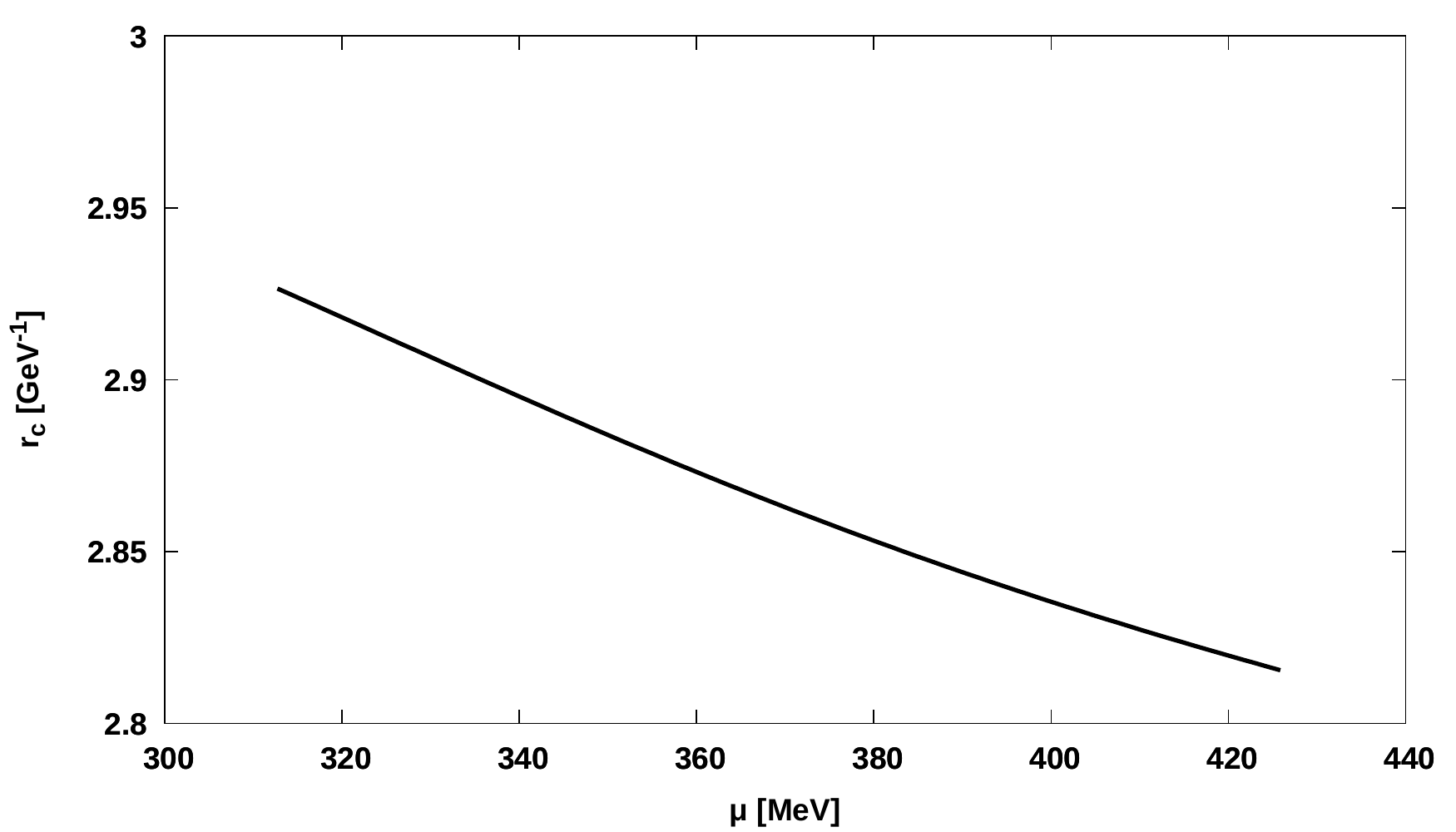}
\caption{\label{fig:holo:holo1smearedrc}Location of the discontinuity in the bulk as a function of baryon chemical potential $\mu$, with $b = 10$\@.}
\end{figure}

Before moving on to the equation of state, it is a good check to look at the location of the discontinuity.
This is shown in figure \ref{fig:holo:holo1smearedrc}\@.
One can see that the discontinuity sits at a `natural' location, as in units of $\Lambda$, which was introduced in chapter \ref{ch:imc}, this distance is $\mathcal{O}(1)$\@.

\begin{figure}[ht]
\centering
\includegraphics[width=0.9\textwidth]{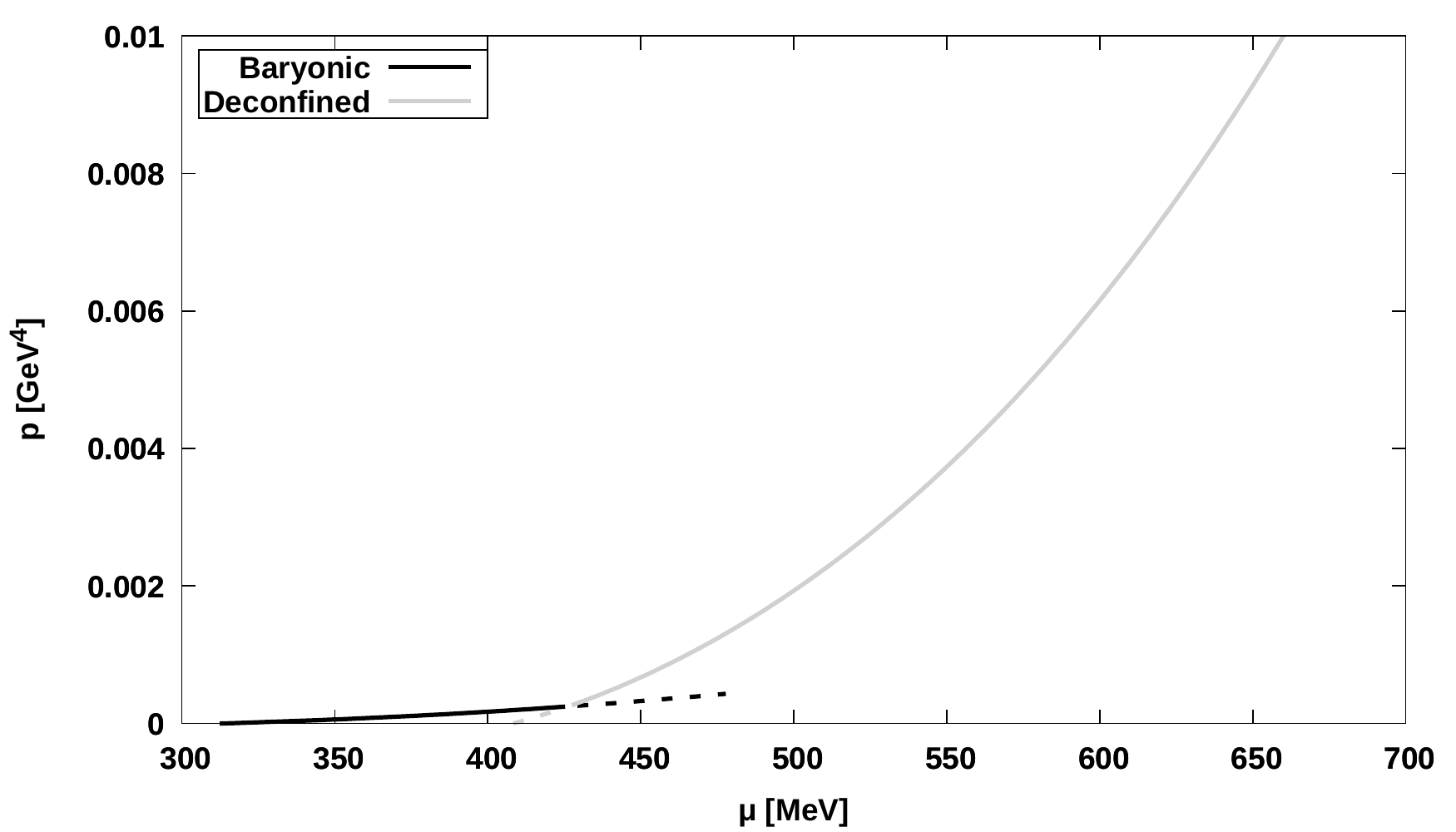}
\caption{\label{fig:holo:holo1freeenergy}Pressure as a function of chemical potential at $T = 0$, and using $b = 10$\@. Unstable branches are denoted with dashed lines.}
\end{figure}

\begin{figure}[ht]
\centering
\includegraphics[width=0.9\textwidth]{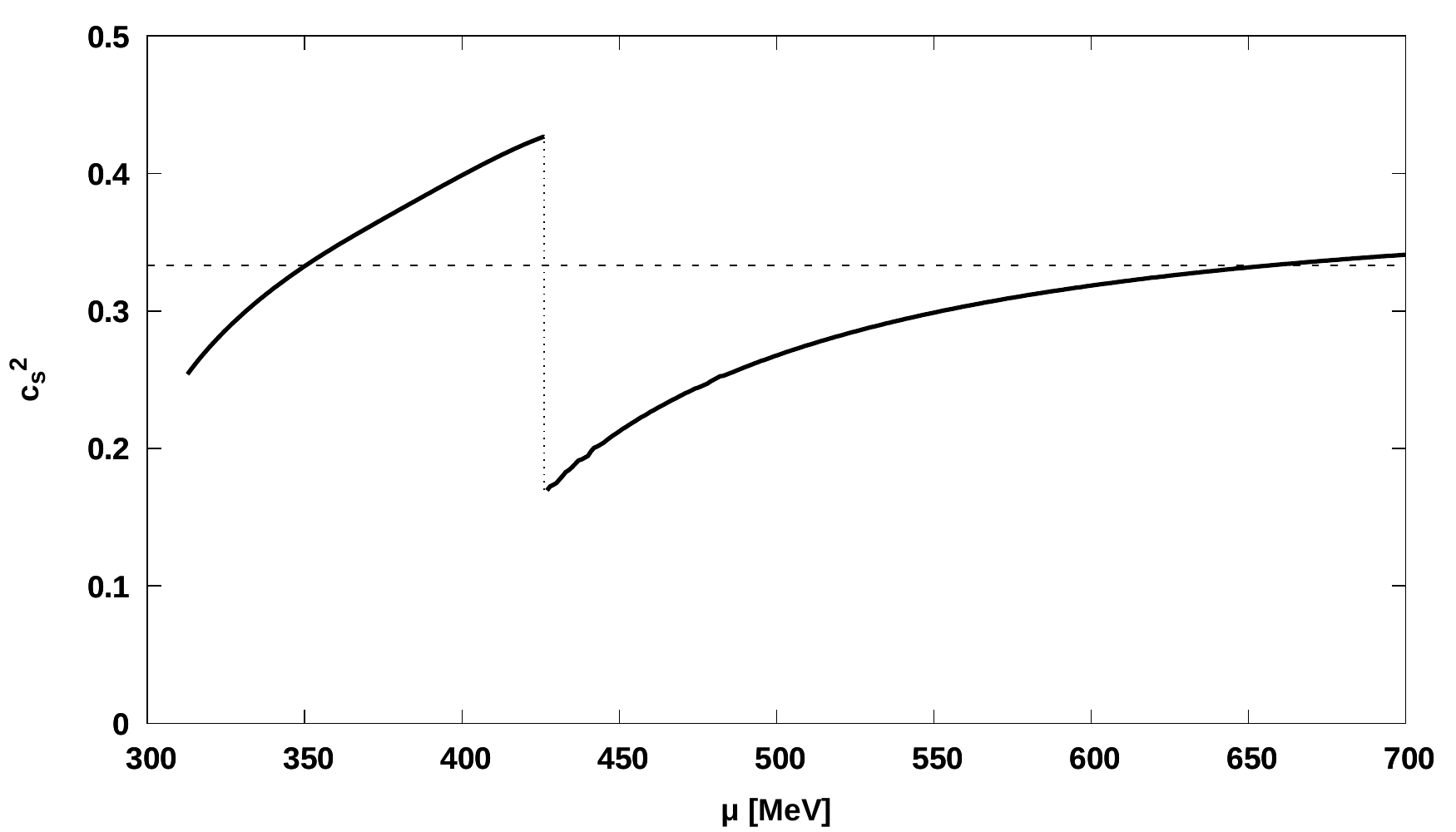}
\caption{\label{fig:holo:holo1cssq}Speed of sound squared at $T = 0$\@. For reference the conformal value of $c_s^2 = 1/3$ is shown as a dashed line.}
\end{figure}

To conclude this section, we will examine the equation of state at zero temperature.
The pressure as a function of chemical potential is shown in figure \ref{fig:holo:holo1freeenergy}\@.
The phase transitions can clearly be seen, and the latent heat associated to the vacuum to baryon transition is $\Delta\epsilon \approx 51\,\text{MeV}/\text{fm}^3$, while the latent heat for the baryonic to deconfined phase transition is $\Delta\epsilon \approx 687\,\text{MeV}/\text{fm}^3$\@.
In figure \ref{fig:holo:holo1cssq}, we show the isothermal speed of sound at zero temperature, given by
\[
c_s^2 = \left(\frac{\partial p}{\partial \epsilon}\right)_{T=0},
\]
with $p$ and $\epsilon$ the pressure and energy density, respectively.
While this is equivalent to figure \ref{fig:holo:holo1freeenergy} up to an integration constant, this is a convenient additional way of visualizing the equation of state.
In the thermal gas phase, the pressure and energy density are both zero, so the speed of sound vanishes.
In the baryonic and deconfined phases, it can be seen that the speed of sound exceeds the conformal value in two places.
This indicates a very stiff equation of state.
As it turns out, a stiff equation of state is likely needed to pass astrophysical constraints \cite{Bedaque:2014sqa,Tews:2018kmu}\@.

Summarizing, in this section we have constructed baryonic solutions in V-QCD using a homogeneous approximation.
In the next section, we will use the resulting equation of state to compute neutron star properties, and thereby investigate the potentially observable consequences from such a strongly coupled approach.
\section{Hybrid neutron star equations of state}\label{sec:holo:hybrideos}
In the previous section, we constructed an equation of state in V-QCD using a homogeneous approach.
This equation of state is likely to be most accurate in the regime of large chemical potential, as the homogeneous ansatz assumes a large baryon density.
Also, we reasoned that because the homogeneous ansatz only really looks at the tails of the soltions, the on-shell action, and hence the pressure, is probably not correct by an overall factor.
Motivated by this, in this section, based on \cite{Ecker:2019xrw}, we will construct a hybrid equation of state, which is equal to a nuclear matter model at low densities, and has speed of sound equal to the one constructed in the previous section at high densities.
We will then examine the potentially observable consequences from this hybrid equation of state, namely the mass-radius relation, the tidal deformability, and post-merger gravitational waveforms resulting from a binary neutron star merger.
\subsection{Matching procedure}
As briefly mentioned above, there are two ingredients that go into the hybrid equation of state, namely a low density nuclear matter model, and V-QCD\@.
For the nuclear matter model, we will use the SLy equation of state \cite{Haensel:1993zw,Douchin:2001sv}\@.
This EoS is based on effective Skyrme interactions between nucleons.
For the V-QCD part of the EoS, we develop a family of solutions, parameterized by varying the parameter $b$ as defined in \eqref{eq:holo:bdefinition}\@.

To perform the matching, we take the two equations of state,
\[
p_\text{SLy}(\mu), \qquad p_\text{V-QCD}(\mu,b),
\]
with $b$ fixed.
We then multiply the pressure belonging to the baryonic phase\footnote{Note that we do not modify the pressure of the deconfined phase, as this has been matched to lattice data in \cite{Jokela:2018ers}\@.} in V-QCD with a constant $c_b$ to reflect that we take only the speed of sound as input from V-QCD, and then we demand that
\begin{align*}
p_\text{SLy}(\mu^*) & = c_b\,p_\text{V-QCD}(\mu^*,b),\\
\partial_\mu p_\text{SLy}(\mu^*) & = c_b\,\partial_\mu p_\text{V-QCD}(\mu^*,b),
\end{align*}
by choosing appropriate $\mu^*$ and $c_b$\@.
These two conditions together guarantee that the phase transition between the SLy and V-QCD parts of the equation of state is second order.
The final equation of state is then given by
\[
p_\text{hybrid}(\mu) = \begin{cases}
p_\text{SLy}(\mu) & \mu < \mu^*, \\
c_b\,p_\text{V-QCD}(\mu) & \mu \geq \mu^*.
\end{cases}
\]
Of course, these two parts should both just be different descriptions of the same phase, so ideally there would be no phase transition at all.
However, a second order phase transition is the best one can do if one is only allowed to change the location of the matching $\mu^*$ and the normalization $c_b$\@.
One could probably make the transition even smoother, but that would require more ad hoc modifications of both EoS ingredients, for which we have no justification.

\begin{figure}[ht]
\centering
\includegraphics[width=0.9\textwidth]{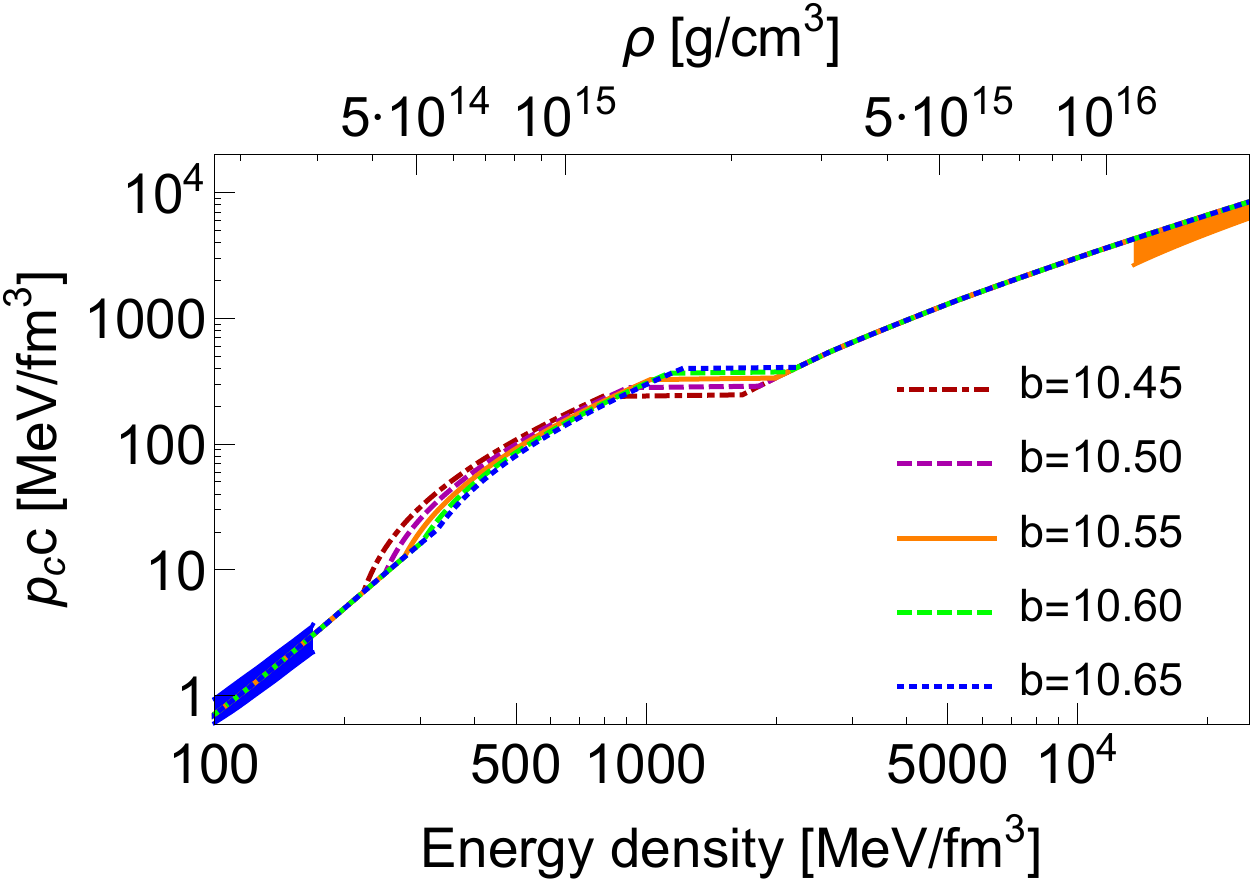}
\caption{\label{fig:holo:holo2EOSfinal}Hybrid equation of state for different values of $b$\@. The blue and orange bands show the constraints coming from effective low-density methods and perturbative QCD, respectively. It can be seen that the hybrid construction satisfies both of these constraints.}
\end{figure}

In this way, for every chosen value of $b$, one obtains a hybrid EoS\@.
This resulting hybrid EoS is shown for several choices of $b$ in figure \ref{fig:holo:holo2EOSfinal}\@.
At low densities, one can see that the hybrid EoS agrees with constraints coming from effective low-density methods, indicated by the blue band.
This is of course true by construction, as this part of the EoS is equal to the SLy EoS, which is itself constrained to satisfy this constraint.
At high densities, the hybrid EoS can be seen to agree with constraints from perturbative QCD at extremely large densities, indicated by the orange band.
This is also true by construction, because the asymptotics of the V-QCD potentials are constrained such that the deconfined phase of V-QCD satisfies this constraint.
Another interesting feature that one can see is that as $b$ increases, the matching chemical potential $\mu^*$ increases as well.
This effectively means that as $b$ increases, the proportion of the neutron stars described by SLy increases, and equations of state with smaller $b$ describe neutron stars with a relatively larger part described by holography.
\subsection{Mass-radius relation and tidal deformability}
\begin{figure}[ht]
\centering
\includegraphics[width=0.9\textwidth]{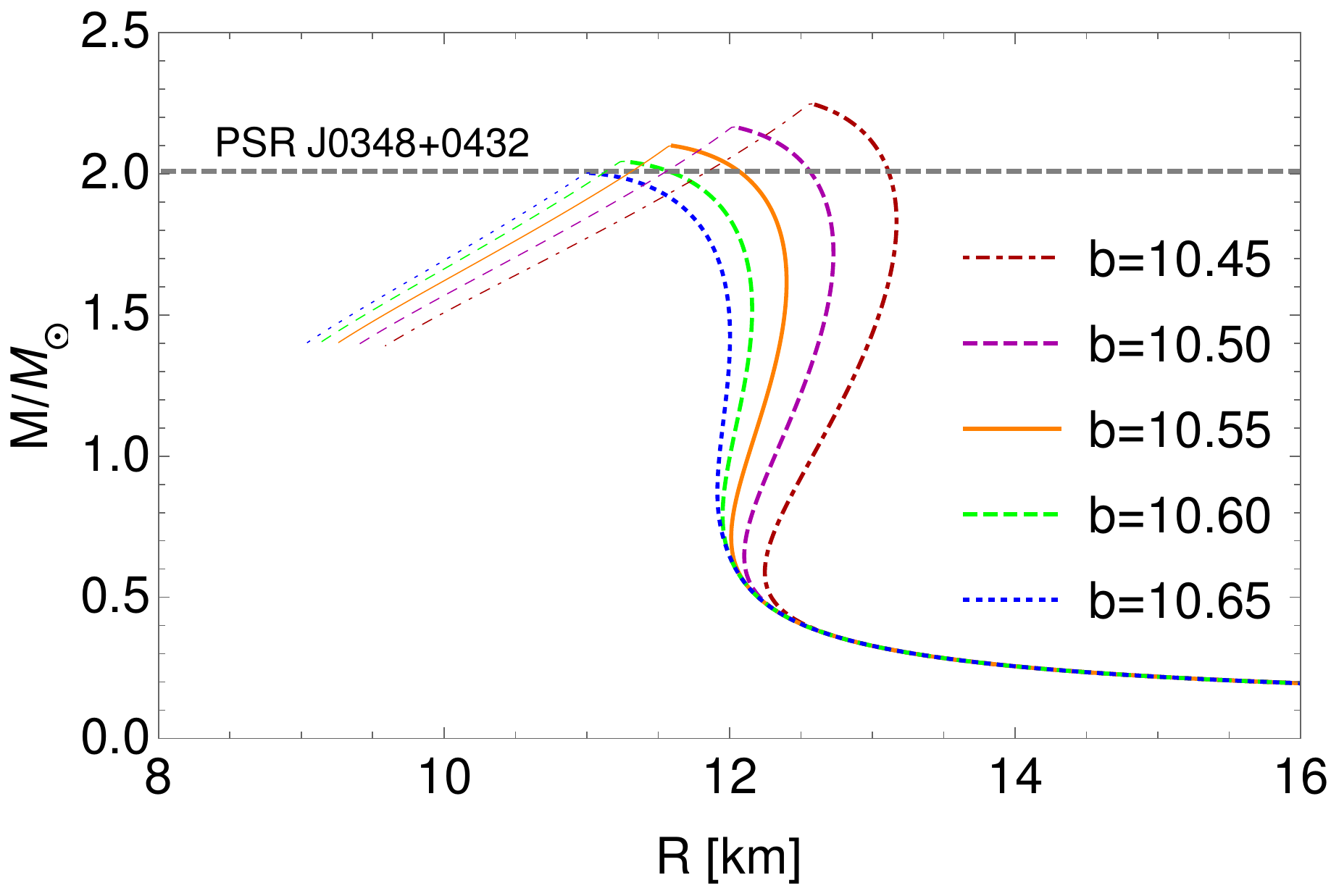}
\caption{\label{fig:holo:holo2MR}Mass-radius relation for the hybrid equation of state, for the values of $b$ also shown in figure \ref{fig:holo:holo2EOSfinal}\@. Also shown as a dashed line is the observational lower bound on the maximum neutron star mass given by \cite{Antoniadis:2013pzd}\@.}
\end{figure}

Now that we have a family of equations of state with input from holography, one can start computing its observable consequences to compare how well the equations of state compare to presently known constraints.
One of these observables is the mass-radius relation.
Given an equation of state, the Tolman-Oppenheimer-Volkov equations can be solved to yield a relation between the mass and radius of a non-rotating neutron star.
This mass-radius relation is shown for the hybrid equations of state in figure \ref{fig:holo:holo2MR}\@.
At some central density, the matter inside the neutron star undergoes a transition to deconfined matter, corresponding to the horizontal lines in figure \ref{fig:holo:holo2EOSfinal}\@.
The kink that is visible in figure \ref{fig:holo:holo2MR} corresponds to the central density at which this happens, and the solutions to the left of this kink have a deconfined quark matter core.
The negative slope of the $M$--$R$ relation for these neutron stars with quark matter cores implies that these solutions are unstable though, so the family of hybrid equations of state does not allow for stable quark matter cores.

One observable that can be read off from the mass-radius relation is the maximum mass of neutron stars, which for our family of equations of state decreases with increasing $b$\@.
This is an important observable, as the mass of a neutron star is a quantity that can be experimentally measured with a reasonably good accuracy.
Indeed, \cite{Antoniadis:2013pzd} measured the mass of the pulsar J0348+0432 to be $2.01 \pm 0.04\,M_\odot$\@.
This immediately implies that any equation of state that does not permit a stable neutron star of at least about $2\,M_\odot$ is incompatible with this observation.
Of course, due to the uncertainty in the measurement of \cite{Antoniadis:2013pzd} it is not precisely clear where one should draw the line, but for our equations of state this means that the EoS corresponding to $b = 10.65$ is near the edge of exclusion at $M_\text{max} = 2.00\,M_\odot$\@.
In 2019, \cite{Cromartie:2019kug} measured a potentially even heavier neutron star, J0740+6620, at $M = 2.14_{-0.09}^{+0.10}\,M_\odot$\@.
Including this measurement, the hybrid EoS for $b = 10.6$ is about one standard deviation lighter than the one measured in \cite{Cromartie:2019kug}, so we will take $b = 10.6$ as our upper bound for $b$\@.

As it turns out, we can also obtain a lower bound for $b$\@.
The reason for this is that using the equation of state one can also compute the tidal deformability $\Lambda$\@.\footnote{This is \emph{not} the same lambda as the one used in chapter \ref{ch:imc}.}
The tidal deformability is a dimensionless number, which for our family of hybrid equations of state decreases as a function of $b$\@.
The tidal deformability, as the name suggests, describes how easy it is for another object to gravitationally deform the star, and in fact this quantity has an influence on the gravitational wave signal of the inspiral phase of binary neutron star mergers.
From GW170817 as measured by LIGO/VIRGO, it was not possible to measure the magnitude of $\Lambda$, but it was possible to extract an upper bound of $\Lambda \lesssim 580$ at 90\% confidence level for a neutron star of mass $1.4\,M_\odot$ \cite{Abbott:2018exr}\@.
For the hybrid equations of state, it turns out the $b = 10.45$ corresponds to $\Lambda \simeq 680$ for such a neutron star, which would be ruled out, but $b = 10.5$ has $\Lambda \simeq 550$\@.

In this way we obtain a family of equations of state with input from holography with $10.5 \lesssim b \lesssim 10.6$ which are compatible with current (2020) observational data.
It is possible that future observations can constrain these values further though.
In particular, future results from NICER, which recently published its first results \cite{Raaijmakers:2019qny}, could potentially constrain the radius, thereby putting another constraint on the equation of state.
Also, with more data, it is possible that the maximum mass constraint becomes more stringent, and also tighter constraints on $\Lambda$ could potentially rule out more, or even perhaps all, values of $b$\@.
\subsection{Holographic neutron star mergers}
We saw in the previous subsection that the hybrid EoS does not exhibit quark matter cores.\footnote{This is not expected to change for rotating neutron stars.}
An interesting question is whether there are any circumstances in which the phase transition would lead to observable consequences.
As there are no static solutions with deconfined matter, such circumstances would have to be fleeting moments in a dynamical process.
One such process is a binary neutron star merger.
The gravitational waves from such a merger can be detected in LIGO/VIRGO, and as these detectors are upgraded, and as new ones are built, detection capabilities continue to grow.
The gravitational waveform contains information about the equation of state, both in the inspiral, as in the post-merger signal.
In the inspiral part of the waveform, the equation of state manifests itself mainly through the tidal deformability $\Lambda$, which in fact we have already used to constrain the parameter $b$\@.
Right after the merger, the single object is in a highly excited state, and vibrates with characteristic frequencies, which depend on the equation of state.
These frequencies can in principle be detected in the corresponding part of the waveform, though as of 2020 the post-merger signal has not been detected.

In this subsection, we will perform simulations of equal mass neutron star mergers, with the aim of extracting observable consequences of the hybrid EoS from the resulting waveforms.
We will restrict ourselves to equal mass mergers, and the stars will be initialized on quasi-circular orbits with a diameter of 45 km.
Here the reason why the orbits are only quasi-circular instead of circular is that the system continuously loses energy due to gravitational wave emission, and the reason to assume vanishing eccentricity is that the same gravitational wave emission tends to have circularized the orbits well before the merger takes place \cite{Radice:2020ddv,Aasi:2013wya}\@.
To generate these initial conditions, we will use the publicly available LORENE pseudospectral code \cite{Gourgoulhon:2000nn}\@.
As explained in the introduction, the subsequent merger is described by general relativity plus general relativistic hydrodynamics \cite{Rezzolla:2013book}\@.
To solve these equations, we use the Einstein Toolkit \cite{Schnetter:2003rb,Thornburg:2003sf,Goodale:2002a}, where we use the high-order, high-resolution shock-capturing code WhiskyTHC \cite{Radice:2012cu,Radice:2013hxh,Radice:2013xpa,Radice:2015nva,Radice:2012THCCode}, which solves the ideal general relativistic hydrodynamics equations in conservative form \cite{Banyuls:1997}\@.
For the solution of the Einstein equations themselves, we use the fourth-order finite differencing McLachlan code \cite{PhysRevD.79.044023,Loffler:2012}, which solves the Einstein equations in the CCZ4 formulation \cite{PhysRevD.85.064040}\@.
Here we use a ``$1 + \log$'' slicing condition, and a ``Gamma-driver'' shift condition \cite{PhysRevD.67.084023,PhysRevD.76.124002}\@.

\begin{figure}[ht]
\centering
\includegraphics[width=0.9\textwidth]{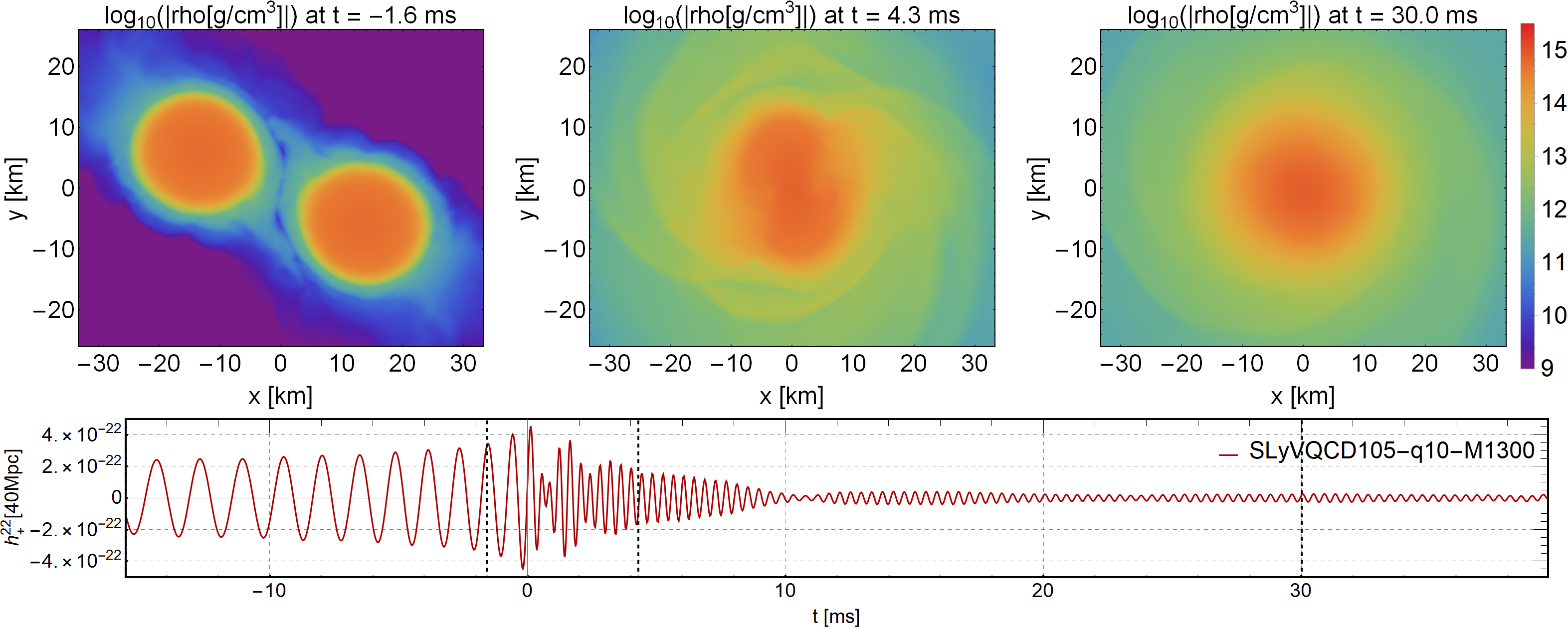}
\caption{\label{fig:holo:holo2moviesnapshots1300}Snapshots of the baryon number density at different times during the merger of two $M = 1.3\,\text{M}_\odot$ neutron stars, together with the resulting gravitational wave signal. The gravitational wave signal is extrapolated to an assumed distance of $40\,\text{Mpc}$\@.}
\end{figure}

In figure \ref{fig:holo:holo2moviesnapshots1300}, one can see the result of such a simulation for a binary neutron star merger where the two stars have mass equal to $1.3\,\text{M}_\odot$\@.
The equation of state used for this simulation is the hybrid EoS with $b = 10.5$\@.
One can see that the signal starts with an inspiral phase where the stars get increasingly close together as they radiate away energy in the form of gravitational waves.
Subsequently, when the stars touch, a highly excited object forms, which continues to emit gravitational waves until it settles down to a stable state.
In this simulation, 40 ms after the merger no horizon has yet formed, but it is unclear whether this simulation would eventually collapse to a black hole.
To investigate this, one would have to continue the simulation for much longer past 40 ms, which is computationally very expensive.
For this reason, it is not clear whether the final state of this merger simulation would be a neutron star or a black hole.

\begin{figure}[ht]
\centering
\includegraphics[width=0.9\textwidth]{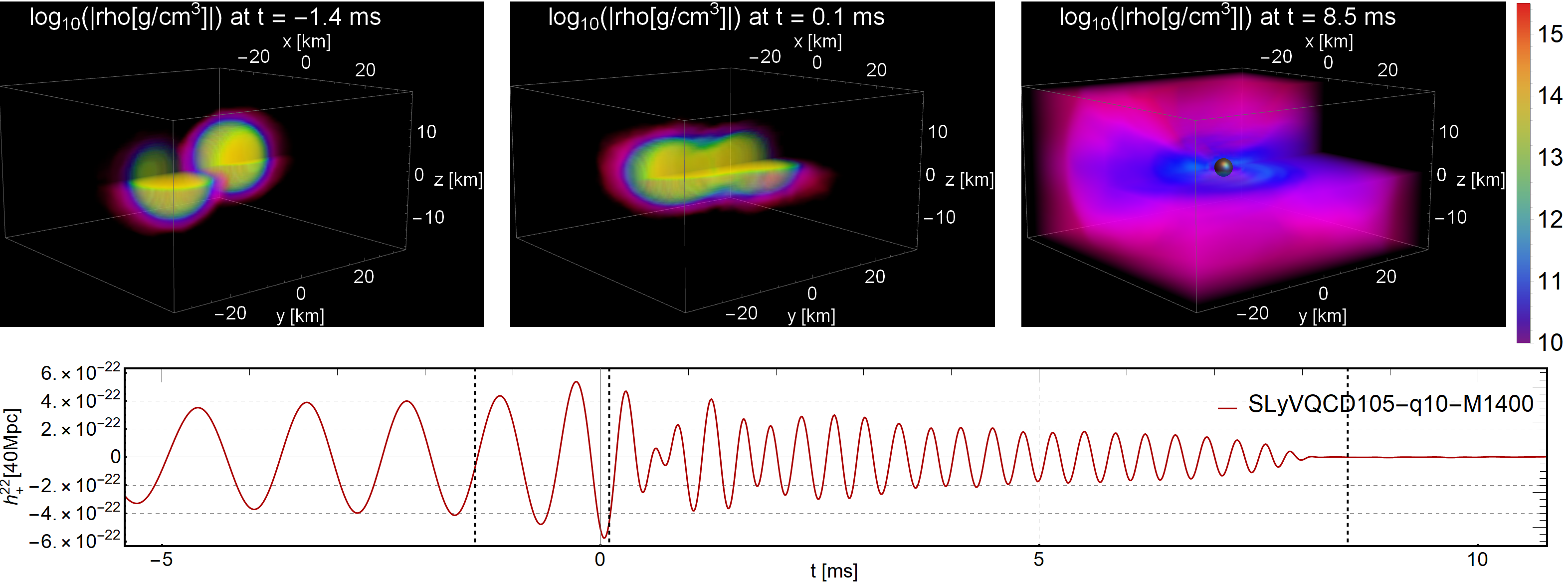}
\caption{\label{fig:holo:holo2moviesnapshots1400}3D snapshots of the baryon number density at different times during the merger of two $M = 1.4\,\text{M}_\odot$ neutron stars, where one quadrant has been omitted for visibility, together with the gravitational wave signal extrapolated to an assumed distance of $40\,\text{Mpc}$\@.}
\end{figure}

If the initial neutron stars are taken to be much heavier, an event horizon will form immediately after the stars touch, and in this case the gravitational wave signal dies down quickly after that.
For the hybrid equation of state with $b = 10.5$, we see this happening when the stars both have mass $M = 1.5\,\text{M}_\odot$\@.
In the intermediate case, something very interesting happens.
In figure \ref{fig:holo:holo2moviesnapshots1400}, the result of a simulation of two neutron stars with masses equal to $1.4\,\text{M}_\odot$ is shown, again with $b = 10.5$\@.
In this case, after the merger a highly excited object forms just as with the lighter stars, but now at $7.6\,\text{ms}$, some of the matter in the middle of the resulting object reaches densities large enough to cross the phase transition into quark matter, which immediately causes a collapse to a black hole.
This is an example of phase transition triggered collapse \cite{Weih:2019xvw}\@.

\begin{figure}[ht]
\centering
\includegraphics[width=0.9\textwidth]{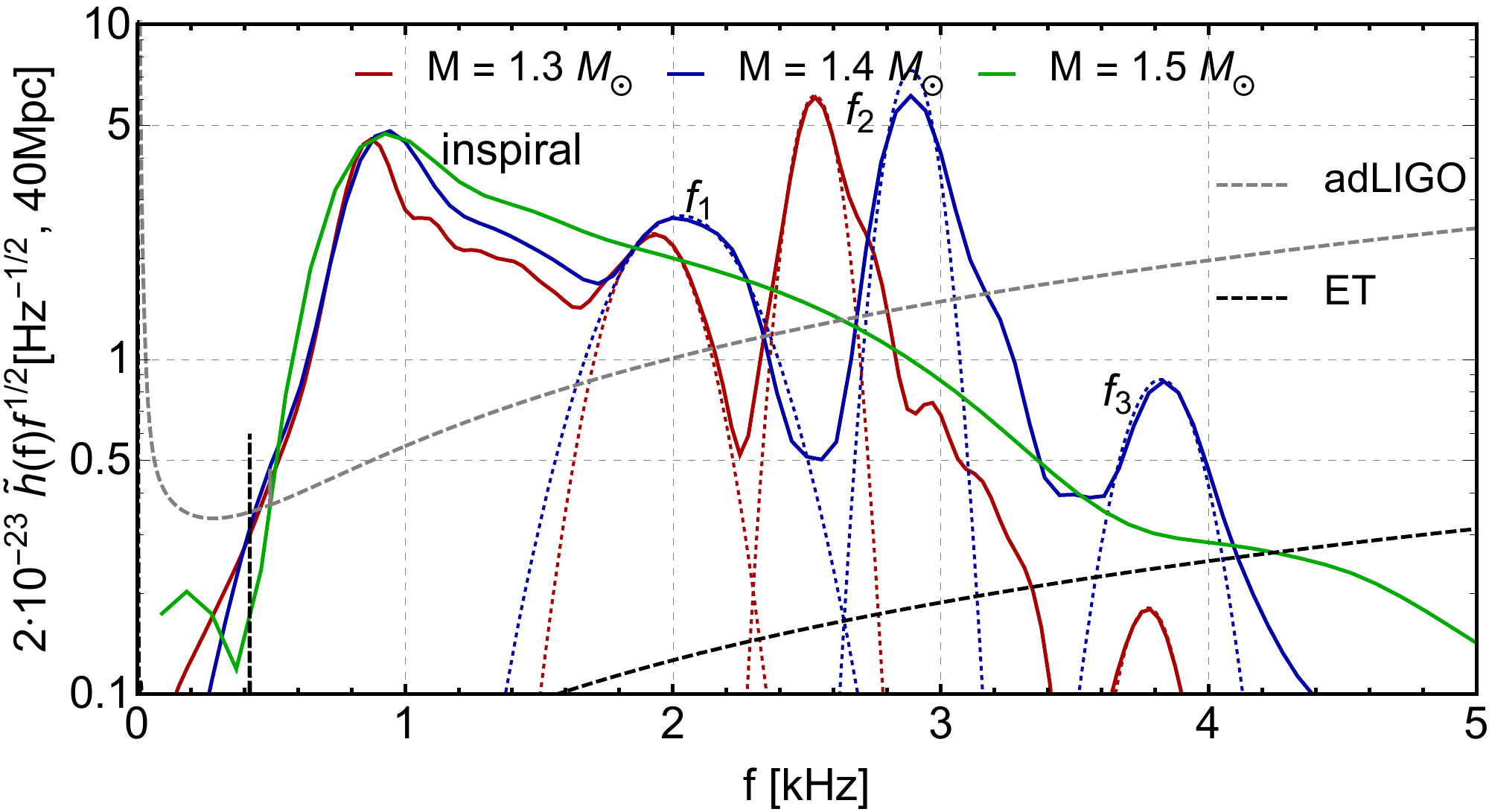}
\caption{\label{fig:holo:holo2psdcomparemass}Power spectrum density (PSD) for the hybrid EoS with $b = 10.5$\@. The dotted lines indicate Gaussian fits to the curves, which determine the characteristic frequencies $f_1$, $f_2$ and $f_3$\@.}
\end{figure}

Together, these three cases describe the possible outcomes from a neutron star merger for the hybrid equations of state.\footnote{In \cite{Weih:2019xvw} there are also options which form remnants with quark matter cores, but this is not possible for our equations of state, as these remnants would be unstable.}
A quantity which contains a lot of information is the power spectral density (PSD) \cite{Takami:2014zpa}:
\[
\tilde h(f) \equiv \sqrt{\frac{|\int h_+(t)e^{-i2\pi ft}\,dt|^2 + |\int h_\times(t)e^{-i2\pi ft}\,dt|^2}{2}},
\]
where the integrations are performed from $-7$ to $24\,\text{ms}$ around the merger, where $t = 0$ is defined as the maximum of the gravitational wave amplitude.
In figure \ref{fig:holo:holo2psdcomparemass}, the PSD is shown for the three possible cases discussed above, each with the hybrid equation of state with $b = 10.5$\@.
Generically, the post-merger signal contains three characteristic frequencies, labeled $f_1$, $f_2$ and $f_3$ \cite{Baiotti:2016qnr}\@.
Here $f_1$ is a universal value which is determined by the compactness $M/R$ \cite{Takami:2014zpa}\@.
This implies that from $f_1$ no information from the equation of state can be obtained that couldn't already be obtained from the mass-radius relation.
The other frequencies $f_2$ and $f_3$ depend on the equation of state in a more non-trivial way, and therefore these frequencies would in principle allow one to obtain additional information regarding the EoS\@.
Note also that the $M = 1.5\,\text{M}_\odot$ case does not contain the characteristic frequencies in its PSD\@.
The reason for this is that the characteristic frequencies are caused by oscillations of the dense matter, which are absent in this case, because the configuration promptly collapses to a black hole.

\begin{figure}[ht]
\centering
\includegraphics[width=0.9\textwidth]{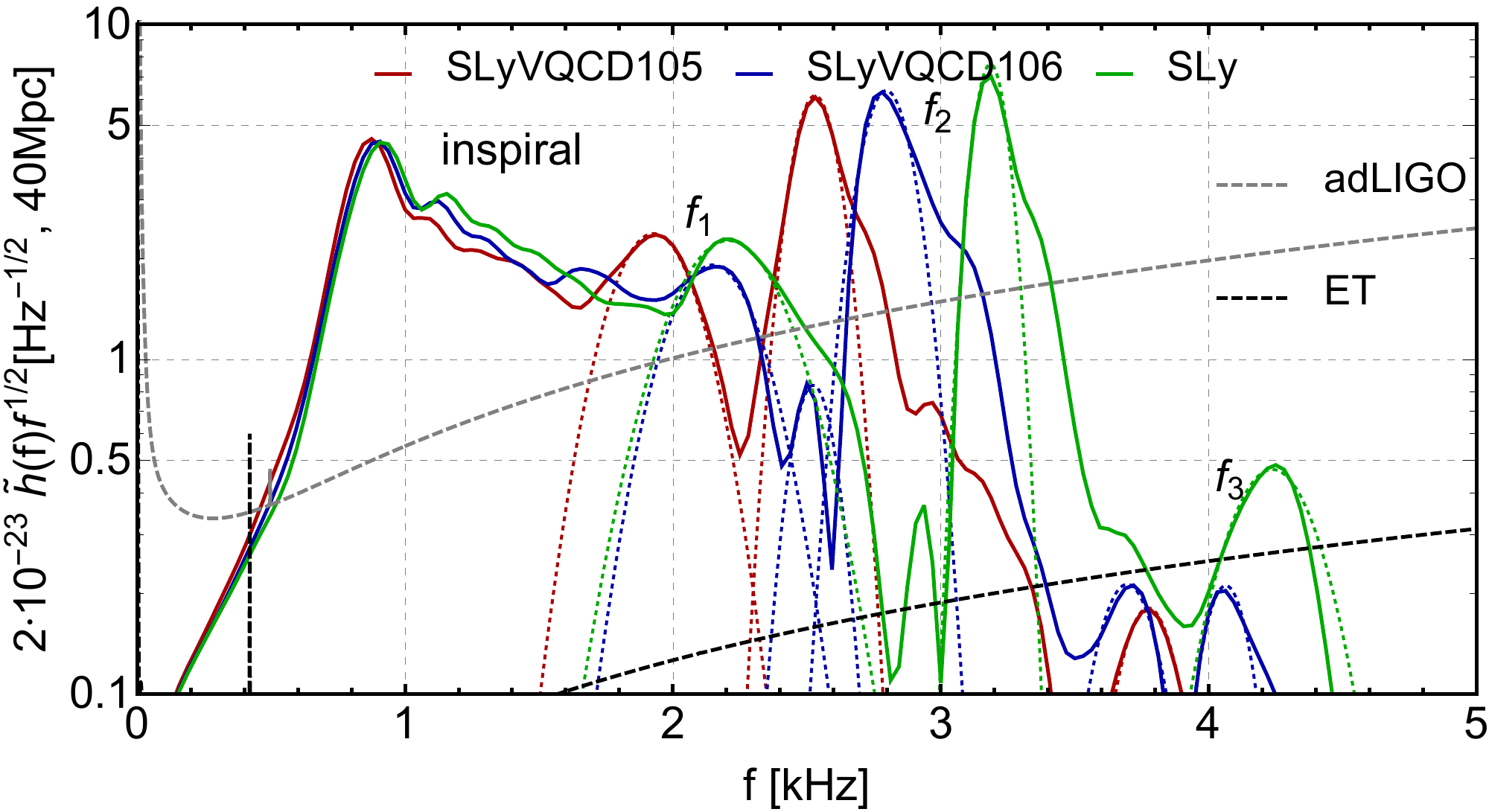}
\caption{\label{fig:holo:holo2psdcompareeos}Power spectrum density (PSD) for mergers of two $M = 1.3\,\text{M}_\odot$ stars, with two hybrid equations of state, with $b = 10.5$ (red) and $b = 10.6$ (blue), and one with the SLy equation of state (green) for comparison. The dotted lines indicate Gaussian fits to the curves, which determine the characteristic frequencies $f_1$, $f_2$ and $f_3$\@.}
\end{figure}

A final interesting thing one can do, is to compare mergers of equal mass neutron stars where one keeps the mass fixed at $1.3\,\text{M}_\odot$, while changing the equation of state.
In figure \ref{fig:holo:holo2psdcompareeos}, the resulting PSD of such a comparison is shown.
Here we show the resulting PSD for the hybrid equations of state with $b = 10.5$ and $b = 10.6$, as well as the SLy equation of state for comparison, so that we can clearly see what the impact is of including input from holography into the equation of state.
The resulting $f_1$ frequencies lie on the universal curve proposed in \cite{Takami:2014zpa}\@.
The $f_2$ peak shifts to significantly lower frequencies for the hybrid equation of state.
Note that this shift is larger for the $b = 10.5$ case, which has a smaller matching density, and therefore contains more input from holography.
Also, the $f_3$ peak has a smaller amplitude for the hybrid equations of state as compared to the pure SLy case.

In conclusion, the hybrid equation of state makes testable predictions for various neutron star observables.
The interesting aspect of this is that holography gives us for the first time an approach to the EoS which is inherently strongly coupled.
In the future, one could refine the analysis from both sections of this chapter to perhaps improve the reliability of the result.

%% file: chapters/trajectum.tex
In the last chapter, we studied the low-temperature, large chemical potential region of the phase diagram using holography.
This gave us insights on how an explicitly strongly coupled framework such as holography can yield observationally testable predictions.
In this chapter, we will focus on another region in the phase diagram, namely that of high temperature and close to vanishing chemical potential, where a quark-gluon plasma (QGP) exists (see \cite{Heinz:2013th} and references therein)\@.
To be precise, we will take the chemical potential to be zero, but a small chemical potential could in principle be added to the analysis presented here in the future.
QCD matter in this region of the phase diagram can be created in heavy ion collisions, such as the ones performed at RHIC and LHC\@.
At the LHC, the collisions are performed at $2.76\,\text{TeV}$ and $5.02\,\text{TeV}$, using lead-208 and xenon-129 nuclei, as well as protons.
In the future it is considered to collide oxygen-16 and argon-40 as well, but as of 2020 no decision has been made.

Heavy ion collisions are able to provide us a unique insight into QCD, and many quantities of theoretical interest can in principle be observed in this way.
Examples of these are the following questions:
\begin{itemize}
\item What does the initial state of a heavy ion collision look like?
\item How can the dynamics of the matter created at the collision be described before it can be described by hydrodynamics? After which time does hydrodynamics apply?
\item What are the values of QGP transport coefficients like the shear viscosity $\eta$ and the bulk viscosity $\zeta$\@? Can higher order transport coefficients and relaxation times like $\tau_\pi$ and $\tau_\Pi$ be measured?
\end{itemize}
One first thing to note is that the question as to the nature of the equation of state at vanishing chemical potential is not included.
The reason for this is that the finite temperature equation of state is well known from lattice QCD computations \cite{Borsanyi:2013bia,Bazavov:2014pvz}\@.
While it is true that the dependence of the EoS on the chemical potential is less well known,\footnote{This is only known on the lattice for relatively small values of the chemical potential \cite{deForcrand:2010ys,Borsanyi:2011sw}\@.} this is beyond the scope of this chapter, since we will not consider finite chemical potential.
A second thing to note is that out of equilibrium properties, like how the initial state reaches local equilibrium, and even near-equilibrium properties like transport coefficients, are very difficult to compute on the lattice.
In holography, however, many of these quantities can be computed, and often give surprising results.
A famous example of this is the prediction that the ratio of the shear viscosity over the entropy density of the QGP is given by \cite{Policastro:2001yc,Kovtun:2004de,CasalderreySolana:2011us,Maldacena:1997re}
\[
\frac{\eta}{s} = \frac{1}{4\pi}.
\]
Another result from holography is that the matter created in the collision can be described using hydrodynamics a very short time after the collision \cite{vanderSchee:2013pia}\@.
However, all of these results are computed under the assumption of infinite coupling and infinite number of colors, so it is an interesting question how well these results match the properties of an actual QGP\@.

Measuring these quantities of interest, however, is not as easy as it may seem.
Most observables in the final state of a heavy ion collision depend on more than just one of the quantities of interest, making it difficult to measure an observable and subsequently inferring something about the quantity of interest.
What \emph{is} possible though is to create a detailed model which simulates heavy ion collisions for different choices of the quantities of interest, which we will from now on call `input parameters'\@.
Such a model can then produce predictions for many different observables for each particular set of input parameters.
By using such a model in a Bayesian analysis, it is then possible to fit the input parameters to real experimental data, allowing one to learn something about which input parameters fit the data best \cite{Bernhard:2016tnd,Ke:2016jrd,Bernhard:2017vql,Moreland:2017kdx,Bass:2017zyn,Ke:2017xyi,Moreland:2018jos,Moreland:2018gsh,Bernhard:2019bmu}\@.

In this chapter, we will construct such a model, called \emph{Trajectum}, which also includes code to analyze the results\@.\footnote{Trajectum is the Roman name for the city of Utrecht, where this code was developed. It also means bridge, which is appropriate, as the aim of \emph{Trajectum} is to bridge the gap between theory and experiment.}
While it will not perform any computations using holography, many of its inputs will be inspired by holographic results of which we would like to test the validity in experiments.
The actual Bayesian analysis will not be performed in this chapter, but the results will be compared to section 5.3 of \cite{Bernhard:2018hnz} as a check that \emph{Trajectum} is able to reproduce known results.
In the next section, we will explain the overall design of \emph{Trajectum}, after which we will explain the various components that work together to create the simulation of the heavy ion collisions.
Finally, we will explain the observables which can be extracted from the resulting collisions, and compare them to previous work.
\section{Overall design of \emph{Trajectum}}
Of course, \emph{Trajectum} is not the first code to perform the tasks described above.
Many codes exist which perform parts of the computation necessary to simulate a collision \cite{Shen:2014vra,Schenke:2010nt,Schenke:2010rr,Paquet:2015lta,Miller:2007ri,Barej:2017kcw,Loizides:2016djv,Bozek:2016kpf,Welsh:2016siu,Weller:2017tsr,Moreland:2014oya,Schenke:2012wb,Gale:2012rq,Eskola:2000xq,Du:2019obx,DelZanna:2013eua,Habich:2014jna,Huovinen:2012is,Cooper:1974mv,Bozek:2009dw,Dusling:2011fd,Teaney:2003kp,Pratt:2010jt,Bernhard:2018hnz,Bass:1998ca,Bleicher:1999xi,Weil:2016zrk}\@.
These codes then need to made to work together though, where care must be taken to hand over the result of each step to the next step in the computation in the right format.\footnote{Recently, a framework \cite{Putschke:2019yrg,Kumar:2020vkx} has been developed which automates this process, but this was not known to the author during the development of \emph{Trajectum}\@.}
The aim of \emph{Trajectum} is to collect all of these steps and to reimplement them into a framework which provides a standard interface between the various required components.
In this way, one ends up with two executables, \emph{collide} and \emph{analyze}\@.
The first of these computes some desired number of heavy ion collisions with input parameters according to the user's preference.
At the time of writing, \emph{collide} does not provide a hadronic afterburner, but the output of \emph{collide} is in the correct format for use in UrQMD \cite{Bass:1998ca,Bleicher:1999xi}\@.
The output of UrQMD can subsequently be used in \emph{analyze}, which is able to compute a range of observables, which will be discussed below.

The advantage of enforcing a standard interface between the components has one very clear advantage.
For each component required for the simulation, several choices are available.
As an example, the simulation requires initial conditions, for which several models exist.
Of these, the following are implemented in \emph{collide}:
\begin{itemize}
\item Monte Carlo Glauber \cite{Miller:2007ri,Barej:2017kcw,Loizides:2016djv,Bozek:2016kpf},
\item Ohio State University \cite{Welsh:2016siu,Weller:2017tsr},
\item T\raisebox{-0.5ex}{R}ENTo \cite{Moreland:2014oya},
\item Gubser flow \cite{Gubser:2010ze,Gubser:2010ui,Marrochio:2013wla}.
\end{itemize}
When using \emph{collide}, the user can choose which of these to use, and as they adhere to the standard interface, they are guaranteed to work together correctly with the other components.
Similarly, all the other components of \emph{collide} can be interchanged for various different options.
The standard interface is implemented in C++ using polymorphism.

As was mentioned above, many of the components implemented in \emph{Trajectum} are reimplementations of existing earlier work.
In the sections below, we will discuss each of the various components in detail.
In certain places, the original works have been extended or modified in order to address some of the holography-inspired questions posed above.
Wherever the implementation of an algorithm in \emph{Trajectum} differs significantly from the one in the corresponding earlier work, this will be clearly stated.
\section{\emph{collide} executable}\label{sec:trajectum:collide}
As was mentioned in the introduction, a heavy ion collision is simulated in several stages.
This consists of a pre-equilibrium stage which provides initial conditions for the hydrodynamical simulation of the QGP, followed by a hydrodynamical evolution.
From the hydrodynamical evolution a freeze-out surface is computed, which is defined as an isotherm called $T_\text{fr}$\@.
At this freeze-out surface, the code translates from a continuous fluid description to a discrete particle description.
The particles must subsequently be written to a file, so that they can be further processed by the hadronic afterburner UrQMD, which simulates interactions between the particles produced up to the point in time where they can be taken to be non-interacting.
The afterburner also simulates the decay of unstable particles.

These stages each translate into one or more components in \emph{collide}\@.
An event is simulated in \emph{collide} as follows:
The component responsible for the pre-equilibrium stage (called `initial conditions') generates initial conditions for the hydrodynamical evolution.
This is then passed to the component for the hydrodynamical evolution.
The component for the hydrodynamical evolution (called `hydrodynamics model') then evolves the evolution in time step by step.
For this it depends on two auxiliary components:
\begin{itemize}
\item A component (called `transport coefficients') containing the equation of state, as well as first and (if needed) second order transport coefficients.\footnote{Note the slight abuse of terminology here. Usually the equation of state is not considered a transport coefficient, but for all purposes one could consider it a 0th order transport coefficient, and we will do so throughout this chapter.}
\item A component (called `PDE solver') which implements an algorithm for solving partial differential equations.
\end{itemize}
By splitting off these components from the hydrodynamical evolution itself, it becomes easier to alter the simulation, as one can easily implement a new solver or set of transport coefficients.
In principle, to do this it is not even necessary to understand how each of the other components works.
One only needs to know how for example the transport coefficients should interface with the framework provided by \emph{collide}, which then guarantees that the new set of transport coefficients correctly works together with the other components.

After each time step is computed, the hydrodynamics model hands the new state of the fluid over to the last component (called `hadronizer'), which is responsible for computing the freeze-out surface and generating particles from the fluid.
When the hadronizer determines that after the last computed time step there is no new addition to the freeze-out surface, the hadronizer causes the entire computation to terminate, and \emph{collide} will move on to the next event.
An important note here is that even though the wording in the last sentence may suggest that \emph{collide} waits for each event to be completed before moving on to the next event, \emph{collide} is actually multithreaded, and will compute 20 simulations simultaneously.\footnote{If desired, multithreading can be turned off when compiling \emph{Trajectum}\@.}

Summarizing, we have 5 different components which together simulate a collision:
\begin{itemize}
\item Initial conditions,
\item Hydrodynamics models,
\item Transport coefficients,
\item PDE solvers,
\item Hadronizers.
\end{itemize}
In the following subsections, each of these components will be explained.
In particular, all currently available options that the user can choose from will be covered.
Also, an important point to mention at this point is that we will assume boost invariance, i.e.~we solve hydrodynamics in $2 + 1$D\@.
This is done by taking the following metric (called the Milne metric):
\begin{equation}
ds^2 = d\tau^2 - dx^2 - dy^2 - \frac{d\eta^2}{\tau^2},\label{eq:trajectum:milnemetric}
\end{equation}
which is related to the Minkowski metric by the coordinate transformation
\[
\tau = \sqrt{t^2 - z^2}, \qquad \eta = \frac{1}{2}\log\left(\frac{t + z}{t - z}\right),
\]
with $\tau$ the proper time, and $\eta$ the pseudorapidity.
With this metric, boosts correspond to shifts in $\eta$, and we can implement boost invariance by assuming that all of the variables constituting the fluid have no $\eta$-dependence.
Also, for the rest of this section, we define the canonical order of variables $(\tau,x,y,\eta) = (0,1,2,3)$\@.
\subsection{Initial conditions}
As was briefly mentioned before, the aim of the initial conditions component is to provide initial conditions to the hydrodynamical evolution.
In \emph{collide}, there are four different sets of initial conditions implemented:
\begin{itemize}
\item Monte Carlo Glauber \cite{Miller:2007ri,Barej:2017kcw,Loizides:2016djv,Bozek:2016kpf},
\item Ohio State University \cite{Welsh:2016siu,Weller:2017tsr},
\item T\raisebox{-0.5ex}{R}ENTo \cite{Moreland:2014oya},
\item Gubser flow \cite{Gubser:2010ze,Gubser:2010ui,Marrochio:2013wla}.
\end{itemize}
In \cite{Gubser:2010ze,Gubser:2010ui}, an analytical solution of viscous hydrodynamics was derived under the assumption of a conformal equation of state, constant $\eta/s$ and $\zeta/s = 0$\@.
Gubser flow initial conditions implement initial conditions as a time slice of this analytical solution.
This provides a good non-trivial check on the implementation of the numerical code to solve the hydrodynamical evolution \cite{Marrochio:2013wla}\@.
The other three initial conditions are phenomenological models attempting to describe the initial state of a heavy ion collision as well as possible.
They also include a model of the evolution from the proper time of collision ($\tau = 0^+$) to the moment that the hydrodynamical evolution is initiated (denoted $\tau_\text{fs}$)\@.
In the remaining paragraphs of this subsection, we will discuss the various steps involved in all of these models.
As the models are rather similar in setup, this will be done for all three models simultaneously, where we point out the differences where they occur.

The first step in all three remaining models is to determine the positions of nucleons in both nuclei.
For protons, this is simple.
As there is only one nucleon inside the nucleus, the single nucleon just sits in the center of the nucleus.
For the other nuclei impemented in \emph{collide}, we assume that the nucleons are distributed according to a Saxon-Woods distribution \cite{Loizides:2014vua,Sievert:2019zjr}:\footnote{Note that this is a probability density in Cartesian coordinates. When sampling from the distribution, one can either sample $x$, $y$, $z$, then covert to $r$, $\theta$, $\phi$ and accept with a probability given by \eqref{eq:trajectum:saxonwoods}, or one can introduce a Jacobian to \eqref{eq:trajectum:saxonwoods} and sample from it directly.}
\begin{equation}
\rho(r,\theta,\phi) = \rho_0\frac{1 + w(r/R(\theta))^2}{1 + \exp\left(\frac{r - R(\theta)}{a}\right)},\label{eq:trajectum:saxonwoods}
\end{equation}
where $\rho$ is the probability density to find a nucleon at radial distance $r$, polar angle $\theta$ and azimuthal angle $\phi$, $\rho_0$ is a normalization factor, $w$ and $a$ are parameters, and
\[
R(\theta) = R\left(1 + \beta_2Y_{20}(\theta) + \beta_4Y_{40}(\theta) + \cdots\right).
\]
Here $R$, $\beta_2$ and $\beta_4$ are more parameters, and $Y_{nm}$ are spherical harmonics.
The values for these parameters for the nuclei implemented in \emph{collide} are given in table \ref{tab:trajectum:saxonwoods}\@.
\begin{table}[ht]
\centering
\begin{tabular}{r|c|c|c|c|c|}
Nucleus & $R$ (fm) & $a$ (fm) & $w$ (fm) & $\beta_2$ & $\beta_4$ \\
\hline
\hline
${}^{16}$O & 2.608 & 0.513 & -0.051 & 0 & 0 \\
${}^{40}$Ar & 3.73 & 0.62 & -0.19 & 0 & 0 \\
${}^{129}$Xe & 5.36 & 0.59 & 0 & 0.161 & -0.003 \\
${}^{208}$Pb & 6.62 & 0.546 & 0 & 0 & 0
\end{tabular}
\caption{\label{tab:trajectum:saxonwoods}Saxon-Woods parameters for the different nuclei available in \emph{collide}\@.}
\end{table}
The nucleons constituting the two nuclei are sampled from the corresponding Saxon-Woods distribution.
In doing this care must be taken that the axis of symmetry of the non-spherically symmetric nuclei needs to point in a random direction.
In T\raisebox{-0.5ex}{R}ENTo initial conditions, the nucleons also have to satisfy the property that the distance between two nucleons has to be larger than a minimal distance $d_\text{min}$\@.
This is enforced as follows:
One samples coordinates $r$, $\theta$ and $\phi$ according to \eqref{eq:trajectum:saxonwoods}\@.
Then, starting from the nucleons with the smallest $r$, $\phi$ is resampled in such a way that the nucleon satisfies the minimal distance requirement with respect to all nucleons with smaller $r$ than itself.
For all initial conditions, the sampled nucleons are subsequently converted into Cartesian coordinates, where the $z$-coordinate which points along the beam is discarded.
The remaining coordinates are given a random offset, reflecting the fact that the nuclei collide with a random impact parameter.

The next step is to determine the locations of the constituents of each nucleon.
Here there is a difference between the initial conditions models we are considering, as the Monte Carlo glauber model does not have nuclear substructure, the Ohio State University model has substructure with 3 constituents per nucleon, and the T\raisebox{-0.5ex}{R}ENTo model only optionally has substructure, but with an arbitrary number of nucleons $N_\text{const}$\@.
For the initial conditions without substructure, one can simply skip this step, and consider a single constituent of each nucleon, located at the location of the corresponding nucleon which was computed in the previous step.
The procedure below describes how constituents are sampled in the T\raisebox{-0.5ex}{R}ENTo model for initial conditions.
The procedure for Ohio State University initial conditions is parameterized differently, but is physically equivalent.
In the T\raisebox{-0.5ex}{R}ENTo model without substructure, nucleons are modelled as Gaussian blobs of density with width $w$\@.\footnote{This is the standard notation. Note that this $w$ is not the same as the one appearing in the Saxon-Woods distribution.}
In the version with substructure, there are $N_\text{const}$ constituents, which are modelled as Gaussians blobs of density with width $v$, sampled from a Gaussian distribution around the center of the nucleon, with width $r$\@.
As it is convenient that the width parameter $w$ from the model without substructure corresponds roughly to the one with substructure, the model with substructure is parameterized in terms of $w$ and $v$, where $v < w$\@.
The width of the distribution of constituents $r$ is then chosen as
\begin{equation}
r^2 = \frac{w^2 - v^2}{1 - \frac{1}{N_\text{const}}}.\label{eq:trajectum:constituentradius}
\end{equation}
For each constituent in each nucleon, the position is sampled from a gaussian with width $r$ around the center of the corresponding nucleon.
However, the average of these sampled constituents does not necessarily correspond to the center of the corresponding nucleon.
Therefore, the constituents are moved by such an amount so that the center \emph{does} coincide with the center of the corresponding nucleon.
Then, because of the choice of $r$ from \eqref{eq:trajectum:constituentradius}, the nucleons will on average have width $w$\@.\footnote{Here width is taken to be the RMS of the nucleon. This is not the same RMS of the deposited energy in the fluid though, as we will later fluctuate how much density each constituent deposits according to a Gamma distribution.}

Now that we have the positions of all the nucleons and (depending on the choice of initial conditions) their constituents, we have to determine which nucleons become `wounded'\@.
Wounded nucleons are precisely the nucleons that participate in the collision.
To determine which nucleons participate, we examine all pairs of nucleons, where we take the first nucleon (nucleon A) from the first nucleus (which we will call nucleus A), and the second nucleon (nucleon B)) from the second nucleus (which we will call nucleus B)\@.
For each such pair, we then compute the overlap function
\[
T_{AB} = \sum_{i=1}^{N_\text{const}}\sum_{j=1}^{N_\text{const}}\frac{1}{4\pi v^2}\exp\left(\frac{-(x_{A,i} - x_{B,j})^2 - (y_{A,i} - y_{B,j})^2}{4v^2}\right),
\]
where $x_{A,i}$ are the $x$-coordinates of the constituents of nucleon A, $y_{A,i}$ the $y$-coordinates of the constituents of nucleon A, and similarly for $x_{B,j}$ and $y_{B,j}$\@.
For the initial conditions without substructure, this overlap function is to be interpreted to have a single constituent for each nucleon, where we also have to replace $v$ by $w$\@.
Now that we have the overlap function, the probability that these two nucleons participate in the collision is
\[
P_\text{coll} = 1 - \exp(-\sigma_{gg}T_{AB}),
\]
where $\sigma_{gg}$ is a constant.
In accordance with this probability, the nucleons are both marked as being `wounded' with probability $P_\text{coll}$\@.\footnote{Note that they are either both marked as wounded, or both not. They are not separately marked with probability $P_\text{coll}$\@.}
If the two nucleons are marked as wounded, the corresponding nucleon pair is also marked as being a `binary collision pair'\@.
We have here introduced the parameter $\sigma_{gg}$\@.
In principle, we could leave this as a parameter to be determined by the user.
Instead of this, however, we choose $\sigma_{gg}$ precisely such that the average cross-section of proton-proton collisions is the same as the parameter $\sigma_{NN}$, which is available to the user to specify.

Knowing which nucleons participate in the collision, we can start constructing a `density' to initialize the plasma.
Note that this density is just a function $T(x,y)$\@.
Depending on the model, it can be interpreted as an entropy density, energy density, or specific component of the stress-energy tensor.
How exactly the different models interpret $T$ will be discussed towards the end of this section.
To compute $T$, let us first define the `thickness functions' $T_A(x,y)$, $T_B(x,y)$ and $T_{AB}(x,y)$:
\[
T_A(x,y) = \sum_{i\in\text{wounded nucleons}}\sum_{j=0}^{N_\text{const}}\frac{\gamma_{ij}}{2\pi v^2N_\text{const}}\exp\left(\frac{-(x_{ij} - x)^2 - (y_{ij} - y)^2}{2v^2}\right),
\]
where $x_{ij}$, and $y_{ij}$ denote the $x$ and $y$-coordinates of constituent $j$ of nucleon $i$, and the outer sum goes over all wounded nucleons in nucleus A\@.
What the weight $\gamma_{ij}$ is, depends on the initial conditions.
For the Monte Carlo Glauber model, it is equal to 1, whereas for both the Ohio State University model and the T\raisebox{-0.5ex}{R}ENTo model it is a random number, which for each constituent is drawn from a Gamma distribution with mean equal to 1, and standard deviation equal to a parameter called $\sigma_\text{fluct}$\@.
An analogous expression holds for $T_B(x,y)$ in terms of the wounded nucleons in nucleus B\@.
The last function, $T_{AB}(x,y)$, is not determined in terms of the wounded nucleons, but in terms of the binary collision pairs.
As $T_{AB}$ will only be used by the Monte Carlo Glauber, and since this model does not have constituents, the expression below is given in terms of the nucleon positions themselves:
\[
T_{AB} = \sum_{(i,j)\in\text{binary collision pairs}}\frac{1}{2\pi v^2}\exp\left(\frac{-(\frac{1}{2}(x_i + x_j) - x)^2 - (\frac{1}{2}(y_i + y_j) - y)^2}{2w^2}\right),
\]
where $(x_i,y_i)$ are the coordinates of the nucleon from nucleus A participating in the binary collision, and $(x_j,y_j)$ are the coordinates of the nucleon from nucleus B\@.
In words, the wounded nucleons deposit density into the thickness functions $T_A$ and $T_B$ at the location of the constituents, and each binary collision pair deposits density into the thickness function $T_{AB}$\@.

How the three thickness functions are subsequently combined into one density $T$ again depends on the specific model.
In the Monte Carlo Glauber model, we have
\[
T(x,y) = \frac{1 - \alpha}{2}\left(T_A(x,y) + T_B(x,y)\right) + \alpha T_{AB}(x,y),
\]
with $\alpha$ a parameter.
In the Ohio State University model, the $T_{AB}$-contribution is ignored, and we have
\[
T(x,y) = T_A(x,y) + T_B(x,y).
\]
In the T\raisebox{-0.5ex}{R}ENTo model, we have
\[
T(x,y) = \left(\frac{T_A^p(x,y) + T_B^p(x,y)}{2}\right)^{1/p},
\]
with $p$ a parameter.
Note that in the limit $p = 0$, this reduces to the geometric mean
\[
T(x,y) = \sqrt{T_A(x,y)T_B(x,y)}.
\]

At this point, for all models considered, we have a function $T(x,y)$ with the dimension of inverse area.
We now have several options for how to interpret this function.
In both the Monte Carlo Glauber and Ohio State University models, $T$ is interpreted as an entropy density as follows:
\[
s(x,y) = \frac{nT(x,y)}{\tau_\text{fs}},
\]
where $n$ is a dimensionless number called the norm, and $\tau_\text{fs}$ is the proper time at which the hydrodynamical evolution is initialized.
Using the equation of state, this can then be used to determine energy density and pressure.
This does not completely fix the initial condition for the hydrodynamical evolution though, as the stress-energy tensor has 7 independent components, and fixing the energy density only fixes one.\footnote{Note that assuming boost invariance reduces the number of components by 3, as $T^{\mu\eta} = 0$ for $\mu = \tau,x,y$\@.}
In \emph{collide}, there are two options to determine the other coefficients.
The first sets the velocities to $u^\mu = (1,0,0,0)$, sets bulk pressure $\Pi = 0$, and subsequently computes the shear tensor as\footnote{Note that in Milne coordinates \eqref{eq:trajectum:milnemetric}, $\sigma^{\mu\nu}$ can be non-zero even if $u^\mu = (1,0,0,0)$\@.}
\begin{equation}
\pi^{\mu\nu} = 2\eta\sigma^{\mu\nu},\label{eq:trajectum:holographicshearinit}
\end{equation}
where the code gets the shear viscosity $\eta$ from the transport coefficients model chosen by the user, and $\sigma^{\mu\nu}$ is defined as in equation \ref{eq:intro:sigmadef}\@.
The second option sets $\pi^{\mu\nu} = 0$, $\Pi = 0$, and \cite{vanderSchee:2013pia}:
\[
u^i = -\frac{\tau_\text{fs}}{3}\frac{d\log e(x,y)}{dx^i},
\]
with $e$ the energy density, $x^1 = x$ and $x^2 = y$\@.

In the T\raisebox{-0.5ex}{R}ENTo model, a different approach is taken.
Here, the stress-energy tensor is initialized as \cite{Bernhard:2018hnz}:
\begin{equation}
T^{\mu\nu}(x,y) = \frac{n}{\tau_\text{fs}}\int_0^{2\pi}d\phi\,\hat p^\mu\hat p^\nu\,T(x - v_\text{fs}\Delta\tau_\text{fs}\cos\phi,y - v_\text{fs}\Delta\tau_\text{fs}\sin\phi),\label{eq:trajectum:freestreaming}
\end{equation}
where $n$ is now a parameter, also called the norm, with dimension of inverse length.
For $\hat p^\mu\hat p^\nu$, we have
\[
\hat p^\mu\hat p^\nu = \left(
\begin{array}{ccc}
1 & v_\text{fs}\cos\phi & v_\text{fs}\sin\phi \\
v_\text{fs}\cos\phi & v_\text{fs}^2\cos^2\phi & v_\text{fs}^2\cos\phi\sin\phi \\
v_\text{fs}\sin\phi & v_\text{fs}^2\cos\phi\sin\phi & v_\text{fs}^2\sin^2\phi
\end{array}
\right),
\]
where the rows and columns of the matrix correspond to $\mu,\nu = 0,1,2$\@.
At this point, we deviate from the original implementation of T\raisebox{-0.5ex}{R}ENTo\@.
In the original implementation, the parameter we call $v_\text{fs}$ is set to 1, in which case \eqref{eq:trajectum:freestreaming} corresponds some density of massless particles determined by $T$ free streaming from time $\tau = 0^+$ to $\tau_\text{fs}$\@.
Indeed, this reflects in the name of the initialization time, where the fs stands for free streaming.
The parameter $v_\text{fs} \in [0,1]$, which we call the free streaming velocity, generalizes this ansatz by allowing free streaming to occur with a speed lower than that of the speed of light.

Another deviation from the original T\raisebox{-0.5ex}{R}ENTo is in the initialization of the shear tensor and bulk pressure.
In \cite{vanderSchee:2013pia}, it was observed that using holography to model the initial stage leads to a smooth transition to hydrodynamics.
The energy density and velocity profile is identical to first order in proper time to that obtained by free streaming, but the shear tensor obtained from holography is given by \eqref{eq:trajectum:holographicshearinit}\@.\footnote{The model in \cite{vanderSchee:2013pia} is conformal, and therefore has no bulk pressure.}
In order so that a future Bayesian analysis can potentially be used to decide whether free streaming or holography fits experimental data best, we introduce a parameter $\alpha \in [0,1]$ with which we can interpolate between free streaming and holography.

The procedure to obtain the stress tensor then becomes the following.
We first construct the stress tensor as given by free streaming \eqref{eq:trajectum:freestreaming}, and decompose the stress tensor in terms of $e$, $u^\mu$, $\pi_\text{fs}^{\mu\nu}$ and $\Pi_\text{fs}$\@.
We then compute $\pi_\text{holographic}^{\mu\nu}$ and $\Pi_\text{holographic}$ as
\[
\pi_\text{holographic}^{\mu\nu} = 2\eta\sigma^{\mu\nu}, \qquad \Pi_\text{holographic} = -\zeta\nabla\cdot u.
\]
We subsequently compute
\[
\pi^{\mu\nu} = \alpha\pi_\text{holographic}^{\mu\nu} + (1 - \alpha)\pi_\text{fs}^{\mu\nu}, \qquad \Pi = \alpha\Pi_\text{holographic} + (1 - \alpha)\Pi_\text{fs},
\]
and reconstruct $T^{\mu\nu}$ using $e$, $u^\mu$, $\pi^{\mu\nu}$ and $\Pi$\@.
In this way, by choosing $\alpha$, we can interpolate between weakly coupled initial conditions ($\alpha = 0$) and strongly coupled initial conditions ($\alpha = 1$)\@.

A final feature that is new in \emph{collide}, which is available for all initial conditions models discussed above, is the ability to optionally bias the distribution of events generated by the initial conditions.
To illustrate what this is and why one would want this, consider proton-lead collisions.
Typically, for such collisions, we are interested in collisions which generate a lot of particles in the final state.
These events correlate fairly well with the initial entropy deposited in the plasma by the initial conditions.\footnote{The correlation is not as good as for lead-lead collisions because the total number of particles produced is lower, and therefore statistical fluctuations from Poisson statistics make the correlation less pronounced.}
However, events which generate a lot of particles in the final state are rare.
This leads to the inconvenient situation that to get a good amount of data to be able to achieve small statistical uncertainties for the observables one is interested in, one has to generate even more events, the vast majority of which will not be useful.
The bias that we introduce now is based on the following idea.
Of the entire collision simulation, the initial conditions are by far computationally the cheapest to compute.
Since the initial entropy correlates well with the number of particles in the final state, we can compute the initial conditions for the hydrodynamical evolution, and then accept the event for further computation with a probability $P(s)$, which depends on its initial entropy.
By choosing the acceptance function $P$ appropriately, we can selectively compute events which will have a large amount of particles in the final state, thereby improving statistics for observables relating to such events without needing to spend much more in computation time.
Of course, when computing these observables, we need to be careful that our bias in the events that we generate does not translate into a bias in the final computed observable.
This can easily be achieved by making sure that each event analyzed by \emph{analyze} is weighted with weight $1/P(s)$\@.
Also, care has to be taken when computing centrality classes, as these weights also have to be taken into account when determining them.
\subsection{Hydrodynamics models}\label{sec:trajectum:hydromodels}
The next component we will describe is the hydrodynamics model itself.
In \emph{collide}, two models are available:
\begin{itemize}
\item First order hydrodynamics, which solves the first order Israel-Stewart equations with only first order transport coefficients \cite{Israel:1979wp,PhysRevC.69.034903,Heinz:2005bw,Song:2007ux},
\item Second order hydrodynamics, which also includes some second order coefficients.
\end{itemize}
Of course, one can just obtain the first order hydrodynamics equations by setting all the second order coefficients to zero.
However, there is still a good reason to have two separate classes, namely that setting the second order coefficients to zero when running the code is a lot slower than explicitly leaving those coefficients out of the code to be evaluated, and solving the simplified equations.
Since the equations which determine the hydrodynamical evolution are used by the solver every time step at every grid point, their speed of execution greatly impacts the overall execution time of the whole program.
However, to explain how these two models work, we will just explain how the second order hydrodynamics model works, as the workings of the first order model can be easily determined by setting the second order coefficients to zero.

In hydrodynamics, one of the first things we have to decide when solving the equations numerically is which variables to choose.
To illustrate this, one could imagine solving for each of the 7 components of the stress-energy tensor.
However, one could equivalently solve for the energy density $e$, two velocities $v^x$ and $v^y$, the bulk pressure $\Pi$ and the components $\pi^{xx}$, $\pi^{xy}$, $\pi^{yy}$\@.\footnote{Note that by tracelessness and orthogonality one can reconstruct $\pi^{\mu\nu}$ completely from these components.}
A convenient choice of variables turns out to be \cite{Song:2007ux}
\begin{equation}
\tilde T^{\tau\tau}, \qquad \tilde T^{\tau x}, \qquad \tilde T^{\tau y}, \qquad \Pi, \qquad \tilde\pi^{\mu\nu},\label{eq:trajectum:hydrovars}
\end{equation}
where we define $\tilde T^{\mu\nu} = \tau T^{\mu\nu}$, and $\tilde\pi^{\eta\eta} = \tau^2\pi^{\eta\eta}$, with $\tilde\pi^{\mu\nu} = \pi^{\mu\nu}$ for the other components.
Just by counting, one can see that this number of variables is larger than the number of independent variables, as we are evolving every non-zero component of the shear tensor.
It turns out however, that for the first step of solving the hydrodynamical equations, which will be discussed below, it is easier if these redundant components are evolved as well, so that they can be used in that step.
Another question one might ask is why certain variables are defined by incorporating a factor $\tau$ into their definition.
The reason for this is twofold.
On the one hand, it reduces the number of Christoffel symbols entering the equations, and on the other hand it removes some general behavior of the variables.
To see what is meant by this, consider $T^{\tau\tau}$\@.
By energy-momentum conservation, $T^{\tau\tau}$ will behave roughly like $1/\tau$\@.
By instead evolving $\tilde T^{\tau\tau}$, this dependence is removed, making it easier to obtain a good accuracy in the numerics.

Solving the hydrodynamical equations themselves involves two steps.
The second step depends for its computation on quantities like the velocity and the energy density, which are quantities that are not listed in \eqref{eq:trajectum:hydrovars}\@.
This immediately explains the need for the first step, which is to reconstruct these quantities, which we will call `auxiliary variables' from now on, from the variables in \eqref{eq:trajectum:hydrovars}, which we will call the `primary variables'\@.
What the second step does depends on which solver is used.
The finite difference solver requires for its algorithm a function which takes in all current values of the primary variables, as well as the auxiliary variables mentioned above, and outputs the proper time derivatives of the primary variables.
The MUSCL solver instead requires the equations to be put in the form
\begin{equation}
\partial_\tau U^k + \partial_x(v_xU^k) + \partial_y(v_yU^k) = S(U^k),\label{eq:trajectum:muscldecomposition}
\end{equation}
where $U^k$ is an array of all variables in \eqref{eq:trajectum:hydrovars}, $v_x = u^x/u^\tau$ and $v_y = u^y/u^\tau$ are the fluid velocities in the $x$ and $y$ directions, and where we call $S$ the source term.
What the MUSCL solver then needs for its evaluation is the function $S$\@.
The hydrodynamics models hence expose three functions to the \emph{collide} framework: one function which reconstructs the auxiliary variables from the primary variables, and two functions which compute the the time derivative, and the source term, respectively.
In the remainder of the discussion of the hydrodynamics model, we will only discuss the equations for source terms, as the equations for the proper time derivatives can be easily obtained from these.

As mentioned above, the first function necessary to solve the hydrodynamical evolution is one that takes in the primary variables and computes from them the auxiliary variables.
To do this, we define \cite{Shen:2014vra}
\[
M^0 = T^{\tau\tau} - \pi^{\tau\tau}, \qquad M^1 = T^{\tau x} - \pi^{\tau x}, \qquad M^2 = T^{\tau y} - \pi^{\tau y}.
\]
Using the decomposition of the stress tensor \ref{eq:intro:viscoushydro}, we can then obtain
\begin{align}
M^0 & = (e + P(e) + \Pi)(u^\tau)^2 - P(e) - \Pi,\nonumber\\
M^1 & = (e + P(e) + \Pi)u^\tau u^x,\label{eq:trajectum:Midecomp}\\
M^2 & = (e + P(e) + \Pi)u^\tau u^y,\nonumber
\end{align}
where $P(e)$ is the pressure.
These equations can be rearranged to give
\[
e = M^0 - \frac{(M^1)^2 + (M^2)^2}{M^0 + P(e) + \Pi}.
\]
This equation can be solved iteratively for $e$ as follows.
The function
\[
f(e) = (M^0 - e)(M^0 + P(e) + \Pi) - (M^1)^2 - (M^2)^2
\]
has the property that $f(0) \geq 0$, whereas $f(M^0) \leq 0$\@.
The physical solution for $e$ corresponds to the solution of $f(e) = 0$, which lies between those values, and can be solved by an algorithm like Brent's method, which was also used in chapter \ref{ch:imc}\@.
Subsequently, once $e$ is known, one can derive from \eqref{eq:trajectum:Midecomp} that
\[
u^\tau = \sqrt{\frac{M^0 + P(e) + \Pi}{e + P(e) + \Pi}}, \qquad u^i = \frac{M^i}{\sqrt{M^0 + P(e) + \Pi}\sqrt{e + P(e) + \Pi}},
\]
where $u^1 = u^x$ and $u^2 = u^y$, completing the task of computing the energy density and the velocity.
Note that during both the iterative solving of $f(e) = 0$ and the computation of $u^\mu$ requires the pressure to be known as a function of the energy density, which is a task which is performed by the transport coefficients component.

Next, we describe the computation of the source terms.
For $\tilde T^{\tau\mu}$, we have \cite{Song:2007ux}
\begin{align*}
S(\tilde T^{\tau\tau}) & = -p - \Pi - \tilde\pi^{\eta\eta} - \tau\partial_x\left((p + \Pi)v_x + \pi^{\tau x} - v_x\pi^{\tau\tau}\right) \\
& \quad - \tau\partial_y\left((p + \Pi)v_y + \pi^{\tau y} - v_y\pi^{\tau\tau}\right), \\
S(\tilde T^{\tau x}) & = -\tau\partial_x\left(p + \Pi + \pi^{xx} - v_x\pi^{\tau x}\right) - \tau\partial_y\left(\pi^{xy} - v_y\pi^{\tau x}\right), \\
S(\tilde T^{\tau y}) & = -\tau\partial_x\left(\pi^{xy} - v_x\pi^{\tau y}\right) - \tau\partial_y\left(p + \Pi + \pi^{yy} - v_y\pi^{\tau y}\right).
\end{align*}
For the bulk pressure $\Pi$, we have used not all the second order expressions, but just the ones also kept by \cite{Bernhard:2018hnz} (The full second order expressions can be found in \cite{Denicol:2010xn,Denicol:2012cn,Denicol:2014vaa}):
\[
S(\Pi) = \Pi\nabla\cdot v - \frac{1}{u^\tau\tau_\Pi}\left(\Pi + \zeta\nabla\cdot u + \delta_{\Pi\Pi}\nabla\cdot u\Pi - \lambda_{\Pi\pi}\tilde\pi^{\mu\nu}\tilde\sigma_{\mu\nu}\right),
\]
where $\tilde\sigma^{\eta\eta} = \tau^2\sigma^{\eta\eta}$, with $\tilde\sigma^{\mu\nu} = \sigma^{\mu\nu}$ for the other components similar to the definition of $\tilde\pi^{\mu\nu}$, and
\[
\nabla\cdot v = \partial_xv_x + \partial_yv_y, \qquad \nabla\cdot u = \partial_\mu u^\mu.
\]
Lastly, for the shear tensor, we have
\begin{align*}
S(\tilde\pi^{\mu\nu}) & = \tilde\pi^{\mu\nu}\nabla\cdot v - \left(u^\mu\tilde\pi^\nu_\rho + u^\nu\tilde\pi^\mu_\rho\right)\left(\partial_\tau + v_x\partial_x + v_y\partial_y\right)u^\rho \\
& \quad - \frac{1}{u^\tau\tau_\pi}\big(\tilde\pi^{\mu\nu} - 2\eta\tilde\sigma^{\mu\nu} + \delta_{\pi\pi}\tilde\pi^{\mu\nu}\nabla\cdot u \\
& \qquad - \phi_7\tilde\pi_\alpha^{\langle\mu}\tilde\pi^{\nu\rangle\alpha} + \tau_{\pi\pi}\tilde\pi_\alpha^{\langle\mu}\tilde\sigma^{\nu\rangle\alpha} - \lambda_{\pi\Pi}\Pi\tilde\sigma^{\mu\nu}\big).
\end{align*}
Here the second order coefficients, which are set to zero for the first order model, are the coefficients $\delta_{\Pi\Pi}$, $\lambda_{\Pi\pi}$, $\delta_{\pi\pi}$, $\phi_7$, $\tau_{\pi\pi}$ and $\lambda_{\pi\Pi}$\@.
\subsection{Transport coefficients}
The task of the transport coefficients model is twofold.
It needs to compute the pressure by means of an equation of state, and it should compute the transport coefficients appearing at the end of the last subsection.
In \emph{collide}, there are three available transport coefficient models:
\begin{itemize}
\item Ideal gas equation of state with `constant' transport coefficients.
\item Lattice QCD equation of state with `constant' transport coefficients.
\item Lattice QCD equation of state with temperature dependence in some of the transport coefficients.
\end{itemize}
Here `constant' means that the transport coefficients are described by a single parameter.
For example, for the shear viscosity $\eta$, `constant' transport coefficients do not imply that $\eta(T)$ itself is constant as a function of temperature, but rather that $\eta(T)/s(T)$ is, where $s(T)$ is the entropy density specified by the equation of state.
We will now describe each of the available models in some detail, starting with the ideal gas equations of state with `constant' transport coefficients.

For the ideal gas equation of state, we assume an equation of state of the form
\[
P(T) = \alpha\cdot T^4,
\]
with $P$ the pressure, and $\alpha$ a constant which can be specified by the user.
Additionally, the shear and bulk viscosities can be specified by the user by means the following constant combinations:
\[
\frac{\eta}{s}, \qquad \frac{\zeta}{s}.
\]
The shear and bulk relaxation times $\tau_\pi$ and $\tau_\Pi$, respectively, are specified as the following constant combinations:
\begin{equation}
\frac{\tau_\pi sT}{\eta}, \qquad \frac{\tau_\Pi sT}{\zeta}.\label{eq:trajectum:idealrelaxtimes}
\end{equation}
The second order coefficients are given by the following constant combinations:\footnote{If the user specified that the first order hydrodynamics model should be used, \emph{collide} does not request the second order coefficients of the user.}
\begin{equation}
\frac{\delta_{\pi\pi}}{\tau_\pi}, \qquad \frac{\phi_7}{P}, \qquad \frac{\tau_{\pi\pi}}{\tau_\pi}, \qquad \frac{\lambda_{\pi\Pi}}{\tau_\pi}, \qquad \frac{\delta_{\Pi\Pi}}{\tau_\Pi}, \qquad \frac{\lambda_{\Pi\pi}}{\tau_\Pi},\label{eq:trajectum:idealsecondorder}
\end{equation}
with $P$ the pressure.
Note that if one specifies $\zeta/s = 0$, $\tau_\pi sT/\eta$ small enough, and all other coefficients zero, one can use this model for transport coefficients in combination with the Gubser flow initial conditions to obtain the analytical solution \cite{Gubser:2010ze,Gubser:2010ui,Marrochio:2013wla}, with which one can check the accuracy of the numerical solution.

For both models which use a lattice QCD equations of state, we actually use a hybrid of a hadron resonance gas (HRG) for temperatures below $165\,\text{MeV}$, an analytical fit to a numerically constructed lattice QCD equation of state for temperatures above $200\,\text{MeV}$, and a polynomial interpolation of the trace anomaly in the intermediate temperature regime \cite{Bernhard:2018hnz,Huovinen:2009yb}\@.
The polynomial interpolation is taken such that at the matching points, $165\,\text{MeV}$ and $200\,\text{MeV}$, the trace anomaly, as well as its first 4 derivatives, is continuous.
The hadron resonance gas equation of state can be computed from
\begin{align*}
e & = \sum_{i\in\text{species}}g_i\int\frac{d^3p}{(2\pi)^3}E_i(p)\frac{1}{\exp(E_i(p)/T) \pm 1}, \\
P & = \sum_{i\in\text{species}}g_i\int\frac{d^3p}{(2\pi)^3}\frac{p^2}{3E_i(p)}\frac{1}{\exp(E_i(p)/T) \pm 1},
\end{align*}
with the $+$ for fermions, and the $-$ for bosons.
Also, $E_i(p) = m_i^2 + p^2$, where $m_i$ and $g_i$ are the mass\footnote{In the actual implementation, the particle masses of the unstable particles are taken from a modified Breit-Wigner distribution, and quantities like the energy density and the pressure are averaged over all masses, where the average is weighted according to the distribution. This is implemented properly in \emph{collide}, but to simplify the discussion, we leave this detail out of the discussion. An excellent explanation can be found in section 3.4 of \cite{Bernhard:2018hnz}\@.} and number of degrees of freedom of species $i$, respectively.
The species used in the sum are precisely the particle content of UrQMD, which is necessary for consistency between the different stages of both \emph{collide} and UrQMD itself.

The lattice QCD part of the equation of state is described by the following parameterization \cite{Bazavov:2014pvz}:
\[
\frac{P}{T^4} = \frac{1}{2}\left(1 + \tanh\left[c_t(t - t_0)\right]\right)\left(\frac{p_\text{id} + a_n/t + b_n/t^2 + d_n/t^4}{1 + a_d/t + b_d/t^2 + d_d/t^4}\right),
\]
with
\begin{equation}
t = T/T_c, \qquad T_c = 154\,\text{MeV}, \qquad p_\text{id} = 95\pi^2/180,\label{eq:trajectum:Tc}
\end{equation}
and the fit coefficients
\[
c_t = 3.8706, \qquad a_n = -8.7704, \qquad b_n = 3.9200, \qquad d_n = 0.3419,
\]
\[
t_0 = 0.9761, \qquad a_d = -1.2600, \qquad b_d = 0.8425, \qquad d_d = -0.0475.
\]

The discussion so far fixes the equation of state of the two lattice EoS based transport coefficient models.
Let us next discuss the first and second order transport coefficients, where we start with the shear and bulk viscosities.
These two transport coefficients are the only difference between the two lattice EoS based models.
In the model without temperature dependence, the two viscosities are given by the same constant combinations as for the ideal gas based model:
\[
\frac{\eta}{s}, \qquad \frac{\zeta}{s}.
\]
In the model with temperature dependence, we have \cite{Bernhard:2018hnz}:
\begin{align*}
\frac{\eta}{s}(T) & = \begin{cases}
\left(\eta/s\right)_\text{hrg} & T < T_c, \\
\left(\eta/s\right)_\text{min} + \left(\eta/s\right)_\text{slope}\cdot\left(T - T_c\right)\cdot\left(\frac{T}{T_c}\right)^{(\eta/s)_\text{crv}} & T > T_c,
\end{cases} \\
\frac{\zeta}{s}(T) & = \frac{\left(\zeta/s\right)_\text{max}}{1 + \left(\frac{T - \left(\zeta/s\right)_{T_0}}{\left(\zeta/s\right)_\text{width}}\right)^2},
\end{align*}
with $T_c$ as given in \eqref{eq:trajectum:Tc}, and
\[
\left(\eta/s\right)_\text{hrg}, \qquad \left(\eta/s\right)_\text{min}, \qquad \left(\eta/s\right)_\text{slope}, \qquad \left(\eta/s\right)_\text{slope},
\]
\[
\left(\zeta/s\right)_\text{max}, \qquad \left(\zeta/s\right)_{T_0}, \qquad \left(\zeta/s\right)_\text{width}
\]
parameters to be specified by the user.
The shear and bulk relaxation times are defined by the following constants \cite{Denicol:2014vaa}:\footnote{To ensure the stability of the numerics, it is important that timescales, like $1/\tau_\pi$ and $1/\tau_\Pi$, are larger (with some margin) than the time step size $\Delta\tau$ used by the PDE solver. To ensure this, \emph{collide} always enforces $1/\tau_\pi > 2\Delta\tau$ and $1/\tau_\Pi > 2\Delta\tau$\@.}
\[
\frac{\tau_\pi sT}{\eta}, \qquad \frac{\tau_\Pi sT\left(\frac{1}{3} - c_s^2\right)^2}{\zeta},
\]
which differ slightly from those in \eqref{eq:trajectum:idealrelaxtimes}\@.
The second order coefficients are then specified by the following constants \cite{Denicol:2014vaa}:
\[
\frac{\delta_{\pi\pi}}{\tau_\pi}, \qquad \frac{\phi_7}{P}, \qquad \frac{\tau_{\pi\pi}}{\tau_\pi}, \qquad \frac{\lambda_{\pi\Pi}}{\tau_\pi}, \qquad \frac{\delta_{\Pi\Pi}}{\tau_\Pi}, \qquad \frac{\lambda_{\Pi\pi}}{\tau_\Pi\left(\frac{1}{3} - c_s^2\right)},
\]
which again differ slightly from those in \eqref{eq:trajectum:idealsecondorder}\@.
\subsection{PDE solvers}
The PDE solver performs the task of solving the hydrodynamics equations.
To do this it interfaces with the hydrodynamics model, which provides the necessary functions, as discussed above.
The PDE solver then updates the state of the fluid from proper time $\tau$ to $\tau + \Delta\tau$, where the size of $\Delta\tau$ can be chosen by the user.
In \emph{collide}, there are two main types of PDE solvers:
\begin{itemize}
\item Finite difference solver,
\item MUSCL solver.
\end{itemize}
Both of these solvers have slight variations implemented, which allow the user even more flexibility.
Each of the solvers has their advantages and disadvantages, so one should pick carefully when using \emph{collide} on a problem.
The finite difference solver uses a very simple algorithm, as we will see below.
This has the advantage that, due to its low complexity, it is faster by about a factor 2 as compared to the MUSCL solver.
The disadvantage of the finite difference solver is that it is not guaranteed to be stable.
Under certain circumstances, like the presence of shocks or when using transport coefficients with extremely small viscosities, numerical instabilities may appear, which grow exponentially.
The MUSCL solver, instead, is guaranteed to be stable.
Indeed, the algorithm that it uses is stable by construction, as it was designed with stability in mind.
For this reason, one can use MUSCL even for an ideal fluid, with all viscosities set to zero.
The disadvantage that accompanies this stability, however, is increased execution time.
Depending on the type of problem, one can choose which of these benifits outweigh the associated costs.
In the following few paragraphs, we will discuss both of these solvers, pointing out the available variations along the way.

Now let us first discuss the finite difference solver.
One can write the hydrodynamics equations for the `primary variables' defined in section \ref{sec:trajectum:hydromodels} as follows:
\[
\partial_\tau U^k_{i,j}(\tau) = f\left(U^k_{i,j}(\tau),\partial_xU^k_{i,j}(\tau),\partial_yU^k_{i,j}(\tau)\right),
\]
with $f$ a function, and where $U^k_{ij}$ is the collection of primary variables at grid site $(i,j)$\@.
Here the first entry corresponds to the $x$-coordinate, and the second corresponds to the $y$-coordinate.
One can then perform the following discretization \cite{Luzum:2008cw}:
\begin{align*}
U^k_{i,j}(\tau + \Delta\tau) & = U^k_{i,j}(\tau) + \Delta\tau f\Big(U^k_{i,j}(\tau),\\
& \quad \frac{U^k_{i+1,j}(\tau) - U^k_{i-1,j}(\tau)}{2a},\frac{U^k_{i,j+1}(\tau) - U^k_{i,j-1}(\tau)}{2a}\Big),
\end{align*}
where $a$ is the spacing between grid points.
In other words, one simply replaces the spatial derivatives by the second order accurate discrete derivatives, and one replaces the time derivative by a linear approximation.
While this may seem a little naive, this works quite well under most circumstances, and as mentioned before, due to its simplicity, is very fast.
In \emph{collide} there are also variations implemented which replace the second order accurate spatial derivatives with fourth or even eighth order accurate ones.
This is however slightly smaller, and seems to be less numerically stable than the second order spatial derivatives, so one should not use it.

The MUSCL solver has a more sophisticated algorithm, which is based on the idea that the equations can `almost' be written as a set of conservation equations \eqref{eq:trajectum:muscldecomposition}\@.
A remarks are in order here.
First of all, it may seem a bit strange to write for example $\partial_\tau(T^{\tau\tau}) + \partial_x(v_xT^{\tau\tau}) + \partial_y(v_yT^{\tau\tau}) = S$, while we know that $\partial_\mu T^{\mu\tau} = 0$\@.
However, the latter equation is a tensor conservation equation.
Numerically, it is simpler to implement the first option, because this guarantees that the same velocities $(v_x,v_y)$ can be used for all primary variables.
How the algorithm updates the primary variables is only different from how the finite difference solver performs the update is in the $\partial_x(v_xU^k)$ and $\partial_y(v_yU^k)$ terms.
The source term is added in the same way as for the finite difference solver, where when needed, we also replace derivatives by their second order accurate approximations.
The two terms mentioned are replaced following the Kurganov-Tadmor (KT) algorithm \cite{Bazow:2016yra,Du:2019obx}:\footnote{An alternative to KT, known as the HLL two-state formula, is also available in \emph{collide} \cite{DelZanna:2013eua}\@.}
\[
\partial_x(v_xU^k_{i,j}) \mapsto \frac{H^x_{i+1/2,j} - H^x_{i-1/2,j}}{a},
\]
with a similar expression for the $y$-derivative term.
We will continue the discussion only for the $x$-derivative term, the formulas for $y$ are analogously defined.
In the above expression, $H^x_{i+1/2,j}$ is defined as
\[
H^x_{i+1/2,j} = \frac{F^x(U^{k,+}_{i+1/2,j}) + F^x(U^{k,-}_{i+1/2,j})}{2} - a^x_{i+1/2,j}\frac{U^{k,+}_{i+1/2,j} - U^{k,-}_{i+1/2,j}}{2},
\]
where $a^x$ is equal to $v_x$, and where $F^x(U^k) = v_xU^k$ is the flux of $U^k$ at the cell's interface.
At this point, an optional shortcut is available in \emph{collide}\@.
To properly evaluate the flux for a specific set of primary variables, one needs to find $v_x$ and $v_y$, which implies that one has to find the auxiliary variables, which is numerically expensive.
The shortcut taken is that we compute $v_x^+$ and $v_x^-$ in an analogous way to the computation of $U^{k,+}$ and $U^{k,-}$, thereby limiting the amount of times one has to solve for auxiliary variables.\footnote{We checked that neither the accuracy nor the stability is negatively impacted by this trick.}
For $U^{k,+}$ and $U^{k,-}$, we have:
\[
U^{k,+}_{i+1/2,j} = U^k_{i+1,j} - \frac{a}{2}\tilde U^k_{i+1,j}, \qquad U^{k,-}_{i+1/2,j} = U^k_{i,j} + \frac{a}{2}\tilde U^k_{i,j},
\]
with the quantity $\tilde U^k$ being roughly equal to the second order accurate derivative with respect to $x$\@.
If one would take this rough equality to be exact, the algorithm would become identical to the finite difference method described above.
This would then however also inherit the stability issues of that method.
It turns out that the crucial element to achieve stability is the addition of a flux limiter.
In the MUSCL implementation in \emph{collide}, we use the minmod flux limiter:
\[
\tilde U^k_{i,j} = \minmod\left(\frac{U^k_{i,j} - U_{i-1,j}}{a},\frac{U^k_{i+1,j} - U^k_{i-1,j}}{2a},\frac{U^k_{i+1,j} - U^k_{i,j}}{a}\right),
\]
where the minmod flux limiter function is defined as
\begin{align*}
\minmod(x,y,z) & = \minmod(x,\minmod(y,z)), \\
\minmod(x,y) & = \left[\sign(x) + \sign(y)\right]\cdot\frac{\min(|x|,|y|)}{2}.
\end{align*}
Finally, note that in \emph{collide}, one has the option to either integrate the time derivatives using the forward Euler algorithm, which is first order accurate, or with the midpoint method, which is second order accurate.
In the results shown in section \ref{sec:trajectum:analyze}, we used the KT solver with the shortcut for the solving of auxiliary variables.
\subsection{Hadronizers}
After each update that the PDE solver makes to the fluid, the hydrodynamics model hands the new fluid state to the hadronizer.
The hadronizer is then tasked with generating particles, which are output in a format compatible with UrQMD\@.
In \emph{collide}, at the time of writing, only one hadronizer is available, namely Cooper-Frye \cite{Cooper:1974mv}\@.
The hadronization procedure consists of several steps.
First, the hadronizer computes a freeze-out surface.
Subsequently, particles are produced at the freeze-out surface according to a modified thermal distribution, where the modifications encode the presence of shear stress and bulk pressure into the final state particles.
In the paragraphs below, we will explain each of these steps in more detail.

The first step to turn the fluid into particles is to compute the freeze-out surface.
This is an isotherm of a specific temperature $T_\text{fr}$ that the user can specify.
At the freeze-out surface, a number of particles are produced according to a Poisson distribution, which has the property that the sum of Poisson distributed processes again follows a Poisson distribution.
As a consequence of this, we can subdivide the freeze-out surface into triangles, and sample particles from each triangle individually.
After this, we can even discard the freeze-out surface from which particles have already been sampled, as it is no longer necessary.
This is exactly how \emph{collide} tackles the problem.
When the hydrodynamics model computes the state of the fluid at proper time $\tau + \Delta\tau$, it gives the state of the fluid to the Cooper-Frye hadronizer, which still has in its memory the state of the fluid at proper time $\tau$\@.
Using the Cornelius algorithm, it then computes a triangulation of the freeze-out surface \cite{Huovinen:2012is}\@.\footnote{Note that for a $2 + 1$D description like the one in \emph{collide}, freeze-out surface elements are triangles. A more general treatment for $3 + 1$D can be found in \cite{Huovinen:2012is}\@.}
For each triangle, it also computes a surface normal $\Delta\sigma_\mu$ \cite{Bernhard:2018hnz}\@.

The next step is to generate particles for each triangle.
This is done using the Cooper-Frye formula \cite{Cooper:1974mv}:
\begin{equation}
E\frac{dN_i}{d^3p} = \frac{g_i}{(2\pi)^3}\int_\sigma\frac{1}{\exp(p\cdot u/T) \pm 1}p^\mu\,d^3\sigma_\mu,\label{eq:trajectum:cooperfrye}
\end{equation}
where $g_i$ is the number of degrees of freedom for species $i$\@.
The idea behind this formula is that the hadronization procedure should be not a change in physical process, but rather a change in description of that process.
The particle species used in the formula are precisely the ones that are used in the HRG equation of state which describes the fluid, and because they are sampled from a thermal distribution, they indeed describe precisely the HRG\@.\footnote{The freeze-out temperature is required to be below $165\,\text{MeV}$, i.e.~in a temperature regime where the equation of state is that of the hadron resonance gas.}
This ensures that, in the absence of shear stress or bulk pressure, the stress energy tensor of the fluid will be on average the same as that of the sampled particles.\footnote{The sampled particles are, of course, subject to Poisson statistics.}
In this sense, the sampled particles form a different description of the fluid, and when one then only keeps particles which move out of the fluid, one thereby obtains a consistent transition between the fluid and the HRG\@.

Everything described so far assumes that $\pi^{\mu\nu} = 0$ and $\Pi = 0$\@.
If this is not the case, the sampling of the particles needs to be adjusted so that the stress-energy tensor of the fluid still matches the average stress-energy tensor of the sampled particles.
The way this is done in \emph{collide} is by rescaling the momentum of each sampled particle in the fluid rest frame \cite{Bernhard:2018hnz}:
\[
p_i \mapsto p_i + \sum_j\lambda_{ij}p_j, \qquad \lambda_{ij} = (\lambda_\text{shear})_{ij} + \lambda_\text{bulk}\delta_{ij},
\]
where $(\lambda_\text{shear})_{ij} \propto \pi_{ij}$ with a proportionality constant that can be computed by a hadron resonance gas computation, and $\lambda_\text{bulk}$ depends in a non-trivial way on $\Pi$\@.\footnote{For details on how to obtain the proportionality constant for $(\lambda_\text{shear})_{ij}$ as well as for how precisely to obtain $\lambda_\text{bulk}(\Pi)$, see \cite{Bernhard:2018hnz}\@.}
For bulk corrections, also the particle density obtained from the Cooper-Frye formula \eqref{eq:trajectum:cooperfrye} needs to be modified, as with the rescaling of the particle momenta described above the energy density is also modified.
The way to correct this is to modify the number density \cite{Bernhard:2018hnz}\@.

We now have produced particles in a way such that the stress-energy tensor is on average continuous across the freeze-out surface.
There is one final step, however, namely decaying the $f_0(500)$ particle.
This particle is not treated by UrQMD, but given its light mass is produced in large quantities during an event \cite{Pelaez:2015qba,Bernhard:2018hnz}\@.
In \emph{collide}, the $f_0(500)$ can be optionally taken into account when compiling \emph{Trajectum}\@.
If the user includes the $f_0(500)$, it is immediately decayed into pions, which is justified given its short lifetime.
\section{\emph{analyze} executable}\label{sec:trajectum:analyze}
The \emph{analyze} executable is tasked with computing the observables according to their proper definitions from the result of the hadronic afterburner.
Additionally, it obtains from the output of \emph{collide} each event's weight.
The executable does this in two passes over the data file.
The first pass reads every event, and computes quantities that will allow us to group the particles into centrality and $N_\text{trk}^\text{off}$ bins during the second pass.
During the second pass, each event is read again, and based on its $N_\text{trk}^\text{off}$ value is placed in a bin together with similar events, after which each event is analyzed by the observables that the user requested.
In the next few paragraphs, each of these two passes will be described in some more detail, after which a selection of observables will be shown.

The first pass over the data has the task to determine the centrality classes.
This is done by counting $N_\text{trk}^\text{off}$ for each event, which is a measure of the number of particles produced in the event.
The definition of $N_\text{trk}^\text{off}$ differs from experiment to experiment, and is usually defined in such a way that only particles are counted which are charged, and have kinematics such that the detector is able to detect them efficiently.
Of course, in the output of \emph{collide} and UrQMD, we are able to detect any particle whatsoever with 100\% efficiency, but for comparison with the experiments it is still important to make the same cuts, so that we are comparing the exact same observable.
This will be a recurring theme throughout this section.
As briefly mentioned already, $N_\text{trk}^\text{off}$ is defined by the number of charged particles in an event, and the kinematic cuts usually require that for a particle to be counted, the transverse momentum $p_T$ needs to be between a certain lower bound and a certain upper bound, and that the pseudorapidity $\eta$ needs to have an absolute value below some upper bound.
For example, the cuts used in this section are
\[
0.4\,\text{GeV} \leq p_T \leq 10\,\text{GeV}, \qquad |\eta| \leq 2.4.
\]
After $N_\text{trk}^\text{off}$ has been counted for each event, the events are sorted from the largest $N_\text{trk}^\text{off}$ to the smallest.
This ordering then determines the centrality bins, and for each centrality bin, \emph{analyze} then saves between which values $N_\text{trk}^\text{off}$ should be such that the event is within that particular bin.
Of course, when events are weighted according to their initial entropy, it is important to take these weights into account when computing the centrality bins, so as not to bias the binning.
For example, 5\% central would no longer mean that 5\% of the events have a larger $N_\text{trk}^\text{off}$, but instead it means that 5\% of the total weights belong to events with a larger $N_\text{trk}^\text{off}$\@.

After the centrality bins\footnote{In \emph{analyze} it is also possible to bin particles based on $N_\text{trk}^\text{off}$ directly. For notational convenience, we will just refer to both of these binning options as `centrality' bins, as for the remainder of the discussion the distinction is not important.} have been determined, a second pass over all events is made, in which each observable that the user requested can examine each event.
Each observable again, typically, makes some cuts on which particles it counts.
All of these cuts, namely cuts on transverse momentum, pseudorapidity, but also which type of particles to count, can be set by the user, where different settings can be chosen per observable.
This may seem like a lot of flexibility, but this is necessary, as different experimental measurements often have different cuts, and as was mentioned before, to be able to compare to experimental values it is extremely important to use the exact same cuts as the experiment in question.
In this way, the flexibility given to the user allows to compute a large number of observables at the same time, where each observable is precisely defined, by a single use of \emph{analyze},i.e.~there is no need to run the executable multiple times.
Another advantage of this flexibility is that the cuts on which types of particles to count makes it possible to compute observables defined in terms of `identified particles', i.e.~one can compute for example the mean transverse momentum of only charged pions, or of only protons.

In the framework provided by \emph{analyze} during the second pass, each observable is allowed to output an array of what we will define `intermediate quantities', along with a weight for each such quantity.
The framework provided by \emph{analyze} will then compute a weighted average for each element in the array, where not only the weights provided by the observable are taken into account, but also the weights for each event based on its initial entropy.
The weighted averages are then given back to the observable, along with errors and correlations between the elements in the array, which provide the observable with all the information it needs to compute its final outputs.
Also, the individual values for each event are given, which allows to, for example, compute event-by-event $v_2$ distributions \cite{Aad:2013xma}\@.
To illustrate this, let us take the quantities $v_2\{2\}$ and $v_2\{4\}$ as examples \cite{Adam:2016izf}\@.
These are both computed by the `v2' component in \emph{analyze}\@.
For each event, `v2' computes quantities known as $\langle2\rangle$ and $\langle4\rangle$, which come with weights $w_2 = M(M - 1)$ and $w_4 = M(M - 1)(M - 2)(M - 3)$, respectively, where $M$ is the number of particles which satisfy the various cuts.\footnote{Precise definitions of $\langle2\rangle$ and $\langle4\rangle$ will be given in section \ref{sec:trajectum:flow}\@.}
After each event has been computed, \emph{analyze} computes the averages
\[
\langle\langle2\rangle\rangle = \frac{\sum_{i\in\text{events}}\tilde w_iw_{2,i}\langle2_i\rangle}{\sum_{i\in\text{events}}\tilde w_iw_{2,i}}, \qquad \langle\langle4\rangle\rangle = \frac{\sum_{i\in\text{events}}\tilde w_iw_{4,i}\langle4_i\rangle}{\sum_{i\in\text{events}}\tilde w_iw_{4,i}},
\]
along with the standard deviations of the averages and correlations between the two measured averages.
Here $\tilde w_i$ stands for the event weight of event $i$ based on its initial entropy, i.e.~$\tilde w_i = 1/P(s_i)$\@.
The averages are then given back to the `v2' component, which computes the final outputs
\[
v_2\{2\} = \sqrt{\langle\langle2\rangle\rangle}, \qquad v_2\{4\} = \sqrt[4]{2\cdot\langle\langle2\rangle\rangle^2 - \langle\langle4\rangle\rangle},
\]
where care is taken to estimate the final errors by standard error propagation methods.

In the remainder of this section, we will use \emph{analyze} to compute a selection of available observables, which we will compare to available experimental data.\footnote{At the time of writing, there are 3132 observables available in \emph{analyze}\@. This may seem like a very large number, but this is mainly due to the fact that many observables have a large number of possible variations. However, this large number does mean that not all observables can be shown here.}
The aim of this is twofold.
Firstly, as \emph{Trajectum} is a new code, checking its predictions against known results is a good check that the code is relatively free of serious mistakes.
For this purpose, we choose the maximum a posteriori (MAP) values obtained for $\sqrt{s} = 2.76\,\text{TeV}$ in \cite{Bernhard:2018hnz}\@.
For certain choices of parameters of \emph{collide}, the predictions of \emph{Trajectum} should be the same as, or at least compatible with, to the result from \cite{Bernhard:2018hnz}\@.
As we will see below, this is the case.
The MAP values we are using for the computation have been obtained by means of a Bayesian fit to experimental data, and as such represent some of the most state-of-the-art knowledge concerning which inputs to the simulation fit the data best.
However, we would like to improve this analysis further by adding additional experimental data to the analysis.
This brings us to the second aim of the remainder of this section.

Any additional experimental data should have the following properties to be useful in the Bayesian analysis, where by useful we mean that the additional observable provides an additional constraint to data.
Firstly, the observable should have small statistical uncertainties for the number of events that we can feasibly generate for each set of input parameters to \emph{collide}\@.
The results shown below have been generated using 10000 minimal-bias events, and with limited computation time available it is not feasible to perform the Bayesian analysis with significantly more events.
This means that the error bars shown in the plots below are what we can reasonably expect to achieve in the Bayesian analysis.
In such an analysis, we need to be able to decide whether a particular observable is well-described by a particular set of inputs to \emph{collide}, or whether there is enough tension to exclude a particular parameter set.
If the error bars are too large, this can not be achieved, and hence the particular observable will not give an additional constraint.

Another requirement for an observable to be useful in the Bayesian analysis is that the constraints it gives on the input parameters should not coincide with the constraints which are already given by other  observables.
To give an obvious example of this, imagine we would like to use the mean transverse momentum of positive pions in the analysis.
This gives a constraint on the input parameters.
In particular, it will constrain the bulk viscosity $\zeta$\@.
However, given that the mean transverse momentum of \emph{all} pions is already being used in the fit, and since there is no significant difference in particles and their antiparticles in collisions at $2.76\,\text{TeV}$, the constraint given by the positive pions will completely coincide with the constraint from all pions.
With regards to this requirement, the most interesting outcome of the analysis of the MAP values would be to find an observable which is completely incompatible with the prediction from \cite{Bernhard:2018hnz}\@.
Such an outcome could mean two things.
Either there is a subset of input parameters to \emph{collide} which is compatible with the data used in \cite{Bernhard:2018hnz}, or the model used to simulate the events is missing an ingredient which is essential to describe the new variable.
\subsection{Charged particle multiplicities and spectra}
\begin{figure}[ht]
\centering
\includegraphics[width=\textwidth]{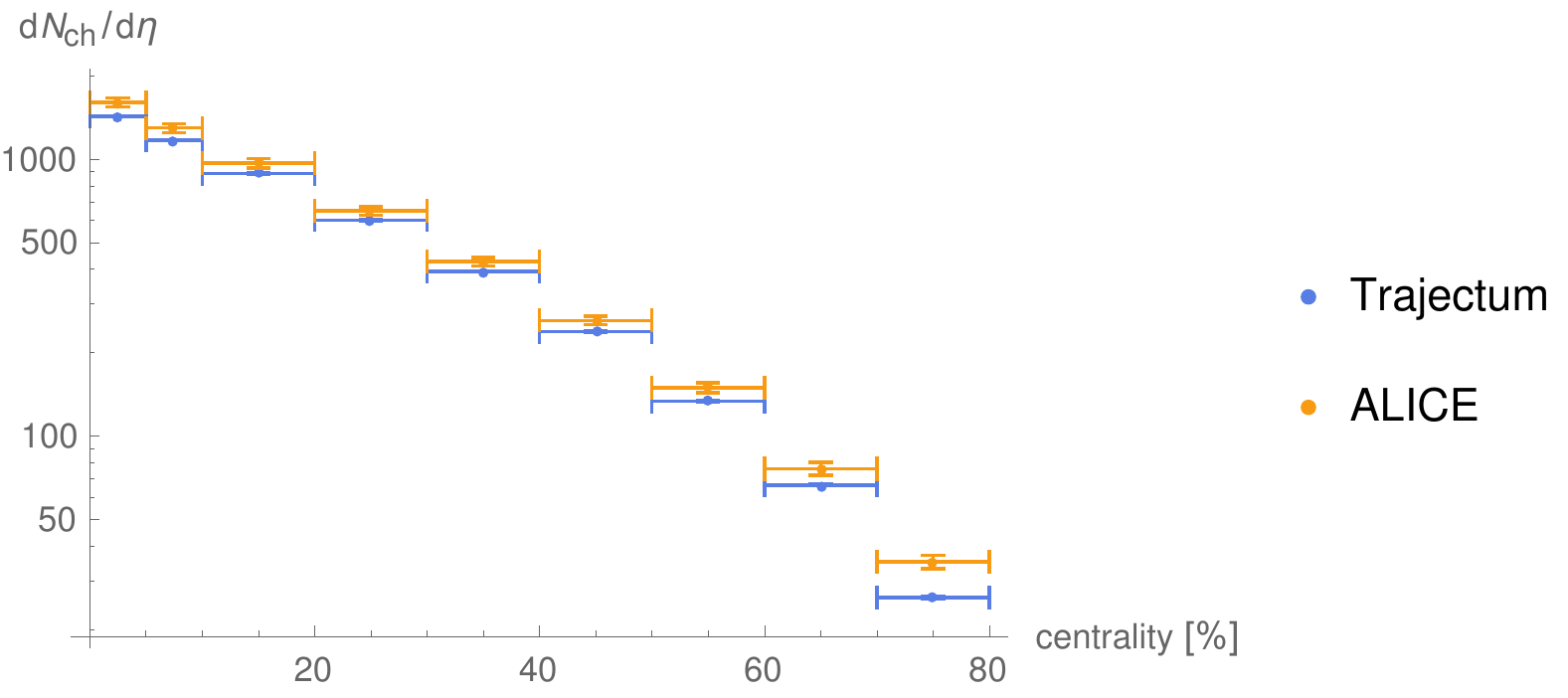}
\caption{\label{fig:trajectum:trajectummultiplicity}Multiplicity of charged particles with $|\eta| \leq 0.5$\@. Both the result from \emph{Trajectum} using the $2.76\,\text{TeV}$ MAP values from \cite{Bernhard:2018hnz}, and experimentally measured data from ALICE \cite{Aamodt:2010cz}, are shown.}
\end{figure}

Let us now start with the comparison of \emph{Trajectum} to experimental data.
In figure \ref{fig:trajectum:trajectummultiplicity}, the charged particle multiplicity is shown, where we only include particles which satisfy $|\eta| \leq 0.5$\@.
The resulting multiplicity is then divided by the size of this pseudorapidity range, so that we arrive at a quantity which is relatively independent of the particular pseudorapidity cut.
In figure \ref{fig:trajectum:trajectummultiplicity}, also the experimental result from ALICE is shown \cite{Aamodt:2010cz}\@.
One can see that \emph{Trajectum} underestimates the experimental result by about 5--10\%\@.
This is no cause for alarm though, as \cite{Bernhard:2018hnz} shows a compatible underestimation.

\begin{figure}[ht]
\centering
\includegraphics[width=\textwidth]{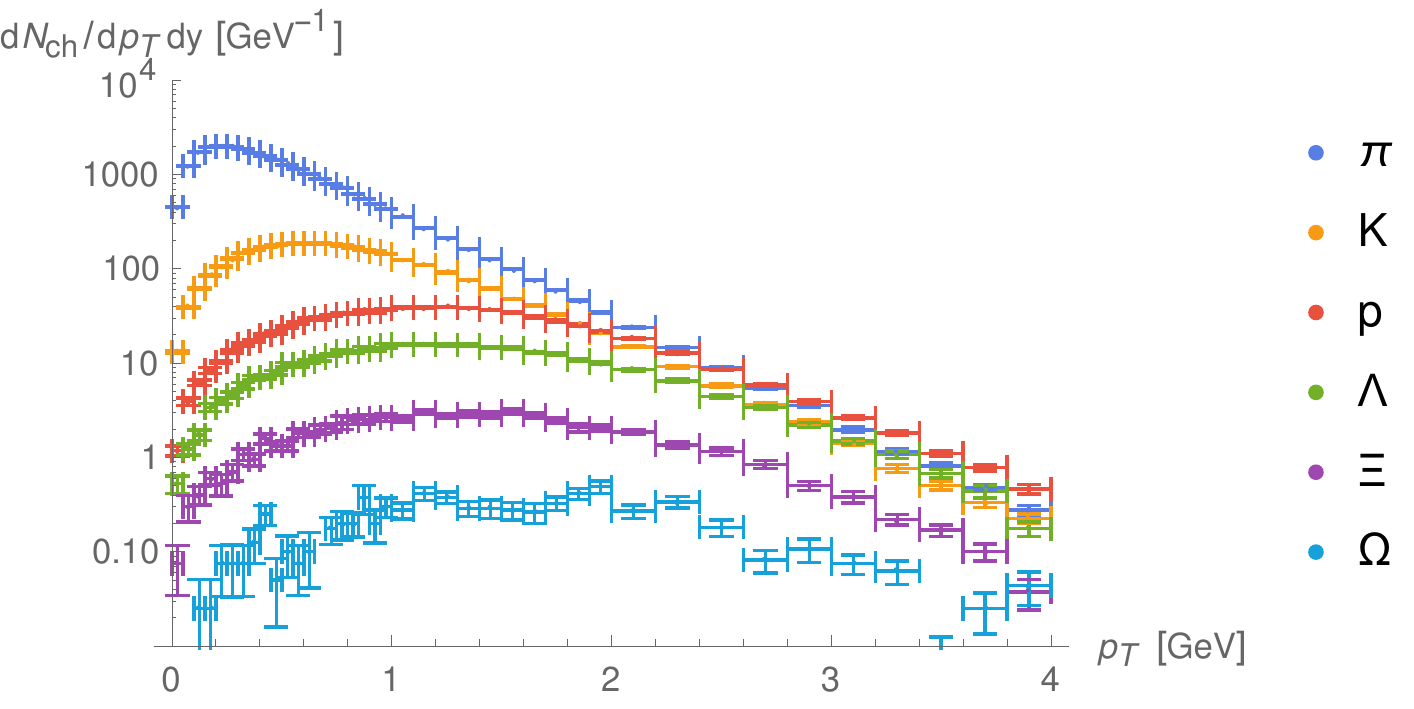}
\caption{\label{fig:trajectum:trajectumspectrum}Transverse momentum spectrum of various identified charged particles with $|y| \leq 0.8$ for events in the 0--5\% centrality interval, computed with \emph{Trajectum} using the $2.76\,\text{TeV}$ MAP values from \cite{Bernhard:2018hnz}\@.}
\end{figure}

An observable which was not used in \cite{Bernhard:2018hnz} is the transverse momentum spectrum, which is shown in figure \ref{fig:trajectum:trajectumspectrum} for events in the 0--5\% centrality interval.
Note that this figure shows the transverse momentum spectrum of identified particles, which reflects in the fact that not a cut in pseudorapidity, but in rapidity (see \eqref{eq:intro:rapiditydef}) has been made.
Also, in order to define the spectrum in such a way that it does not scale significantly with the size of the transverse momentum bins and the rapidity cut, the multiplicity in each bin has been divided by both the bin size and the rapidity range.
Even though we don't compare the transverse momentum spectrum with experimental data, one can see that for heavier masses the momenta shift to higher values, something we will see quantatively in the next subsection.
Also, one can clearly see that even for rare particles like $\Xi$ or maybe even $\Omega$, it is possible to obtain reasonable statistics for the transverse momentum spectra of identified particles.
\subsection{Mean transverse momentum}
\begin{figure}[ht]
\centering
\includegraphics[width=\textwidth]{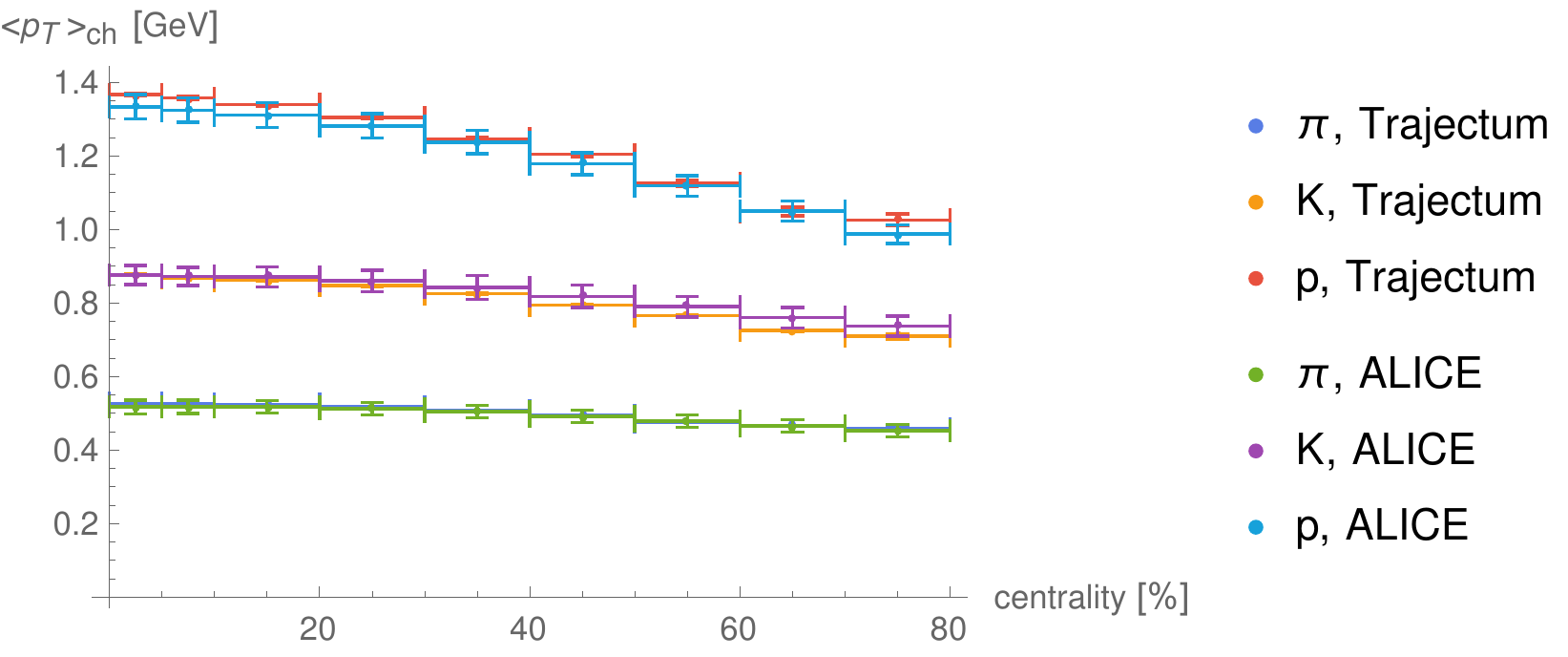}
\caption{\label{fig:trajectum:trajectummeanpt}Mean transverse momentum of various identified charged particles with $|y| \leq 0.5$\@. Both the result from \emph{Trajectum} using the $2.76\,\text{TeV}$ MAP values from \cite{Bernhard:2018hnz}, and experimentally measured data from ALICE \cite{Abelev:2013vea}, are shown.}
\end{figure}

In figure \ref{fig:trajectum:trajectummeanpt}, we show the mean transverse momentum of charged identified particles, where we have a rapidity cut of $|y| \leq 0.5$\@.
The mean transverse momentum is quite simply defined as the mean of the transverse momentum of all particles satisfying the cuts in all events.
To mold this definition into a weighted average over events, we define for each event
\[
\langle p_T\rangle_\text{event} = \frac{1}{M}\sum_{i=1}^Mp_{T,i}, \qquad w_\text{event} = M,
\]
with $w$ the weight associated to the event, and $M$ the number of particles in the event which satisfy the cuts.
This definition guarantees that if we then average $\langle p_T\rangle_\text{event}$ over all events with the given weights, the method agrees with the experimental one.
Figure \ref{fig:trajectum:trajectummeanpt} shows excellent agreement with \cite{Abelev:2013vea}\@.
This is no surprise however, as the result from \cite{Bernhard:2018hnz} agrees with the experimental data equally well, because $\langle p_T\rangle$ was used in their Bayesian analysis.

\begin{figure}[ht]
\centering
\includegraphics[width=\textwidth]{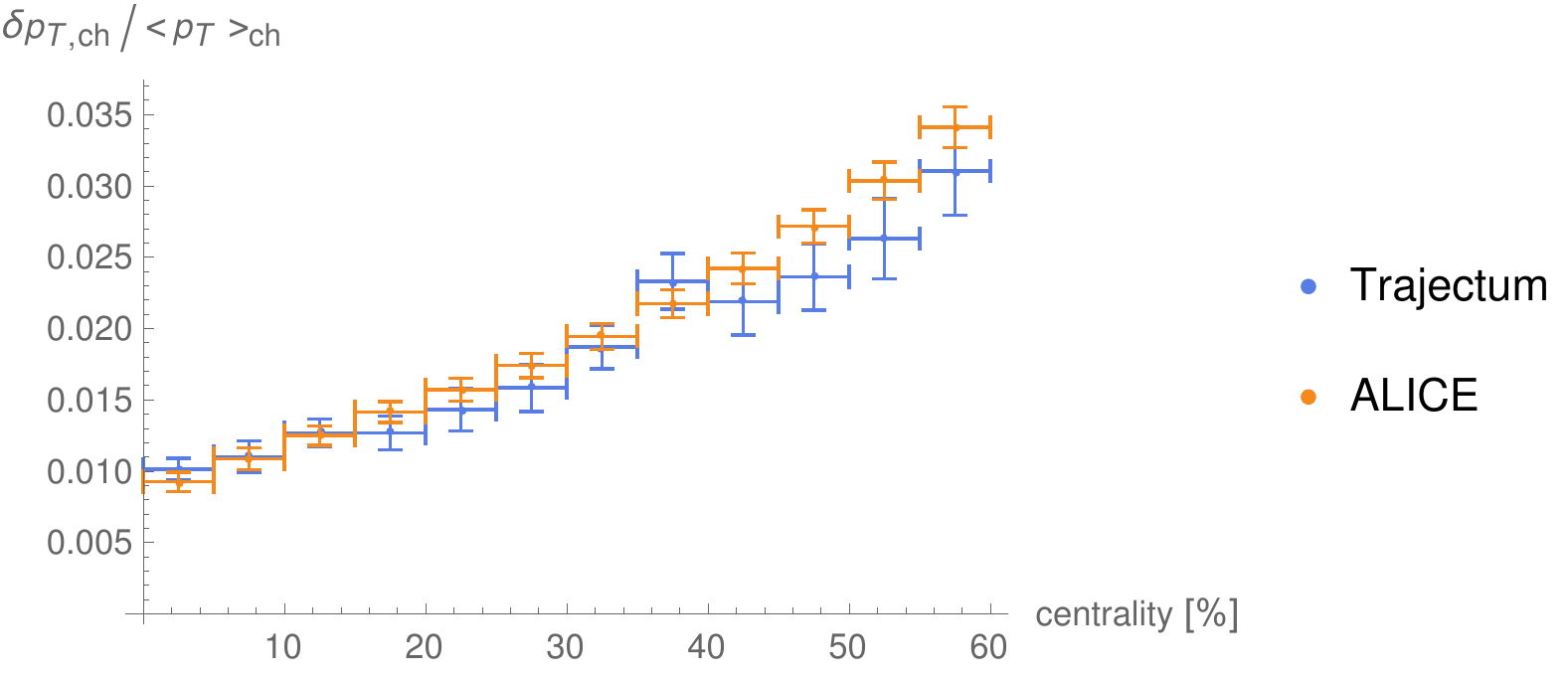}
\caption{\label{fig:trajectum:trajectumptfluct}Fluctuations of mean transverse momentum of charged particles with $|\eta| \leq 0.8$\@. Both the result from \emph{Trajectum} using the $2.76\,\text{TeV}$ MAP values from \cite{Bernhard:2018hnz}, and experimentally measured data from ALICE \cite{Abelev:2014ckr}, are shown.}
\end{figure}

Another observable we will examine is the mean transverse momentum fluctuations, which is shown in figure \ref{fig:trajectum:trajectumptfluct}, where we include only charged particles with pseudorapidity $|\eta| \leq 0.8$\@.
The mean transverse momentum fluctuations are defined as \cite{Abelev:2014ckr}:
\[
\left(\delta p_T\right)^2 = \left\langle\left\langle\left(p_{T,i} - \langle p_T\rangle\right)\left(p_{T,j} - \langle p_T\rangle\right)\right\rangle\right\rangle,
\]
where we average over all particle pairs $(i,j)$ occuring in the same event.
To write this in the form required by \emph{analyze}, we define the following two quantities for each event, in addition to the mean transverse momentum:
\begin{align*}
C_1 & = \frac{1}{M(M - 1)}\left[\left(\sum_{i=1}^Mp_{T,i}\right)^2 - \sum_{i=1}^Mp_{T,i}^2\right], \\
C_2 & = \frac{2}{M}\sum_{i=1}^Mp_{T,i},
\end{align*}
where each of these quantities have weight $M(M - 1)$ and where $M$ is the number of particles in the event which satisfy the cuts.\footnote{Note that while it may seem that $\langle C_2\rangle = 2\langle p_T\rangle$, the weighting for $C_2$ is different compared to that of the mean transverse momentum, invalidating this equality.}
With these definitions, we now have that
\[
\left(\frac{\delta p_T}{\langle p_T\rangle}\right)^2 = \frac{\langle C_1\rangle}{\langle p_T\rangle^2} - \frac{\langle C_2\rangle}{\langle p_T\rangle} + 1,
\]
where the angle brackets are the properly weighted averages of $C_1$ and $C_2$ over all events.
The result from \emph{Trajectum} shown in figure \ref{fig:trajectum:trajectumptfluct} is in good agreement with \cite{Abelev:2014ckr}, which is again unsurprising, because this observable was also used in the Bayesian analysis of \cite{Bernhard:2018hnz}, with good results.
\subsection{Anisotropic flow}\label{sec:trajectum:flow}
\begin{figure}[ht]
\centering
\includegraphics[width=\textwidth]{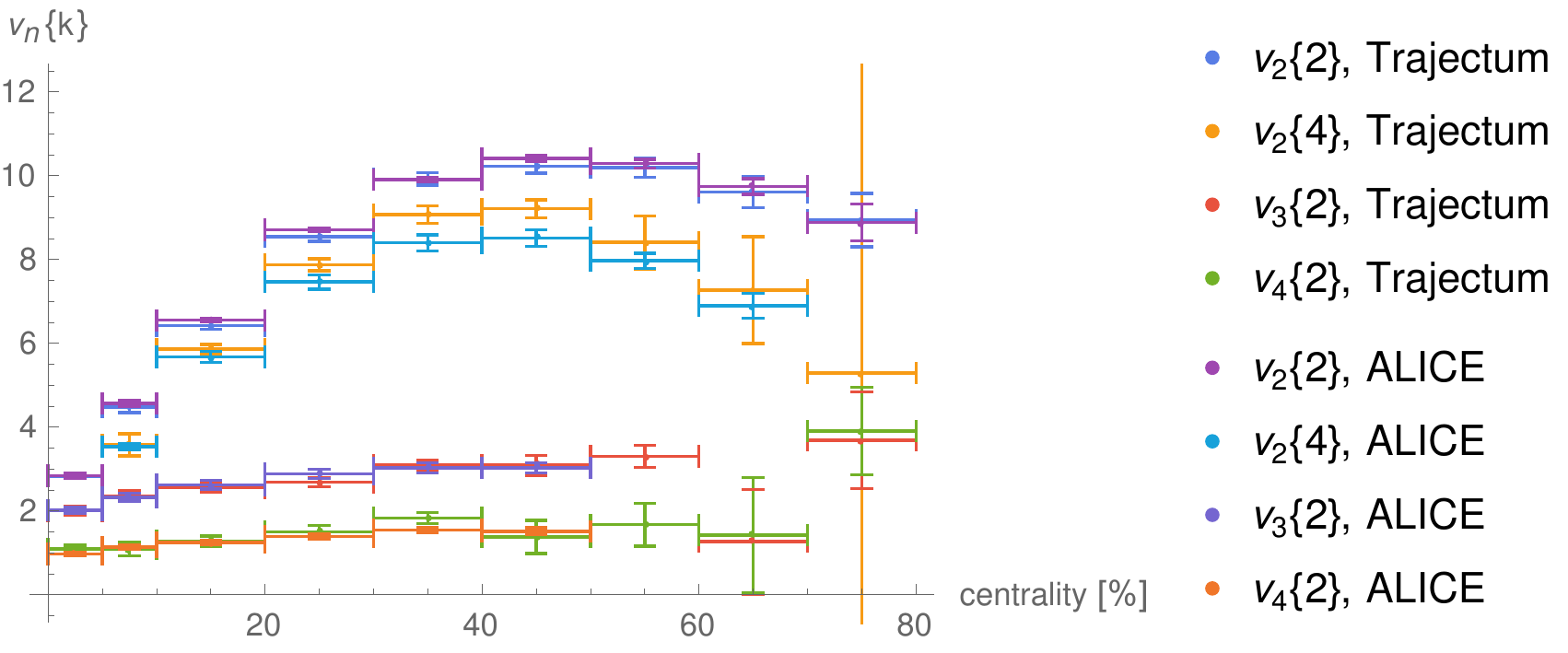}
\caption{\label{fig:trajectum:trajectumvn}Anisotropic flow coefficients $v_2\{2\}$, $v_2\{4\}$, $v_3\{2\}$ and $v_4\{2\}$ for charged particles with $0.2\,\text{GeV} \leq p_T \leq 5\,\text{GeV}$ and $|\eta| \leq 0.8$\@. Both the result from \emph{Trajectum} using the $2.76\,\text{TeV}$ MAP values from \cite{Bernhard:2018hnz}, and experimentally measured data from ALICE \cite{Adam:2016izf}, are shown.}
\end{figure}

The anisotropic flow measures how anisotropic the azimuthal distribution of particles emitted from an event is.
As this can vary substantially on an event-by-event basis, one usually defines observables as an average over a large number of events in a centrality class.
Additional observables look at the amount of variation of anisotropic flow within a centrality class, which will be covered in the next subsection.
In this section, we define the flow coefficients $v_n\{k\}$\@.
Several of these are shown in figure \ref{fig:trajectum:trajectumvn} for charged particles with $0.2\,\text{GeV} \leq p_T \leq 5\,\text{GeV}$ and $|\eta| \leq 0.8$\@.
To define the flow coefficients, let us first define for each event the cumulants \cite{Bilandzic:2010jr}:
\begin{equation}
Q_n = \sum_{i=1}^Me^{in\phi_i},\label{eq:trajectum:cumulants}
\end{equation}
where $M$ is again the number of particles in the event which satisfy the cuts, and $\phi_i$ is the azimuthal angle of particle $i$\@.
We subsequently define for each $n$:
\begin{align*}
\langle2\rangle & = \frac{|Q_n|^2 - M}{M(M - 1)}, \\
\langle4\rangle & = \frac{|Q_n|^4 + |Q_{2n}|^2 - 2\cdot\mathfrak{Re}\left[Q_{2n}Q_n^*Q_n^*\right] - 4(M - 2)\cdot|Q_n|^2 + 2M(M - 3)}{M(M - 1)(M - 2)(M - 3)},
\end{align*}
and we also define the weights
\[
w_{\langle2\rangle} = M(M - 1), \qquad w_{\langle4\rangle} = M(M - 1)(M - 2)(M - 3).
\]
If we now average $\langle2\rangle$ and $\langle4\rangle$ with these weights, then for each $n$ we obtain $\langle\langle2\rangle\rangle$ and $\langle\langle4\rangle\rangle$, in terms of which are defined
\[
v_n\{2\} = \sqrt{\langle\langle2\rangle\rangle}, \qquad v_n\{4\} = \sqrt[4]{2\cdot\langle\langle2\rangle\rangle^2 - \langle\langle4\rangle\rangle}.
\]
In figure \ref{fig:trajectum:trajectumvn}, one can see that, as with the previous observables, there is excellent agreement between \emph{Trajectum} and experimental data \cite{Adam:2016izf}\@.
Again, this is in agreement with \cite{Bernhard:2018hnz} as well, who used exactly these experimental findings in their analysis.
\subsection{Event-by-event anisotropic flow}
\begin{figure}[ht]
\centering
\includegraphics[width=\textwidth]{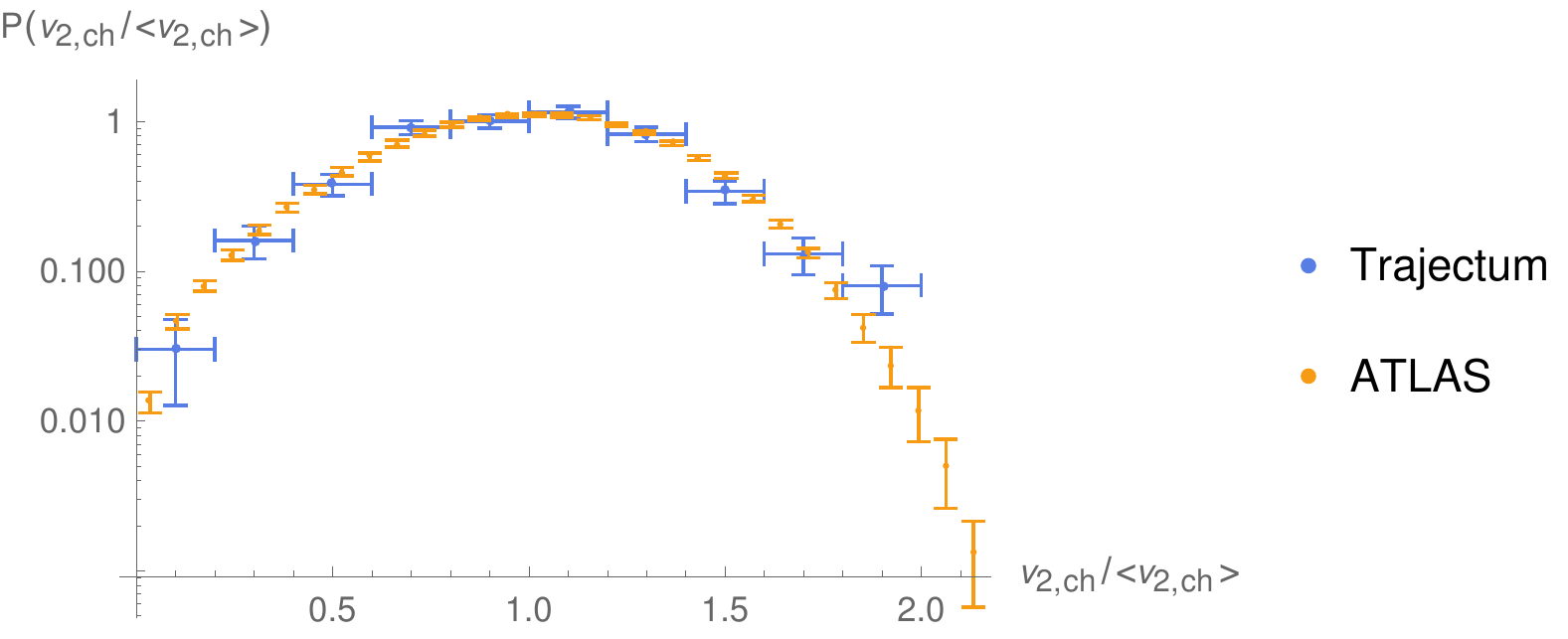}
\caption{\label{fig:trajectum:trajectumebev2}Event-by-event $v_2$ of charged particles with $|\eta| \leq 2.5$ for events in the 30--35\% centrality interval, using the single particle method. Both the result from \emph{Trajectum} using the $2.76\,\text{TeV}$ MAP values from \cite{Bernhard:2018hnz}, and experimentally measured data from ATLAS \cite{Aad:2013xma}, are shown.}
\end{figure}

As briefly mentioned above, one can also look at the amount by which the anisotropic flow, in this case specifically $v_2$, fluctuates event by event.
This event-by-event $v_2$ is shown in figure \ref{fig:trajectum:trajectumebev2} for charged particles with $|\eta| \leq 2.5$ for events in the 30--35\% centrality interval.
The method used in this figure for computing the $v_2$ of a single event is called the single particle method \cite{Aad:2013xma}\@.
It can be defined in terms of the cumulants \eqref{eq:trajectum:cumulants} as
\begin{equation}
v_2 = \frac{|Q_2|}{M},\label{eq:trajectum:v2perevent}
\end{equation}
where $M$ is again the number of particles in the event which satisfy the cuts.
We can then store the $v_2$ values for each event, and then compute a probability distribution around the average value as in figure \ref{fig:trajectum:trajectumebev2}\@.
As can be seen, even though this observable was not used in the Bayesian fit in \cite{Bernhard:2018hnz}, there is agreement with the experimental value \cite{Aad:2013xma}, but the error bars are quite large.
Also, the tail of the distribution is expected to be the most constraining \cite{Niemi:2015qia}, especially to the initial condition, but this is entirely missing from the \emph{Trajectum} prediction, and would require more events to compute.
Maybe one could still achieve a non-trivial constraint on the inputs by choosing larger centrality bins though.

Another potential issue with this observable is that the experimental value \cite{Aad:2013xma} has been modified using an unfolding procedure, which attempts to remove the bias inherent in defining $v_2$ through \eqref{eq:trajectum:v2perevent}\@.
This implies that the two observables shown in figure \ref{fig:trajectum:trajectumebev2} are not the same, and should not be compared.
However, in the limit of the number of particles per event going to infinity, this mentioned bias in defining $v_2$ through \eqref{eq:trajectum:v2perevent} vanishes.
Therefore, by studying the dependence of the \emph{Trajectum} result shown in figure \ref{fig:trajectum:trajectumebev2} on the acceptance cuts, particularly on the pseudorapidity, it may be possible to extrapolate the result to this desired limit.
This would allow the comparison to the ATLAS data, but we leave this for future work.

%% file: chapters/discussion.tex
In this thesis, we have studied QCD from three different angles, corresponding to the three main chapters.
A recurring theme throughout this has been the idea to use holography to study QCD-like theories in a genuinely strongly coupled setup.
We saw that this works rather well in many cases.
In the case of observables at zero chemical potential, one can obtain qualitatively reasonable results which match results from lattice QCD, in which case holography can be used to try to understand the mechanisms behind those phenomena seen on the lattice.
In the case of finite chemical potential, one can use holography in a regime where no results from other strongly coupled methods exist.
This also gives qualitatively reasonable results, which do not directly contradict known experimental results.
In this chapter, we will briefly summarize the main results, and provide an outlook for how the work in this thesis could be improved or extended in the future.
\section{(Inverse) magnetic catalysis in holo\-graph\-ic QCD}
In chapter \ref{ch:imc}, we studied the V-QCD model in the presence of a magnetic field $B$ and anisotropy $a$, which could be either parallel or perpendicular to $B$\@.
This model is constructed such that it matches qualitative features of QCD, like the running of certain coupling constants in the UV, glueball spectra, and the phase structure at finite temperature.
In particular, V-QCD displays confinement and spontaneous chiral symmetry breaking at low temperatures, while these features are lacking at high temperatures.
This makes V-QCD a good tool to study the behavior of the chiral condensate associated to chiral symmetry breaking in the presence of a magnetic field.
At zero temperature in a perturbative setting, one expects the chiral condensate to increase as a function of $B$\@.
However, in lattice QCD studies, the opposite effect was seen, called inverse magnetic catalysis.

In section \ref{sec:imc:b}, we saw that with a judicious choice of the parameter $c$ occuring in the $w$-potential of V-QCD, it is possible to obtain the same behavior of the chiral condensate as seen on the lattice.
Indeed, we saw that at small temperatures, the condensate increases as a function of $B$, whereas in a region between the chiral phase transition and the deconfinement transition, the condensate decreases as a function of $B$\@.
We also saw that at larger values of $B$, eventually inverse magnetic catalysis disappears, and the condensate starts increasing again.
Furthermore, we were able to identify two competing effects, namely a direct effect of $B$ on the condensate which tends to increase the condensate, and an indirect effect which tends to decrease it.
This was in agreement with earlier studies on the lattice.

We subsequently extended the analysis to include a finite baryon chemical potential $\mu$ in section \ref{sec:imc:mub}\@.
In this region of the phase diagram, lattice QCD is affected by the sign problem.
In holography, however, there are no major technical issues preventing us from exploring this region.
We found that the region of chirally broken deconfined plasma which exists at zero $B$ extends in seze at finite $B$, particularly at large values of the chemical potential.
In contrast, in the region between the chiral transition and the deconfinement transition, the condensate decreases as seen earlier in section \ref{sec:imc:b}, but now we can also conclude that this region of inverse magnetic catalysis only extends to a finite $\mu$, after which the region of inverse magnetic catalysis disappears.

Section \ref{sec:imc:a} explores a different approach, by adding an anisotropy $a$, which is dual to a space-dependent theta term.
This way we can examine the effects that anisotropy has on QCD, without that anisotropy being a magnetic field.
In other words, we can investigate whether the effects that a magnetic field has on the plasma are due to the magnetic field specifically, or whether they apply more generally for other sources of anisotropy.
We saw several interesting effects.
The chiral condensate behaved in much the same way as it did in the presence of a magnetic field, giving evidence for the conjecture first proposed in \cite{Giataganas:2017koz} that inverse magnetic catalysis is caused by the anisotropy induced by the magnetic field.
We named this conjectured effect `inverse anisotropic catalysis'\@.

Finally in section \ref{sec:imc:ba}, we explored the interplay between a magnetic field $B$ and anisotropy $a$, where we considered both the $B \parallel a$ and $B \perp a$ cases.
We find that the presence of $a$ seems to effectively decrease the value of $B$, especially for the parallel case.
We also see that the chiral transition temperature for $B \parallel a$ seems to cross that of $B \perp a$ roughly where $B \sim a$\@.

In the future it would be very interesting to determine whether it is possible to find an explanation for the observed behavior in section \ref{sec:imc:ba}\@.
Another interesting avenue of research would be to see if other sources of anisotropy can be added to lattice studies, thereby testing the conjecture of `inverse anisotropic catalysis'\@.
Additionally, the holographic model itself can be improved.
For the studies presented in this thesis, the quark flavors are assumed to all be identical.
In particular, the baryon number is assumed to be equal to the electric charge.
In the future, one could attempt to couple the magnetic field to the quarks in a different way for different flavors, thereby getting closer to a model for QCD\@.
This is however likely a very challenging extension.
A future study which is simpler to achieve is to use the existing model to study transport coefficients in the presence of a magnetic field.
In \cite{Grozdanov:2016tdf,Hernandez:2017mch}, the complete set of hydrodynamic transport coefficients for relativistic magnetohydrodynamics was derived to first order, but the values that these coefficients take are model-dependent.
In principle, by studying perturbations around the background space-times constructed in section \ref{sec:imc:model}, one can derive the values for the transport coefficients in V-QCD\@.
\section{Holographic baryons and neutron stars}
In neutron star physics, one of the most basic questions is one of the biggest question marks.
This question is what the neutron star equation of state is.
The equation of state governs observables like the maximum possible mass of a neutron star, the mass to radius relation and tidal deformability, and the post-merger spectrum of a binary neutron star merger.
On the other hand the answer to the question what the equation of state is would give us valuable information about QCD, including potentially the density at which a phase transition to deconfined matter occurs.
Many models exist which derive an equation of state, but none so far have used explicitly strongly coupled methods.
For this reason, in chapter \ref{ch:holographicns}, we used holography to study the equation of state and the resulting model for neutron stars.

In the Witten-Sakai-Sugimoto model, several approximations exist to incorporate baryons into the holographic model.
In section \ref{sec:holographicns:baryons}, we adapted one such approximation into V-QCD\@.
As it turns out, this gives a qualitatively reasonable phase diagram, with baryons condensing in the region in the phase diagram where one would expect this to happen.
We also saw that the distance into the holographic direction where the baryons are located is $\mathcal{O}(1)$, giving confidence that indeed the approximation is reasonable.
Looking at the zero temperature equation of state itself, we see that it exceeds the conformal value for the speed of sound in two places, a feature that is difficult to obtain in weakly coupled theories.
This lends confidence to the idea that the assumption of strong coupling leads to genuinely different predictions for the equation of state as compared to weak coupling.

In section \ref{sec:holo:hybrideos}, we used the speed of sound of the resulting equation of state, and used it as the high-density part of a hybrid equation of state, the low density part coming from the SLy equation of state.
This hybrid equation of state was shown to be compatible with currently existing constraints, including the maximum mass, mass to radius relation, and tidal deformability.
Subsequently a merger simulation using the hybrid equation of state was shown, which included an example of phase transition induced collapse.
We also examined the post-merger spectrum, finding a value for the frequency $f_1$ which is compatible with a universal relation, and a value for $f_2$ which shifts to lower frequencies as the hybrid equation of state is changed to incorporate the holographic part of the equation of state up to lower densities.

There are many opportunities in the future to improve the analysis from chapter \ref{ch:holographicns}\@.
One example is ongoing work to repeat the analysis for different choices of potentials for V-QCD, where the potentials are taken from \cite{Jokela:2018ers}\@.
This analysis will also include different choices for the part of the equation of state used for the low-density part.
Another extension already being done is to extend the analysis to the presence of a magnetic field, to see whether the equation of state changes as a function of $B$\@.
Further in the future, it is important to investigate how good the various approximations made in section \ref{sec:holographicns:baryons} are and, where possible, to improve them.
This will likely be a large effort, but this is worth doing if it increases the reliability of the results.
The merger simulation can also be improved, for example by including neutrinos and increasing the resolution.
One particular effect which is very important for the outcome of a merger simulation is the temperature dependence of the equation of state.
This is hard to do in bottom-up holographic models such as V-QCD since this requires stringy corrections, which are hard to obtain in a bottom-up model.
It may be possible though to obtain a reasonable approximation, which can then be used in a merger simulation.
\section{Simulation of heavy ion collisions with \emph{Trajectum}}
In chapter \ref{ch:trajectum}, the new \emph{Trajectum} framework was introduce, which provides a consistent interface between the various components necessary to simulate heavy ion collisions.
This framework consists of two executables, \emph{collide} and \emph{analyze}, to simulate collisions and to analyze the result, respectively.
Also, several examples of components which fit into the \emph{Trajectum} framework.
Most of these are reimplementations of existing models, but some have been modified in non-trivial ways.
The aim of these extensions is to incorporate results from AdS/CFT, such as the way the fluid should be initialized at the start of the hydrodynamical evolution.
Subsequently, a Bayesian analysis will be performed to attempt to infer whether experimental data shows a preference for AdS/CFT, or whether it points into another direction.

We then tested the code using the maximum a posteriori (MAP) values obtained from a Bayesian analysis in \cite{Bernhard:2018hnz}\@.
This yielded excellent agreement.
We also added new observables, which were not used in the analysis of \cite{Bernhard:2018hnz}\@.
This therefore potentially adds new constraints to the Bayesian analysis to be performed, which can then be used to learn something about the various parameters used in the simulation in \emph{collide}\@.

In the future, there are several ways in which \emph{Trajectum} can be improved.
An interesting option is to include thermal photon emission by the fluid.
This could give interesting new constraints, as it is directly sensitive to fluid quantities in the center of the collision, in contrast to the quantities we are using now, which are all produced at the edge of the quark-gluon plasma.
Other obvious extensions are to extend the code to optionally work in $3 + 1$D, which provides a better physical description of the collisions, as well as adding conserved quantities other than the stress-energy tensor.
The latter include baryon number density and electric charge.
With the inclusion of electric charge, it would also become possible to further extend the hydrodynamics into full magnetohydrodynamics, which could be used for various interesting problems \cite{Gursoy:2014aka,Gursoy:2018yai,Inghirami:2019mkc}\@.
A final interesting possibility is to include critical fluctuations which occur around a potential critical point \cite{Glorioso:2018wxw,Rajagopal:2019xwg}\@.
This would allow us to study what the effect of a critical point on observables is, and these observables could then also be used in a future Bayesian analysis.

Throughout these chapters, the overarching theme has been that holography can be used as a tool to gain qualitative insight into strongly coupled theories such as QCD\@.
This results both in new explanations of existing observations and in new qualitative predictions in areas where holography is so far the only way to gain insight in an explicitly strongly coupled setting.
Furthermore, by simulating neutron star mergers and heavy ion collisions, the ideas coming from holography can be compared to experimental observations.
All of these avenues taken together have taught us a lot about strongly coupled physics, and will surely teach us much more in the future.

%% file: chapters/acknowledgements.tex
First of all, I would like to thank my supervisors, Raimond and Umut, for hiring me and giving me this great opportunity to conduct research in so many different areas.
You have both helped a lot whenever I was stuck at something, provided interesting projects, but you also allowed me to work in other collaborations, something I've learned a lot from.
You've also always been encouraging to go to conferences and summer schools to present our work and learn new things.

My supervisors were not the only people I have had the pleasure to work with during my time as a PhD.
Therefore I would also like to thank my other collaborators, Ioannis, Matti, Juan, Takaaki, Christian and Wilke.
Without you, the work presented in this thesis could not have been done.

I would also like to thank my officemates, Bernardo and Sonja, for insightful discussions both within and outside physics.
I've had a good time in our office the past four years, and I think that you've always made sure there is a good atmosphere inside the office.

Also the entire string group has been great to discuss both physics and non-physics related topics with, so I'd like to give thanks to Brice, Chongchuo, Chris, Damian, Domingo, Eric, Huibert, Kilian, Koen, Miguel, Natale, Nava, Phil, Pierre, Ronnie, Sebastian, Stefan and Thomas.

Moving outside of physics itself, I would like to thank my two paranymphs, Lotte and Stephanie, both for the support they've given me during the writing of this thesis, and for the invaluable support they will give me during the promotion ceremony itself.
Of my two paranymphs, I would like to especially thank my girlfriend Lotte, who has supported me throughout my time as a PhD\@.
She has encouraged me to never give up when things went less well, and I thank her enormously for her support.

Also my parents plus their spouses, Annelies, Arie, Jacobien and Eveline, I would like to thank for their support.
In addition, I'd like to thank the rest of my family, my in-laws, and my friends in Ceros, USBC and SUF\@.
Lastly, let me thank two very special friends that Lotte and I have made, namely Domingo and Danna.
We have both very much enjoyed our dinner and board game evenings together, and we hope to come and visit you in your new home.

%% file: chapters/potentials.tex
In this appendix, we discuss the potentials which appear in the V-QCD action defined by \eqref{eq:intro:Sg} and \eqref{eq:intro:DBI}\@.
Furthermore, we will discuss the various constants \cite{Alho:2012mh} which appear in the near-boundary expansions discussed in \ref{sec:imc:eomandbc}\@.
First, we will discuss some general statements which apply to both of the sets of potentials that will be considered here, before moving on two the two specific cases used in chapters \ref{ch:imc} and \ref{ch:holographicns}\@.

In V-QCD, there are 5 potentials which enter the action, namely $V_g(\lambda)$, $V_f(\lambda,\tau)$, $\kappa(\lambda)$, $w(\lambda)$ and $Z(\lambda)$\@.
Firstly, let us restrict ourselves to $V_f$ of the form
\[
V_f(\lambda,\tau) = V_{f0}(\lambda)e^{-a(\lambda)\tau^2}.
\]
Using this, we then define
\[
V_\text{eff}(\lambda) = V_g(\lambda) - x_fV_{f0}(\lambda),
\]
and we obtain the following expansion:
\[
V_\text{eff}(\lambda) = \frac{12}{\mathcal{L}_\text{UV}^2}\left[1 + V_1\lambda + V_2\lambda^2 + \mathcal{O}(\lambda^3)\right],
\]
which defines the constants $\mathcal{L}_\text{UV}$, $V_1$ and $V_2$\@.
In terms of these constants, one can also define
\[
b_0 = \frac{9}{8}V_1, \qquad b_1 = -\frac{9}{256}\left[23V_1^2 - 64 V_2\right].
\]
Another expansion we can obtain is:
\[
\frac{\kappa(\lambda)}{a(\lambda)} = \frac{2\mathcal{L}_\text{UV}^2}{3}\left[1 + \kappa_1\lambda + \kappa_2\lambda^2 + \mathcal{O}(\lambda^3)\right],
\]
from which we can obtain
\[
\frac{\gamma_0}{b_0} = -\frac{4}{3} - \frac{4\kappa_1}{3V_1}.
\]
For both of the sets of potentials considered below, we will have
\[
b_0 = \frac{11 - 2x_f}{3}, \qquad b_1 = \frac{34 - 13x_f}{6}, \qquad \frac{\gamma_0}{b_0} = \frac{9}{22 - 4x_f},
\]
which matches the perturbative QCD beta function, as well as the perturbative anomalous dimension of the quark mass in QCD\@.
Since the potentials from section \ref{sec:potentials:baryons} are only meant to be used for $x_f = 1$, these potentials only satisfy these properties at this precise value of $x_f$\@.
\section{Inverse magnetic catalysis}\label{sec:potentials:imc}
In chapter \ref{ch:imc}, we use the following potentials \cite{Alho:2012mh,Alho:2013hsa,Gursoy:2016ofp,Gursoy:2018ydr}:
\begin{align*}
V_g(\lambda) & = {12\over \mathcal{L}_0^2}\left[1+{88\lambda\over27}+{4619\lambda^2 \over 729}{\sqrt{1+\log(1+\lambda)}\over(1+\lambda)^{2/3}}\right],\\
V_{f}(\lambda,\tau) & = {12\over x_f \mathcal{L}_{UV}^2}\left[{\mathcal{L}_{UV}^2\over\mathcal{L}_0^2}-1+{8\over27}\left(11{\mathcal{L}_{UV}^2\over\mathcal{L}_0^2}-11+2x_f \right)\lambda\right.\\
& \quad + \left.{1\over729}\left(4619{\mathcal{L}_{UV}^2\over \mathcal{L}_0^2}-4619+1714x_f - 92x_f^2\right)\lambda^2\right]\,e^{-a_0\tau^2},\\
\kappa(\lambda) & = {[1+\log(1+\lambda)]^{-1/2}\over[1+\frac{3}{4}(\frac{115-16x_f}{27}-{1\over 2})\lambda]^{4/3}},\\
w(\lambda) & = \kappa(c\lambda),\\
Z(\lambda) & = 1+\frac{\lambda^4}{10},
\end{align*}
where
\[
a_0 = \frac{3}{2\mathcal{L}_{UV}^2}, \qquad \mathcal{L}_{UV}^3 = \mathcal{L}_0^3 \left(1+{7 x_f \over 4} \right),
\]
and $c$ is a free constant which has different values in different parts of chapter \ref{ch:imc}\@.
\section{Baryons}\label{sec:potentials:baryons}
In chapter \ref{ch:holographicns}, we use the potential which were fitted to lattice data in \cite{Jokela:2018ers}\@.
In particular, we use potentials 7a, where the notation `7a' follows \cite{Jokela:2018ers}\@.
As we will not use the $Z$ potential in chapter \ref{ch:holographicns}, we will also not define it.
The potentials are:\footnote{Note that there is a factor $8\pi^2$ difference in the definition of $\lambda$ with respect to \cite{Jokela:2018ers}\@.}
\begin{align*}
V_g(\lambda) & = 12\left[1 + \frac{88}{27}\lambda + \frac{4619\lambda^2}{729(1 + \lambda/\lambda_0)}\right.\\
& \quad + \left.V_\text{IR}e^{-\lambda_0/\lambda}\left(\lambda/\lambda_0\right)^{4/3}\sqrt{\log\left(1 + \lambda/\lambda_0\right)}\right], \\
V_f(\lambda,\tau) & = e^{-\tau^2}\left[W_0 + W_1\lambda + \frac{W_2\lambda^2}{1 + \lambda/\lambda_0} + W_\text{IR}e^{-\lambda_0/\lambda}\left(\lambda/\lambda_0\right)^2\right], \\
\frac{1}{\kappa(\lambda)} & = \left(\frac{3}{2} - \frac{W_0}{8}\right)\left[1 + \frac{11}{3}\lambda + \tilde\kappa_0\left(1 + \frac{\tilde\kappa_1\lambda_0}{\lambda}\right)e^{-\lambda_0/\lambda}\frac{\left(\lambda/\lambda_0\right)^{4/3}}{\sqrt{\log\left(1 + \lambda/\lambda_0\right)}}\right], \\
\frac{1}{w(\lambda)} & = w_0\left[1 + \frac{w_1\lambda/\lambda_0}{1 + \lambda/\lambda_0} + \tilde w_0e^{-\lambda_0/\lambda w_s}\frac{\left(w_s\lambda/\lambda_0\right)^{4/3}}{\log\left(1 + w_s\lambda/\lambda_0\right)}\right],
\end{align*}
with
\[
\lambda_0 = \frac{1}{3}, \qquad V_\text{IR} = 2.05, \qquad W_0 = 2.5, \qquad W_1 = \frac{64 + 24W_0}{9},
\]
\[
W_2 = \frac{6488 + 999W_0}{243}, \qquad W_\text{IR} = 0.9, \qquad \tilde\kappa_0 = 1.5, \qquad \tilde\kappa_1 = -0.047,
\]
\[
w_0 = 0.83, \qquad 3w_s = 0.925, \qquad w_1 = 2, \qquad \tilde w_0 = 45.
\]
Compared to the potentials from appendix \ref{sec:potentials:imc}, $b_0$, $b_1$ and $\gamma_0/b_0$ are unchanged, provided we compare at $x_f = 1$\@.
However, we have
\[
\mathcal{L}_\text{UV}^2 = \frac{12}{12 - W_0} \approx 1.26.
\]
These potentials will be used in a setting where the planck mass $M$ enters non-trivially into the computation through \eqref{eq:holographicns:Vrho}\@.
For these potentials, we have
\[
M^3 = \frac{11\cdot1.32}{180\pi^2\mathcal{L}_\text{UV}^3}.
\]